\documentclass[reqno,12pt]{nuthesis}
\usepackage{epsfig}
\usepackage{amsmath}
\usepackage{amssymb}
\usepackage{amsthm}
\usepackage{longtable,lscape,setspace}

\author{James Phearl Jenkins Jr.}

\title{Model Independent Explorations of Majorana Neutrino Mass Origins}

\degree{DOCTOR OF PHILOSOPHY}  % Default: DOCTOR OF PHILOSOPHY

\field{Physics}            % Default: Mathematics

\graduationmonth{June}         % The default is June or December
\graduationyear{2008}          % Default: current year.

\begin{document}

\frontmatter        % Preliminary pages start here.

\maketitle      % Produces the title page.

%\copyrightpage      % Creates the copyright page.

\abstract       % Abstract.

The recent observation of nonzero neutrino mass is the first
concrete indication of physics beyond the Standard Model. Their
properties, unique among the other fermions, leads naturally to the
idea of a Majorana neutrino mass term.  Despite the strong
theoretical prejudice toward this concept, it must be tested
experimentally.  This is indeed possible in the context of next
generation experiments.  Unfortunately, the scale of neutrino mass
generation may be too large to explore directly, but useful
information may still be extracted from independent experimental
channels.  Here I survey various model independent probes of
Majorana neutrino mass origins.  A brief introduction   %100
 to the
concepts relevant to the analysis is followed by a discussion of the
physical ranges of neutrino mass and mixing parameters within the
context of standard and non-standard interactions.  Armed with this,
I move on to systematically analyze the properties of radiatively
generated neutrino masses induced by nonrenormalizable lepton number
violating effective operators of mass dimensions five through eleven.
By fitting these to the observed light mass scale, I extract
predictions for neutrino mixing as well as neutrinoless double beta
decay, rare meson/tau decays and collider phenomenology.  I find
that many such models are already constrained by current data %200
 and
many more will be probed in the near future.  I then move on
demonstrate the utility of a low scale see saw mechanism via a
viable $3+2+1$ sterile neutrino model that satisfies all oscillation
data as well as solves problems associated with supernova kicks and
heavy element nucleosynthesis. From this I extract predictions for
tritium  and neutrinoless double beta decay searches. This is
supplemented throughout by descriptions of practical limitations in
addition to suggestions for future work. %277

\acknowledgements   % Acknowledgements (optional).

I would like to acknowledge the help of many friends and colleges,
without whom this work could never have begun.  First, I thank my
advisor Andre de Gouvea for his advice, support and patience.  On the
same note, I thank the remainder of my thesis committee, Heidi
Schellman and Robert Oakes for their kind remarks and encouragement
throughout my years of graduate study. Finally, I would like to
thank Tina Jakubosky for editing services.

The success of my graduate career also depended heavily on my vast
exposure to scholarly dialog/instruction.  For this reason I also
take this opportunity to acknowledge the faculty of Northwestern
University's high energy physics group as well as the Fermilab and
Argonne theory groups.  Additionally, I thank the organizers,
participants, and lecturers of the TASI 2006 and PITP 2007 summer
schools for many useful discussions and insights into this thesis
topic.

\newpage
\begin{center}
\begin{Large}\textbf{Dedication}\end{Large}
\\
\vspace{1cm}
This thesis is dedicated to my best friend and companion Tina
Jakubosky for her unending love and support.
\end{center}

\preface        % Preface (optional).

While there exists many astrophysical hints of physics beyond the
Standard Model, the observation of nonzero neutrino mass is the only
undisputed terrestrial evidence.  It is set apart from the rest
because it is not (directly) related to poorly understood
gravitational interactions.  If it were just the fact that neutrinos
have mass, this observation would be important but not earth
shattering... After all, the quarks and charged leptons have nonzero
mass values that span nearly six orders of magnitude (this is
puzzling by itself).  The neutrinos' mystique is enhanced because
they are so different from the charged fermions in both structure
and magnitude.  Specifically, we find that the neutrinos possess
unprecedented tiny masses, nearly twelve orders of magnitude smaller
then the largest quark mass, or equivalently the electroweak scale
(another puzzle).  Additionally, we observe large (perhaps maximal)
mixing which is quite distinct from the observed quark mixings.  It
is likely not a coincidence that neutrinos are the only known
electrically neutral fundamental fermion.  They are not protected by
this $U(1)_Q$ gauge symmetry and consequently may have new types of
interactions of the Majorana type.  On the grander scale, neutrino
data unambiguously confirms that the pattern of fermion masses begs
for an explanation.  Fitting the known Yukawa couplings as well as
neutrino masses (of either Dirac or Majorana type) into the Standard
Model can be easily accomplished, but there is likely something
more.  Flavor constrains are some of the strongest in all of
particle physics and, enhanced by information in the neutrino
sector, are currently poised to select the next set of fundamental
laws that govern the universe.  The next step on this journey is the
exploration of neutrino mass origins.

It is possible that this goal is beyond the reach of next generation
experiments due to the possibility of potentially ultra high energy
scales involved in neutrino mass generation.  Still, it is my biased
hope that, by combining information from distinct terrestrial,
astrophysical and cosmological probes, the physics community can at
least narrow down the set of model possibilities, if not choose the
correct one.  Of course completeness dictates that such analysis
should be as model independent as possible.  This spirit is the
inspiration for the majority of my research interests and, in
particular, the present analysis.  Here, I embark on a model
independent survey of (potentially) testable models of Majorana
neutrino mass.

This thesis is taken from three previously published papers done in
collaboration with Andre de Gouvea and Nirmala Vasudevan \cite{de Gouvea:2007xp,MySeeSaw,MyParameterPaper}.  They were
selected and modified to reflect the unifying themes outlined
above.  While this work does not cover all possible models of
neutrino mass, it is easily broad enough to aid in the analysis of
current and future neutrino data.  This analysis will help constrain
large classes of new physics models as well as lead the way for
other theoretical/phenomenological studies.

\tableofcontents    % Table of Contents will be automatically
            % generated and placed here.

\listoftables       % List of Tables and List of Figures will be placed
\listoffigures      % here, if applicable (optional).

\mainmatter             % Actual text starts here.

%%%%%%%%%%%%%%%%%%%%%%%%%%%
% Actual text starts here %
%%%%%%%%%%%%%%%%%%%%%%%%%%%
\chapter{Introduction} \label{chap:Introduction}

The Standard Model (SM) of particle physics has enjoyed tremendous
success in recent decades.  Indeed, consistency between
experiments and theoretical calculations has been demonstrated in a
variety of terrestrial channels over a wide range of energy scales
exceeding twelve orders of magnitude\footnote{This conservative range is taken from the sub-eV scale associated with atomic fine structure to the ultra-GeV scale accessable to present day colliders.}.  Despite extensive
searches, the majority of experiments have not yet uncovered a shred
of (convincing) evidence of new physics Beyond the Standard Model
(BSM). On the contrary, most results agree with SM predictions to
an unprecedented and often puzzling degree.  See appendix
\ref{app:SM} for a brief review of the SM.

Despite its success, there is still strong reason to believe that
there exists new physics BSM.  These are of both an indirect and
direct nature.  Indirectly, one is naturally led to the idea of new
physics above the electroweak scale to stabilize the scalar Higgs
mass to radiative corrections \cite{Mahbubani:2006kq}.  Additionally, the running and
approximate intersection of the SM gauge couplings at high energies
hint at a possible grand unification of particle physics within a
single (broken) gauge group \cite{Langacker:1980js}.  The details of this unification requires the existance of supersymmetry for full consistency.  Direct evidence for phenomenon BSM
is also available from astrophysical/cosmological observations.
Specifically, an analysis of the energy budget of the universe
reveals that only a small fraction of the universe is composed of
the SM particles with the majority residing in dark energy (70\%) and
dark matter (26\%) \cite{Trodden:2004st}.  Finally, as will become clear shortly, the
discovery of neutrino mass via flavor oscillations is a clear sign
of physics BSM and is the motivation of this thesis. Originally, the
evidence of this phenomenon was also of astrophysical origin but has
been convincingly confirmed by terrestrial experiments.  All of
these ``problems'' suggest the existence of new physics for full
resolution.

 The observation of nonzero neutrino mass is the first
unambiguous evidence for physics BSM.  In other words, neutrino mass
can not be accommodated within the SM as currently formulated. This
is easy to understand by considering the ``types'' of masses
available to the neutrino.  First, neutrinos may be Dirac particles
and thus posses a mass term similar to that of the charged
fermions,  of the general form
\begin{equation}
\mathcal{L}_{Dirac} = M_D \left(\bar{\nu}N + \bar{N}\nu\right)
\end{equation}
where $N$ is a \emph{new} right-handed SM singlet dubbed ``right
handed neutrino.''  This mass term arises after electroweak
symmetry breaking from the Yukawa coupling of the $\bar{\nu}N$
combination to the neutral component of the SM Higgs doublet. Notice
that these terms involve two different fields, the familiar
left-handed neutrino and the new right-handed state.  The addition
of $N$ is already an indication of new physics, but as is often
overlooked in such discussions, it is not the only BSM effects
guaranteed to arise from Dirac neutrinos. Since the singlet $N$
carries no unbroken charge, it should combine to form a mass term
$\overline{N^c}N$, which ultimately spoils the Dirac nature of the
physical neutrino state upon diagonalization.  A new global symmetry
must be imposed on the system by hand to forbid it whereas it is
accidental within the SM. Indirectly, the Dirac nature of the
neutrino suggests more substantial modifications to the SM in order
to explain the tiny Yukawa coupling constant needed to yield the
observed sub-eV neutrino mass, roughly $10^{-12}$.

 The second type of neutrino mass term is allowed since the neutrino carries no unbroken gauge quantum numbers.  This is of the
  Majorana form written as
 \begin{equation}
\mathcal{L}_{Majorana} = M_M \left(\bar{\nu^c}\nu +
\bar{\nu}\nu^c\right). \label{eq:MajoranaMassTerm}
 \end{equation}
Notice that this mass is written entirely in terms of a single field
$\nu$ and so no new field content is required for its construction.
Physically, such an interaction states that left-handed neutrinos
have a small but nonzero probability (proportional to $M_M^2$) of
being observed as right-handed antineutrinos.  In this case, neutrinos
and antineutrinos are equivalent.  The only reason the distinction
between them is convenient is that the numerical coefficient $M_M$
is so small.  This mass term, however, is not invariant under the
full unbroken electroweak symmetry of SM $SU(2)_L \times U(1)_Y$
where the neutrino is charged under both weak isospin and
hypercharge. The only means of such a construction is via the Higgs
boson, but unlike the Dirac case, the Higgs quantum numbers do not
allow the required renormalizable interaction.  This necessarily
requires the addition of new physics of various types.  One can form
the required term by coupling to a pair of Higgs doublets as
$(1/\Lambda)HH\overline{\nu^c}\nu$, but this is of dimension five
and as such, nonrenormalizable.  It requires an ultraviolet
completion at or below the scale $\Lambda$ to preserve unitarity.
This is in fact the effective operator induced by the seesaw
mechanisms \cite{SeeSaw} and yields a mass of order $v^2/\Lambda$
after electroweak symmetry breaking.  The seesaw neutrino mass
mechanism will be discussed at length in chapter
\ref{chap:SeeSawLSND}.  The case of neutrino mass generation by more
general high dimensional operators is discussed in chapter
\ref{chap:LNV}.  Alternatively, to construct
Eq.~(\ref{eq:MajoranaMassTerm}), one could simply introduce a new
scalar $\phi$, in addition to the SM Higgs that transforms as a
triplet under $SU(2)_L$ to form the SM singlet $\phi
\bar{\nu^c}\nu$.  It will yield a Majorana mass after the neutral
component of  $\phi$ acquires a vacuum expectation value either
directly, or indirectly via a trilinear Higgs coupling $\phi H H$.  In the first case the vacuum expectation of $\phi$ must be unnaturally small to accomidate the tiny observed neutrino masses.  However, in the latter case, provided the mass of $\phi$ is large, the system
reduces down to the previously mentioned dimension five seesaw
operator after electroweak symmetry breaking with the scalar mass acting as the suppression scale
$\Lambda$.

It is natural to wonder about the consequences of introducing both
Dirac and Majorana masses.  It is well known that any such
combination will yield physical Majorana neutrinos after
diagonalization to the mass basis.  This is the principle behind the
type-I see-saw mechanism.  From this it seems that the Majorana
nature of the neutrino is the most natural choice and indeed it is a
majority prejudice within the theoretical neutrino physics
community.  I emphasize that the construction of Dirac-type neutrinos
\emph{requires} the existence of a new symmetry imposed to protect
lepton number (L) or more precisely Baryon minus Lepton Number
(B-L). While B-L is accidental within the SM there is no reason to
believe that it will be so with the addition of new physics.
Ultimately, the nature of the neutrino is one we must answer
experimentally.  The overall purpose of my thesis is to aid in
this endeavor by exploiting the connection between Majorana
neutrinos, Lepton Number Violation (LNV), and mixing phenomena and
thereby extract predictions for a variety of physics experiments.

\chapter{The Physical Range of Neutrino Parameters}

In order to properly study Majorana neutrino interactions, it is
important to understand how to fully parameterize neutrino mixing
both within the SM, minimally extended to include neutrino mass, and
in the context of more general BSM interactions.  This involves
questions that arise regarding both the number of free parameters
and their physical ranges.  Parameter counting arguments are well
known in both the Dirac and Majorana cases.  In words, these go as
follows for the general case of $n$ Majorana neutrinos.  The
transformation from the flavor bases, where the gauge interactions
are diagonal, to the mass basis is accomplished by an $n\times n$
complex matrix. These $2n^2$ degrees of freedom minus $n^2$
unitarity conditions and $n$ field rephasings leaves room for
$n(n-1)$ parameters.  For Dirac neutrinos, one is allowed to rephase
both the left-handed lepton doublet and right handed neutrino fields
separately, thus reducing the parameter count by $n-1$ where the
additional ``1'' arises from the universal ability to rephase the
entire system.  The freedom is shared between real mixing angles,
Majorana phases, and Dirac phases.

While this much is known, the complex relationship between the
physical ranges of these parameters has not been explored fully.  It
turns out that one may limit the range of a particular parameter set
by simply extending the range of another in a non-trivial way.
Furthermore, this relationship is modified in the presence of new
physics.  In what follows, I develop a framework in which to study
these ranges via symmetries of the mass matrix.  The reasoning
behind this is simple: If a symmetry exists between the mixing
values, it implies the existence of two or more degenerate regions
within the parameter space.  One may choose to exclusively populate any one of
them while at the same time taking heed of the consequences imposed
by this choice on other parameters.  This is accomplished using an
extension of the $SU(2)$ algebra defined in Appendix
\ref{app:notation}.  Admittedly, the use of this system is not
optimal for a specified neutrino system.  It is natural to use the
well known algebra of $SU(n)$ to study the symmetries of an $n$
neutrino system.  The drawback of this approach is that one must
use a different algebra to analyze each different system.  The
utility of the approach adopted here is that it works for all cases,
as will become apparent.

In this chapter, I explore the physical ranges of neutrino mixing
parameters within the SM for the minimal case of two neutrinos, the
realistic case of three neutrinos, and the general case of $n$
neutrinos.  Special emphesis is placed on the parameter implications of additional sterile neutrinos and their mixing.  This is followed by a discussion of the modifications to
these relationships under the influence of new physics.

\section{Within the Minimally extended standard model}
\label{sec:PhyRange}

With the introduction offered in Appendix \ref{app:SM}, it is easy to
see that the neutrino sector of the SM extended to accommodate
Majorana neutrino mass may be expressed by the Lagrangian
\begin{equation}
\mathcal{L}_{\nu SM} \supset \frac{g}{\sqrt{2}}
\left(\bar{\nu_\alpha} \gamma^\mu \ell_\alpha W^+_\mu +
\bar{\ell_\alpha}\gamma^\mu\nu_\alpha W^-_\mu \right) +
\frac{g}{2\cos\theta_W} \bar{\nu_\alpha}\gamma^\mu \nu_\alpha Z_\mu
+ \frac{1}{2} \bar{\nu_\alpha^c} M_{\alpha\beta} \nu_\beta.
\label{eq:LSM}
\end{equation}
Where $M_{\alpha\beta}$ is the Majorana neutrino mass matrix.  This
expression is written in the flavor bases where the charged lepton
masses and neutrino interaction terms are diagonal.  I now determine
the transformations of the neutrino mixing matrix that leaves this
invariant.  The reader should note for future use that an overall
sign change of any neutrino field, either in the flavor or mass
basis, is unphysical, in that it will not affect any observable
process and is a common property of all quantum field theories.  This followes directly from the rephasing freedom already discussed in the previous section.
Using this I analyze the symmetries of the mass matrix.

\subsection{The Two Neutrino Case} \label{subsec:TwoNu}

In the case of only two Majorana neutrinos, the mass eigenstates
$\nu_1$ and $\nu_2$ are related to the flavor eigenstates
$\nu_\alpha$ and $\nu_\beta$ by the unitary transformation
$U^\dagger$. This can be parameterized as
\begin{eqnarray}
\left(
  \begin{array}{c}
    \nu_1 \\
    \nu_2 \\
  \end{array}
\right) &=& {\bf P^2}(\phi) {\bf R^{12}}(\theta) \left(
\begin{array}{c}
    \nu_\alpha \\
    \nu_\beta \\
  \end{array}
\right)  \\
 &=& \nonumber \left(
            \begin{array}{cc}
              1 & 0 \\
              0 & e^{i\phi} \\
            \end{array}
          \right)\left(
            \begin{array}{cc}
              \cos\theta & -\sin\theta \\
              \sin\theta & \cos\theta \\
            \end{array}
          \right)\left(
  \begin{array}{c}
    \nu_\alpha \\
    \nu_\beta \\
  \end{array}
\right) \label{eq:TwoNuMix}
\end{eqnarray}
where the phase matrix ${\bf P^2}$ and rotation matrix ${\bf R^{12}}$ are defined in Appendix \ref{app:notation} Here, all physics is contained in four parameters expressible as two
real and positive masses $m_1,m_2$, one real mixing angle $\theta$,
and one CP violating Majorana phase $\phi$.  The symmetries of
Eq.~(\ref{eq:TwoNuMix}) may be used to limit the physical ranges of
these quantities.  To this end, I must find transformations up to an
overall rephasing that renders
\begin{equation}
{\bf N_m P^2}(\phi) {\bf R^{12}}(\theta) {\bf N_f} = {\bf S^{12}
P^2}(\phi^\prime) {\bf R^{12}}(\theta^\prime),
\label{eq:2NuCondition}
\end{equation}
where ${ \bf N_m}$ and ${ \bf N_f}$ are unphysical shifts of the
mass and flavor eigenstates respectively, and ${\bf S^{12}}$ is a discrete permutation matrix that flips 1-2 vector elements. Assuming the Standard
Model Lagrangian augmented by Majorana neutrino masses, these are just
simple field sign redefinitions.  All possible physical
transformations are contained in the continuous shifts of mixing
parameters ${\bf P^2}(\phi + \delta\phi) {\bf
R^{12}}(\theta+\delta\theta)$, as well as the discrete interchange
of mass eigenstates.  This is accomplished by the operation ${\bf
S^{12}}={\bf P^2}(\pi){\bf R^{12}}(-\pi/2)$ which changes the sign
of the mass squared difference $\Delta m^2 = m_2^2 - m_1^2$.  To
perform a systematic symmetry search, I put both sides of
Eq.~(\ref{eq:2NuCondition}) into the same form for easy comparison.
Beginning on the right hand side, I find that the case without the
mass eigenstate flip is trivially ${\bf P^2}(\phi) {\bf
P^2}(\delta\phi) {\bf R^{12}}(\delta\theta) {\bf R^{12}}(\theta)$.
Adding the $\nu_1 \leftrightarrow \nu_2$ operation I commute to find
\begin{eqnarray}
{\bf S P^2}(\phi + \delta\phi) {\bf R^{12}}(\theta + \delta\theta)
&=& {\bf P^2}(\pi){\bf R^{12}}(-\frac{\pi}{2}){\bf P^2}(\phi) {\bf
P^2}(\delta\phi) {\bf R^{12}}(\delta\theta) {\bf R^{12}}(\theta)\\
\nonumber &=& {\bf P^2}(\pi){\bf P^1}(\phi) {\bf P^1}(\delta\phi)
{\bf
R^{12}}(-\frac{\pi}{2}){\bf R^{12}}(\delta\theta) {\bf R^{12}}(\theta)\\
\nonumber&=& {\bf P^2}(\phi){\bf P^2}(\pi-\phi){\bf P^1}(\phi +
\delta\phi)
{\bf R^{12}}\left(\delta\theta - \frac{\pi}{2}\right) {\bf R^{12}}(\theta)\\
\nonumber &=& {\bf P^2}(\phi){\bf P^2}(\pi-2\phi - \delta\phi){\bf
R^{12}}\left(\delta\theta - \frac{\pi}{2}\right) {\bf
R^{12}}(\theta).
\end{eqnarray}
The last step utilizes a total rephasing by $e^{-i(\phi +
\delta\phi)}$ to yield the requisite form.  On the left of
Eq.~(\ref{eq:2NuCondition}) I must consider eight cases defined by
the matrix pairs
\begin{eqnarray} ({\bf N_m},{\bf N_f})&=&\left\{({\bf I},{\bf I}),({\bf
I},{\bf -I}),({\bf P^2}(\pi),{\bf I}),({\bf P^1}(\pi),{\bf I}),({\bf
P^2}(\pi),{\bf P^2}(\pi)),({\bf P^2}(\pi),{\bf P^1}(\pi)), \right. \nonumber \\
& & \left. ({\bf
I},{\bf P^2}(\pi)),({\bf I},{\bf P^1}(\pi)) \right\}.
\end{eqnarray}
The only difficulty arises from a nontrivial ${\bf N_f}$ commuting
with the rotation matrix. Since ${\bf P^1}(\pi) = -{\bf P^2}(\pi)$
in this two dimensional case, it is enough to consider only
\begin{eqnarray}
{\bf N_m P^2}(\phi) {\bf R^{12}}(\theta) {\bf P^1}(\pi) &=& {\bf
P^2}(\phi) {\bf N_m}{\bf R^{12}}(\theta) {\bf P^1}(\pi){\bf
R^{12}}(-\theta){\bf R^{12}}(\theta)\\ \nonumber &=& {\bf P^2}(\phi)
{\bf N_m}{\bf P^1}(\pi){\bf R^{12}}(-2\theta){\bf R^{12}}(\theta) \nonumber \\
 &=& {\bf P^2}(\phi) {\bf N_m}{\bf P^2}(\pi){\bf
R^{12}}(-2\theta+\pi){\bf R^{12}}(\theta)
\end{eqnarray}
and change overall signs as needed, with the operation ${\bf
R^{12}}(\pi)$, to match the intended structure.  Once this is done
to both sides of Eq.~(\ref{eq:2NuCondition}), the outer factors of
${\bf P^2}(\phi)$ and ${\bf R^{12}}(\theta)$ cancel, revealing
simplified equations that may be easily solved for the physical
parameter shifts $\delta\phi$ and $\delta\theta$.

\begin{table}
{\SMALL \begin{tabular}{|c c|c|c|}
  \hline
  {\bf Matrix} & & $\bf{P^2}(\delta\phi)\bf{R^{12}}(\delta\theta)$ & $\bf{P^2}(\pi - 2\phi - \delta\phi)\bf{R^{12}}(\delta\theta - \pi/2)$ \\
 \hline
   ${\bf P^2}(0){\bf R^{12}}(0)$ & None & $(\phi,\theta) \rightarrow (\phi,\theta)$  & $(\phi,\theta) \rightarrow (-\phi + \pi,\theta + \pi/2)$   \\
   ${\bf P^2}(0){\bf R^{12}}(\pi)$ & $\nu_\alpha \rightarrow -\nu_\alpha,\nu_\beta \rightarrow -\nu_\beta$  & $(\phi,\theta) \rightarrow(\phi,\theta + \pi)$ & $(\phi,\theta) \rightarrow(-\phi + \pi,\theta - \pi/2)$ \\
   ${\bf P^2}(\pi){\bf R^{12}}(0)$ & $\nu_2 \rightarrow -\nu_2$ & $(\phi,\theta) \rightarrow(\phi+\pi,\theta)$ & $(\phi,\theta) \rightarrow(-\phi,\theta + \pi/2)$ \\
   ${\bf P^2}(\pi){\bf R^{12}}(\pi)$ & $\nu_1 \rightarrow -\nu_1$ & $(\phi,\theta) \rightarrow(\phi+\pi,\theta+\pi)$ & $(\phi,\theta) \rightarrow(-\phi,\theta - \pi/2)$ \\
   ${\bf P^2}(0){\bf R^{12}}(-2\theta)$ & $\nu_2 \rightarrow -\nu_2,\nu_\beta \rightarrow -\nu_\beta$ & $(\phi,\theta) \rightarrow(\phi,-\theta)$ & $(\phi,\theta) \rightarrow(-\phi + \pi,-\theta+\pi/2)$ \\
   ${\bf P^2}(0){\bf R^{12}}(-2\theta + \pi)$ & $\nu_2 \rightarrow -\nu_2,\nu_\alpha \rightarrow -\nu_\alpha$ & $(\phi,\theta) \rightarrow(\phi,-\theta+\pi)$ & $(\phi,\theta) \rightarrow(-\phi+\pi,-\theta-\pi/2)$ \\
   ${\bf P^2}(\pi){\bf R^{12}}(-2\theta)$ & $\nu_\beta \rightarrow -\nu_\beta$ & $(\phi,\theta) \rightarrow(\phi+\pi,-\theta)$ & $(\phi,\theta) \rightarrow(-\phi,-\theta+\pi/2)$ \\
   ${\bf P^2}(\pi){\bf R^{12}}(-2\theta + \pi)$ & $\nu_\alpha \rightarrow -\nu_\alpha$ & $(\phi,\theta) \rightarrow(\phi+\pi,-\theta+\pi)$ & $(\phi,\theta) \rightarrow(-\phi,-\theta-\pi/2)$ \\
  \hline
\end{tabular}}
\label{tab:2NuInv}
\caption[Exhaustive summary of two neutrino mixing
symmetries]{Exhaustive summary of two neutrino mixing symmetries.
See text for details.}
\end{table}

Table \ref{tab:2NuInv} shows the solutions to these equations for
every possible case.  The first column and row displays the
simplified matrix equations obtained from manipulations of the left
and right side of Eq.~(\ref{eq:2NuCondition}), respectively.  For
convenience, the second column displays the unphysical field
redefinitions associated with each corresponding row.  The remaining
columns contain an exhaustive list of physical transformations that
leave the two neutrino Majorana mixing matrix invariant.  All of
these take the general form $(\phi,\theta) \rightarrow (\pm \phi +
\delta\phi, \pm \theta + \delta\theta)$. Notice that many of the
listed transformations yield equivalent information.  In particular,
I find that the phase $\phi \rightarrow -\phi$ for all cases where
the mass eigenstates are exchanged since $\nu_1 \leftrightarrow
\nu_2\Rightarrow {\bf P^2}(\phi) \rightarrow {\bf P^1}(\phi)$, which
must be countered by a total rephasing by $-\phi$. Additionally, I
see that $\phi$ is invariant under translations by $\pi$ in cases
defined by the number of unphysical field redefinitions.  In a
similar way, it is clear that the mixing angle $\theta$ is only
invariant under discrete sign changes as well as shifts by $\pi$ and
$\pi/2$. Each invariance listed in the table may be interpreted as a
single constraint on the physical parameter ranges.  I follow \cite{de Gouvea:2000cq}, choosing
independent entries, and limit these as:

\begin{enumerate}
\item  $\nu_2 \rightarrow
-\nu_2$, and $\phi \rightarrow \phi+\pi$.\\
Here I find that the phase $\phi$ is invariant under shifts by
$\pi$, given a nonphysical mass eigenstate redefinition. This
suggests degeneracies between the $\phi$ parameter regions of size
$\pi$.  I choose $\phi \in [-\pi/2,\pi/2]$ without loss of
generality. This result conforms via direct symmetry arguments to
the common conception of Majorana phase ranges.

\item $\theta \rightarrow \theta + \pi$, $\nu_1 \rightarrow
-\nu_1$, and $\nu_2 \rightarrow -\nu_2$.\\
The mixing angle shift suggests parameter space degeneracies of size
$\pi$.  Hence, $\theta$ is limited to span $\pi$ radians, half its
original range, to the interval $\theta \in [-\pi/2,\pi/2]$, chosen
for convenience.

\item $\theta \rightarrow -\theta$, $\nu_\beta \rightarrow
-\nu_\beta$, and $\nu_2 \rightarrow -\nu_2$.\\
Thus, neutrino mixing is invariant under reflections about
$\theta=0$, implying a degeneracy between positive and negative
values.  This suggests a halving of the mixing angle physical range
to $\theta \in [0,\pi/2]$.

\item $\theta \rightarrow \theta + \pi/2$, $\nu_2 \rightarrow
-\nu_2$, $\nu_1 \leftrightarrow \nu_2$ and $\phi \rightarrow
-\phi$.\\
This symmetry relates three distinct physical transformations, and
it is clear from Table \ref{tab:2NuInv} that no smaller set will
yield the same results.  Hence, this relation may be interpreted in
one of three ways.  First, if $\theta$ and $\phi$ are allowed to
move within their full range as constrained above, one may choose a
particular ordering of mass eigenstates, or equivalently, the sign
of the neutrino mass squared difference.  This is typically done in
full, two neutrino oscillation analysis where $\Delta m^2$ is taken
positive. If both positive and negative $\Delta m^2$ are allowed,
one has the freedom to limit \emph{either} $\theta \in [0,\pi/4]$ or
$\phi \in [0,\pi/2]$.  In the case of standard vacuum oscillations,
the sign of $\Delta m^2$ is unphysical, leading to the notion that
the mixing parameter ranges are automatically limited as described
here.  While true for pure neutrino mixing, the SM charged current
interactions shown in Eq.~(\ref{eq:LSM}), relevant for matter
effects, break this degeneracy and one must explicitly choose how to
limit and interpret the resulting parameter space.
\end{enumerate}

I point out that the Majorana neutrino mass matrix, which governs
such lepton number violating processes as neutrinoless double beta
decay, is invariant under these transformations by construction. It
is also easy to check that other processes such as neutrino flavor
oscillations and neutrino-antineutrino oscillations share this
property since their amplitudes are strongly related to the mass
matrix.  I illustrate some of the above observations by example in
this simple two neutrino context.  Most conclusions are apparent by
inspection here, but carry over to the more general case of $n$
neutrino flavors in a much less obvious way.

First, consider the two flavor neutrino oscillation probability of a
$\nu_\alpha$ being measured as a $\nu_\beta$ after traversing a
baseline $L$ with energy $E$.  In vacuum, this is given by
\begin{equation}
P_{\nu_\alpha\rightarrow \nu_\beta}^{\rm vacuum}(L,E) = \sin^2
2\theta \sin^2 \left( \frac{\Delta m^2 L}{4E} \right).
\label{eq:VacNuOss}
\end{equation}
Here, the Majorana phase, as well as the sign of $\Delta m^2$, is
unphysical.  It is clear that the physical range of $\theta$ is only
in the $[0,\pi/4]$ interval from the factor $\sin^2 2\theta$.  To
introduce the $\Delta m^2$ sign as a physical degree of freedom, I
must introduce matter effects.  For neutrinos propagating in a
constant electron density background, the mixing parameters of
Eq.~(\ref{eq:VacNuOss}) must be replaced by effective matter
quantities.  In terms of the dimensionless parameter $A = 2\sqrt{2}E
G_F N_e/\Delta m^2$, where $N_e$ is the local electron number
density, the modified oscillation probability is
\begin{equation}
P_{\nu_\alpha\rightarrow \nu_\beta}^{\rm matter}(L,E) = \frac{\sin^2
2\theta}{1+A^2-2A\cos 2\theta} \sin^2 \left( \frac{\Delta m^2
L}{4E}\sqrt{1+A^2-2A\cos 2\theta} \right). \label{eq:MatNuOss}
\end{equation}
Since $\cos 2\theta$ now appears independently, I have lost the
freedom to constrain $\theta$ beyond $[0,\pi/2]$.  However, if I
allow for a mass eigenstate flip, which induces an $A \rightarrow
-A$ transformation, the angular degeneracy is restored.  I see that
one may choose to confine either $\Delta m^2$ positive \emph{or}
$\theta \in [0,\pi/4]$.

To see how the Majorana phases enter into this discussion, I must
move to lepton number violating processes where they contribute to
important physical effects.  These are much harder to
observe/constrain since all rates must be proportional to neutrino
mass values, as opposed to the interferometric dependence in
standard neutrino oscillations. Consider the rate for
neutrino-antineutrino oscillation which goes like \cite{deGouvea:2002gf,Langacker:1998pv}
\begin{equation}
\Gamma_{\nu \rightarrow \bar{\nu}} \propto \frac{\sin^2
2\theta}{4E^2}\left\{m_1^2 + m_2^2 - 2m_1 m_2 \cos
\left(\frac{\Delta m^2 L}{2E} - 2\phi \right) \right\}.
\end{equation}
This is similar to the neutrino vacuum flavor oscillation formulae,
except that this rate is directly proportional to the neutrino mass
scale and shows a Majorana phase dependence.  Here the physical
range of $\theta$ is still of size $\pi/4$, but now the sign of
$\Delta m^2$ and $\phi$ are physical. I may limit the range for one
of these by noting that a negative $\phi$ value can be compensated
by mass eigenstate flip.  This is the physical manifestation of the
ambiguity noted in the final entry of the above symmetry list.  The
next logical step is to consider neutrino-antineutrino oscillations
in matter, where, as in the flavor cases, the mass matrix
diagonalization is modified by an effective matter potential.  The
process carries through as before, yielding no new information
except for an illustration of the interplay between $\theta$,
$\Delta m^2$ and $\phi$ at the same time.  I leave this to the
reader and move on to the more realistic three neutrino case.

\subsection{The Three Neutrino Case} \label{subsec:ThreeNu}

The case of three Majorana neutrinos is more complicated than the
two neutrino analysis performed in Subsection \ref{subsec:TwoNu}.
Nevertheless, the methodology and many features of the previous
example carry over directly.  Here, neutrino mixing is defined in
terms of nine parameters, conventionally chosen as three mass
eigenvalues $\{m_1,m_2,m_3\}$ taken real and positive, three real
mixing angles $\{\theta_{12}, \theta_{13}, \theta_{23}\}$, and three
CP violating phases $\{\delta, \phi_2, \phi_3 \}$.  The Majorana
phases $\phi_2$ and $\phi_3$ are not physical, meaning that they may
be phased away, when neutrinos are Dirac particles.  The mixing
between the mass eigenstates $\nu_1$, $\nu_2$ and $\nu_3$ and the
flavor eigenstates $\nu_e$, $\nu_\mu$, and $\nu_\tau$ may be
parameterized by
\begin{eqnarray}
\left(
  \begin{array}{c}
    \nu_1 \\
    \nu_2 \\
    \nu_3 \\
  \end{array}
\right) &=& \left[{\bf P^2}(\phi_2){\bf P^3}(\phi_3)\right] \left[ {\bf
R^{12}}(\theta_{12}) \right] \left[{\bf P^1}(-\delta){\bf P^3}(\delta){\bf
R^{13}}(\theta_{13}){\bf P^1}(\delta){\bf P^3}(-\delta)\right] \nonumber \\ \nonumber &~~~& \times \left[{\bf
R^{23}}(\theta_{23})\right] \left(
\begin{array}{c}
    \nu_e \\
    \nu_\mu \\
    \nu_\tau \\
  \end{array}
\right) \nonumber \\
 &=&  \left(
            \begin{array}{ccc}
              1 & 0 & 0\\
              0 & e^{i\phi_2} & 0 \\
              0 & 0 & e^{i\phi_3} \\
            \end{array}
          \right)\left(
            \begin{array}{ccc}
              c_{12} & -s_{12} & 0 \\
              s_{12} & c_{12} & 0 \\
              0 & 0 & 1 \\
            \end{array}
          \right)
          \left(
            \begin{array}{ccc}
              c_{13} & 0 & -s_{13}e^{-i\delta} \\
              0 & 1 & 0 \\
              s_{13}e^{i\delta} & 0 & c_{13} \\
            \end{array}
          \right) \\ \nonumber
&~~& \times
          \left(
            \begin{array}{ccc}
              1 & 0 & 0 \\
              0 & c_{23} & -s_{23} \\
              0 & s_{23} & c_{23} \\
            \end{array}
          \right) 
          \left(
  \begin{array}{c}
    \nu_e \\
    \nu_\mu \\
    \nu_\tau
  \end{array}
\right) \label{eq:ThreeNuMix},
\end{eqnarray}
where, to conserve space, I employ the shorthand $c_{ij}\equiv \cos
\theta_{ij}$ and $s_{ij} \equiv \sin \theta_{ij}$.

As in the previous case, I search for physical transformations of
the mixing variables that leave Eq.~(\ref{eq:ThreeNuMix}) invariant
and interpret these as degeneracies within their parameter spaces,
which lead directly to range limitations.  Here, it is not useful to
list every possible symmetry due to the overwhelming number of
possibilities, many of which yield equivalent physical information.
Rather, I enumerate in Table \ref{tab:3NuInv} a subset of simple
cases that best reveal the underlying physics. The first column
labels the transformation number referred to in the text, followed
by the unphysical field redefinitions and the physical shifts in
columns two and three, respectively.  For convenience, the symmetry
ordering is set such that the first physical transformation in each
row may be used to limit a particular parameter provided the
constraints listed above it are imposed.  In this way, one may run
down the list systematically to reveal the most constrained
parameter spaces available to the system. One should note a few key
features of this table. First, for simplicity, I only include those
transformations that lead to parameter space degeneracies and skip
nontrivial parameter redefinitions of the form $P_{i} \rightarrow
P_i(P_1,P_2...P_n)$ for arbitrary mixing variables $P_i$. Such cases
that mix parameters are addressed separately. Furthermore, the
entries listed here are in no way unique, but they do constitute a
complete set, since combining them in various ways will produce all
other elements. These variations, however, yield no new physical
insight and are therefore neglected.

\begin{table}
 \label{tab:3NuInv}
\begin{tabular}{|c|c|c|}
  \hline
   \# & Unphysical & Physical \\
    \hline
 1 & $\nu_2 \rightarrow -\nu_2$ & $\phi_2 \rightarrow \phi_2 + \pi$ \\
 2 & $\nu_3 \rightarrow -\nu_3$ & $\phi_3 \rightarrow \phi_3 + \pi$ \\
 3 & & $ \theta_{13}\rightarrow -\theta_{13},\delta \rightarrow \delta + \pi$ \\
 4 & $\nu_\tau \rightarrow -\nu_\tau, \nu_3 \rightarrow -\nu_3$ & $ \theta_{23}\rightarrow -\theta_{23}, \theta_{13}\rightarrow -\theta_{13}$ \\
 5 & $\nu_\mu \rightarrow -\nu_\mu, \nu_2 \rightarrow -\nu_2$ & $ \theta_{12}\rightarrow -\theta_{12}, \theta_{23}\rightarrow -\theta_{23}$ \\
 6 & $\nu_1 \rightarrow -\nu_1, \nu_2 \rightarrow -\nu_2$ & $\theta_{12} \rightarrow \theta_{12}+\pi$   \\
% 3 & $\nu_1 \rightarrow -\nu_1$ & $\phi_2 \rightarrow -\phi_2,\phi_3 \rightarrow -\phi_3$ \\
 7 & $\nu_2 \rightarrow -\nu_2, \nu_3 \rightarrow -\nu_3$  & $\theta_{23} \rightarrow \theta_{23}+\pi, \theta_{13}\rightarrow -\theta_{13}, \theta_{12}\rightarrow -\theta_{12}$   \\
 8 & & $\theta_{13} \rightarrow \theta_{13}+\pi, \theta_{12} \rightarrow \theta_{12}+\pi, \theta_{23} \rightarrow \theta_{23}+\pi, \delta \rightarrow \delta + \pi$   \\
  \hline
\end{tabular}
\caption[Summary of selected three neutrino mixing
symmetries]{Summary of selected three neutrino mixing symmetries.
Although this list is not exhaustive it is a complete representation
of the symmetry structure. See text for details.}
\end{table}

From this, the physical mixing parameters may be limited in the
following way.  The first two entries tell us that I may halve the
ranges of the Majorana phases to $\pi$ with impunity. %provided that I allow the
%mass eigenstates to take on both positive and negative values.
The next entry constrains the $\theta_{13}$ mixing angle and its
associated phase $\delta$.  Here, one may chose either variable to
be limited to a range of $\pi$ provided the other occupies the full
$2\pi$ region.  This suggests that one may have either $\delta \in
[0,\pi]$ and $\theta_{13}\in [-\pi,\pi]$ \emph{or} $\delta \in
[0,2\pi]$ and $\theta_{13}\in [0,\pi]$, but not both. Typically, the
latter option is chosen by experimental analyses to maintain
consistency with the two neutrino case where the Dirac phase, and
therefore this ambiguity, is not present. It will become clear that, since these two symmetries are
independent of unphysical transformations they are thus robust to
the effects on non-standard interactions.  Entries $4$ and $5$ allow for the similar range limitation of the
mixing angles $\theta_{12}$ and $\theta_{23}$ to $[0,\pi]$. Notice
that while naively requiring both positive and negative
$\theta_{13}$ values, this is independent of the previously
described ambiguity. This is because $\theta_{13} \rightarrow
-\theta_{13}$ and $\delta \rightarrow \delta + \pi$ are
interchangeable transformations, such that the mixing angle sign
change may be compensated by a phase shift. Entries $6$, $7$ and $8$
permit one to further limit each mixing angle to a range of $\pi/2$,
which I take to be $[-\pi/2, \pi/2]$ without loss of generality. I
see here that these actions utilize the nonphysical sign change of
the mass eigenstates. Finally, making these standard choices, I see
that the physical ranges of each mixing angle is limited to
$\theta_{ij} \in [0,\pi/2]$ and the Majorana phases are limited to
$\phi_i \in [-\pi/2,\pi/2]$, while the Dirac phase is unconfined
within its full $\delta \in [0,2\pi]$ range.

Unfortunately, without an exhaustive search, it is impossible to
know if and when I include all relevant transformations.  To this
end, I must submit to plausibility arguments supplemented by
numerical examples.  The invariance criterion is simple if I
consider each of the six parameters of the mixing matrix Eq.~
(\ref{eq:ThreeNuMix}) to transform as $P_i \rightarrow P_i + \delta
P_i$ and formally commute the result into the form ${\bf
N_m}U^{\dagger\prime}(p_i,\delta P_i) = U^\dagger(P_i) M(P_i,\delta
P_i)$.  One must then find all possible $\delta P_i=\delta P_i(P_j)$
that renders the matrix $M$ proportional to some product of ${\bf
P^k}(\pi)$s, which are just nonphysical field redefinitions.
Working through this manipulation, I make the observation that only
factors of ${\bf P^k}(\pi)$ commute through the rotation matrices
trivially, only inducing parameter sign changes.  All other attempts
do not commute and lead to additional terms directly proportional to
some of the mixing parameters. It follows that $M$ is a product of
such factors, so it is natural to conclude that solutions $\delta
P_i(P_j)$ can not exist in general for all parameter values. This
statement holds true in the case of three parameter variations,
which may be checked analytically, but is not obvious when all six
parameters are allowed to shift.  To establish confidence that no
symmetries are missed in the general case, I perform a comprehensive
numerical scan for solutions.  Due to the large number of possible
variations, this operation is coarse by necessity.  That being said,
our methodology did find the class of solutions already explored
with no indication of other possibilities.  I therefore conclude
that the symmetries spanned by the transformations listed in Table
\ref{tab:3NuInv} are complete.

In addition to the symmetries outlined in Table \ref{tab:3NuInv},
physically distinct transformations also exist that do not limit
parameter spaces, but should rather be interpreted as variable
redefinitions. These are cases where parameters mix with each other
in a nontrivial way.  The simplest example that still supplies
physical insight is the three neutrino analog of symmetry four in
the two neutrino case involving the exchange of mass eigenstates.
Here I see that the Lagrangian is invariant under $\theta_{12}
\rightarrow \theta_{12} - \pi/2, ~\phi_2 \rightarrow -\phi_2,~
\phi_3 \rightarrow \phi_3 - \phi_2$, $\nu_1 \leftrightarrow \nu_2$
provided the unphysical sign redefinition $\nu_2 \rightarrow
-\nu_2$. No phase space limitations may be extracted from this due
to the mixing of the Majorana phases $\phi_2$ and $\phi_3$.  It
requires a nontrivial relabeling of the $\phi_3$ parameter.  The
utility of this transformation is clear when the Majorana phases are
unphysical, as is the case for Dirac neutrinos, where the
troublesome shifts may be phased away, or in all lepton number
conserving processes where $\phi_2$ and $\phi_3$ simply do not show
up.  Under these circumstances, one may use this symmetry to limit
parameters in much the same way as in the two neutrino scenario.
Namely, one may choose either positive $\Delta m_{12}^2$, or
$\theta_{12} \in [0,\pi/4]$. The central position of the $1-2$
parameters in this discussion is unambiguously chosen by our
standard mixing matrix parameterizations. The general conclusion is
that the $i-j$ plane may enjoy this constraint if it contains the
\emph{first} Euler rotation of the mass eigenstates in the
transformation to the flavor basis. The noncommutative nature of
orthogonal rotations renders the effects of any other
$\nu_{i^\prime} \leftrightarrow \nu_{j^\prime}$ much more complex.
In particular, all other mass eigenstate exchange symmetries require
mixing/relabeling of the mixing angles and Dirac phase that cannot
be phased away in any reasonable situation. Therefore, most such
transformations are of little practical importance and are not
considered further.

\begin{figure}[t]
\begin{center}
\includegraphics[scale=.60]{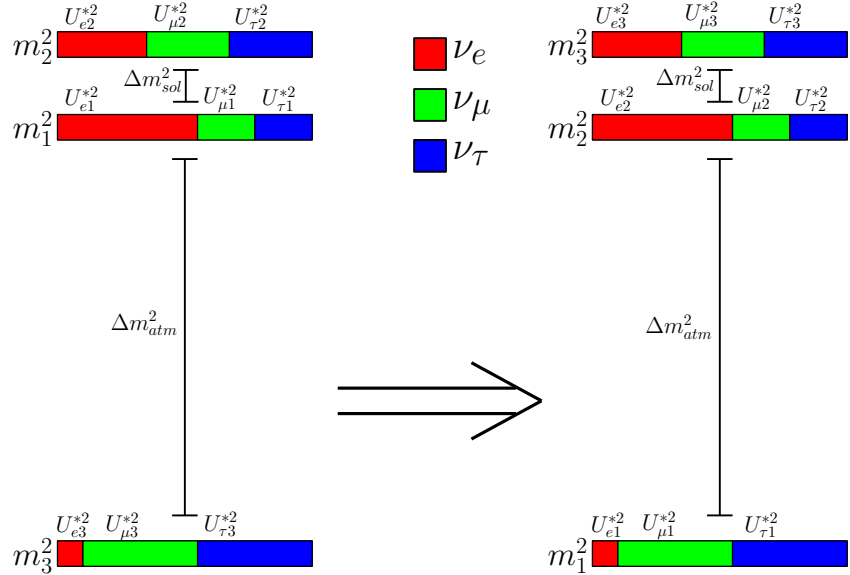}
\caption[Mapping the inverted neutrino hierarchy between popular
schemes]{Mapping of the inverted neutrino mass hierarchy between the
popular (312) and (123)schemes.  The relationship between mixing
elements is clear from this plot, but the mapping between the
specific mixing parameters is highly nontrivial as given in the
text.}\label{fig:InvMap}
\end{center}
\end{figure}

The only exception to this involves the permutations of two
eigenstate pairs $\nu_1 \leftrightarrow \nu_2$ followed by $\nu_3
\leftrightarrow \nu_2$.  This transformation is of interest because
it defines the mapping between two popular parameterizations of
three neutrino mixing within the context of the inverted mass
hierarchy displayed in Figure \ref{fig:InvMap}.  On one hand, as
shown in the leftmost spectrum, one may assume the mass states
ordered, from smallest to largest, as $m_3, m_1, m_2$ (henceforth
the $312$-case), so as to maintain the mixing angle definitions
obtained for the normal hierarchy.  On the other hand, one may wish
to maintain the mass ordering scheme $m_1, m_2, m_3$ (the
$123$-case) as shown in the rightmost spectrum \cite{Petcov:2001sy}.
It is natural to explore the mapping between the mixing parameters
in these two cases. Figure \ref{fig:InvMap}, which indicates the
flavor composition of each state by color coding, represents this
transformation between the mixing matrix elements $U^*_{\alpha i}$
via explicit labeling.  From this alone I see, for example, that
$U^{(312)*}_{e3} = U^{(123)*}_{e1}$ which implies that
$s_{12}^{(312)}s_{23}^{(312)} +
c_{12}^{(312)}c_{23}^{(312)}s_{13}^{(312)}e^{-i\delta^{(312)}} =
c_{12}^{(123)}c_{13}^{(123)}$, where the parameter superscripts
indicate the mass ordering.  This is one of the simplest cases taken
from the figure, but it still provides a very nontrivial relation
between the variables.  I proceed formally, using the language of
Appendix \ref{app:notation}, to complete the mapping.  Specifically,
I study the effect of the following operation
\begin{equation}
{\bf H}\left(
  \begin{array}{c}
    \nu_3 \\
    \nu_1 \\
    \nu_2 \\
  \end{array}
\right) = {\bf R^{23}}(\frac{\pi}{2}){\bf P^3}(\pi) {\bf
R^{13}}(\frac{\pi}{2}){\bf P^3}(\pi)\left(
  \begin{array}{c}
    \nu_3 \\
    \nu_1 \\
    \nu_2 \\
  \end{array}
\right) = \left(
  \begin{array}{c}
    \nu_1 \\
    \nu_2 \\
    \nu_3 \\
  \end{array}
\right)
\end{equation}
on the mixing matrix.  Here, one of the phase rotations ${\bf
P^3}(\pi)$ may commute through the real rotations to cancel out,
leaving only a simple sign change.  Acting on the mixing matrix,
this has the following effect:
\begin{eqnarray}
{\bf H}^{-1}{\bf U}^\dagger &=& {\bf R^{13}}(\frac{\pi}{2}){\bf
R^{23}}(-\frac{\pi}{2}){\bf P^2}(\phi_2){\bf P^3}(\phi_3) {\bf
R^{12}}(\theta_{12}){\bf P^1}(-\delta){\bf P^3}(\delta) {\bf
R^{13}}(\theta_{13})\\ \nonumber &~~~~~& \times{\bf P^1}(\delta){\bf P^3}(-\delta){\bf
R^{23}}(\theta_{23}) \\ \nonumber &=& {\bf
R^{13}}(\frac{\pi}{2}){\bf P^3}(\phi_2){\bf P^2}(\phi_3) {\bf
R^{13}}(\theta_{12}){\bf P^1}(-\delta){\bf P^2}(\delta){\bf
R^{12}}(\theta_{13}){\bf P^1}(\delta)\\ \nonumber &~~~~~& \times{\bf P^2}(-\delta){\bf
R^{23}}(\theta_{23}-\frac{\pi}{2})\\ \nonumber &=& {\bf
P^1}(\phi_2){\bf P^2}(\phi_3) {\bf
R^{13}}(\theta_{12}+\frac{\pi}{2}){\bf P^1}(-\delta){\bf
P^2}(\delta){\bf R^{12}}(\theta_{13}){\bf P^1}(\delta) \times{\bf
P^2}(-\delta)\\ \nonumber &~~~~~&{\bf R^{23}}(\theta_{23}-\frac{\pi}{2}).
\label{eq:InvFormalMap}
\end{eqnarray}
Notice that the entire transformation may be absorbed into shifts of
the physical mixing parameters leaving no nonphysical sign change
requirements.  Formally, the transformations shown in
Eq.~(\ref{eq:InvFormalMap}) are very simple.  The largest change
arises from the relabeling of the mixing planes which leads to a
situation where the complex rotation governed by $\theta_{13}$ and
$\delta$ now acts in the $1-2$ plane, while $\theta_{12}$ governs
the $1-3$ rotation.  As intuitively expected, it exchanges the roles
of $\theta_{12}$ and $\theta_{13}$.  The main difficulty arises
operationally while massaging the Euler rotations into the
conventional order, since they do not commute.  To put the result of
Eq.~(\ref{eq:InvFormalMap}) into the form of
Eq.~(\ref{eq:ThreeNuMix}) one must commute the $1-2$ and $1-3$
rotations.  A little thought reveals that this alone is not enough,
as no consistent parameter mapping may exist for all relevant angles
in this case. Thus, to proceed, one must also search for highly
nontrivial transformations of $\theta_{23}$.  Commuting the matrices
formally for the full transformation, I find the following mapping of
mixing parameters from the $312$ to the $123$ spectral cases.
\begin{equation}
\begin{array}{lll}
\Delta m_{13}^2 & \rightarrow & -\Delta m_{12}^2 \\
\Delta m_{23}^2 & \rightarrow & -\Delta m_{31}^2 \\
\Delta m_{12}^2 & \rightarrow & \Delta m_{23}^2 \\
\end{array} \Rightarrow \left\{
\begin{array}{c}
c_{12}^2 \rightarrow \frac{s_{13}^2}{1-c_{13}^2s_{12}^2}\\
s_{13}^2 \rightarrow c_{13}^2s_{12}^2\\
c_{23}^2 \rightarrow \frac{c_{23}^2s_{12}^2s_{13}^2 + c_{12}^2s_{23}^2 + 2\cos\delta c_{12}c_{23}s_{12}s_{23}s_{13}}{1 - c_{13}^2s_{12}^2}\\
\delta \rightarrow \cos^{-1}\left( \frac{c_{12}s_{23} + \cos\delta
c_{12}s_{23}s_{13}}{\sqrt{c_{23}^2s_{12}^2s_{13}^2 +
c_{12}^2s_{23}^2 + 2\cos\delta c_{12}c_{23}s_{12}s_{23}s_{13}}}
\right)\\
\phi_3 \rightarrow \phi_2 - \phi_3 - \delta - \cos^{-1}\left(
\frac{c_{12}s_{23} + \cos\delta
c_{12}s_{23}s_{13}}{\sqrt{c_{23}^2s_{12}^2s_{13}^2 +
c_{12}^2s_{23}^2 + 2\cos\delta c_{12}c_{23}s_{12}s_{23}s_{13}}}\right) \\
\phi_2 \rightarrow -\phi_3 - \delta \\
\end{array}
\right.
\end{equation}
These are unique up to trivial sign changes, provided the overall
nonphysical rephasing by $e^{i(\delta + \phi_3)}$.  The
transformations are relatively simple for the $\theta_{12}$ and
$\theta_{13}$ parameters, but the redefinition is less attractive
for $\theta_{23}$ due to its induced dependence on $\delta$.  In the
case of Dirac CP conservation, the properties $\delta=0$ and
$\delta=\pi$ are maintained in the mapping (as they should) and the
troublesome transformations are simplified to $\phi_2 \rightarrow
-\phi_3 + (-\pi)$, $\phi_3 \rightarrow \phi_2-\phi_3 + (-\pi)$ and
$c_{23}^2 \rightarrow \left(c_{23}s_{12}s_{13} + c_{12}s_{23}
\right)^2/\left(c_{13}^2s_{12}^2 - 1 \right)$.

\subsection{Comments on the $n$ Neutrino case} \label{subsec:NNu}

The results of the previous subsections may be directly generalized
to the case of $n$ neutrinos.  Here, all physics may be
parameterized by $n$ real and positive masses $m_i$ and $n(n-1)$
mixing angles/phases.  This arises from the $2n^2$ degrees of
freedom in an arbitrary complex $n\times n$ matrix minus $n^2$
unitarity conditions and $n$ field rephasing. Choosing the standard
parameterizations, the remaining freedom may be categorized into
$n(n-1)/2$ real mixing angles $\theta_{ij}$, $(n-1)(n-2)/2$ Dirac
phases $\delta_{ij}$ and $n-1$ Majorana phases $\phi_i$.  Due to the
limited rephasing freedom, I may define select elements of the
mixing matrix real.  Following \cite{Fritzsch:1986gv} I choose to
remove all Dirac phases from the three row main diagonal band so
that rotations between the neighboring $(a-1)-a$ planes are taken to
be real, while the rest are not, and thus, include a CP violating
Dirac phase. Taking these complex rotations in the $i-j$ plane to be
${\bf \widetilde{R}^{ij}}(\theta_{ij},\delta_{ij}) \equiv {\bf
P^i}(-\delta_{ij}/2) {\bf P^j}(\delta_{ij}/2) {\bf
R^{ij}}(\theta_{ij}{\bf P^i}(\delta_{ij}/2) {\bf
P^j}(-\delta_{ij}/2)$, I find that an arbitrary $n\times n$ Majorana
mixing matrix may be written as:

\begin{eqnarray}
U_n^\dagger &=& \prod_{i=2}^n \left[{\bf P^i}(\phi_i)
\prod_{j=1}^{i-2}\left[ {\bf
\widetilde{R}^{ji}}(\theta_{ji},\delta_{ji})\right]
{\bf R^{i-1,i}}(\theta_{i-1,i}) \right]\\
\nonumber &=& {\bf P^n}(\phi_n) U_{n-1}^\dagger
\prod_{j=1}^{n-2}\left[ {\bf
\widetilde{R}^{jn}}(\theta_{jn},\delta_{jn})\right] {\bf
R^{n-1,n}}(\theta_{n-1,n}), \label{eq:NNuMix}
\end{eqnarray}
where the products are taken on the right such that $\prod_{i=1}^n
A_i = A_1 A_2 A_3 ... A_n$.  The $U_{n-1}^2$ in the second line is
actually of dimension $n\times n$, but is suggestively written to
indicate that the upper $n-1 \times n-1$ diagonal block is the
reduced $n-1$ neutrino mixing matrix. This reproduces the standard
mixing matrices for the previous $n=2$ and $n=3$ cases and one may,
in principle, compute this structure for an arbitrary number of
states. For example, assuming $n=4$ Eq.~(\ref{eq:NNuMix}) takes the
form
\begin{equation}
U_4^\dagger = {\bf P^2}(\phi_2){\bf R^{12}}(\theta_{12}){\bf
P^3}(\phi_3){\bf \widetilde{R}^{13}}(\theta_{13},\delta_{13}){\bf
R^{23}}(\theta_{23}){\bf P^4}(\phi_4){\bf
\widetilde{R}^{14}}(\theta_{14},\delta_{14}){\bf
\widetilde{R}^{24}}(\theta_{24},\delta_{24}){\bf
R^{34}}(\theta_{34}). \label{eq:fourNuMix}
\end{equation}

%%%%%%%%%%%%%%%%%%%%beginsterilestuff%%%%%%%%%%%%%%%%%%%%%%%%%%%%%%%%%%%%%%%%%%%%%
In order for this analysis to make contact with observed reality, I now describe the parameterization of the $3+n$ sterile neutrino scenario.  Assuming the standard three light neutrinos that mix among themselves via the $3\times 3$ matrix $U_\ell$ and $n$ sterile states that mix via the approximately unitary $n \times n$ matrix $U_H$.  The total mixing matrix $[U_{tot}]_{n+3}$ to first order in the light-heavy mixing is

\begin{equation}
% use packages: array
[U_{tot}]_{n+3} = \left(\begin{array}[c]{cc}
[U_\ell]_3 & [\Theta^\dagger]_{3\times n} \\ 
-[U_H]_n [\Theta]_{n\times 3} [U_\ell]_3 & [U_H]_3
\end{array} \right),
\end{equation}
where
\begin{equation}
[\Theta]_{n\times 3} = \sum_{j=1}^n [U_{H,j-1}]_n [\Theta_j]_{n\times 3}.
\end{equation}
Here, $[U_{H,0}]_n = [U_{H,1}]_n = 1$ and $\Theta_j$ is zero for all elements $a,b$ except the row $a=i$ where it is
\begin{equation}
\theta_{b,a+3}e^{i\delta_{b,a+3}}.
\end{equation}
Note that all Dirac phases are present, per my outlined convention, except $\delta_{34}=0$ (since I hold rotations between adjacent planes real).  

In the special $3+1$ sterile neutrino case I can work out the mixing matrix to second order.

\begin{equation}
\left(
\begin{array}[c]{cc}
U_\ell^\dagger & {\bf 0}_{3\times 1} \\ 
{\bf 0}_{1\times 3} & 1
\end{array}
\right)
 \left( \begin{array}[c]{cccc}
1-\frac{1}{2}\theta_{14}^2 & -\theta_{14}\theta_{24}e^{-i(\delta_{14} - \delta_{24})} & -\theta_{14}\theta_{34}e^{-i\delta_{14}} & -\theta_{14}e^{-i\delta_{14}} \\ 
0 & 1-\frac{1}{2}\theta_{24}^2 & -\theta_{24}\theta_{34}e^{-i\delta_{24}} & -\theta_{24}e^{-i\delta_{24}} \\ 
0 & 0 & 1-\frac{1}{2}\theta_{34}^2 & -\theta_{34} \\ 
\theta_{14}e^{i\delta_{14}} & \theta_{24}e^{i\delta_{24}} & \theta_{34} & 1-\frac{1}{2}\left(\theta_{14}^2 + \theta_{24}^2 + \theta_{34}^2 \right)×
\end{array} \right)
\end{equation}
The schematic form of the sterile-active mixing 
\begin{equation}
\left(
\begin{array}[c]{ll}
C \approx 1 - \frac{1}{2}\theta^2 & S^\dagger \approx \theta^\dagger \\ 
S \approx \theta & C \approx 1 - \frac{1}{2}\theta^2
\end{array}
\right)
\end{equation}
is clear in all cases and will be very useful in Chapter \ref{chap:SeeSawLSND}.

%%%%%%%%%%%%%%%%%%%%endsterilestuff%%%%%%%%%%%%%%%%%%%%%%%%%%%%%%%%%%%%%%%%%%%%%%%%%

I may now search for symmetries of $U_n^\dagger$.  To limit the
parameter space in a systematic fashion, I present the symmetries in
the same order as in the three neutrino case.  The transformations
outlined here are equally complete.  First, notice that all Majorana
phases may be trivially commuted to the left of the expression.  As
such, for all integers $i\in (2,n)$, a phase shift $\phi_i
\rightarrow \phi_i + \pi$ may be absorbed by a sign redefinition of
the mass eigenstate $\nu_i \rightarrow -\nu_i$. Hence, as was recently shown in \cite{Jenkins:2007ip} using different methods, each Majorana
phase may be limited to a range of $\pi$. Next, I find that
$\theta_{ij}\rightarrow -\theta_{ij}$, followed by $\delta_{ij}
\rightarrow \delta_{ij}+\pi$ is also an invariance for integers $i+1
\neq j$.  Thus, for the complex rotations one may compensate for
negative angle values, provided a corresponding shift $\pi$ of the
Dirac phase.  In a similar way, the angles governing real rotations
participate in this invariance via the symmetries $\theta_{a-1,a}
\rightarrow -\theta_{a-1,a}, \theta_{ia} \rightarrow -\theta_{ia},
\theta_{aj} \rightarrow -\theta_{aj}$ for all integers $i\in
(1,a-2)$ and $j\in (a+1,n)$ provided the unphysical redefinitions
$\nu_a \rightarrow -\nu_a$ and $ \nu_{\alpha_a} \rightarrow
-\nu_{\alpha_a}$ of the mass and flavor states, respectively.  Here,
one may systematically limit the angles $\theta_{a-1,a}$ since all
other angles are either constrained by the previous transformations,
or show up to the right of the real rotation in question.  In this
latter case one may proceed inductively from the rightmost rotation
${\bf R^{n-1,n}}(\theta_{n-1,n})$. Finally, I see that our system
obeys the symmetry $ \theta_{ab} \rightarrow \theta_{ab} + \pi,~
\theta_{ib} \rightarrow -\theta_{ib}, ~\theta_{aj} \rightarrow
-\theta_{aj}$ for all integers $i < a$ and $a < j<b$, provided the
unphysical redefinitions $\nu_a \rightarrow -\nu_a$ and $ \nu_b
\rightarrow -\nu_b$ of the mass and flavor states, respectively.
Hence, I obtain the expected result that the physical range of
\emph{all} mixing angles, Majorana phases, and Dirac phases are
$\theta_{ij} \in [0,\pi/2]$, $\phi \in [-\pi/2,\pi/2]$, and
$\delta_{ij} \in [0,2\pi]$, respectively.  As in the three neutrino
case, no simple invariance may be formed with mass eigenstate
permutations.

That being said, some relations still hold among the mixing angles
and phases in these most general cases.  Such instances are highly nontrivial and a
full exploration is beyond the scope of this analysis and not very
enlightening.  In any case, all of the examples outlined in
subsections \ref{subsec:TwoNu} and \ref{subsec:ThreeNu} are still
valid in the full $n$ neutrino case.  To see this, I show that if
$U_{n-1}^\dagger$ is invariant under some symmetry up to allowed
state sign changes, then $U_n^\dagger$ is also invariant. In this
context, invariant is taken to mean $U_{n-1}^{\dagger\prime} = {\bf
N_m} U_{n-1}^{\dagger} {\bf N_f}$ where, as in subsection
\ref{subsec:TwoNu}, ${\bf N_m}$ and ${\bf N_f}$ are simply sign
change matricies acting on the mass and flavor eigenstates,
respectively.  Both of these may be written as products of ${\bf
P^i}(\pi)$ and handled with the algebra outlined in Appendix
\ref{app:notation}.  To begin, I note that ${\bf N_m}$ can always be
absorbed by mass eigenstate sign changes, as can a $\pi$ phase shift
of ${\bf P^n}(\phi_n)$.  Thus, it is enough to show that an
arbitrary ${\bf P^a}(\pi)$ from ${\bf N_f}$ commutes through the
remaining rotations to the flavor state vector.  Since $a \leq n-1$,
only the commutation with ${\bf R^{an}}$ is nontrivial. For $a <
n-1$ I have ${\bf P^a}(\pi){\bf
\widetilde{R}^{an}}(\theta_{an},\delta_{an}) = {\bf
\widetilde{R}^{an}}(-\theta_{an},\delta_{an}){\bf P^a}(\pi)={\bf
\widetilde{R}^{an}}(\theta_{an},\delta_{an}+\pi){\bf P^a}(\pi)$,
which yields only a $\delta_{an}$ shift by $\pi$. For the case
$a=n-1$ I am left with ${\bf P^{n-1}}(\pi){\bf
R^{n-1,n}}(\theta_{n-1,n})={\bf P^{n-1}}(\pi){\bf P^{n-1}}(\pi){\bf
P^{n}}(\pi){\bf R^{n-1,n}}(\theta_{n-1,n}+\pi)={\bf P^{n}}(\pi){\bf
R^{n-1,n}}(\theta_{n-1,n}+\pi)$, where the remaining phase shift is
absorbed by a $\nu_n$ sign change after commuting through each ${\bf
\widetilde{R}^{in}}$, which induces a $\delta_{in} \rightarrow
\delta_{in} +\pi$ for all $i < n$.  The $\theta_{n-1,n} \rightarrow
\theta_{n-1,n} + \pi$ is absorbed by a separate $\nu_{\alpha_{n-1}}
\rightarrow -\nu_{\alpha_{n-1}},\nu_{\alpha_{n}} \rightarrow
-\nu_{\alpha_{n}}$ transformation.  Hence, I see that the invariance
of $U_{n-1}^\dagger$ is preserved, provided each $\delta_{in}$
occupy its full physical range of $2\pi$.

\section{Nonstandard Interactions} \label{sec:NonStandardInt}

Many conclusions derived in the previous sections will be modified
when nonstandard neutrino interactions are introduced into the SM
Lagrangian of Eq.~(\ref{eq:LSM}).  These can take the form of lepton
number conserving interactions given by
\begin{equation}
\mathcal{L}_{NS}^{LNC} = \xi^{\rm n}_{\alpha\beta}
\overline{\nu}_\alpha \gamma_\mu \nu_\beta Z^{\prime\mu} + \xi^{\rm
c}_{\alpha\beta} \overline{\nu}_\alpha \gamma_\mu \ell_\beta
W^{\prime\mu} + {\rm h.c.} \label{eq:LNC_NSI}
\end{equation}
and lepton number violating interactions such as
\begin{equation}
\mathcal{L}_{NS}^{LNV} = \eta^{\rm n}_{\alpha\beta}
\overline{\nu^c}_\alpha \nu_\beta S_n + \eta^{\rm c}_{\alpha\beta}
\overline{\nu^c}_\alpha \ell_\beta S_c + {\rm h.c.}
\label{eq:LNV_NSI}
\end{equation}
Here, each line contains a neutral current type and charged current
type interaction denoted by the superscripts on the coupling
constants $\xi$ and $\eta$.  To be concrete, these are written in
terms of renormalizable expressions where neutrinos couple to vectors
$Z^\prime$ and $W^\prime$, as well as charged and neutral scalars
denoted by $S_c$ and $S_n$, respectively.  All that is needed for
my purpose is the neutrino structure, so one may substitute a
general fermion current in place of the bosons to yield effective
operators if desired.  All four of the couplings listed are general
complex matrices in flavor space defined by the flavor subscripts
$\alpha$ and $\beta$. These are not diagonal in either the flavor or
mass basis.  I point out that terms in Eq.~(\ref{eq:LNV_NSI})
violate lepton number by two units and as such, will themselves lead
to radiatively generated Majorana neutrino masses at some order in
perturbation theory \cite{de Gouvea:2007xp}.  If this is the
dominant source of neutrino masses generation, $\eta^{n(c)}$ would
then be approximately aligned with the mass basis.

Upon transforming to the mass basis with the unitary matrix $U$, I
find that the coupling constants shift to
\begin{eqnarray}
\xi^{\rm n} &\rightarrow& U^\dagger \xi^{\rm n}U \\
\nonumber \xi^{\rm c} &\rightarrow& U^\dagger \xi^{\rm n} \\
\nonumber \eta^{\rm n} &\rightarrow& U^T \eta^{\rm n}U \\
\nonumber \eta^{\rm c} &\rightarrow& U^T \eta^{\rm n}.
\end{eqnarray}
Using the invariance condition $U^\dagger \rightarrow {\bf
N_m}U^\dagger {\bf N_f}$ as before, I search for the symmetries of
these interactions.  First, by recalling that both ${\bf N}$
matrices are Hermitian, it is clear that both the lepton number
violating and conserving interactions yield the same conditions, so
it is enough to discuss the transformation of $\xi$.  For the
neutral current term I want ${\bf N_m}U^\dagger {\bf N_f} \xi^{\rm
n} {\bf N_f}U {\bf N_m} = U^\dagger \xi^{\rm n}U$ up to field sign
redefinitions which can clearly absorb the ${\bf N_m}$ factors.
Thus, a given symmetry of the mixing matrix is also a symmetry of
neutral current nonstandard interactions if the resulting ${\bf
N_f}$ commutes with $\xi^{\rm n}$.  Since ${\bf N_f}$ is some
product of ${\bf P^{a_i}}(\pi)$, for some set of integers $a_i$,
this translates to the condition that $\xi^{\rm n}_{a_i j}=\xi^{\rm
n}_{j a_i}=0$ for all $j\neq a_i$. The charged current case is even
less restrictive in that all that is required is ${\bf N_m}U^\dagger
{\bf N_f} \xi^{\rm c}  = U^\dagger \xi^{\rm n}$ up to field sign
redefinitions.  Here, this reduces to ${\bf N_f} \xi^{\rm c} =
\xi^{\rm c}{\bf N_f}^\prime$, where ${\bf N_f}^\prime$ is some other
product of ${\bf P^{b_i}}(\pi)$.  In component form, this translates
to the condition that for each $b_i$ there exists a $c_i$ such that
$\xi^{\rm c}_{b_i,j}=0$ for all $j \neq c_i$ and $\xi^{\rm
c}_{k,c_i}=0$ for all $k \neq b_i$.  This is a much looser set of
conditions than for the neutral current case. I point out that any
symmetries of the mixing matrix broken by the nonstandard
interactions may be restored by allowing for coupling constant sign
changes of the form $\xi^{\rm n} \rightarrow {\bf N_f} \xi^{\rm n}
{\bf N_f}$ or $\xi^{\rm c} \rightarrow {\bf N_f} \xi^{\rm c}$.

In terms of the specific symmetries found in Section
\ref{sec:PhyRange}, I see that most parameter limitations still hold
in the face of general new physics.  These are transformations whose
${\bf N_f}$ may be commuted and absorbed entirely into a
redefinition of the mass eigenstates.  In particular, within the $n$
neutrino scenario, the invariance involving $\theta_{a-1,a}
\rightarrow -\theta_{a-1,a}$ is broken while the following
symmetries still hold:
\begin{enumerate}
\item $\phi_i \rightarrow \phi_i +\pi, \nu_i \rightarrow -\nu_i$ for
all integers $1 \leq i$.
\item $\theta_{ij} \rightarrow -\theta_{ij}, \delta_{ij} \rightarrow
\delta_{ij} + \pi$ for all integers $i \neq j-1$.
\item $\theta_{ij} \rightarrow \theta_{ij} + \pi, \theta_{i,i+1} \rightarrow -\theta_{i,i+1}, \theta_{ia} \rightarrow -\theta_{ia}, \theta_{bj} \rightarrow -\theta_{bj}, \nu_i \rightarrow -\nu_i, \nu_j \rightarrow
-\nu_j$ for all integers $b<i$ and $i+1 < a < j$.
\end{enumerate}
In the last entry the $\theta_{i,i+1} \rightarrow -\theta_{i,i+1}$
transformation is factored out of the other operations, as it may
not be independently addressed by another unbroken symmetry.
Therefore, in the case where the $\theta_{i,i+1}$ positive-negative
symmetry is broken, one must choose which degeneracy to exploit when
limiting the parameter space. In either case, this leads to an
expanded mixing angle range $\theta_{ij} \in [-\pi/2, \pi/2]$.  This
occurs when the nonstandard interactions, written in the flavor
basis, have a structure where there exists at least one nondiagonal
$\xi^{\rm n}$ entry in each of the $i^{\rm th}$ and $(i+1)^{\rm th}$
row or column for the neutral current case.  The situation is
similar for charged current new terms, except that here one must
have a nondiagonal $\xi^{\rm n}$ entry in both the $i^{\rm th}$ and
$(i+1)^{\rm th}$ rows, or at least one in each column.

\begin{figure}[t]
\begin{center}
\includegraphics[scale=1]{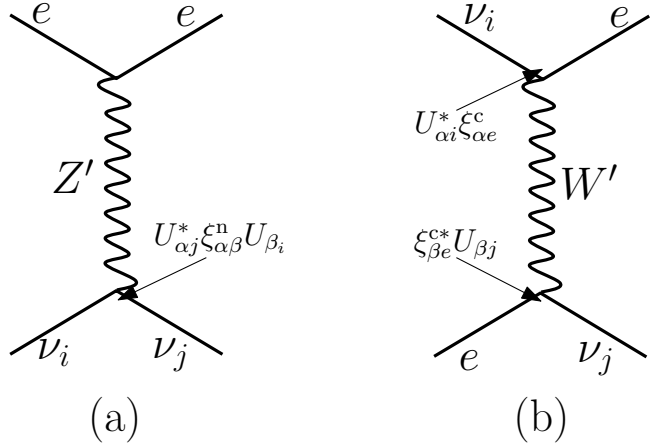}
\caption[Neutrino electron quasi-elastic scattering diagrams via
$W^\prime$ and $Z^\prime$]{Neutrino electron quasi-elastic scattering
diagrams via the exchange of massive $W^\prime$ and $Z^\prime$
vector bosons.}\label{fig:NonStandMatter}
\end{center}
\end{figure}

It is useful to explore this result with examples taken from the two
neutrino scenario.  Of course many processes may be used to
illustrate these points, but for simplicity, I focus on neutrino
electron scattering $\nu_i e \rightarrow \nu_j e$.  The Standard
Model contribution to this via t-channel $Z$ and $W$ exchange is
easy to understand and is responsible for coherent forward neutrino
scattering in dense media that leads to matter effects in neutrino
oscillations.  For this, one may subtract the diagonal $Z$ couplings
from the effective matter potential as a common factor and consider
only the charged current interactions. In a similar way, nonstandard
neutrino couplings will modify matter affected neutrino
oscillations. One may calculate the modified probabilities, but for
my purpose it is enough to simply consider the relative
amplitudes.  I consider the SM contributions as well as new neutral
and charged current interactions mediated by $Z^\prime$ and
$W^\prime$, respectively.  The relevant diagrams are shown in Figure
\ref{fig:NonStandMatter} along with explicitly labeled vertex
couplings. In this example, I assume that new physics has nonzero
coupling to first generation charged leptons, but this is not a
general requirement. The amplitude for this process is
\begin{equation}
\mathcal{A}_{ij} = F_Z\delta_{ij} + F_W U_{ie}^* U_{ej} +
F_{Z^\prime}U^*_{\alpha j}\xi^n_{\alpha \beta} U_{\beta i} +
F_{W^\prime} U^*_{\alpha i}\xi^c_{\alpha e} \xi^{c*}_{\beta e}
U_{\beta j},
\end{equation}
where $F_i$ is some function containing process-dependent
kinematical factors and irrelevant coupling constants.  For my
purpose, it is enough to know that these functions are independent
of neutrino mixing parameters and couplings.  The full expression
may be calculated within the context of a specific model if needed.
I point out that all Majorana phase dependency washes out of this
expression since it conserves lepton number.  For illustrative
purposes, I evaluate the $i=j=1$ elastic scattering case since it has
contributions from all four interaction types identifiable by their
$F_i$ factors. In component form, the amplitude is
\begin{eqnarray}
\mathcal{A}_{11} &=& F_Z + F_W \cos^2 \theta + \frac{1}{2}F_{Z^\prime}
\left(\xi^n_{ee} + \xi^n_{\mu\mu} + (\xi^n_{ee}-\xi^n_{\mu\mu})\cos
2\theta -  (\xi^n_{e\mu}+\xi^n_{\mu e})\sin 2\theta\right) \nonumber \\ &~~~~~& +
F_{W^\prime}\left(\xi^c_{ee}\cos\theta - \xi^c_{\mu e}\sin\theta
\right)^2.
\end{eqnarray}
I explore this term by term, beginning with the SM contributions.
Clearly, the $Z$ exchange terms do not effect the symmetry structure
of neutrino mixing, but $W$ mediation breaks the $\theta$ degeneracy
between $[0,\pi/4]$ and $[\pi/2,\pi/2]$ due to the $\cos^2 \theta$
factor. This is the reason behind the physical interpretation of the
$\Delta m^2$ sign discussed by example in Subsection
\ref{subsec:TwoNu}.  Moving to the new physics contrabutions I see
that $W^\prime$ mediation maintains $\theta \rightarrow \theta +
\pi$ while breaking the $\theta \rightarrow -\theta$ symmetry,
unless compensated by a corresponding sign flip of the off diagonal
coupling constant $\xi^c_{\mu e}$.  The same is true for the
$Z^\prime$ interaction which breaks $\theta \rightarrow -\theta$
unless $\xi^n_{e\mu}+\xi^n_{\mu e} \rightarrow
-(\xi^n_{e\mu}+\xi^n_{\mu e})$.  Therefore, I see explicitly that
flavor off diagonal nonstandard interactions expand the physical
mixing angle range to $\theta \in [-\pi/2,\pi/2]$, as expected.

\section{Conclusions} \label{sec:Par_Conclusions}

I conclude with a brief summary of the key issues explored in this
chapter.  From exploiting the symmetries of the neutrino mixing
matrix, I explore degeneracies within the neutrino parameter space.
There is a significant amount of freedom in choosing the physical
ranges of these quantities and care must be taken to ensure
consistent analysis of neutrino data.  This is particularly true
when comparing the results of experiments that rely on different
``physical'' parameter combinations as in, for example, lepton number
violating and lepton number conserving phenomena where Majorana
neutrino phases are physical and non-physical, respectively.  With
this in mind, it is best to choose a parameter scheme and stick with
it.  Following the majority of the literature, I chose to restrict
all Majorana phases $\phi_i \in [-\pi/2,\pi/2]$, Dirac phases
$\delta_{ij} \in [0,2\pi]$, and mixing angles $\theta_{ij} \in
[0,\pi/2 ]$ for the remainder of this thesis.  Such a convention can
accommodate all of the physics of neutrino mixing, assuming SM
interactions.  Interestingly, this also holds in the case of
arbitrary BSM effects, provided that some select new physics coupling
constants are allowed to take on both positive and negative values.

\chapter{A Survey of Lepton Number Violation Via Effective Operators} \label{chap:LNV}
\setcounter{equation}{0} \setcounter{footnote}{0}

As previously emphasized, the discovery of neutrino masses via their
flavor oscillations over long baselines constitutes the first solid
evidence of physics beyond the standard model (SM) of particle
physics \cite{NeutrinoReview}. While this is an important first step
toward a deeper understanding of nature, it poses many more
questions than it answers. A number of theoretically well-motivated
models have been proposed and explored to address the origin of the
neutrino mass but, strictly speaking, these represent only a handful
out of an infinite set of possibilities.  The question of how well
future experiments can probe and distinguish different scenarios
arises naturally and is quite relevant given the current state of
high energy physics.  The coming years promise detailed explorations
of the terascale with the Large Hadron Collider (LHC) and the more
distant International Linear Collider (ILC) or variants thereof.
Expectations are that combined information from these two
facilities, coupled with high precision, low energy results and
cosmological observations will shed light on some of the current
mysteries of physics, including that of the neutrino mass.

Here, I concentrate on the possibility that the neutrino masses are
generated at some high energy scale $\Lambda$ where $U(1)_{B-L}$,
the only non-anomalous global symmetry of the standard model, is
broken. Such a scenario is
 well-motivated by the observed properties of the light neutrinos including tiny
 masses, large mixings and the fact that neutrinos are the only electrically
 neutral fundamental fermions. More specifically, once $U(1)_{B-L}$ is broken, neutrinos
 are not protected from getting non-zero Majorana masses after electroweak symmetry breaking.
 On the other hand, since the renormalizable minimal standard model\footnote{Throughout, I will
 assume that the weak scale degrees of freedoms are the known standard model fields, plus a minimal Higgs sector. Hence, I assume that there are no gauge singlet ``right-handed neutrino'' fermions or higher $SU(2)_L$ Higgs boson representations, such as Higgs boson triplets.} preserves $U(1)_{B-L}$, $B-L$ breaking
 effects will only manifest themselves at low energies through higher dimensional operators. This
 being the case, one generically expects neutrino masses to be suppressed with respect to charged
 fermion masses by $(v/\Lambda)^n$, $n\ge 1$, where $v$ is the Higgs boson vacuum expectation value.

By further assuming that all new degrees of freedom are much heavier
than the weak scale, I am guaranteed that, regardless of the details
of the new physics sector, all phenomena below the weak scale are
described by irrelevant, higher dimensional operators. In this
spirit, the observable consequences of all high energy models that
lead to small Majorana neutrino masses can be catalogued by
understanding the consequences of irrelevant operators that break
$B-L$ by two units. With this in mind, I will survey all such
non-renormalizable effective operators for phenomenological
signatures at future and current experiments.  I restrict myself
to operators that will lead to lepton number violation (LNV), as
these will be directly connected to the existence of small Majorana
neutrino masses. This means that I do not consider operators that
conserve $L$ but violate $B$, and hence also $B-L$, by two units
(such operators lead to, for example, neutron--antineutron
oscillations), nor do I include operators that respect $B-L$. Most
of the time, the latter will not mediate any observable consequences
for large enough $\Lambda$, except for operators of dimension-six
and above that can mediate proton decay.

To begin, I systematically name and classify all relevant LNV
operators. Fortunately, this has already been done\footnote{The
authors of \cite{Operators} discuss all possible effective operators
of dimensions up to and including eleven, but only explicitly list
those deemed unique in the sense that they cannot be written as the
product of any previous operator with a Standard Model interaction.
I append their list and naming scheme to include these into our
analysis.} in \cite{Operators} up to and including operators of mass
dimension eleven.\footnote{I will argue later that irrelevant
operators with mass dimension thirteen and higher, if related to
neutrino masses, will require new physics below the electroweak
scale so that I would have already observed new physics if neutrino
masses were generated in this way. Furthermore, from a model
building perspective, it is difficult to develop models that
predominantly yield effective operators of very high mass dimension.
The probability that such scenarios are both theoretically
well-motivated and evade all observations appears to be slim.} For
each operator I then calculate/estimate the analytic form of the
radiatively generated neutrino mass matrix.  Upon setting this
expression equal to the experimentally measured neutrino masses, I
extract the energy scale $\Lambda$ associated to the new LNV
physics. Armed with these scales, I proceed to calculate each
operator's phenomenological signatures at a variety of experimental
settings. Additionally, having explicitly calculated the
operator-induced neutrino mass \emph{matrices}, I may also verify,
under some generic assumptions, whether one can account for the
observable lepton mixing pattern.  After such a general survey, one
is adequately equipped to take a step back and select
phenomenologically/theoretically interesting operators for further
detailed study by ``expanding'' effective vertices to reveal
particular ultraviolet completions.  In this way, one can use the
results presented here as a means of systematically generating
renormalizable models with well-defined experimental predictions.

This chapter is organized as follows.  Sec.~\ref{sec:Scale} is
devoted to an introduction to the effective operators and methods.
In Sec.~\ref{subsec:NuMass}, I derive and comment on the scales
$\Lambda$ of new physics that are used throughout the remainder of
the text.  In Sec. \ref{sec:constraints}, I survey various
experimental probes of LNV for each operator, and address if and
when our analysis breaks down due to added model structure or
additional assumptions. Specifically, I study both current
constraints and future prospects for neutrinoless double-beta decay
experiments in Sec.~\ref{subsec:bb0nu}, followed,  in
Sec.~\ref{subsec:OtherProcesses},  by a similar analysis of other
rare decay modes, including those of various mesons and W/Z gauge
bosons. In Sec.~\ref{subsec:Collider}, I present collider signatures
of LNV as they apply to future linear collider facilities running in
the $e^-e^-$ collision mode, and describe extensions of our analysis
to include associated $\gamma\gamma$ collisions. I also comment on
searches for LNV in future hadron machines.
Sec.~\ref{sec:Oscillations} describes current constraints from
neutrino oscillation phenomenology due to the general structure of
the derived neutrino mass matrices. In Sec.~\ref{sec:interesting}, I
highlight a number of ``interesting'' operators, defined by low
cutoff scales and prominence of experimental signatures, which are
still allowed by current constraints on LNV. I undertake a slightly
more detailed discussion of their characteristics and signatures and
present some sample ultraviolet completions. I conclude in
Sec.~\ref{sec:conclusion} with a summary of our assumptions and
results, augmented by commentary on future prospects for LNV
searches. Our results are tabulated by operator name in
Table~\ref{tab:AllOps} for easy reference.

I hope that this  analysis will prove useful to various audiences on
a number of distinct levels. In the most superficial sense, the
casual reader should note the general features of LNV as well as the
diversity of model variations.  Such information is best expressed
in terms of the operator distribution histograms scattered
throughout the text. These are color-coded by operator dimension or
cutoff scale, and typically contain additional information,
including current experimental prospects. On the more technical
side, those interested in specific neutrino mass generating models
will find detailed, operator specific, information that may be
utilized as crude model predictions. Additionally, as already
alluded to, one may even ``hand-pick'' operators for model
development based on specific phenomenological criteria. Finally, I
urge experimentalists to search for new physics in all accessible
channels. It is our ultimate goal to provide motivation for
experimental considerations of non-standard LNV effects, beyond
neutrinoless double-beta decay.

\setcounter{footnote}{0} \setcounter{equation}{0}
\section{The lepton number violating scale}
\label{sec:Scale}

Here I analyze $SU(3)_c \times SU(2)_L \times U(1)_Y$ invariant
$\Delta L = 2$ non-renormalizable effective operators of mass
dimension up to and including eleven.  They are composed of only the
SM field content as all other, presumably heavy, degrees of freedom
are integrated out. As already emphasized, I do not allow for the
existence of SM singlet states (right-handed neutrinos) or any other
``enablers'' of renormalizable neutrino masses, such as Higgs
$SU(2)_L$ triplet states. I therefore assume that all lepton number
violation originates from new ultraviolet physics and that neutrino
masses are generated at some order in perturbation theory.

A $d$-dimensional operator ${\mathcal O}^d$ is suppressed by $d-4$
powers of a mass scale $\Lambda$ that characterizes the new physics,
in addition to a dimensionless coupling constant $\lambda$:
\begin{equation}
{\mathcal L}\in \sum_i\frac{\lambda_i{\mathcal
O}_i^d}{\Lambda^{d-4}},
\end{equation}
where I sum over all possible flavor combinations that make up the
same ``operator-type,'' as defined below. For each operator,
$\Lambda/\lambda$ is approximately the maximum energy scale below
which the new perturbative ultraviolet physics is guaranteed to
reside, and $\Lambda$ is used as a hard momentum cutoff in the
effective field theory. Among all $d$-dimensional operators, I
define $\Lambda$ so that the largest dimensionless coupling
$\lambda$ is equal to unity. Unless otherwise noted, I will assume
that all other $\lambda$ are of order one.

In the first two columns of Table \ref{tab:AllOps}, I exhaustively
enumerate all possible lepton number violating operators of mass
dimension less than or equal to eleven. All together, this amounts
to 129 different types of operators, most of which, 101 to be exact,
are of dimension eleven and consist of six fermion and two Higgs
fields.  Remaining are 21, 6 and 1 operator of dimension nine,
seven, and five, respectively.  The dimension-nine operators can be
of two different kinds, as defined by their respective field
content.  They either contain four fermion and three Higgs fields or
simply six fermion fields with no
 Higgs field content.  For consistency, I use the notation of
reference \cite{Operators}, where such a listing was first
introduced. Our operator naming scheme is also derived from the same
list, which I trivially extend to include $21$ elements only
mentioned in that analysis. These are the dimension-nine
 and dimension-eleven LNV operators that can be constructed from the ``product''
of the previously listed dimension-five and dimension-seven
operators with the SM Yukawa interactions. I individually identify
those operators with the same field content but different $SU(2)_L$
gauge structure with an additional roman character subscript added
onto the original designation from \cite{Operators}.  This is done
in order to render our discussion of the various operators clearer,
since specific gauge structures can play an important role in the
derived energy scale and predictions of a given operator. Note that
I neglect effective operators that contain SM gauge fields, since,
as argued in \cite{Operators}, these are not typically generated by
renormalizable models of new physics.

Our notation is as follows. \begin{equation}  L = \left(
                                                            \begin{array}{c}
                                                              \nu_L \\
                                                              e_L \\
                                                            \end{array}
                                                          \right)~~~
                                                         {\rm and}~~~
                                                         Q=\left(
                                                            \begin{array}{c}
                                                              u_L \\
                                                              d_L \\
                                                            \end{array}
                                                          \right)
                                                          \end{equation}
are the left-handed lepton and quark $SU(2)_L$ doublets,
respectively. $e^c$, $u^c$ and $d^c$ are the charge-conjugate of the
$SU(2)_L$ singlet right-handed charged lepton and quark fermion
operators, respectivelty. Conjugate fields are denoted with the
usual ``bar'' notation ($\bar{L}$, $\bar{Q}$, $\bar{e}^c$). For
simplicity, I am omitting flavor indices, but it is understood that
each matter fermion field comes in three flavors. All matter fields
defined above are to be understood as flavor eigenstates: all SM
gauge interactions, including those of the $W$-boson, are diagonal.
Without loss of generality, I will also define the $L$ and $e^c$
fields so that the charged-lepton Yukawa interactions are
flavor-diagonal.

I take the $SU(2)_L$ doublet Higgs scalar to be
\begin{equation}H = \left(
                                                            \begin{array}{c}
                                                              H^+ \\
                                                              H^0 \\
                                                            \end{array}
                                                          \right),
                                                          \end{equation}
and assume that, after electroweak symmetry breaking, its neutral
component acquires a vacuum expectation value (vev) of magnitude $v
\approx 0.174$~TeV,\footnote{Our numerical value for $v$ is distinct
from many treatments of the SM where $v$ is taken to be
$0.246~\rm{TeV}$.  These are equivalent up to a factor of $\sqrt{2}$
and are both valid provided a consistent treatment of the
interaction Lagrangian.} thus spontaneously breaking the electroweak
gauge symmetry $SU(2)_L \times U(1)_Y \rightarrow U(1)_{\rm em}$. In
Table~\ref{tab:AllOps}, the components of the $SU(2)_L$ doublets are
explicitly listed and labeled with $i,j,k,\ldots=1,2$. In order to
form gauge singlets, operators are contracted either by the
antisymmetric tensor $\epsilon_{ij}$, defined such that
$\epsilon_{12} = 1$, or by trivial contractions with a conjugate
doublet field.  Different gauge contractions are partially
responsible for the wide variety of operator structures encountered
in this study.

In order to avoid unnecessarily messy expressions, several features
are missing from the operators as listed in Table \ref{tab:AllOps}.
To begin, $SU(3)_c$ color indices are suppressed in these
expressions.
 Color contractions are only implied here because $SU(3)_c$ is an unbroken
symmetry of the SM and hence there is no sense in distinguishing the
various quark field components.  I assume that the parent
ultraviolet completion to each operator treats the color gauge
symmetry properly by introducing appropriately chosen heavy colored
particles to render the theory gauge invariant.  Slightly more
serious is the omission of flavor indices to label the fermion
generations.  For most of this analysis, I assume that all new
physics effects are generation universal and thus, flavor
independent.  This is not guaranteed to be the case, as is painfully
obvious within the SM.  One will also note that, depending on the
$SU(2)$ structure of the effective operator, different
flavor-dependent coefficients will be strictly related. For example,
including flavor dependent couplings $\lambda^1_{\alpha\beta}$,
${\mathcal O}_1$ should read
$\lambda^1_{\alpha\beta}L_{\alpha}^iL_{\beta}^jH^kH^l\epsilon_{ik}\epsilon_{jl}$,
where $\lambda^1_{\alpha\beta}=\lambda^1_{\beta\alpha}$ (symmetric)
for all $\alpha,\beta=e,\mu,\tau$. On the other hand, ${\mathcal
O}_{3a}$ should read (for fixed $Q$ and $d^c$ flavors)
$\lambda^{3_a}_{\alpha\beta}L_{\alpha}^iL_{\beta}^jQ^kd^cH^l\epsilon_{ij}\epsilon_{kl}$,
where $\lambda^{3_a}_{\alpha\beta}=-\lambda^{3_a}_{\beta\alpha}$
(antisymmetric) for all $\alpha,\beta=e,\mu,\tau$. Large differences
among the various flavor structures of each operator may very well
exist. Flavor is an important facet of LNV phenomenology, and is
addressed where relevant within the text.

The final feature missing from our notation is explicit Lorentz
structure. Each operator must, of course, form a Lorentz scalar, but
there are numerous field configurations that can bring this about.
The Higgs field is a scalar, and as such, transforms trivially under
the Lorentz group and is thus of no relevance to this discussion.
The fermions, however, transform non-trivially and their
contractions must be accounted for in each operator.  Simple
combinatorics dictate that there are at most $45$ such possibilities
for the six-fermion operators that comprise the bulk of our sample,
3 in the four-fermion case and only $1$ for the lone dimension five
operator. Additionally, each contraction can be made in a variety of
ways, corresponding to the bilinear Dirac operators $\mathbf{1}$,
$\gamma^\mu$ and $\sigma^{\mu\nu} =
\frac{i}{2}[\gamma^\mu,\gamma^\nu]$ of the scalar, vector and tensor
types, respectively.  Since I am dealing with chiral fields, the
addition of the $\gamma_5$ matrix to form the pseudoscalar and
axial-vector bilinears is redundant. While this helps reduce the
number of possibilities, the task of listing, categorizing, and
analyzing all possible Lorentz structures for each operator is still
quite overwhelming and is not undertaken in this general survey.
Fortunately, different Lorentz structures for the same operator-type
lead to the same predictions up to order one effects. This is
especially true for the ``interesting'' operators characterized by
TeV $\Lambda$ scales. I shall quantify this statement and mention
specific structures when relevant.  That being said, the Lorentz
structure of an effective operator can suggest a lot of information
about its parent renormalizable model. For example, it can suggest
the spin of the heavy intermediate states and the forms of various
vertices.

Armed with these operators, I can calculate the amplitude of any
$\Delta L=2$ LNV process. It is important to emphasize that when
addressing the phenomenological consequences of any particular
operator ${\mathcal O}$, I assume that it characterizes the dominant
tree-level effect of the new heavy physics, and that all other
effects -- also characterized by other LNV effective operators of
lower mass dimension -- occur at higher orders in perturbation
theory. Our approach is purely diagrammatic, in that I begin with an
operator-defined vertex and then proceed to close  loops and add SM
interactions as needed to yield the correct external state
particles.  In this sense, special care must be taken to respect the
chiral structure as defined by each operator.  In order to reach the
intended external states, to couple to particular gauge bosons, or
to close fermion loops, one must often induce a helicity flip with a
SM mass insertion.  I express these inserted fermion masses in terms
of the respective Yukawa couplings, $y_f$ ($f=\ell,u,d$) and the
Higgs vev, $v$. The Higgs field can be incorporated into this
procedure in a number of ways. I treat the two charged and single
neutral Nambu-Goldstone Higgs bosons, $H^\pm$ and $H^0$,
respectively, within the Feynman-'t Hooft gauge as propagating
degrees of freedom with electroweak scale masses. The physical
neutral Higgs, $h_0$, can be either chosen to propagate as a virtual
intermediate state, or couple to the vacuum with amplitude $v$.

In order to avoid the task of explicitly evaluating a huge number of
multiloop Feynnman diagrams, I succumb to approximate LNV amplitudes
based on reasonable assumptions and well-motivated rules. Our
methodology is motivated by exact computations with one-loop,
dimension 7 operators where the work is analytically tractable, as
well as on general theoretical grounds.  For select operators, I
have also checked our assumptions against predictions from
ultraviolet complete models with success.  In order to perform a
particular calculation, I draw the appropriate diagram(s), taking
care that no momentum loop integral vanishes by symmetry reasons.
This step is potentially quite involved, as multiple diagrams can
give sizable amplitude contributions depending on the characteristic
energy transfer in the system, not to mention the cumbersome Dirac
algebra within the respective loops.  Given the high, often
super-TeV, mass scale associated with our calculations, it is often
convenient to work in the gauge field basis where each boson state
is associated with a single SM group generator, as is natural before
electroweak symmetry breaking.  In a similar sense, all fermions,
including those of the third generation, are taken to be massless to
zeroth order. All masses are included perturbatively where needed
via mass insertions. At first guess, it would seem that our results
are only valid in the rather subjective limit $\Lambda \gg v$. By
direct comparison with other more complete approximations, however,
I find that our predictions are very reasonable at all scales above
$0.5~\rm{TeV}$. Keeping all of this in mind, I apply the following
``rules'' to obtain
 approximate amplitude expressions.

\begin{enumerate}
\item
\textit{Trivial numerical factors}:  A number of numerical factors
can be read off trivially from the Feynman diagrams.  Specifically,
one can extract the presence of the suppression scale
$\Lambda^{-(d-4)}$ directly from the dimension $d$ operator, as well
as the dimensionless coupling constants $\lambda$. Generally,
$\lambda$ is a generation dependent quantity, but for lack of any
experimental evidence to the contrary, I take $\lambda = 1$
universally unless stated otherwise. In  the case of scenarios
already constrained by current data, I will relax this assumption to
``save'' the operator and comment on the phenomenological
consequences of the change.

Furthermore, various factors of the electroweak scale $v$ may be
extracted from the operator's Higgs field content, in addition to
fermion/gauge boson mass terms.  In this way, I may also include the
various Yukawa and gauge coupling factors $y_f$ ($f=\ell,u,d$) and
$g_i$, respectively, where $i$ runs over the three SM gauge groups.
For simplicity, I neglect the gauge subscript $i$ in further
analytic expressions. Finally, a color factor of $3$ associated with
each quark loop should also be included in our computations, but can
(and will) be neglected for simplicity from algebraic expressions
where order one factors are irrelevant and only serve to render
expressions more cumbersome. I note that all coupling constants are
subject to renormalization group running.  In particular, those
occurring within a loop should be evaluated at the scale $\Lambda$.
I neglect this order one effect since it is most important at large
$\Lambda$ scales where operators tend to have less of a
phenomenological impact due to the $(1/\Lambda)^n$ suppression.
\item
\textit{Loop factors}:  In all of our calculations, I assume that
each operator defines an effective field theory, characterized by
the scale $\Lambda$. This implies that all momentum integrals are
effectively cut off at $\Lambda$, above which new states will emerge
to regularize the theory. Divergences in such loops tend to cancel
the large scale suppressions inherent to the bare operators, and
thus enhance predicted LNV rates.  Specific divergences can be
determined by simple power counting of momentum factors.  Of course,
multiple loop integrals are often convoluted to the point where
substantial simplification is needed to determine the dominant
divergent term.  Such a complication is in part due to the numerator
of the Dirac propagators, which include single momentum factors and
must therefore be present in pairs to contribute effectively to an
ultraviolet divergence.  The process of adding loops to induce
$\Lambda$ power law divergences should only be pursued to the point
where the suppression of the induced effective term is no less than
$\Lambda^{-1}$.  Any further divergent contribution must be treated
as a renormalization to lower order terms, and hence, can only add
small finite corrections to the total amplitude.  In any case, those
diagrams with the smallest scale suppressions are not always the
most dominant, as will become clear later when I discuss specific
results.

 In addition to power-law divergences, each
loop is also associated with a numerical suppression factor.  This
arises from the proper normalization of the loop four-momentum
integral as a factor of $(2\pi)^{-4}$, the characteristic phase
space ``volume'' of a quantum state.  It allows one to view the
integral as a coherent sum over all possible intermediate
configurations in a consistent way.  Partially evaluating these
integrals for a number of examples, one quickly finds that two
powers of $\pi$ cancel with the four dimensional Euclidian space
solid angle $\int d\Omega_4$.  I introduce a suppression factor of
$(16\pi^2)^{-1} \sim 0.0063$ for each diagram loop, which tends to
offset enhancements from associated divergent factors. A
quadratically divergent loop diagram is often proportional to the
lowest order contribution times $(1/16\pi^2)(\Lambda/v)^{2}$ to the
power $n$ (number of loops in the diagram). This contribution is
larger than the leading order one if
 $\Lambda > 4\pi v \sim
2~\rm{TeV}$ for any number of loops.  The situation is often more
involved, as many loops turn out to be logarithmically divergent or
even convergent.  The important conclusion is that adding loops is
not an efficient way to enhance LNV rates at the low scales
accessible to future experiments.  This fact is demonstrated by
example in Sec.~\ref{sec:constraints}.

Finally, as already alluded to, many diagram loops will exhibit
logarithmic divergences, as is the standard case in renormalizable
theories involving fermion and vector fields. This occurrence
typically reflects the differences between the two characteristic
scales inherent to the system, namely $\Lambda$ and $v$, and are of
the general form $\xi\log^n(\Lambda/v)$ for some power $n$. $\xi$ is
a small, loop suppressed, dimensionless coupling coefficient.
Numerically, these logarithms are much softer than their
quadratically divergent counterparts seen elsewhere in the diagrams
and can safely be neglected.
\item
\textit{Intermediate states}:  I treat all virtual intermediate
states, outside of loops, as if they carry the characteristic
momentum of the interaction $Q$ and neglect Dirac structure, unless
stated otherwise.  In particular, goldstone bosons are assigned the
propagator $(Q^2 - M_g^2)^{-1}$ and fermions are assigned $(Q -
M_f)^{-1}$.  In the case of an intermediate neutrino, this reduces
to a simple factor of $Q^{-1}$ for all realistic $Q$ values.  Hence,
for very low energy processes ($Q\ll 100$~MeV), neutrino exchange
diagrams tend to dominate LNV rates.
\item
\textit{Lorentz structure}:  For the purposes of our analysis, I
assume that all Lorentz contractions between fermions are
scalar-like.  As previously mentioned, the absolute magnitude of
most LNV amplitudes is robust under this assumption up to order one
factors.  The only qualitative exception to this occurs in some
cases involving fermion bilinear terms with a tensor Lorentz
structure ($\bar{\psi}^{\prime}\sigma_{\mu\nu}\psi$). This factor,
when coupled between two fermions contracted in a loop, will yield a
vanishing rate due to its antisymmetry inside of a trace, since
$Tr(\sigma_{\mu\nu})=0$.  This can be bypassed by introducing a new
momentum vector into the trace, implying the addition of another
loop.  In most cases, this  is most efficiently accomplished with a
new gauge boson line, which is accompanied by a logarithmic
divergence.  The combination of both factors leads to a marginal
amplitude suppression (with respect to the same operator where all
fermion bilinears are Lorentz scalars) for all energies of interest.
\end{enumerate}

With these approximations in hand, it is a simple matter to estimate
the amplitude associated to any given diagram.  Still, one must
wonder about the uncertainty induced onto the calculations by such
varied assumptions.  Can results obtained by such methods supply
valid physical predictions?  The answer, of course, depends on the
question that is being asked. Here, I will only be interested in
estimating order of magnitude effects, including what value of
$\Lambda$ is required in order to explain the observed neutrino
masses and, once $\Lambda$ is so constrained, what is the order of
magnitude of other related observable effects.

One may wonder whether a more detailed estimate of the effects of
each individual operator would lead to more reliable results. The
answer is negative. It is easy to show that different renormalizable
theories that lead to the same effective operator at tree-level will
mediate different processes at the loop-level with order one
different relative strengths. Furthermore, the derived cutoff scales
inherit the uncertainty from the absolute value of the heaviest
neutrino mass, which is only loosely bounded between $0.05~\rm{eV}$
and $1~\rm{eV}$ by the extracted atmospheric mass squared difference
\cite{OscBestFit} and tritium beta decay kinematic measurements
\cite{Mainz,Troitsk}. This is an order of magnitude uncertainty that
cannot be avoided even if one were to perform a detailed computation
within a well-defined ultraviolet complete theory.

In summary, given all approximations and uncertainties, our results
are only valid up to $\pm$ an order of magnitude. In this spirit,
one need not explicitly consider order one factors that will
necessarily yield negligible corrections by these standards. Such a
large error tolerance supplies the need for care when interpreting
results. In particular, one should not place too much emphasis on
any one bound or prediction, unless it is very robust, {\it i.e.},
able to withstand variations of at least a factor of ten.  Of
course, for those operators constrained by several different
independent sources one can, and should, take more marginal results
seriously.

\subsection{Neutrino Masses and the Scale of New Physics}
\label{subsec:NuMass}

Having defined the  set of LNV operators, I now extract the scale of
new physics from the direct comparison of radiatively generated
neutrino mass expressions to their observed values. Since there are
three light neutrino masses, I will use the heaviest of these to set
the overall mass scale. Neutrino oscillation data, currently
providing the only evidence for neutrino masses, constrain the
relative magnitudes of the mass eigenstates but not the overall
scale \cite{NeutrinoReview}. Such data only supply a lower bound on
the heaviest neutrino mass, derived from the largest observed mass
squared difference $\Delta m^2_{13} = |m_3^2-m_1^2| \approx
0.0025~\rm{eV}^2$, the atmospheric mass squared difference
\cite{OscBestFit}.  At least one neutrino mass must be greater than
$\sqrt{\Delta m^2_{13}}\approx 0.05$~eV. Neutrino oscillations also
teach us that the next-to-heaviest neutrino weighs at least
$\sqrt{\Delta m^2_{12}}\approx 0.009$~eV (the solar mass-squared
difference), in such a way that the ratio of the heaviest to the
next-to-heaviest neutrino masses is guaranteed to be larger than,
approximately, 0.2. No lower bounds can be placed on the lightest
neutrino mass. An upper bound on the heaviest neutrino mass is
provided by several non-oscillation neutrino probes. Cosmology
provides interesting constraints on the sum of light neutrino
masses, but these are quite dependent on unconfirmed details of the
thermal history of the universe and its composition \cite{CosmoSum,Fogli:2006yq,seljak,Hannestad:2006mi}.
Most direct are kinematic measurements of the tritium beta decay
electron endpoint spectrum \cite{TritNuEffMass}.  Both types of
probes provide upper bounds near $1~\rm{eV}$, likely to improve in
coming years. I choose to perform our calculations assuming the mass
scale $m_\nu \approx 0.05$~eV, corresponding to the experimental
lower bound.  In this way, each extracted operator scale $\Lambda$,
inversely related to the neutrino mass, represents a loose upper
bound. Since most rates for LNV observables are proportional to some
inverse power of $\Lambda$, this choice implies the added
interpretation that, all else remaining equal, our results for such
rates should conservatively reflect lower limit predictions.

\begin{figure}
\begin{center}
\includegraphics[scale=1]{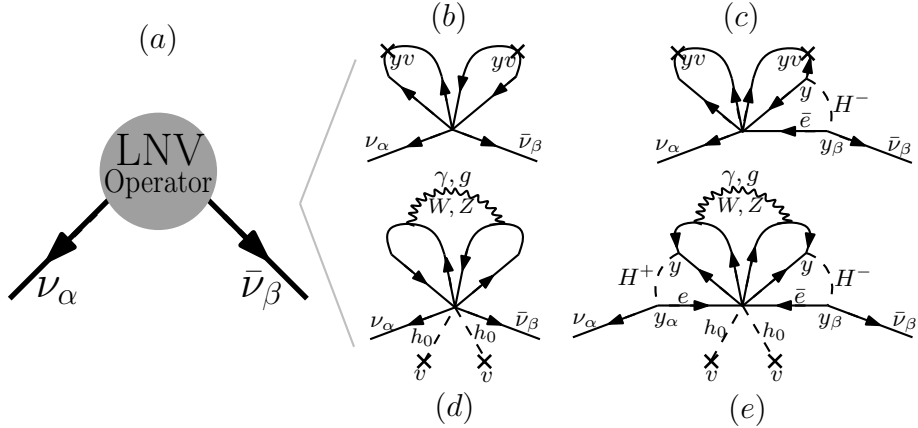}
\caption[Sample diagrams that radiatively generate Majorana neutrino
masses]{Sample diagrams that radiatively generate Majorana neutrino
masses. Diagram $(a)$ is representative of all operators that can
generate the needed external state neutrinos. This usually proceeds
via loop contractions and other couplings, hidden within the light
gray region. Diagrams $(b)-(e)$ help illustrate the methodology of
this analysis. Despite obvious differences, all of these generate
effective dimension-five interactions of calculable strength. See
the text for more details.} \label{fig:MassDiagrams}
\end{center}
\end{figure}

LNV  neutrino masses are nothing more than self-energy diagrams
evaluated at vanishing momentum transfer.  These must couple
together the left-handed neutrino state $\nu_\alpha$ with the
right-handed anti-neutrino state $\nu_\beta$, as shown schematically
in diagram $(a)$ of Fig.~\ref{fig:MassDiagrams}.  Here the flavor
indices $\alpha$ and $\beta$ can accommodate any of the three lepton
flavors ($\alpha,\beta=e,\mu,\tau$).  The derived Majorana masses
$m_{\alpha\beta}=m_{\beta\alpha}$ are generally complex.
 The large grey circle in this diagram represents all
 possible contributions to the neutrino mass.  Specifically, it
 contains the underlying $\Delta L=2$ operator along with all
 modifications needed to yield the correct external state structure.
  This includes such objects as loops, additional gauge boson propagators and SM coupling
  constants.  Generally, several diagrams can contribute to this
  mass generation, but special care must be taken that these are not
  proportional to any positive power of $Q$, the momentum carried by the
  neutrino legs, as this would not lead to a nonzero rest mass
  correction.

Diagrams $(b)-(e)$ of Fig.~\ref{fig:MassDiagrams} are examples that
serve to illustrate some typical features encountered in our
effective operator induced self-energy calculations.  The underlying
LNV operators shown in each diagram contain six fermion fields and
are therefore of dimension nine or eleven, as are the majority of
the analyzed operators. Each of these diagrams generates an
effective dimension-five interaction
\begin{equation}
{\mathcal L}_5= \xi_{\alpha\beta}^{(5)}
\frac{(L^{\alpha}H)(L^{\beta}H)}{\Lambda}, \label{eq:Gen5}
\end{equation}
where $\xi_{\alpha\beta}^{(5)}=\xi_{\beta\alpha}^{(5)}$ is a
generation dependent coupling constant that is calculable, given the
structure of the original operator.  This is  easily verifiable via
direct power counting, despite differences in dimension, loop
number, field content, and helicity structure.  It turns out that
most operators, especially those characterized by super-TeV scales,
possess this property.

I describe each sample diagram in turn to point out important
features. A subset of the subtleties described below is encountered
when estimating the neutrino masses $m_{\alpha\beta}$ for the entire
effective operator set.
 Diagram $(b)$ is a simple two-loop
radiatively generated mass term proceeding from dimension-nine
operators, such as
$\mathcal{O}_{11_b}=L^iL^jQ^kd^cQ^ld^c\epsilon_{ik}\epsilon_{jl}$,
containing the fermion structure $f_L f_R^c f_L^\prime f_R^{\prime
c}$, where the fields $f$ and $f^\prime$ are contracted into loops
with mass insertions that supply the needed field chirality flip.
Masses arising from such operators are proportional to two powers of
fermion Yukawa couplings.  Strictly speaking, allowed fermions from
all three generations traverse the closed loops and contribute to
the mass. However, assuming universal new physics coupling
constants, third generation fermions will strongly dominate the
induced neutrino mass.  In cases such as these, one can freely
suppress couplings to the lighter two generations without modifying
the expected value of $\Lambda$. Since diagrams arising from the
majority of our operator set contain at least one loop of this kind,
this property proves quite useful when attempting to avoid low
energy nuclear physics constraints, as will be discussed in more
detail later.

Diagram $(d)$ involves an operator of dimension eleven, such as \newline
$\mathcal{O}_{22}=L^iL^jL^ke^c \bar{L}_k
\bar{e^c}H^lH^m\epsilon_{il}\epsilon_{jm}$, but has a similar
structure to Diagram $(b)$ since both neutral Higgs fields $h_0$
couple to the vacuum, yielding a $v^2$ factor. In this case, the
parent underlying operators contain the fermion structure $f_{L(R)}
f_{L(R)}^c f_{L(R)}^\prime f_{L(R)}^{\prime c}$, or simple variants
thereof. From this, it is clear that such operators will create and
annihilate the \emph{same} field and one can close the fermion loops
without mass insertions. A little thought reveals that such loops,
if left on their own, will vanish by symmetry, since $\int d^4k
[k^\mu/g(k^2)] = 0$ for all functions $g(k^2)$. Hence, non-zero
neutrino masses only appear at a higher order in perturbation theory
({\it i.e.}, I need to add another loop).  To maintain the chiral
structure of the diagram a gauge boson line insertion is always the
most effective. The specific gauge field required in this step
depends critically on the quantum numbers of the fermions $f$ and
$f^\prime$ contained in the operator itself and must be determined
on a case-by-case basis. The absence of Yukawa dependence renders
the estimated value of the cutoff scale $\Lambda$ insensitive to the
values of the dimensionless operator couplings ($\lambda$), given
the way $\Lambda$ is defined.
%In
%fact, the extracted scale $\Lambda$ is invariant to these
%modifications, provided that at least one coupling $\lambda$ is
%order one, a property that is already garenteed by our methodology
%of scale extraction.
% since, given the array of coupling constants
%$\lambda_{\alpha\beta...\zeta}$, one may normalize to its maximum
%value by scaling $\Lambda$ by the appropriate amount
Notice that this three-loop diagram, like Diagram $(b)$, predicts an
anarchic neutrino Majorana mass matrix, currently allowed by the
neutrino oscillation data \cite{anarchy}. That is, up to order one
corrections, all entries are of the same magnitude, $m_\nu \approx
0.05$~eV.  This is in contrast to the remaining sample diagrams
($(c)$ and $(e)$), which both suggest flavor-structured mass
matrices.

Dimension-nine operators, such as
$\mathcal{O}_{19}=L^iQ^jd^cd^c\bar{e^c}\bar{u^c}\epsilon_{ij}$,
yielding diagram $(c)$ have the peculiar property that, upon
expanding out the various $SU(2)_L$ contractions in terms of
component fields, no $\nu_\alpha\nu_\beta$ content is present to
form the external legs of a mass diagram. Here the LNV is introduced
via the fermion structure $\nu_\alpha e_{R\beta}$, which annihilates
a left-handed neutrino and creates a left-handed positron.  Hence,
to tie in the needed antineutrino line, one must both flip the
charge lepton helicity and carry away the excess charge with some
bosonic state. Of course, such a charged boson is guaranteed by
charge conservation to be needed elsewhere in the system to close
some $f/f^\prime$ loop.  In this particular example, the process is
illustrated by the exchange of a charged Higgs goldstone boson $H^-$
to clearly visualize the chiral structure, but of course, one may
equivalently think in terms of a transverse $W^-$ boson. The crucial
point is that this mass is necessarily proportional to a charged
lepton Yukawa coupling $y_{\ell_\beta}$, of the \emph{same} flavor
as the external neutrino, since I am working in the weak eigenbasis
where the gauge couplings and the charged-lepton Yukawa couplings
are flavor diagonal.  By symmetry,
%this diagram, as well as its
%charged conjugate counterpart,
the contribution to the $m_{\alpha\beta}$ entry of the neutrino mass
matrix
%generation
%, implying that any particular mass matrix element
is proportional to $y_{\ell_\alpha} + y_{\ell_\beta}$, which reduces
to the largest coupling $y_{\ell_\beta}$ in the realistic case of
hierarchial charged lepton Yukawa couplings.
%In this case, the
%specific structure of the loops is not important, as their
%properties were explained in the previous examples.

Finally, Diagram $(e)$ yields a five-loop suppressed neutrino
self-energy originating from a dimension-eleven LNV operator such as
$\mathcal{O}_{36} =
\bar{e^c}\bar{e^c}Q^id^cQ^jd^cH^kH^l\epsilon_{ik}\epsilon_{jl}$.
This represents the most complicated structure considered in this
analysis.  As in Diagram $(c)$, no explicit $\nu_\alpha\nu_\beta$
structure is available in the underlying operator, but in this case
all of the LNV arises from $e_{R\alpha}^c e_{R\beta}^c$-type
interactions. Curiously, this interaction already flips helicity as
it annihilates a left-handed positron and creates a right-handed
electron. Unfortunately, being an $SU(2)_L$ singlet, $e_R$ only
couples to the neutrino via a charged Higgs induced Yukawa
interaction; therefore, this amplitude must be proportional to the
product $y_{\ell_\alpha}y_{\ell_\beta}$ to yield a legitimate
neutrino mass contribution.  One might imagine that the Higgs fields
contained in the LNV operator could be used to produce the needed
neutrino legs, but this is not possible since the resulting loop
would have a structure of the form $\int d^4k \frac{k^\mu +
Q^\mu}{g(k^2)} \propto Q^\mu$ which vanishes in the ``rest mass''
limit.  It is clear that both Higgs fields must again couple to the
vacuum and the needed flip must come from the other fermion loops.
The resulting loop integrals can be separated into two convoluted
pieces corresponding to both loop/leg pairs.  A little thought
reveals that each loop set contains three fermion lines whose
associated integral is again proportional to the momentum of the
external neutrino, and thus is not a valid mass correction.  To fix
this last problem without further complicating the chiral structure,
one can add a gauge boson exchange between the fermion loops, as was
also done in Diagram $(d)$.

Despite the dominance of the generated dimension-five interactions
described by Eq.~(\ref{eq:Gen5}) for the majority of the studied LNV
operators, I find that this need not be the case for all of them.
For some operators, the dimension-five neutrino-mass effective
operator Eq.~(\ref{eq:Gen5}) occurs at higher order in perturbation
theory than the dimension-seven neutrino-mass effective operator
(schematically, $(LH)^2 H^2$).  For these, the neutrino mass matrix
is generated after electroweak symmetry breaking from
\begin{equation}
{\mathcal L}_{57} =
\frac{\xi_{\alpha\beta}^{(5)}}{16\pi^2}\frac{(LH)(LH)}{\Lambda} +
\xi_{\alpha\beta}^{(7)}\frac{(LH)(LH)H\bar{H}}{\Lambda^3},
\label{eq:Gen57}
\end{equation}
where $\xi_{\alpha\beta}^{(7)}$ are new  calculable coefficients.
This type of structure is present in the following operators
\begin{equation}
\mathcal{O}_7,\mathcal{O}_{21_{a,b}},\mathcal{O}_{22},\mathcal{O}_{23},\mathcal{O}_{25},\mathcal{O}_{26_b},\mathcal{O}_{27_b},\mathcal{O}_{29_a}\mathcal{O}_{30_b},\mathcal{O}_{31},\mathcal{O}_{44_c},\mathcal{O}_{57}.
\label{eq:ODim7Mass}
\end{equation}
In general, they are associated with dimension-eleven
operators\footnote{This also occurs with operator $\mathcal{O}_7$,
which is of dimension nine. This operator is the exception, in that
it explicitly contains three Higgs bosons which naturally aids in
building the needed $v^4$ factors in a way similar to that discussed
in the text.} whose mass diagrams are found trivially by connecting
the external fermion loops and coupling the neutral Higgs fields to
the vacuum.  This adds two factors of the electroweak scale to the
mass expressions. Dimensional analysis dictates that the fermion
loops must conspire to yield an addition factor of $v^2$, usually
from mass insertions utilized to flip helicities. For the
dimension-seven operators in Eq.~(\ref{eq:Gen57}), the resulting
neutrino mass expression is proportional to $v^4/\Lambda^3$.  If I
assume, as is usually the case, that most of the dimensionless
factors of Eq.~(\ref{eq:Gen57}) are common to both
$\xi_{\alpha\beta}^{(5)}$ and $\xi_{\alpha\beta}^{(7)}$, I find
$m_{\alpha\beta} \propto 1/16\pi^2 + v^2/\Lambda^2$.  In such cases,
the dimension-seven contribution is only relevant for operator
cutoff scales $\Lambda\lesssim 4\pi v \approx 2~\rm{TeV}$.  Such low
scales are seldom reached considering that these operators are
efficient at mass generation at low orders and consequently do not
possess the necessary suppression factors. Still, for completeness,
I include these terms when relevant.

\setlength{\LTcapwidth}{9in}

\begin{singlespace}
{\small
\begin{landscape}
\begin{longtable}{|l|c||c|c|c|c|}
\caption[Summary of dimension 5-11 LNV operators surveyed
here]{  Dimension-five through dimension-eleven LNV operators analyzed
in this survey.  The first two columns display the operator name and
field structure, respectively. Column three presents the induced
neutrino mass expressions, followed by the
inferred scale of new physics, $\Lambda_{\nu}$.  Column five lists favorable modes of experimental exploration. Column six describes an operator's current status according to the key U (Unconstrained), C (Constrained) and D (Disfavored).  See text for details. }
 \label{tab:AllOps}   \\ 
\endfirsthead
 \hline \\[-2.3ex]
  \multicolumn{1}{c|}{$\mathcal{O}$} &
  \multicolumn{1}{c||}{Operator} &
  \multicolumn{1}{c|}{$m_{\alpha\beta}$} &
  \multicolumn{1}{c|}{$\Lambda_\nu~\rm{(TeV)}$} &
  \multicolumn{1}{c|}{Best Probed} &
  \multicolumn{1}{c|}{Disfavored}
  \\[.1ex] \hline
  \\[-2ex]
\endhead
$1$ & $L^i L^j H^k H^l \epsilon_{ik} \epsilon_{jl}$ & $\frac{v^2}{\Lambda}$ & $6\times 10^{11}$ & $\beta\beta0\nu$ & U \\
$2$ & $L^i L^j L^k e^c H^l \epsilon_{ij} \epsilon_{kl}$ & $\frac{y_\ell }{16\pi^2}\frac{v^2}{\Lambda}$ & $4\times 10^7$ & $\beta\beta0\nu$ & U \\
$3_a$ & $L^i L^j Q^k d^c H^l \epsilon_{ij} \epsilon_{kl}$ & $\frac{y_d g^2}{(16\pi^2)^2}\frac{v^2}{\Lambda}$ & $2\times 10^5$ & $\beta\beta0\nu$ & U \\
$3_b$ & $L^i L^j Q^k d^c H^l \epsilon_{ik} \epsilon_{jl}$ & $\frac{y_d}{16\pi^2}\frac{v^2}{\Lambda}$ & $1\times 10^8$ &$\beta\beta0\nu$& U \\
$4_a$ & $L^i L^j \overline{Q}_i \bar{u^c} H^k \epsilon_{jk}$ & $\frac{y_u}{16\pi^2}\frac{v^2}{\Lambda}$ & $4\times 10^9$ &$\beta\beta0\nu$& U \\
$4_b$ & $L^i L^j \overline{Q}_k\bar{u^c}H^k \epsilon_{ij}$ & $\frac{y_u g^2}{(16\pi^2)^2}\frac{v^2}{\Lambda}$ & $6\times 10^6$ &$\beta\beta0\nu$& U \\
$5$ & $L^i L^j Q^k d^c H^l H^m \overline{H}_i \epsilon_{jl}\epsilon_{km}$ & $\frac{y_d}{(16\pi^2)^2}\frac{v^2}{\Lambda}$ & $6\times 10^5$ &$\beta\beta0\nu$& U \\
$6$ & $L^i L^j \overline{Q}_k\bar{u^c}H^l H^k \overline{H}_i\epsilon_{jl}$ & $\frac{y_u}{(16\pi^2)^2}\frac{v^2}{\Lambda}$ & $2\times 10^7$ &$\beta\beta0\nu$& U \\
$7$ & $L^iQ^j \bar{e^c}\overline{Q}_kH^k H^l H^m\epsilon_{il} \epsilon_{jm}$ & $y_{\ell _\beta}\frac{g^2}{(16\pi^2)^2}\frac{v^2}{\Lambda}\left(\frac{1}{16\pi^2} + \frac{v^2}{\Lambda^2}\right)$ & $4\times 10^2$ &mix& C \\
$8$ & $L^i \bar{e^c} \bar{u^c} d^c H^j \epsilon_{ij}$ & $y_{\ell _\beta}\frac{ y_d y_u}{(16\pi^2)^2}\frac{v^2}{\Lambda}$ & $6\times 10^3$  &mix& C \\
$9$ & $L^i L^j L^k e^c L^l e^c \epsilon_{ij}\epsilon_{kl}$ & $\frac{y_\ell ^2}{(16\pi^2)^2}\frac{v^2}{\Lambda}$ & $3\times 10^3$ &$\beta\beta0\nu$& U \\
$10$  & $L^i L^j L^k e^c Q^l d^c \epsilon_{ij}\epsilon_{kl}$ & $\frac{y_\ell y_d}{(16\pi^2)^2}\frac{v^2}{\Lambda}$ & $6\times 10^3$  &$\beta\beta0\nu$& U \\
$11_a$& $L^i L^j Q^k d^c Q^l d^c \epsilon_{ij} \epsilon_{kl}$ & $\frac{y_d^2 g^2}{(16\pi^2)^3}\frac{v^2}{\Lambda}$ & $30$ &$\beta\beta0\nu$& U \\
$11_b$& $L^i L^j Q^k d^c Q^l d^c \epsilon_{ik}\epsilon_{jl}$ & $\frac{y_d^2}{(16\pi^2)^2}\frac{v^2}{\Lambda}$ & $2\times 10^4$ &$\beta\beta0\nu$& U \\
$12_a$& $L^iL^j\overline{Q}_i\bar{u^c}\overline{Q_j}\bar{u^c}$ & $\frac{y_u^2}{(16\pi^2)^2}\frac{v^2}{\Lambda}$ & $2\times 10^7$ &$\beta\beta0\nu$& U \\
$12_b$& $L^iL^j\overline{Q}_k\bar{u^c}\overline{Q}_l\bar{u^c}\epsilon_{ij}\epsilon^{kl}$ & $\frac{y_u^2 g^2}{(16\pi^2)^3}\frac{v^2}{\Lambda}$ & $4\times 10^4$ &$\beta\beta0\nu$& U \\
$13$ & $L^i L^j \overline{Q}_i \bar{u^c}L^l e^c\epsilon_{jl}$ & $\frac{y_\ell y_u}{(16\pi^2)^2}\frac{v^2}{\Lambda}$ & $2\times 10^5$ &$\beta\beta0\nu$& U \\
$14_a$& $L^iL^j\overline{Q}_k\bar{u^c}Q^kd^c\epsilon_{ij}$ & $\frac{y_dy_ug^2}{(16\pi^2)^3}\frac{v^2}{\Lambda}$ & $1\times 10^3$ &$\beta\beta0\nu$& U \\
$14_b$& $L^i L^j \overline{Q}_i \bar{u^c}Q^ld^c\epsilon_{jl}$ & $\frac{y_dy_u}{(16\pi^2)^2}\frac{v^2}{\Lambda}$ & $6\times 10^5$ &$\beta\beta0\nu$& U \\
$15$ & $L^i L^j L^k d^c \overline{L}_i \bar{u^c}\epsilon_{jk}$ & $\frac{y_dy_ug^2}{(16\pi^2)^3}\frac{v^2}{\Lambda}$ & $1\times 10^3$ &$\beta\beta0\nu$& U \\
$16$ & $L^i L^j e^c d^c \bar{e^c} \bar{u^c}\epsilon_{ij}$ & $\frac{y_dy_ug^4}{(16\pi^2)^4}\frac{v^2}{\Lambda}$ & $2$ &$\beta\beta0\nu$, LHC& U \\
$17$ & $L^i L^j d^c d^c \bar{d^c} \bar{u^c}\epsilon_{ij}$ & $\frac{y_dy_ug^4}{(16\pi^2)^4}\frac{v^2}{\Lambda}$ & $2$ &$\beta\beta0\nu$, LHC& U \\
$18$ & $L^i L^j d^c u^c \bar{u^c} \bar{u^c}\epsilon_{ij}$ & $\frac{y_dy_ug^4}{(16\pi^2)^4}\frac{v^2}{\Lambda}$ & $2$ &$\beta\beta0\nu$, LHC& U \\
$19$ & $L^i Q^j d^c d^c \bar{e^c} \bar{u^c}\epsilon_{ij}$ & $y_{\ell _\beta}\frac{y_d^2y_u}{(16\pi^2)^3}\frac{v^2}{\Lambda}$ & $ 1 $ &$\beta\beta0\nu$, HElnv, LHC, mix & C \\
$20$ & $L^i d^c \overline{Q}_i \bar{u^c} \bar{e^c} \bar{u^c}$ &$y_{\ell _\beta}\frac{y_dy_u^2}{(16\pi^2)^3}\frac{v^2}{\Lambda}$ & $40$ &$\beta\beta0\nu$, mix & C \\
$21_a$& $L^iL^jL^ke^cQ^lu^cH^mH^n\epsilon_{ij}\epsilon_{km}\epsilon_{ln}$ & $\frac{y_\ell y_u}{(16\pi^2)^2}\frac{v^2}{\Lambda}\left(\frac{1}{16\pi^2} + \frac{v^2}{\Lambda^2}\right)$ & $2\times 10^3$ &$\beta\beta0\nu$& U \\
$21_b$& $L^i L^j L^k e^c Q^l u^c H^m H^n \epsilon_{il} \epsilon_{jm}\epsilon_{kn}$ & $\frac{y_\ell y_u}{(16\pi^2)^2}\frac{v^2}{\Lambda}\left(\frac{1}{16\pi^2} + \frac{v^2}{\Lambda^2}\right)$ & $2\times 10^3$ &$\beta\beta0\nu$& U \\
$22$ & $L^i L^j L^k e^c \overline{L}_k \bar{e^c}H^lH^m\epsilon_{il}\epsilon_{jm}$ & $\frac{g^2}{(16\pi^2)^3}\frac{v^2}{\Lambda}$ & $4\times 10^4$  & $\beta\beta0\nu$ & U \\
$23$ & $L^i L^jL^k e^c \overline{Q}_k\bar{d^c}H^lH^m \epsilon_{il} \epsilon_{jm}$ & $\frac{y_\ell y_d}{(16\pi^2)^2}\frac{v^2}{\Lambda}\left(\frac{1}{16\pi^2} + \frac{v^2}{\Lambda^2}\right)$ & $40$  & $\beta\beta0\nu$ & U \\
$24_a$& $L^i L^j Q^k d^c Q^l d^c H^m \overline{H}_i\epsilon_{jk} \epsilon_{lm}$ & $\frac{y_d^2}{(16\pi^2)^3}\frac{v^2}{\Lambda}$ & $1\times 10^2$  & $\beta\beta0\nu$ & U \\
$24_b$& $L^i L^j Q^k d^c Q^l d^cH^m \overline{H}_i \epsilon_{jm} \epsilon_{kl}$ & $\frac{y_d^2}{(16\pi^2)^3}\frac{v^2}{\Lambda}$ & $1\times 10^2$  & $\beta\beta0\nu$ & U \\
$25$ & $L^i L^j Q^k d^c Q^l u^c H^m H^n \epsilon_{im}\epsilon_{jn} \epsilon_{kl}$ & $\frac{y_dy_u}{(16\pi^2)^2}\frac{v^2}{\Lambda}\left(\frac{1}{16\pi^2} + \frac{v^2}{\Lambda^2}\right)$ & $4\times 10^3$  & $\beta\beta0\nu$ & U \\
$26_a$& $L^i L^j Q^k d^c \overline{L}_i \bar{e^c}H^l H^m \epsilon_{jl} \epsilon_{km}$ & $\frac{y_\ell y_d}{(16\pi^2)^3}\frac{v^2}{\Lambda}$ & $40$ & $\beta\beta0\nu$ & U \\
$26_b$& $L^i L^j Q^k d^c \overline{L}_k \bar{e^c} H^lH^m\epsilon_{il}\epsilon_{jm}$ & $\frac{y_\ell y_d}{(16\pi^2)^2}\frac{v^2}{\Lambda}\left(\frac{1}{16\pi^2} + \frac{v^2}{\Lambda^2}\right)$ & $40$ & $\beta\beta0\nu$ & U \\
$27_a$& $L^i L^j Q^k d^c \overline{Q}_i\bar{d^c} H^lH^m \epsilon_{jl} \epsilon_{km}$ & $\frac{g^2}{(16\pi^2)^3}\frac{v^2}{\Lambda}$ & $4\times 10^4$  & $\beta\beta0\nu$ & U \\
$27_b$& $L^i L^j Q^k d^c \overline{Q}_k\bar{d^c} H^l H^m \epsilon_{il} \epsilon_{jm}$ & $\frac{g^2}{(16\pi^2)^3}\frac{v^2}{\Lambda}$ & $4 \times 10^4$  & $\beta\beta0\nu$ & U \\
$28_a$& $L^i L^j Q^k d^c \overline{Q}_j \bar{u^c}H^l \overline{H}_i \epsilon_{kl}$ & $\frac{y_dy_u}{(16\pi^2)^3}\frac{v^2}{\Lambda}$ & $4\times 10^3$  & $\beta\beta0\nu$ & U \\
$28_b$& $L^i L^j Q^k d^c \overline{Q}_k\bar{u^c} H^l \overline{H}_i \epsilon_{jl}$ & $\frac{y_dy_u}{(16\pi^2)^3}\frac{v^2}{\Lambda}$ & $4\times 10^3$  & $\beta\beta0\nu$ & U \\
$28_c$& $L^i L^j Q^k d^c \overline{Q}_l \bar{u^c} H^l\overline{H}_i\epsilon_{jk}$ & $\frac{y_dy_u}{(16\pi^2)^3}\frac{v^2}{\Lambda}$ & $4\times 10^3$  & $\beta\beta0\nu$ & U \\
$29_a$& $L^i L^j Q^k u^c \overline{Q}_k \bar{u^c}H^l H^m \epsilon_{il} \epsilon_{jm}$ & $\frac{y_u^2}{(16\pi^2)^2}\frac{v^2}{\Lambda}\left(\frac{1}{16\pi^2} + \frac{v^2}{\Lambda^2}\right)$ & $2\times 10^5$  & $\beta\beta0\nu$ & U \\
$29_b$& $L^i L^j Q^k u^c \overline{Q}_l \bar{u^c} H^l H^m\epsilon_{ik} \epsilon_{jm}$ & $\frac{g^2}{(16\pi^2)^3}\frac{v^2}{\Lambda}$ & $4\times 10^4$  & $\beta\beta0\nu$ & U \\
$30_a$& $L^i L^j \overline{L}_i \bar{e^c}\overline{Q}_k\bar{u^c} H^k H^l \epsilon_{jl}$ & $\frac{y_\ell y_u}{(16\pi^2)^3}\frac{v^2}{\Lambda}$ & $2\times 10^3$  & $\beta\beta0\nu$ & U \\
$30_b$& $L^i L^j\overline{L}_m \bar{e^c} \overline{Q}_n \bar{u^c} H^k H^l\epsilon_{ik} \epsilon_{jl} \epsilon^{mn}$ & $\frac{y_\ell y_u}{(16\pi^2)^2}\frac{v^2}{\Lambda}\left(\frac{1}{16\pi^2} + \frac{v^2}{\Lambda^2}\right)$ & $2\times 10^3$  & $\beta\beta0\nu$ & U \\
$31_a$& $L^i L^j \overline{Q}_i\bar{d^c}\overline{Q}_k\bar{u^c} H^k H^l \epsilon_{jl}$ & $\frac{y_dy_u}{(16\pi^2)^2}\frac{v^2}{\Lambda}\left(\frac{1}{16\pi^2} + \frac{v^2}{\Lambda^2}\right)$ & $4\times 10^3$  & $\beta\beta0\nu$ & U \\
$31_b$& $L^i L^j \overline{Q}_m\bar{d^c} \overline{Q}_n\bar{u^c}H^k H^l\epsilon_{ik} \epsilon_{jl} \epsilon^{mn}$ & $\frac{y_dy_u}{(16\pi^2)^2}\frac{v^2}{\Lambda}\left(\frac{1}{16\pi^2} + \frac{v^2}{\Lambda^2}\right)$  & $4\times 10^3$  & $\beta\beta0\nu$ & U \\
$32_a$& $L^i L^j \overline{Q}_j \bar{u^c}\overline{Q}_k \bar{u^c} H^k \overline{H}_i$ & $\frac{y_u^2}{(16\pi^2)^3}\frac{v^2}{\Lambda}$ & $2\times 10^5$ & $\beta\beta0\nu$ & U \\
$32_b$& $L^i L^j\overline{Q}_m \bar{u^c} \overline{Q}_n \bar{u^c} H^k\overline{H}_i \epsilon_{jk} \epsilon^{mn}$ & $\frac{y_u^2}{(16\pi^2)^3}\frac{v^2}{\Lambda}$ & $2\times 10^5$ & $\beta\beta0\nu$ & U \\
$33$ & $\bar{e^c} \bar{e^c} L^i L^j e^c e^c H^kH^l \epsilon_{ik} \epsilon_{jl}$ & $\frac{g^2}{(16\pi^2)^3}\frac{v^2}{\Lambda}$ & $4\times 10^4$  & $\beta\beta0\nu$ & U \\
$34$ & $\bar{e^c} \bar{e^c} L^i Q^j e^c d^c H^kH^l \epsilon_{ik} \epsilon_{jl}$ & $y_{\ell _\beta}\frac{y_dg^2}{(16\pi^2)^4}\frac{v^2}{\Lambda}$ & $< 0.5$  & $\beta\beta0\nu$, mix, ILC, LHC & C \\
$35$ & $\bar{e^c} \bar{e^c} L^i e^c\overline{Q}_j\bar{u^c} H^jH^k\epsilon_{ik}$ & $y_{\ell _\beta}\frac{y_ug^2}{(16\pi^2)^4}\frac{v^2}{\Lambda}$ & $2$ & mix, LHC & C \\
$36$ & $\bar{e^c} \bar{e^c} Q^i d^c Q^j d^c H^k H^l \epsilon_{ik} \epsilon_{jl}$ & $y_{\ell _\alpha}y_{\ell _\beta}\frac{y_d^2g^2}{(16\pi^2)^5}\frac{v^2}{\Lambda}$ & $< 0.5$ & $\beta\beta0\nu$, mix, HElnv, ILC, LHC & D \\
$37$ & $\bar{e^c} \bar{e^c} Q^i d^c\overline{Q}_j\bar{u^c} H^j H^k \epsilon_{ik}$ & $y_{\ell _\alpha}y_{\ell _\beta}\frac{y_dy_ug^2}{(16\pi^2)^5}\frac{v^2}{\Lambda}$ & $< 0.5$  & $\beta\beta0\nu$, mix, HElnv, ILC, LHC & D \\
$38$ & $\bar{e^c} \bar{e^c} \overline{Q}_i\bar{u^c}\overline{Q}_j \bar{u^c} H^i H^j$ & $y_{\ell _\alpha}y_{\ell _\beta}\frac{y_u^2g^2}{(16\pi^2)^5}\frac{v^2}{\Lambda}$ & $< 0.5$ & $\beta\beta0\nu$, mix, HElnv, ILC, LHC & D \\
$39_a$& $L^i L^j L^k L^l \overline{L}_i \overline{L}_jH^m H^n \epsilon_{km} \epsilon_{ln}$\footnote{This operator is modified slightly from its original form as given in reference \cite{Operators} where it appeared as $\mathcal{O}_{39}(a) = L^iL^jL^kL^l\overline{L}_i\overline{L}_jH^mH^n\epsilon_{jm}\epsilon_{kl}$.  I corrected this error. } & $\frac{g^2}{(16\pi^2)^3}\frac{v^2}{\Lambda}$ & $8\times 10^4$  & $\beta\beta0\nu$ & U \\
$39_b$& $L^i L^j L^k L^l\overline{L}_m \overline{L}_n H^m H^n \epsilon_{ij} \epsilon_{kl}$ & $\frac{g^2}{(16\pi^2)^3}\frac{v^2}{\Lambda}$ & $4\times 10^4$  & $\beta\beta0\nu$ & U \\
$39_c$& $L^i L^j L^k L^l \overline{L}_i \overline{L}_m H^m H^n\epsilon_{jk} \epsilon_{ln}$ & $\frac{g^2}{(16\pi^2)^3}\frac{v^2}{\Lambda}$ & $4\times 10^4$  & $\beta\beta0\nu$ & U \\
$39_d$& $L^i L^j L^k L^l \overline{L}_p\overline{L}_q H^m H^n \epsilon_{ij} \epsilon_{km} \epsilon_{ln}\epsilon^{pq}$ & $\frac{g^2}{(16\pi^2)^3}\frac{v^2}{\Lambda}$ & $4\times 10^4$  & $\beta\beta0\nu$ & U \\
$40_a$& $L^i L^j L^k Q^l \overline{L}_i \overline{Q}_jH^m H^n \epsilon_{km} \epsilon_{ln}$ & $\frac{g^2}{(16\pi^2)^3}\frac{v^2}{\Lambda}$ & $4\times 10^4$  & $\beta\beta0\nu$ & U \\
$40_b$& $L^i L^j L^k Q^l \overline{L}_i \overline{Q}_l H^m H^n\epsilon_{jm}\epsilon_{kn}$ & $\frac{g^2}{(16\pi^2)^3}\frac{v^2}{\Lambda}$ & $4\times 10^4$  & $\beta\beta0\nu$ & U \\
$40_c$& $L^i L^j L^k Q^l \overline{L}_l \overline{Q}_i H^m H^n\epsilon_{jm}\epsilon_{kn}$ & $\frac{g^2}{(16\pi^2)^3}\frac{v^2}{\Lambda}$ & $4\times 10^4$  & $\beta\beta0\nu$ & U \\
$40_d$& $L^i L^j L^k Q^l \overline{L}_i \overline{Q}_m H^m H^n\epsilon_{jk}\epsilon_{ln}$ & $\frac{g^2}{(16\pi^2)^3}\frac{v^2}{\Lambda}$ & $4\times 10^4$  & $\beta\beta0\nu$ & U \\
$40_e$& $L^i L^j L^k Q^l \overline{L}_i \overline{Q}_m H^m H^n\epsilon_{jl}\epsilon_{kn}$ & $\frac{g^2}{(16\pi^2)^3}\frac{v^2}{\Lambda}$ & $4\times 10^4$  & $\beta\beta0\nu$ & U \\
$40_f$& $L^i L^j L^k Q^l \overline{L}_m \overline{Q}_i H^m H^n\epsilon_{jk}\epsilon_{ln}$ & $\frac{g^2}{(16\pi^2)^3}\frac{v^2}{\Lambda}$ & $4\times 10^4$  & $\beta\beta0\nu$ & U \\
$40_g$& $L^i L^j L^k Q^l \overline{L}_m \overline{Q}_i H^m H^n\epsilon_{jl}\epsilon_{kn}$ & $\frac{g^2}{(16\pi^2)^3}\frac{v^2}{\Lambda}$ & $4\times 10^4$  & $\beta\beta0\nu$ & U \\
$40_h$& $L^i L^j L^k Q^l \overline{L}_m \overline{Q}_n H^m H^n\epsilon_{ij}\epsilon_{kl}$ & $\frac{g^2}{(16\pi^2)^3}\frac{v^2}{\Lambda}$ & $4\times 10^4$ & $\beta\beta0\nu$ & U \\
$40_i$& $L^i L^j L^k Q^l \overline{L}_m \overline{Q}_n H^p H^q\epsilon_{ip}\epsilon_{jq} \epsilon_{kl} \epsilon^{mn}$ & $\frac{g^2}{(16\pi^2)^3}\frac{v^2}{\Lambda}$ & $4\times 10^4$ & $\beta\beta0\nu$ & U \\
$40_j$& $L^i L^j L^k Q^l \overline{L}_m \overline{Q}_n H^p H^q\epsilon_{ip}\epsilon_{lq} \epsilon_{jk} \epsilon^{mn}$ & $\frac{g^2}{(16\pi^2)^3}\frac{v^2}{\Lambda}$ & $4\times 10^4$  & $\beta\beta0\nu$ & U \\
$41_a$& $L^i L^j L^k d^c \overline{L}_i \bar{d^c}H^l H^m \epsilon_{jl} \epsilon_{km}$ & $\frac{g^2}{(16\pi^2)^3}\frac{v^2}{\Lambda}$ & $4\times 10^4$ & $\beta\beta0\nu$ & U \\
$41_b$& $L^i L^j L^k d^c \overline{L}_l \bar{d^c} H^l H^m\epsilon_{ij}\epsilon_{km}$ & $\frac{g^2}{(16\pi^2)^3}\frac{v^2}{\Lambda}$ & $4\times 10^4$ & $\beta\beta0\nu$ & U \\
$42_a$& $L^i L^j L^k u^c \overline{L}_i\bar{u^c} H^l H^m \epsilon_{jl} \epsilon_{km}$ & $\frac{g^2}{(16\pi^2)^3}\frac{v^2}{\Lambda}$ & $4\times 10^4$ & $\beta\beta0\nu$ & U \\
$42_b$& $L^i L^j L^k u^c \overline{L}_l\bar{u^c} H^l H^m \epsilon_{ij} \epsilon_{km}$ & $\frac{g^2}{(16\pi^2)^3}\frac{v^2}{\Lambda}$ & $4\times 10^4$ & $\beta\beta0\nu$ & U \\
$43_a$& $L^i L^j L^k d^c \overline{L}_l\bar{u^c} H^l\overline{H}_i \epsilon_{jk}$ & $\frac{y_dy_ug^2}{(16\pi^2)^4}\frac{v^2}{\Lambda}$ & $6$ & $\beta\beta0\nu$, LHC & U \\
$43_b$& $L^i L^j L^k d^c \overline{L}_j\bar{u^c} H^l\overline{H}_i\epsilon_{kl}$ & $\frac{y_dy_ug^2}{(16\pi^2)^4}\frac{v^2}{\Lambda}$ & $6$ & $\beta\beta0\nu$, LHC & U \\
$43_c$& $L^i L^j L^k d^c \overline{L}_l\bar{u^c} H^m\overline{H}_n\epsilon_{ij} \epsilon_{km} \epsilon^{ln}$ & $\frac{y_dy_ug^2}{(16\pi^2)^4}\frac{v^2}{\Lambda}$ & $6$ & $\beta\beta0\nu$, LHC & U \\
$44_a$& $L^i L^j Q^k e^c \overline{Q}_i \bar{e^c} H^l H^m\epsilon_{jl} \epsilon_{km}$ & $\frac{g^2}{(16\pi^2)^3}\frac{v^2}{\Lambda}$ & $4\times 10^4$ & $\beta\beta0\nu$ & U \\
$44_b$& $L^i L^j Q^k e^c \overline{Q}_k \bar{e^c} H^l H^m\epsilon_{il}\epsilon_{jm}$ & $\frac{g^2}{(16\pi^2)^3}\frac{v^2}{\Lambda}$ & $4\times 10^4$ & $\beta\beta0\nu$ & U \\
$44_c$& $L^i L^j Q^k e^c \overline{Q}_l \bar{e^c} H^l H^m\epsilon_{ij}\epsilon_{km}$ & $\frac{g^4}{(16\pi^2)^4}\frac{v^2}{\Lambda}$ & $60$ & $\beta\beta0\nu$ & U \\
$44_d$& $L^i L^j Q^k e^c \overline{Q}_l \bar{e^c} H^l H^m\epsilon_{ik} \epsilon_{jm}$ & $\frac{g^2}{(16\pi^2)^3}\frac{v^2}{\Lambda}$ & $4\times 10^4$ & $\beta\beta0\nu$ & U \\
$45$ & $L^i L^j e^c d^c \bar{e^c} \bar{d^c} H^k H^l\epsilon_{ik} \epsilon_{jl}$ & $\frac{g^2}{(16\pi^2)^3}\frac{v^2}{\Lambda}$ & $4\times 10^4$ & $\beta\beta0\nu$ & U \\
$46$ & $L^i L^j e^c u^c \bar{e^c} \bar{u^c} H^k H^l\epsilon_{ik} \epsilon_{jl}$ & $\frac{g^2}{(16\pi^2)^3}\frac{v^2}{\Lambda}$ & $4\times 10^4$ & $\beta\beta0\nu$ & U \\
$47_a$& $L^i L^j Q^k Q^l \overline{Q}_i\overline{Q}_j H^mH^n \epsilon_{km} \epsilon_{ln}$ & $\frac{g^2}{(16\pi^2)^3}\frac{v^2}{\Lambda}$ & $4\times 10^4$ & $\beta\beta0\nu$ & U \\
$47_b$& $L^i L^j Q^k Q^l \overline{Q}_i \overline{Q}_k H^m H^n\epsilon_{jm}\epsilon_{ln}$ & $\frac{g^2}{(16\pi^2)^3}\frac{v^2}{\Lambda}$ & $4\times 10^4$ & $\beta\beta0\nu$ & U \\
$47_c$& $L^i L^j Q^k Q^l \overline{Q}_k\overline{Q}_l H^m H^n\epsilon_{im}\epsilon_{jn}$ & $\frac{g^2}{(16\pi^2)^3}\frac{v^2}{\Lambda}$ & $4\times 10^4$ & $\beta\beta0\nu$ & U \\
$47_d$& $L^i L^j Q^k Q^l \overline{Q}_i\overline{Q}_m H^m H^n\epsilon_{jk}\epsilon_{ln}$ & $\frac{g^2}{(16\pi^2)^3}\frac{v^2}{\Lambda}$ & $4\times 10^4$ & $\beta\beta0\nu$ & U \\
$47_e$& $L^i L^j Q^k Q^l \overline{Q}_i\overline{Q}_m H^m H^n\epsilon_{jn}\epsilon_{kl}$ & $\frac{g^2}{(16\pi^2)^3}\frac{v^2}{\Lambda}$ & $4\times 10^4$ & $\beta\beta0\nu$ & U \\
$47_f$& $L^i L^j Q^k Q^l \overline{Q}_k\overline{Q}_m H^m H^n\epsilon_{ij}\epsilon_{ln}$ & $\frac{g^4}{(16\pi^2)^4}\frac{v^2}{\Lambda}$ & $60$ & $\beta\beta0\nu$ & U \\
$47_g$& $L^i L^j Q^k Q^l \overline{Q}_k\overline{Q}_m H^m H^n\epsilon_{il}\epsilon_{jn}$ & $\frac{g^2}{(16\pi^2)^3}\frac{v^2}{\Lambda}$ & $4\times 10^4$ & $\beta\beta0\nu$ & U \\
$47_h$& $L^i L^j Q^k Q^l \overline{Q}_p\overline{Q}_q H^m H^n\epsilon_{ij}\epsilon_{km} \epsilon_{ln} \epsilon^{pq}$ & $\frac{g^4}{(16\pi^2)^4}\frac{v^2}{\Lambda}$ & $60$ & $\beta\beta0\nu$ & U \\
$47_i$& $L^i L^j Q^k Q^l \overline{Q}_p\overline{Q}_q H^m H^n\epsilon_{ik}\epsilon_{jm} \epsilon_{ln} \epsilon^{pq}$ & $\frac{g^2}{(16\pi^2)^3}\frac{v^2}{\Lambda}$ & $4\times 10^4$ & $\beta\beta0\nu$ & U \\
$47_j$& $L^i L^j Q^k Q^l \overline{Q}_p\overline{Q}_q H^m H^n\epsilon_{im}\epsilon_{jn} \epsilon_{kl} \epsilon^{pq}$ & $\frac{g^2}{(16\pi^2)^3}\frac{v^2}{\Lambda}$ & $4\times 10^4$ & $\beta\beta0\nu$ & U \\
$48$ & $L^i L^j d^c d^c \bar{d^c} \bar{d^c} H^kH^l \epsilon_{ik} \epsilon_{jl}$ & $\frac{g^2}{(16\pi^2)^3}\frac{v^2}{\Lambda}$ & $4\times 10^4$ & $\beta\beta0\nu$ & U \\
$49$ & $L^i L^j d^c u^c \bar{d^c} \bar{u^c} H^kH^l \epsilon_{ik} \epsilon_{jl}$ & $\frac{g^2}{(16\pi^2)^3}\frac{v^2}{\Lambda}$ & $4\times 10^4$ & $\beta\beta0\nu$ & U \\
$50$ & $L^i L^j d^c d^c \bar{d^c} \bar{u^c} H^k\overline{H}_i \epsilon_{jk}$ & $\frac{y_dy_ug^2}{(16\pi^2)^4}\frac{v^2}{\Lambda}$ & $6$ & $\beta\beta0\nu$ LHC & U \\
$51$ & $L^i L^j u^c u^c \bar{u^c} \bar{u^c} H^kH^l \epsilon_{ik} \epsilon_{jl}$ & $\frac{g^2}{(16\pi^2)^3}\frac{v^2}{\Lambda}$ & $4\times 10^4$  & $\beta\beta0\nu$ & U \\
$52$ & $L^i L^j d^c u^c \bar{u^c} \bar{u^c} H^k\overline{H}_i \epsilon_{jk}$ & $\frac{y_dy_ug^2}{(16\pi^2)^4}\frac{v^2}{\Lambda}$ & $6$  & $\beta\beta0\nu$, LHC & U \\
$53$ & $L^i L^j d^c d^c \bar{u^c} \bar{u^c}\overline{H}_i \overline{H}_j$ & $\frac{y_d^2y_u^2g^2}{(16\pi^2)^5}\frac{v^2}{\Lambda}$ & $< 0.5$  & $\beta\beta0\nu$, HElnv, ILC, LHC & D \\
$54_a$& $L^i Q^j Q^k d^c \overline{Q}_i \bar{e^c}H^l H^m \epsilon_{jl} \epsilon_{km}$ & $y_{\ell _\beta}\frac{y_dg^2}{(16\pi^2)^4}\frac{v^2}{\Lambda}$ & $< 0.5$ & $\beta\beta0\nu$, mix, HElnv, ILC, LHC & D \\
$54_b$& $L^i Q^j Q^k d^c \overline{Q}_j \bar{e^c} H^l H^m\epsilon_{il}\epsilon_{km}$ & $y_{\ell _\beta}\frac{y_dg^2}{(16\pi^2)^4}\frac{v^2}{\Lambda}$ & $< 0.5$ & $\beta\beta0\nu$, mix, HElnv, ILC, LHC & D \\
$54_c$& $L^i Q^j Q^k d^c \overline{Q}_l \bar{e^c} H^l H^m\epsilon_{im}\epsilon_{jk}$ & $y_{\ell _\beta}\frac{y_dg^2}{(16\pi^2)^4}\frac{v^2}{\Lambda}$ & $< 0.5$ & $\beta\beta0\nu$, mix, ILC, LHC & D \\
$54_d$& $L^i Q^j Q^k d^c \overline{Q}_l \bar{e^c} H^lH^m \epsilon_{ij} \epsilon_{km}$ & $y_{\ell _\beta}\frac{y_dg^2}{(16\pi^2)^4}\frac{v^2}{\Lambda}$ & $< 0.5$ & $\beta\beta0\nu$,mix, HElnv, ILC, LHC & D \\
$55_a$& $L^i Q^j \overline{Q}_i \overline{Q}_k\bar{e^c} \bar{u^c} H^k H^l \epsilon_{jl}$ & $y_{\ell _\beta}\frac{y_ug^2}{(16\pi^2)^4}\frac{v^2}{\Lambda}$ & $2$ & $\beta\beta0\nu$, mix, LHC & C \\
$55_b$& $L^i Q^j \overline{Q}_j \overline{Q}_k \bar{e^c}\bar{u^c}H^k H^l \epsilon_{il}$ & $y_{\ell _\beta}\frac{y_ug^2}{(16\pi^2)^4}\frac{v^2}{\Lambda}$ & $2$ & $\beta\beta0\nu$, mix, LHC & C \\
$55_c$& $L^i Q^j \overline{Q}_m \overline{Q}_n \bar{e^c}\bar{u^c} H^k H^l\epsilon_{ik} \epsilon_{jl} \epsilon^{mn}$ & $y_{\ell _\beta}\frac{y_ug^2}{(16\pi^2)^4}\frac{v^2}{\Lambda}$ & $2$  & $\beta\beta0\nu$, mix, LHC & C \\
$56$ & $L^i Q^j d^c d^c \bar{e^c} \bar{d^c} H^kH^l \epsilon_{ik} \epsilon_{jl}$ & $y_{\ell _\beta}\frac{y_dg^2}{(16\pi^2)^4}\frac{v^2}{\Lambda}$ & $< 0.5$  & $\beta\beta0\nu$, mix, ILC, LHC & C \\
$57$ & $L^i d^c \overline{Q}_j \bar{u^c} \bar{e^c}\bar{d^c} H^j H^k \epsilon_{ik}$ & $y_{\ell _\beta}\frac{y_ug^2}{(16\pi^2)^4}\frac{v^2}{\Lambda} $ & $2$ & $\beta\beta0\nu$, mix, LHC & C \\
$58$ & $L^i u^c \overline{Q}_j \bar{u^c} \bar{e^c}\bar{u^c} H^j H^k \epsilon_{ik}$ & $y_{\ell _\beta}\frac{y_ug^2}{(16\pi^2)^4}\frac{v^2}{\Lambda}$& $2$ & mix, LHC & C \\
$59$ & $L^i Q^j d^c d^c \bar{e^c} \bar{u^c}H^k \overline{H}_i \epsilon_{jk}$ & $y_{\ell _\beta}\frac{y_d^2y_u}{(16\pi^2)^4}\frac{v^2}{\Lambda}$ & $< 0.5$ & $\beta\beta0\nu$, mix, HElnv, ILC, LHC & D \\
$60$ & $L^i d^c \overline{Q}_j \bar{u^c}\bar{e^c}\bar{u^c} H^j \overline{H}_i$ & $y_{\ell _\beta}\frac{y_dy_u^2}{(16\pi^2)^4}\frac{v^2}{\Lambda}$ & $< 0.5$ & $\beta\beta0\nu$, mix, HElnv, ILC, LHC & D \\
$61$ & $L^i L^j H^k H^l L^r e^c \overline{H}_r \epsilon_{ik} \epsilon_{jl}$ & $\frac{y_\ell }{16\pi^2}\frac{v^2}{\Lambda}\left(\frac{1}{16\pi^2} + \frac{v^2}{\Lambda^2}\right)$ & $2\times 10^{5}$ & $\beta\beta0\nu$ & U \\
$62$ & $L^i L^j L^k e^c H^l L^r e^c \overline{H}_r\epsilon_{ij} \epsilon_{kl}$ & $\frac{y_\ell ^2}{(16\pi^2)^2}\frac{v^2}{\Lambda}\left(\frac{1}{16\pi^2} + \frac{v^2}{\Lambda^2}\right)$ & $20$  & $\beta\beta0\nu$ & U \\
$63_a$& $L^i L^j Q^k d^c H^l L^r e^c \overline{H}_r\epsilon_{ij} \epsilon_{kl}$ & $\frac{y_\ell  y_d}{(16\pi^2)^3}\frac{v^2}{\Lambda}$ & $40$ & $\beta\beta0\nu$ & U \\
$63_b$& $L^i L^j Q^k d^c H^l L^r e^c \overline{H}_r\epsilon_{ik} \epsilon_{jl}$ & $\frac{y_\ell  y_d}{(16\pi^2)^2}\frac{v^2}{\Lambda}\left(\frac{1}{16\pi^2} + \frac{v^2}{\Lambda^2}\right)$ & $40$ & $\beta\beta0\nu$ & U \\
$64_a$& $L^i L^j \overline{Q}_i \bar{u^c} H^k L^r e^c \overline{H}_r\epsilon_{jk}$ & $\frac{y_\ell  y_u}{(16\pi^2)^2}\frac{v^2}{\Lambda}\left(\frac{1}{16\pi^2} + \frac{v^2}{\Lambda^2}\right)$ & $2\times 10^3$ & $\beta\beta0\nu$ & U \\
$64_b$& $L^i L^j \overline{Q}_k\bar{u^c}H^k L^r e^c \overline{H}_r\epsilon_{ij}$ & $\frac{y_\ell  y_u}{(16\pi^2)^3}\frac{v^2}{\Lambda}$ & $2\times 10^3$ & $\beta\beta0\nu$ & U \\
$65$ & $L^i \bar{e^c} \bar{u^c} d^c H^j L^r e^c \overline{H}_r\epsilon_{ij}$ & $\frac{y_d y_ug^2}{(16\pi^2)^4}\frac{v^2}{\Lambda}$ & $6$ & $\beta\beta0\nu$, LHC & U \\
$66$ & $L^i L^j H^k H^l \epsilon_{ik} Q^r d^c \overline{H}_r\epsilon_{jl}$ & $\frac{y_d}{16\pi^2}\frac{v^2}{\Lambda}\left(\frac{1}{16\pi^2} + \frac{v^2}{\Lambda^2}\right)$ & $6\times 10^{5}$ & $\beta\beta0\nu$ & U \\
$67$ & $L^i L^j L^k e^c H^l Q^r d^c \overline{H}_r\epsilon_{ij} \epsilon_{kl}$ & $\frac{y_\ell y_d}{(16\pi^2)^2}\frac{v^2}{\Lambda}\left(\frac{1}{16\pi^2} + \frac{v^2}{\Lambda^2}\right)$ & $40$ & $\beta\beta0\nu$ & U \\
$68_a$& $L^i L^j Q^k d^c H^l Q^r d^c \overline{H}_r\epsilon_{ij} \epsilon_{kl}$ & $\frac{y_d^2 g^2}{(16\pi^2)^3}\frac{v^2}{\Lambda}\left(\frac{1}{16\pi^2} + \frac{v^2}{\Lambda^2}\right)$ & $1$ & $\beta\beta0\nu$, LHC & U \\
$68_b$& $L^i L^j Q^k d^c H^l Q^r d^c \overline{H}_r\epsilon_{ik} \epsilon_{jl}$ & $\frac{y_{q_d^2}}{(16\pi^2)^2}\frac{v^2}{\Lambda}\left(\frac{1}{16\pi^2} + \frac{v^2}{\Lambda^2}\right)$ & $1\times 10^2$  & $\beta\beta0\nu$ & U \\
$69_a$& $L^i L^j \overline{Q}_i \bar{u^c} H^k Q^r d^c \overline{H}_r\epsilon_{jk}$ & $\frac{y_dy_u}{(16\pi^2)^2}\frac{v^2}{\Lambda}\left(\frac{1}{16\pi^2} + \frac{v^2}{\Lambda^2}\right)$ & $4\times 10^3$ & $\beta\beta0\nu$ & U \\
$69_b$& $L^i L^j \overline{Q}_k\bar{u^c}H^k Q^r d^c \overline{H}_r\epsilon_{ij}$ & $\frac{y_dy_u g^2}{(16\pi^2)^3}\frac{v^2}{\Lambda}\left(\frac{1}{16\pi^2} + \frac{v^2}{\Lambda^2}\right)$ & $7$ & $\beta\beta0\nu$, LHC & U \\
$70$ & $L^i \bar{e^c} \bar{u^c} d^c H^j Q^r d^c \overline{H}_r\epsilon_{ij}$ & $y_{\ell _\beta}\frac{ y_d^2 y_u}{(16\pi^2)^3}\frac{v^2}{\Lambda}\left(\frac{1}{16\pi^2} + \frac{v^2}{\Lambda^2}\right)$ & $< 0.5$ & $\beta\beta0\nu$, mix, HElnv, ILC, LHC & D \\
$71$ & $L^i L^j H^k H^l Q^r u^c H^s \epsilon_{rs} \epsilon_{ik} \epsilon_{jl}$ & $\frac{y_u}{16\pi^2}\frac{v^2}{\Lambda}\left(\frac{1}{16\pi^2} + \frac{v^2}{\Lambda^2}\right)$ & $2\times 10^{7}$ & $\beta\beta0\nu$ & U \\
$72$ & $L^i L^j L^k e^c H^l Q^r u^c H^s \epsilon_{rs}\epsilon_{ij} \epsilon_{kl}$ & $\frac{y_\ell y_u}{(16\pi^2)^2}\frac{v^2}{\Lambda}\left(\frac{1}{16\pi^2} + \frac{v^2}{\Lambda^2}\right)$ & $2\times 10^3$  & $\beta\beta0\nu$ & U \\
$73_a$& $L^i L^j Q^k d^c H^l Q^r u^c H^s \epsilon_{rs}\epsilon_{ij} \epsilon_{kl}$ & $\frac{y_dy_u g^2}{(16\pi^2)^3}\frac{v^2}{\Lambda}\left(\frac{1}{16\pi^2} + \frac{v^2}{\Lambda^2}\right)$ & $7$ & $\beta\beta0\nu$, LHC & U \\
$73_b$& $L^i L^j Q^k d^c H^l Q^r u^c H^s \epsilon_{rs}\epsilon_{ik} \epsilon_{jl}$ & $\frac{y_dy_u}{(16\pi^2)^2}\frac{v^2}{\Lambda}\left(\frac{1}{16\pi^2} + \frac{v^2}{\Lambda^2}\right)$ & $4\times 10^3$ & $\beta\beta0\nu$ & U \\
$74_a$& $L^i L^j \overline{Q}_i \bar{u^c} H^k Q^r u^c H^s \epsilon_{rs}\epsilon_{jk}$ & $\frac{y_u^2}{(16\pi^2)^2}\frac{v^2}{\Lambda}\left(\frac{1}{16\pi^2} + \frac{v^2}{\Lambda^2}\right)$ & $2\times 10^5$ & $\beta\beta0\nu$ & U \\
$74_b$& $L^i L^j \overline{Q}_k\bar{u^c}H^k Q^r u^c H^s \epsilon_{rs}\epsilon_{ij}$ & $\frac{y_u^2 g^2}{(16\pi^2)^3}\frac{v^2}{\Lambda}\left(\frac{1}{16\pi^2} + \frac{v^2}{\Lambda^2}\right)$ & $2\times 10^2$ & $\beta\beta0\nu$ & U \\
$75$ & $L^i \bar{e^c} \bar{u^c} d^c H^j Q^r u^c H^s \epsilon_{rs}\epsilon_{ij}$ & $y_{\ell _\beta}\frac{ y_d y_u^2}{(16\pi^2)^3}\frac{v^2}{\Lambda}\left(\frac{1}{16\pi^2} + \frac{v^2}{\Lambda^2}\right)$ & $1$ & $\beta\beta0\nu$, mix & C \\
\hline 
 \end{longtable}
\end{landscape}}
 \end{singlespace}

The third column of Table \ref{tab:AllOps}, labeled
$m_{\alpha\beta}$, presents our estimate for the  operator-induced
Majorana neutrino mass expressions.  These were derived based on the
estimation procedure discussed earlier. Trivial order one factors,
as well as the generation dependent coupling constants $\lambda$
have been omitted, as already advertised.  Flavor specific charged
lepton Yukawa couplings are explicitly denoted $y_{\ell_\alpha}$ and
$y_{\ell_\beta}$ to distinguish them from $y_\ell$, $y_u$ and $y_d$,
meant to represent $\alpha,\beta$-independent Yukawa couplings.  A
summation over all ``internal flavors'' is assumed for each entry.
For order one coupling constants, this sum is strongly dominated by
third generation Yukawa couplings. Upon setting these mass
expressions equal to the observed scale of light neutrino masses
($0.05~\rm{eV}$), I extract the required cutoff scale $\Lambda$ for
each operator.  This quantity, defined to be $\Lambda_{\nu}$, is
listed in column four in units of one TeV. Numerical results were
obtained assuming the current best fit values for all SM parameters.
Associated errors are negligibly small as far as our aspirations are
concerned.

\begin{figure}
\begin{center}
\includegraphics[angle=270,scale=.55]{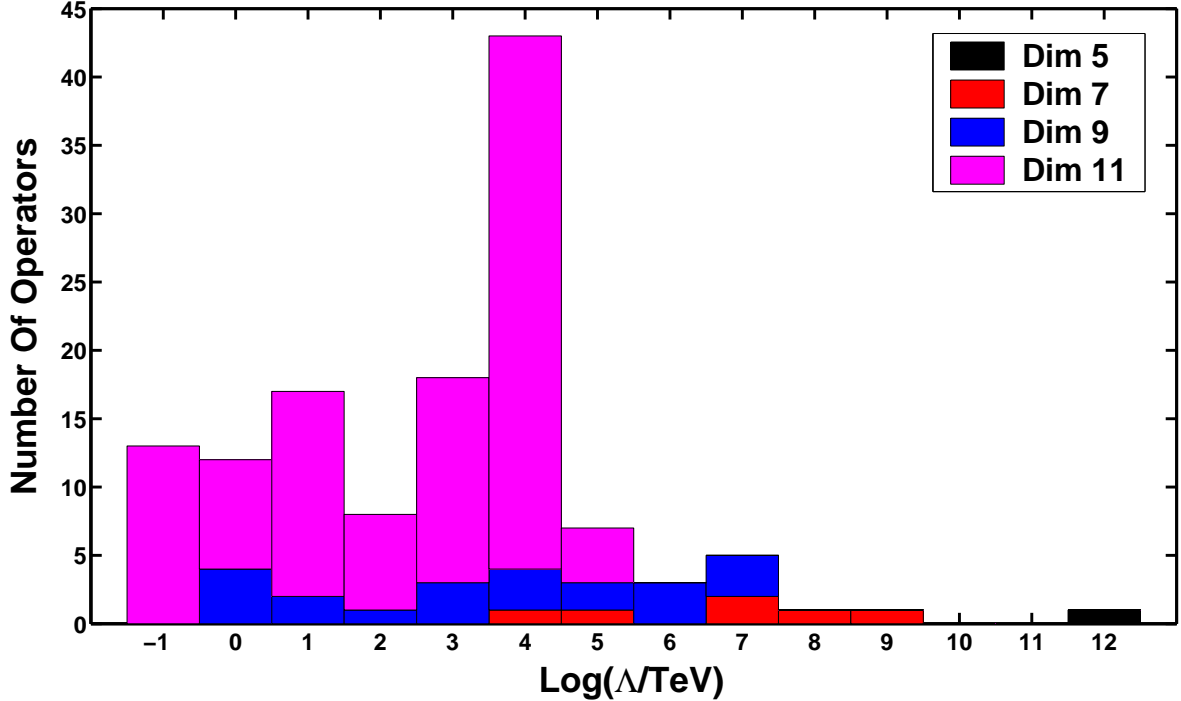}
\caption[Summary histogram of the scale $\Lambda$ extracted from 129
LNV operators]{A summary histogram of the scale of new physics
$\Lambda_{\nu}$ extracted from the 129 LNV operators introduced in
Table \ref{tab:AllOps}.  I assume a radiatively generated neutrino
mass of $0.05~\rm{eV}$ and universal order one coupling constants.
The contributions of operators of different mass dimensions are
associated to different colors (shades of gray), as indicated in the
caption.} \label{fig:OpSum}
\end{center}
\end{figure}

Fig.~\ref{fig:OpSum} displays the distribution of extracted cutoff
scales, $\Lambda_{\nu}$. The histogram bars are color coded to
reflect the different operator mass dimensions.  The distribution
spans thirteen orders of magnitude, from the electroweak scale to
$10^{12}~\rm{TeV}$.  It is interesting to note the general trend of
operator dimension with scale:  as expected, higher dimension
operators are characterized by lower ultraviolet scales.  For
operators associated with the lowest ultraviolet cutoffs, the lepton
number breaking physics occurs at the same energy scale as
electroweak symmetry breaking. In this case, one needs to revisit
some of the assumptions that go into obtaining the bounds and
predictions discussed here. Regardless, it is fair to say that some
of these effective operators should be severely constrained by other
experimental probes, as will be discussed in the next section.

The natural scale for most of the explored operators is well above
$10~\rm{TeV}$, and thus outside the reach of future experimental
efforts except, perhaps, those looking for neutrinoless double-beta
decay. The remainder, however, should yield observable consequences
in next-generation experiments.  This small subset arguably contains
the most interesting cases on purely economic grounds, as they
naturally predict tiny neutrino masses as well as TeV scale new
physics, which is already thought to exist for independent reasons.
It is aesthetically pleasing to imagine that all, or at least most,
of nature's current puzzles can arise from the same source, as
opposed to postulating various solutions at different energy scales.
It is important to note that one can ``push'' more of the operators
into the observable TeV window by modifying the coupling of the new
physics to different fermion generations.  In particular, since many
of the induced neutrino masses depend upon fermion Yukawa couplings,
one can efficiently reduce scales by simply and uniformly decoupling
the third generation. In most cases this can yield a $\Lambda_{\nu}$
reduction of several orders of magnitude; a factor that can be
further enhanced by also decoupling the second generation.  Under
these conditions, the resulting distribution, analogous to
Fig.~\ref{fig:OpSum}, would show the majority of the operators piled
up near and slightly above the electroweak scale. A detailed
exploration of this possibility would be impractical and is not
pursued further. I will, however, like to emphasize that this
strategy of decoupling the new physics from the \emph{heavy}
fermions is very non-standard. In most cases, one is tempted to
decouple \emph{light} fermions from new physics both because these
lead to the strongest constraints and because one tends to believe
that the large Yukawa couplings of third generation fermions are
entangled with the physics of electroweak symmetry breaking.

Not all extracted cutoff scales are subject to a strong dependency
on SM Yukawa couplings. In particular, the $\Lambda_{\nu}$ values
for the majority of dimension-eleven operators in the large
histogram bar near $10^4~\rm{TeV}$ would not shift down at all under
this hypothetical decoupling of the third generation from the new
physics.  These are the operators, as shown in Diagram $(d)$ of
Fig.~\ref{fig:MassDiagrams}, whose induced neutrino mass matrix is
independent of the Yukawa sector. In such cases, $m_{\alpha\beta}$
are only functions of the various gauge couplings. As such, these
constitute the most robust results of our analysis. These operators
all predict an anarchic Majorana neutrino mass matrix of overall
scale given by $m_{\nu} = g^2/(16\pi^2)^3 v^2/\Lambda$, implying an
energy scale $\Lambda_{\nu}\sim 10^5~\rm{TeV}$. The only other
``Yukawa invariant'' cutoff scale estimate arises for the
dimension-five operator $\mathcal{O}_1$. $\mathcal{O}_1$ captures
the physics of all versions of the
 seesaw mechanism \cite{SeeSaw}, and is at the heart of
most of the model building currently done within the neutrino
sector. Its ultraviolet completion can precede in only three
distinct ways \cite{Bajc:2006ia}.  These possibilities are via the
exchange of heavy gauge singlet fermions (type I seesaw), $SU(2)_L$
triplet scalars (type II seesaw) \cite{SeeSaw2}, $SU(2)_L$ triplet
fermions (type III seesaw) \cite{SeeSaw3}, or some combination
thereof.  Its popularity is well-founded for a number of reasons,
including its underlying simplicity in structure as well as the
purely empirical fact that it is the ``lowest order means'' of
neutrino mass, and as such is easily generated by a ``generic'' LNV
model. Additionally, the high scale associated with the seesaw
mechanism can be easily incorporated within existing theoretical
models and serves to help explain the observed baryon antisymmetry
of the universe via leptogenesis \cite{Leptogenesis}.

For the purposes of direct observation, $\mathcal{O}_1$'s high
cutoff scale, nearly $10^{12}~\rm{TeV}$, places it well outside of
the ``detectable region'' ($\Lambda\lesssim 10$~TeV) and renders it
uninteresting for the purposes of our analysis. Of course, there
always remains the possibility that $\mathcal{O}_1$ is generate by
very weakly coupled new physics (or very finely-tuned new physics
\cite{fine_nus}), in which case I expect to run into the new
ultraviolet degrees of freedom at energies well below
$10^{12}~\rm{TeV}$. In the case of $\mathcal{O}_1$, it has been
argued that new physics at almost any energy scale (from well below
the sub-eV realm to well above the weak scale) will lead to light
neutrino masses \cite{LowScaleSeeSaw,nuSM_dark,MySeeSaw,SeeSaw_LSND} without contradicting current
experimental results. Such possibilities -- related to the fact that
the new physics is very weakly coupled -- are not being explored
here, as I always assume that the new degrees of freedom are heavier
than typical experimentally accessible energy scales.

Armed with our derived new physics scales $\Lambda_{\nu}$, I proceed
to plug them back into the different irrelevant LNV operators and
search for possible means of future observation as well as already
existing constraints. Generally, those operators that yield the
largest experimental signals have the lowest cutoff scales. I
conclude that, if associated to neutrino masses, the effective
cutoff scale $\Lambda_{\nu}$ of the following effective operators is
constrained to be less than  $1~\rm{TeV}$:
\begin{equation}
\mathcal{O}_{34},\mathcal{O}_{36},\mathcal{O}_{37},\mathcal{O}_{38},\mathcal{O}_{53},\mathcal{O}_{54_{a,b,c,d}},\mathcal{O}_{56},\mathcal{O}_{59},\mathcal{O}_{60},\mathcal{O}_{70}.
\label{eq:O1TeV}
\end{equation}
These may lead to observable effects at future high energy
accelerator facilities. Additionally, such low scales may also
indirectly lead to observable effects in ``low energy'' (but high
sensitivity) experiments.  There are more operators associated with
slightly higher scales between $(1-10)~\rm{TeV}$ that may manifest
themselves experimentally via virtual effects.  These are
\begin{equation}
\mathcal{O}_{16},\mathcal{O}_{17},\mathcal{O}_{18},\mathcal{O}_{19},\mathcal{O}_{35},\mathcal{O}_{43_{a,b,c}},\mathcal{O}_{50},\mathcal{O}_{52},\mathcal{O}_{55_{a,b,c}},\mathcal{O}_{57},\mathcal{O}_{58},\mathcal{O}_{65},\mathcal{O}_{68_{a,b}},\mathcal{O}_{73_a},\mathcal{O}_{75}.
\label{eq:O1to10TeV}
\end{equation}
These operators yield finite predictions for more than one
observable, such that experimental efforts in seemingly unrelated
fields can help  constrain the class of possible LNV models or even
help identify the true LNV model.

\setcounter{footnote}{0} \setcounter{equation}{0}
\section{General operator constraints and predictions}
\label{sec:constraints}

There are, currently, bounds on LNV processes from a number of
independent experimental sources \cite{LNVUpperBounds,PDG}.  Many of
these are presently too mild to constrain the operators listed in
Table~\ref{tab:AllOps} once their ultraviolet cutoffs $\Lambda$ are
set to the required value indicated by the presence of non-zero
neutrino masses, $\Lambda_{\nu}$. The situation, however,  is
expected to improve in the next several years with increased rare
decay sensitivities and higher collider energies. Here I survey the
experimental signatures of these operators in terms of the minimal
scenarios described above.  Specifically, I address the potential of
neutrinoless double-beta decay (Sec.~\ref{subsec:bb0nu}), rare meson
decays (Sec.~\ref{subsec:OtherProcesses}), and collider experiments
(Sec.~\ref{subsec:Collider}) to constrain the effective operators in
question, assuming that, indeed, they are responsible for the
observed non-zero neutrino masses. As before, I will use the
approximations discussed in Sec.~\ref{sec:Scale}, and warn readers
that all the results presented are to be understood as order of
magnitude estimates. The results, however, are useful as far as
recognizing the most promising LNV probes and identifying different
scenarios that may be probed by combinations of different LNV
searches.

Most of this section will be devoted to  probes of LNV via simple
variants of the following process, which can be written
schematically as
\begin{equation}
\ell _\alpha\ell _\beta \leftrightarrow d_\kappa d_\zeta
\bar{u}_\rho \bar{u}_\omega. \label{eq:GoldenCh}
\end{equation}
Greek subscripts run over all different fermion flavors. Given the
assumed democratic models, coupled with our present lack of
experimental information, one would expect that all flavor
combinations are equivalent to zeroth order.  Any indication to the
contrary would signify important deviations from simple
expectations, and thus begin to reveal the flavor structure of the
new physics.  The above selected ``golden modes'' often yield the
largest LNV rates, but this is not always the case. For example,
some operators do not allow tree-level charged dilepton events, but
rather prefer to include neutrino initial or final states.  LNV
processes with initial and final state neutrinos are extremely
difficult to identify. The only hope of such discovery channels is,
perhaps, via neutrino scattering experiments on either electron or
nucleon targets, using well understood neutrino beams.  I point out
that any neutrino/anti-neutrino cross contamination induces
ambiguity onto the total lepton number of the incident beam and
would serve as a crippling source of background for LNV searches.
This reasoning rules out conventional superbeam \cite{SupBeam}
facilities as well as proposed neutrino factories \cite{NuFact},
which contain both neutrino and anti-neutrino components, but does
suggest modest possibilities for future beta-beams \cite{BetaBeam}.
Given projected beta-beam luminosities and energies along with the
derived cutoff scales $\Lambda_{\nu}$, it seems unlikely that LNV
can be observed in such experiments.  Another possible discovery
mode  involves only two external state quarks and an associated
gauge boson as in the sample process $\ell_{\alpha}  \ell_{\beta}
\rightarrow d \bar{u} + W^-$. It turns out that the rates for such
processes are generally suppressed for the majority of operators
involving six fermion fields, as I am trading a phase space
suppression for a stronger loop suppression. For those operators
with only four fermion fields, the situation is not as
straightforward and, in some cases, the three particle final state
is preferred.  Typically, the neutrino mass induced cutoff scales of
those operators are high $(\Lambda_{\nu} \gg 100~\rm{TeV})$, so it
would be quite difficult to observe such effects.  Of course, any
$W$-boson final state will either promptly decay leptonically,
yielding missing energy and unknown total lepton number, or
hadronically, reducing the reaction back to that of the golden mode.

Another possibility is to replace two or more of the external quark
states in Eq.~(\ref{eq:GoldenCh}) with leptons in such a way as to
preserve charge, baryon number, and $\Delta L = 2$ constraints.
While many operators favor this structure, a little thought reveals
that at least one external neutrino state is always present, which
leaves only a missing energy signature, and little means of lepton
number identification in a detector. Such events would not be clean,
but of course, three final-state charged leptons and missing energy
are enough to extract the existence of at least $\Delta L=1$ LNV,
provided that the number of invisible states is known to be no
greater than one. This last requirement is difficult to achieve in
the presence of the large backgrounds and the limited statistics
expected at future collider facilities, but should still be possible
given a concrete model probed near resonance (see for example
\cite{EmEm2munu}). Therefore, while important and potentially
observable, this mode is not generally the best place to look for
LNV and is neglected in the remainder of our analysis. From this
perspective, the only other relevant channel of LNV discovery is
related to $W$ and $Z$ rare decays into final states with non-zero
total lepton number. This possibility is briefly addressed in
Sec.~\ref{subsec:OtherProcesses}.

\subsection{Neutrinoless Double-Beta Decay} \label{subsec:bb0nu}

Here I probe the expectations for neutrinoless double-beta decay
$(\beta\beta0\nu)$ for each operator listed in Table
\ref{tab:AllOps}.  $\beta\beta0\nu$ is the LNV ($\Delta L=2$)
process where, within a nucleus, two down quarks convert into two up
quarks with the emission of two electrons but no neutrinos, or in
the language of nuclear physics $(A,Z)\rightarrow (A,Z+2)+e^-e^-$.
See \cite{BB0n_review,EVcommon} and references therein for a comprehensive
review. While precise computations of nuclear matrix elements are
essential for making detailed predictions
 \cite{NuMatrixEl}, the minimal parton-level description
given above is adequate for the purposes of this study. There is a
continuing legacy of cutting edge experiments designed to search for
$\beta\beta0\nu$ with no success to date.\footnote{There is
currently a positive report of $\beta\beta0\nu$ at the $4.2\sigma$
level by a subset of the Heidelberg-Moscow collaboration
\cite{BB0nPosSig}. They report a measured half-life of
$1.74^{+0.18}_{-0.16}\times 10^{21}~\rm{years}$ which maps to
$m_{ee}^{\rm eff}\sim (0.2-0.6)~\rm{eV}$. I choose to neglect this
controversial result, which is still awaiting independent
conformation.} Currently the $^{76}$Ge half-life for this process is
bounded to be greater than $1.9\times 10^{25}~\rm{yr}$ and
$1.57\times 10^{25}~\rm{yr}$ at $90\%$ confidence by the
Heidelberg-Moscow \cite{HMoscowCurrent} and IGEX \cite{IGEX}
experiments, respectively. Future experiments are poised to improve
these limits (for several different nuclei) by a couple of orders of
magnitude within the next five to ten years \cite{BB0nFuture}.

If one assumes that $\beta\beta0\nu$ proceeds via the exchange of
light Majorana neutrinos, its amplitude is proportional to the $ee$
element of the Majorana neutrino mass matrix,
\begin{equation}
m_{ee} = \sum_{i=1}^3 m_i U_{ei}^2, \label{eq:std_Mee}
\end{equation}
where $m_i$  are the neutrino masses and $U_{ei}$ are elements of
the leptonic mixing matrix.  With this, one can extract the upper
bound $m_{ee}<0.35~{\rm eV}$ (90\% confidence level bound,
\cite{PDG}) from current experiments while next-generation
experiments are aiming at $m_{ee}\gtrsim 0.05$~eV\footnote{The
parameter change from half-life to $m_{ee}$ depends heavily on
nuclear matrix element calculations.
 Current calculations induce
  an uncertainty of less then a factor of four on $m_{ee}$ for most parent isotopes \cite{NuMatrixEl}.}
   \cite{BB0nFuture}.
 In general, LNV new physics will lead to additional contributions to $\beta\beta0\nu$, most of
 which are not proportional to $m_{ee}$. However, the amplitude for  $\beta\beta0\nu$
can still be expressed in terms of an effective $m_{ee}$,
$m_{ee}^{\rm eff}$, which is an operator-specific quantity that will
be used to analyze  new models of LNV.

\begin{figure}
\begin{center}
\includegraphics[scale=0.8]{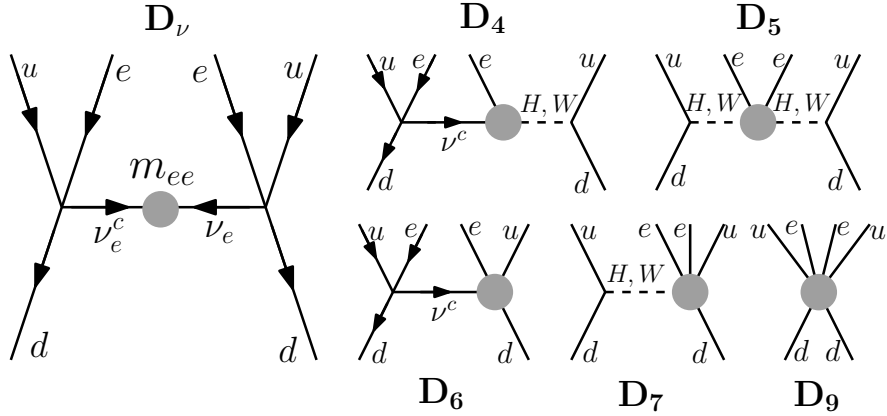}
\caption[Feynman diagrams contributing to neutrinoless double-beta
decay]{The parton level Feynman diagrams contributing to
neutrinoless double-beta decay, labeled by the dimension of the
underlying lepton number violating interaction, indicated by gray
dots.  Each diagram is generated, at some order in perturbation
theory, by all analyzed interactions, but estimates of their
magnitudes depend heavily on the details of the operators, including
their associated scale $\Lambda_{\nu}$, fermion content and helicity
structure.} \label{fig:Bb0nOps}
\end{center}
\end{figure}

Here, I define six different ``classes'' of diagrams one can
construct out of LNV irrelevant operators that contribute to
$\beta\beta 0\nu$ at the parton level. These are illustrated in
Fig.~\ref{fig:Bb0nOps}, and classified by the dimension of the
generated LNV interaction, depicted by large gray dots. In order to
unambiguously separate the different classes, note that the grey
circles are defined in such a way that all fermion and Higgs legs
that come out of it are part of the ``parent'' operator ${\mathcal
O}$ (and not attached on via reducible SM vertices), while all other
interactions are SM vertices. The dots should be viewed as hiding
the underlying LNV interactions. In general, they contain a mixture
of coupling constants and loop factors that must be evaluated
explicitly for each diagram.  It is important to emphasize that the
contribution of a  generic operator ${\mathcal O}$ to
$\beta\beta0\nu$ will consist of contributions from all different
classes, while usually dominated by one of them. I show the lepton
number conserving electroweak vertices (point-like) as effective
four-fermion interactions, justified by the low energy scale of
nuclear beta decays. The dotted lines indicate the exchange of
$W$-bosons, labeled by $W$ and $H$ (charged Higgs goldstone boson).
Helicity arrows are explicitly included where uniquely determined,
implying that the arrowless legs can have any helicity.

${D_\nu}$ describes the standard scenario of $\beta\beta0\nu$
mediated by light Majorana neutrinos. It is simply two electroweak
vertices held together by a Majorana mass term on which two
neutrinos are annihilated.  The amplitude for this diagram is
proportional to $m_{ee}$, as defined in Eq.~(\ref{eq:std_Mee}). The
dependence on such a neutrino mass is intuitively clear considering
the need for a helicity flip on the internal neutrino line.  The
remaining diagrams are qualitatively different from this standard
case. Most importantly, none of them require ``helicity flips'' and
are therefore not directly proportional to neutrino masses. They
are, however, proportional to inverse powers of the new mass scale
$\Lambda_{\nu}$, and hence also suppressed.  These effects are not
entirely independent, since the value of $\Lambda_{\nu}$ was
extracted from the requirement that  neutrino masses are small, but
correlations are relaxed enough to allow nontrivial consequences. It
is this partial decoupling from neutrino masses that allows larger
than naively expected contributions to  $\beta\beta0\nu$ from some
of the LNV irrelevant operators. Before proceeding, I make the
trivial observation that the amplitudes following from ${D_\nu}$,
${D_4}$ and ${D_5}$ are additionally proportional to two powers of
CKM matrix elements, namely $|V_{ud}|^2$, whereas ${D_6}$ and
${D_7}$ are only proportional to one power of
$V_{ud}$.\footnote{This is true provided that I assume no flavor
structure for the underlying operator, or, equivalently, that all
dimensionless coupling constants are order one.  If one is motivated
by experiment to postulate a minimally flavor violating scenario, to
perhaps ease constraints from flavor changing neutral currents, the
statement must be modified accordingly.}  The tree-level diagram
${D_9}$ has no CKM ``suppression.''  While this is a purely academic
fact in the case of $\beta\beta0\nu$ ($|V_{ud}|\sim 1$!),  it leads
to important consequences for analogous rare decays that depend on
the much smaller off-diagonal CKM matrix elements. I will return to
these in the next subsection.

For a given operator, the relative size of each diagram's
contribution to the total decay rate depends on many factors
including the operator's dimension, scale, fermion content and
helicity structure.  The dominant contributions must be calculated
on a case by case basis.  Generally, the high scale operators
$(\Lambda_{\nu} \gtrsim 10~\rm{TeV})$ are dominated by the two
dimension-four diagrams ${D_\nu}$ and ${D_4}$ since many factors of
$\Lambda_{\nu}$ will be canceled by divergent loops inside the gray
dots thereby minimizing the $1/\Lambda_{\nu}$ suppression.  All else
being equal, ${D_\nu}$ is the strongest of the pair since it is
enhanced by $\sim Q^{-2}$ from the two propagating neutrino lines as
opposed to only $\sim Q^{-1}$ for the one neutrino case shown in
diagram $D_4$. For those operators with no tree-level $\nu\nu$ field
content, ${D_4}$ can still be very important, but its dominance is
nevertheless rare. As discussed in Sec.~\ref{sec:Scale}, these are
precisely the operators that have the greatest loop suppressions and
consequently lower energy scales suggesting the need for diagrams
beyond ${D_4}$. The effects of low cutoff scale operators
$(\Lambda_{\nu} \lesssim 1~\rm{TeV})$ are not severely suppressed by
$1/\Lambda_{\nu}$ (by definition), so the dominant diagrams will
typically be of the highest dimension allowed by the tree-level
structure of the operator.  For such low scales and for operators of
the following schematic form $dd\bar{u}\bar{u}\bar{e}\bar{e}$
(dimension $9$) or $dd\bar{u}\bar{u}\bar{e}\bar{e}H_0H_0$ (dimension
$11$), ${D_9}$ always dominates the $\beta\beta0\nu$ rate yielding
amplitudes proportional to $1/\Lambda_{\nu}^5$ and
$v^2/\Lambda_{\nu}^7$, respectively. For intermediate scales, and
when the operator's field content does not directly support
$\beta\beta0\nu$ due to lack of quark fields, the situation is not
as straight forward and one must perform the relevant computations
to determine the dominant diagrams.  Still, it should be noted that
diagrams containing internally propagating neutrinos are enhanced by
inverse powers of $Q$ and maintain a slight advantage over their
neutrinoless counterparts.  One can thus generally expect diagram
$D_6$ to dominate the decay rates for low $\Lambda_{\nu}$ scale
operators when $D_9$ is suppressed.  The opposite is true for
interactions taking place at higher energies in, for example,
next-generation colliders, as discussed in
Sec.~\ref{subsec:Collider}.

Since each diagram in Fig.~\ref{fig:Bb0nOps} can have different
external helicity structures, the different  contributions to the
total rate will be added incoherently, thus eliminating the effects
of interference.  There are some case specific coherent
contributions that I neglected in our treatment since most rates are
dominated by a single diagram. Another potential difference among
the different contributions is related to nuclear matrix element
calculations: can the calculations done assuming $\beta\beta0\nu$
via the standard light Majorana neutrino exchange scenario of
diagram ${D_\nu}$ be applied to the more general cases encountered
here?  I have nothing to add to this discussion except to naively
note that there is no obvious reason why such rates should be
severely suppressed or enhanced relative to the standard scenario. I
therefore assume that all nuclear matrix elements are identical and
can be factored out of the incoherent sum. I assume that this
approximation is not more uncertain than the other sources of
uncertainty inherent to our study (likely a very safe assumption).

As drawn, each diagram $D_i$ contributes to the amplitude that
characterizes $\beta\beta0\nu$.  For example, the amplitude
associated with $D_\nu$ is proportional to
\begin{equation}
\mathcal{A}_{D_\nu} \equiv m_{ee}\frac{|V_{ud}|^2 G_F^2}{Q^2},
 \label{eq:ad}
 \end{equation}
where $G_F$ is the Fermi constant. The remaining diagrams will
contribute with $\mathcal{A}_{D_i}\propto\zeta(v,Q)\Lambda^{4-i}$,
up to a dimensionless coefficient containing various numerical/loop
factors, as well as general scale depencies parameterized by some
power of the ratio $v/\Lambda$.  The function $\zeta(v,Q)$ has mass
dimension ${i-9}$ so that all ${\mathcal A}_{D_i}$ have the same
mass dimension.  Note that all aspects of $\mathcal{A}_{D_i}$ are
calculable given a LNV operator and diagram. I can analyze each
operator in terms of an effective $m_{ee}^{\rm eff}$,  defined in
terms of the underlying dimension nine amplitude $\mathcal{A}_{D_i}$
by
\begin{equation}
m_{ee}^{\rm eff} = \frac{Q^2}{G_F^2 |V_{ud}|^2}\sqrt{\sum_i
\mathcal{A}_{D_i}^2},
 \label{eq:MeeEff}
\end{equation}
where $i$ runs over the set $\{\nu,4,5,6,7,9\}$  that labels the
diagrams shown in Fig.~\ref{fig:Bb0nOps}, and $Q\sim 50~\rm{MeV}$ is
the typical momentum transfer in $\beta\beta0\nu$. $m_{ee}^{\rm
eff}$ can be directly compared with experiment and used to make
prediction for future observations. A few comments are in order
regarding this quantity. First, it is a useful derived object that
has no direct connection to a real neutrino mass and is valid to
arbitrarily large values. Note that in the case of Majorana neutrino
exchange, $m_{ee}^{\rm eff}=m_{ee}$ only if $m_{ee}\ll Q$.  When
neutrino masses are greater than Q, $m_{ee}^{\rm eff}\propto 1/m$.
Our definition of $m_{ee}^{\rm eff}$ also conforms to the use of
large effective masses in \cite{LNVUpperBounds}.  The second comment
is that, unlike the case of $m_{ee}$, which is valid for any process
involving the exchange of electron-like Majorana neutrinos,
$m_{ee}^{\rm eff}$ is case specific.  It must be calculated
separately for each process, as each one, in general, is composed of
different diagrams.  In particular, the calculations of the
effective mass for $\beta\beta0\nu$ expressed here are not directly
applicable to other LNV processes and should not be interpreted as
such.
\begin{figure}
\begin{center}
\includegraphics[angle=270,scale=.57]{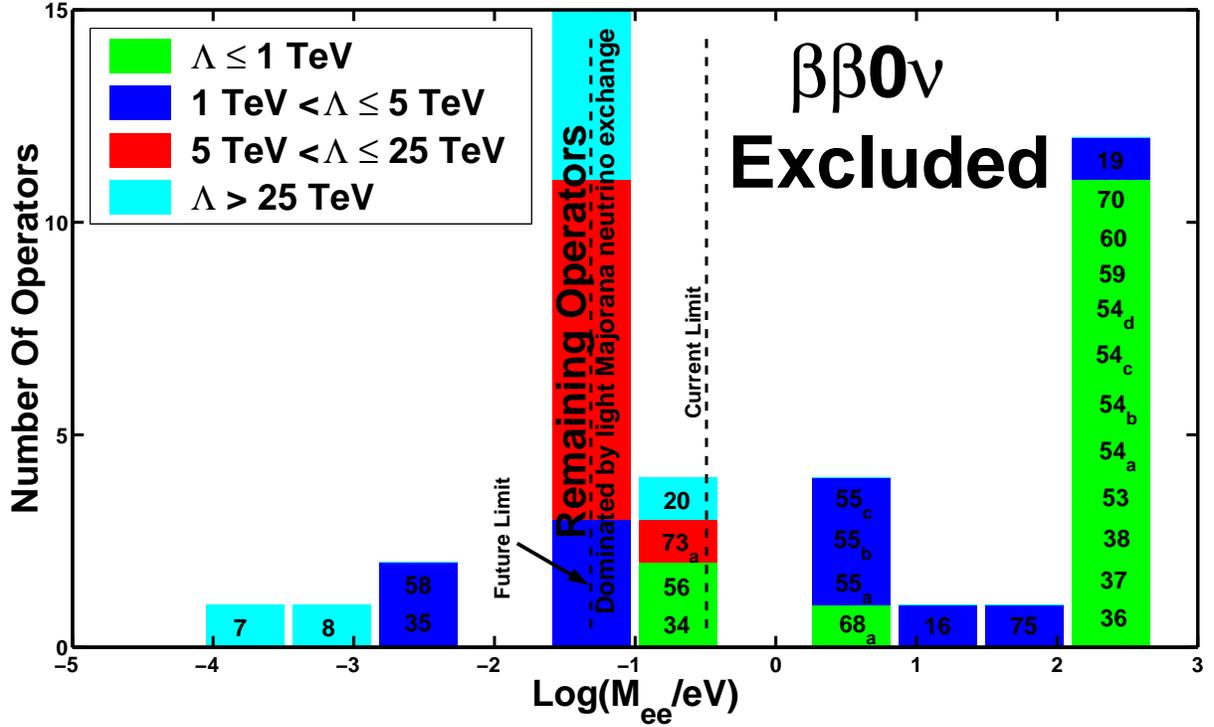}
\caption[$m_{ee}^{\rm eff}$ distribution derived for neutrinoless
double-beta decay process]{$m_{ee}^{\rm eff}$ distribution derived
for the neutrinoless double-beta decay process as described in the
text. The calculations were made assuming the scales $\Lambda_{\nu}$
derived in Sec.~\ref{sec:Scale}, as well as universally order one
coupling constants. The histogram bars are labeled explicitly with
operator names and color-coded by their cutoff scales.  Also shown
in light gray is the region probed by next-generation experiments.
The vertical axis is truncated at 15 operators to best display the
relevant features of the plot.} \label{fig:BBEX}
\end{center}
\end{figure}

The $m_{ee}^{\rm eff}$ distribution extracted from all operators is
shown in Fig.~\ref{fig:BBEX} assuming the scales $\Lambda_{\nu}$
derived in Sec.~\ref{sec:Scale} and color-coded for convenience
within the histogram. Specifically, I indicate in green the
operators that are characterized by sub-TeV scales and thus
accessible to next-generation experiments via direct production. The
blue and red operators are characterized by scales between
$(1-5)~\rm{TeV}$ and $(5-25)~\rm{TeV}$ respectively, where virtual
effects should be most important for collider searches. The majority
of operators, shown in cyan, are suppressed by scales greater than
$25~\rm{TeV}$ and are hence quite difficult to observe in other
search modes.  I also explicitly label each operator within the
histogram bars for easy identification and comparison. One should
notice the expected general trend that increasing $\Lambda_{\nu}$
leads to a decrease in $m_{ee}^{\rm eff}$ and vice-versa. The
vertical axis is truncated at $15$ operators, as the bar near
$0.05~\rm{eV}$, dominated by the light Majorana neutrino exchange
described above, would extend to nearly $100$ operators. With broken
vertical lines, I indicate the current 90\% upper bound \cite{PDG},
 $m_{ee}^{\rm eff} = 0.35~\rm{eV}$,
 and the potential reach of future experiments.

This distribution, which spans over six orders of magnitude (from
$10^{-4}$~eV till $10^{2}$~eV), reveals many important features of
the effective operator set.  Beginning at the largest $m_{ee}^{\rm
eff}$ values, I find that the twelve operators appearing near
$300~\rm{eV}$ all have the expected common feature of low energy
scales, including $\mathcal{O}_{19}$ with $\Lambda_{\nu}$ only just
above the $1~\rm{TeV}$ mark.  Additionally, the contribution of the
majority of these operators to $\beta\beta0\nu$ is dominated by the
tree-level $D_9$ diagram.  The exceptions are
$\mathcal{O}_{54_{c,d}}$ and $\mathcal{O}_{70}$, all of which are
characterized by sub $0.5~\rm{TeV}$ scales and dominated by diagram
$D_6$.  Consequently, these are subject to a loop and
Yukawa/gauge\footnote{As it turns out these are all suppressed by a
single bottom quark Yukawa coupling as well as two powers of the
$SU(2)_L$ gauge coupling $g$, but this fact cannot be deduced from
Fig.~\ref{fig:BBEX} alone.} suppression relative to their $D_9$
dominated cousins, but the difference is not visible given the
resolution of the figure.  It is interesting to note that these
three operators  have the correct quark and lepton content for large
$\beta\beta0\nu$, but their $SU(2)_L$ gauge structures forbid large
tree-level contributions. Similarly, operators $\mathcal{O}_{16}$,
$\mathcal{O}_{55_{a,b,c}}$, $\mathcal{O}_{68_a}$, and
$\mathcal{O}_{75}$ are also dominated by diagram $D_6$ accompanied
by slightly higher cutoff scales.  This drives down $m_{ee}^{\rm
eff}$ significantly considering the leading one-loop scale
suppressions of $\Lambda^{-5}$ and $\Lambda^{-3}$ for the
dimension-eleven and dimension-nine operators respectively.  I point
out that operators $\mathcal{O}_{54_{a,b,c,d}}$ and
$\mathcal{O}_{55_{a,b,c}}$ yield almost identical expressions for
their respective $\beta\beta0\nu$ amplitudes (as well as their
radiatively generated neutrino mass expressions) with up and down
quark Yukawa couplings exchanged. While this action enhances most of
the $\mathcal{O}_{55}$ $\beta\beta0\nu$ couplings relative to those
of $\mathcal{O}_{54}$, it also raises the $\mathcal{O}_{55}$
$\Lambda_{\nu}$ scale by nearly a factor of four and thus drives
$m_{ee}^{\rm eff}$ down by orders of magnitude.

The remaining operators all predict $m_{ee}^{\rm eff} < 1~\rm{eV}$,
close to current experimental bounds. The histogram bar near
$0.1~\rm{eV}$ is composed of operators of very different
$\Lambda_{\nu}$ scales.  $\mathcal{O}_{34}$ and $\mathcal{O}_{56}$
are both characterized by low cutoff energy scales around
$0.5~\rm{TeV}$, but, due to their fermion and helicity structure,
their contributions to $\beta\beta0\nu$ are dominated by two-loop
versions of diagram $D_6$. The neutrino-mass-required cutoff for
$\mathcal{O}_{73_a}$ is around $7~\rm{TeV}$ and its contribution to
$\beta\beta0\nu$ is also dominated by diagram $D_6$. In this case,
however, the two-loop version turns out to be larger than the
allowed one-loop amplitude due to strong scale suppressions (the
added loop reduces the cutoff dependency from $\Lambda^{-5}$ to
$\Lambda^{-3}$). This behavior is characteristic of operators with a
larger value of $\Lambda_{\nu}$.  The $\Lambda_{\nu} = 40~\rm{TeV}$
operator $\mathcal{O}_{20}$, defines the lower edge of this
histogram bar.  It is dominated by the one-loop diagram $D_6$
enhanced by a top quark Yukawa coupling and, being a dimension nine
operator, is only suppressed by $\Lambda^{-3}$ from the start. The
next bar down contains operators dominated by $D_\nu$.  Most of
these are suppressed by a very high energy scale, but a small subset
is characterized by scales $\Lambda_{\nu} < 25~\rm{TeV}$.  In
particular operators $\mathcal{O}_{17}$, $\mathcal{O}_{18}$ and
$\mathcal{O}_{57}$ are all cutoff at $2~\rm{TeV}$ but, due to their
fermion content they cannot participate in any of the non-standard
interactions of Fig.~\ref{fig:Bb0nOps} at a low enough order in
perturbation theory.  Similarly, the intermediately scaled operators
$\mathcal{O}_{43_{a,b,c}}$, $\mathcal{O}_{50}$, $\mathcal{O}_{52}$,
$\mathcal{O}_{62}$, $\mathcal{O}_{65}$, and $\mathcal{O}_{69_b}$
have either the wrong fermion content or gauge structure to enhance
any of the $\beta\beta0\nu$ diagrams (other than $D_\nu$) to an
observable level.  These operators are important because their
minimal forms are experimentally unconstrained yet still potentially
observable to both next-generation $\beta\beta0\nu$ and collider
experiments. The remaining histogram bars with $m_{ee}^{\rm eff} <
10^{-2}~\rm{eV}$ are not accessible to $\beta\beta0\nu$ experiments
in the foreseeable future. Each of these diagrams are dominated by
$D_\nu$, either due to high suppression scales as in the case of
$\mathcal{O}_{7}$ and $\mathcal{O}_{8}$, or, as in
$\mathcal{O}_{35}$ and $\mathcal{O}_{58}$, the operator's fermion
content simply disfavors other contributions to the $\beta\beta0\nu$
amplitude. It is the general form of the neutrino mass matrix
derived in Table \ref{tab:AllOps}, where I see that $m_{ee}\propto
y_e$, that drives these operators away from their peers near
$m_{ee}^{\rm eff} = 0.05~\rm{eV}$.  It is unfortunate that the two
``low'' dimensionality operators $\mathcal{O}_{7}$ and
$\mathcal{O}_{8}$ are cutoff by energy scales $\Lambda_{\nu}$ in
excess of $100~\rm{TeV}$ and are hence invisible to any direct
probe. If either of these operators have anything to do with nature,
it is unlikely that LNV will be observed in the foreseeable future
in \emph{any} experiment.  On the other hand,  any observation of
LNV will rule out these types of scenarios. Additionally, as will
become clear shortly in Sec.~\ref{sec:Oscillations}, current
neutrino oscillation data already marginally disfavor such operators
and have ample room to tighten constraints in the near future.

It is interesting to point out that the lower boundary of the
currently excluded region falls within the $m_{ee}^{\rm eff}$
distribution, suggesting exciting prospects for the future.  That
being said, one should not read too much into current and future
null results as, for most operators, relatively small cancelations
and order one factors, not accounted for here, can push the relevant
rates below the observable level depending on the underlying
ultraviolet theory. On the other hand, one is allowed to interpret
that operators that lead to  $m_{ee}^{\rm eff} \gtrsim 10$~eV are
severely constrained (if not ruled out) as proper explanations for
neutrino masses if one assumes the new physics to be flavor
``indifferent'' -- order one factors cannot be evoked to save the
scenario. Once this assumption is dropped, however, it is quite easy
to ``fix'' these scenarios, since the large $\beta\beta0\nu$ rate is
a direct consequence of the universal order one couplings and the
relatively low cutoff energy scale $\Lambda_{\nu}$.  For example,
one can suppress the coupling of new physics to first generation
fermions (compared to second and third generation fermions), thereby
suppressing the worrisome diagrams of Fig.~\ref{fig:Bb0nOps}. This
will have little effect on the relation between $\Lambda$ and the
neutrino masses, discussed in Sec.~\ref{sec:Scale}, since these are
either generation independent or highly reliant on third generation
Yukawa couplings. Of course, by combining $\beta\beta0\nu$ searches
with other probes I can obtain a much better idea of the origins of
LNV as well as the relevant model(s), if any, chosen by nature.

\setcounter{footnote}{0}
\subsection{Other Rare LNV Decay Processes}
\label{subsec:OtherProcesses}

Most of the qualitative discussions of Sec.~\ref{subsec:bb0nu},
devoted to $\beta\beta0\nu$, can be directly applied to other rare
decay processes with the same underlying kernel interaction
described by Eq.~(\ref{eq:GoldenCh}). For such processes one need
only analyze simple variants of the diagrams listed in
Fig.~\ref{fig:Bb0nOps}, using crossing amplitude symmetries to
account for the needed initial and final state fermions. Other
factors must be added to the various electroweak vertices to account
for quark flavor mixing. The requisite CKM matrix elements can
highly suppress many diagrams for processes involving
cross-generational quark couplings.  In fact, only tree-level $D_9$
diagrams are  safe from such suppressions. Next, and most
importantly, one must include the appropriate characteristic
momentum transfer $Q$ of the new system. Specific rates are highly
dependent on this quantity as effective operator cross-sections
typically grow with some power of $Q$. The particular exponent of
the power law depends on the  diagram, but naive dimensional
analysis dictates that $\Gamma \propto Q^{12}$ for diagram $D_9$,
rendering it highly dependent on a reaction's energy transfer.  The
fact that each diagram varies with $Q$ in a different way implies
that predicting the dominant contributions to a given process is
non-trivial and must be addressed quantitatively.
 Finally, in the cases of hadronic decays, one must also
account for initial/final state matrix elements.  I assume that all
factors can be simply estimated on dimensional grounds.

Unlike the $\beta\beta0\nu$ case, some meson decay modes proceeding
via new LNV tensor interactions are expected to be suppressed.  Such
processes are one instance in our analysis where an operator's
Lorentz structure can qualitatively affect expected LNV decay rates.
One can understand this by considering a meson decay mediated by a
new tensor particle. The parton level interaction has the form
$(\bar{u}\sigma_{\mu\nu}d) T^{\mu\nu}$ where the initial state
quarks are explicitly shown and all other fields are contained in
the tensor $T^{\mu\nu}$. Following the standard procedure I factor
out the hadronic structure in the form of a free decay constant and
write the amplitude as generally allowed by Lorentz invariance in
terms of the external state's four-momentum. Due to the antisymmetry
of $\sigma_{\mu\nu}$, this amplitude vanishes to first order.
Non-zero contributions to this decay mode must necessarily involve
individual parton momenta and are therefore suppressed relative to
the usual vector-like decay calculations. From this, it is clear
that models of LNV containing tensor couplings will often evade the
predictions and bounds of this section. Tensor operators will
mediate LNV meson decays into more complicated final states (one may
include, say, initial/final state radiation). Associated rates are,
however, subject to additional gauge coupling and phase space
suppression that  tend to further reduce the already tiny LNV rates
beyond any hope of detection.
\begin{figure}
\begin{center}
\includegraphics[scale=0.8]{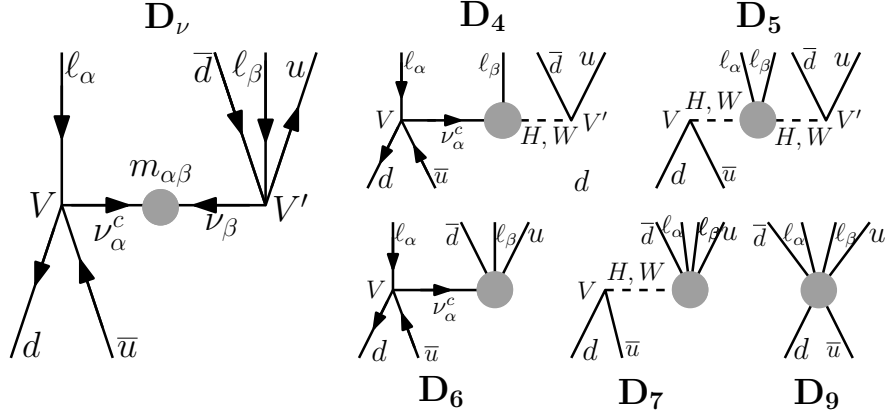}
\caption[Parton level Feynman diagrams contributing to LNV meson
decay]{The parton level Feynman diagrams contributing to rare LNV
meson decay labeled by the dimension of the underlying lepton number
violating interaction, indicated by gray dots.  Each diagram is
generated, at some order in perturbation theory, by all analyzed
interaction, but estimates of their magnitudes depend heavily on the
details of the operators, including their associated scale
$\Lambda_{\nu}$, fermion content, and helicity structure.}
\label{fig:MesonOps}
\end{center}
\end{figure}

Rare LNV meson decays have been experimentally pursued for many
years \cite{PDG}.  Here, I focus on the $\Delta L=2$ processes
$M^\prime \rightarrow M + \ell^{\pm}_{\alpha} \ell^{\pm}_{\beta}$,
where $M^\prime$ and $M$ are the initial and final states mesons
respectively and the $\ell $s represent like-sign lepton pairs of
arbitrary flavor. Electric charge conservation dictates that
$M^\prime$ and $M$  have equal and opposite charge.  Here I take
each meson to consist of a color singlet up-type/antidown-type bound
state\footnote{For simplicity I assume that both the process
$M^\prime \rightarrow M + \ell _\alpha \ell _\beta $ and its
conjugate have similar amplitudes and therefore treat them
symmetrically.  Large CP-violating effects can invalidate this
assumption.} and factor out all long distance hadronic effects.  In
this way I can view the meson decay process as $d\bar{u} \rightarrow
\ell _\alpha \ell _\beta + \bar{d}u$ for all up-type and down-type
quark flavor combinations.  The effective LNV diagrams contributing
to this process are shown in Fig. \ref{fig:MesonOps} with the same
naming scheme as their analogs in Fig.~\ref{fig:Bb0nOps}.  Here, $V$
and $V^\prime$ denote potentially distinct elements of the quark
mixing matrix.  I additionally point out the potential dependency on
all entries of the Majorana neutrino mass matrix elements
$m_{\alpha\beta}$ in diagram $D_\nu$, as opposed to the
$\beta\beta0\nu$ case where $D_{\nu}$ depends only on $m_{ee}$.
These processes probe combinations of the neutrino masses that are
naively unconstrained by $\beta\beta0\nu$ \cite{Flanz:1999ah}. In
general, the varied flavor structures encountered in meson decays
allow for experimental probes into new physics couplings across the
fermion generations. I pointed out earlier that some of the LNV
operators lead to unacceptably large rates for $\beta\beta0\nu$
unless first generation quarks participate in the new interactions
with severely suppressed couplings (compared with second and third
generation quarks). If such a scenario is realized in nature, rare
$D$ or $B$ decays may be much more frequent than naive expectations.
 For
this reason, improving rare decay sensitivities to all channels is
essential to completely constrain models of new LNV physics beyond
the minimal framework analyzed here.

\begin{figure}
%\begin{center}
\includegraphics[scale=.74]{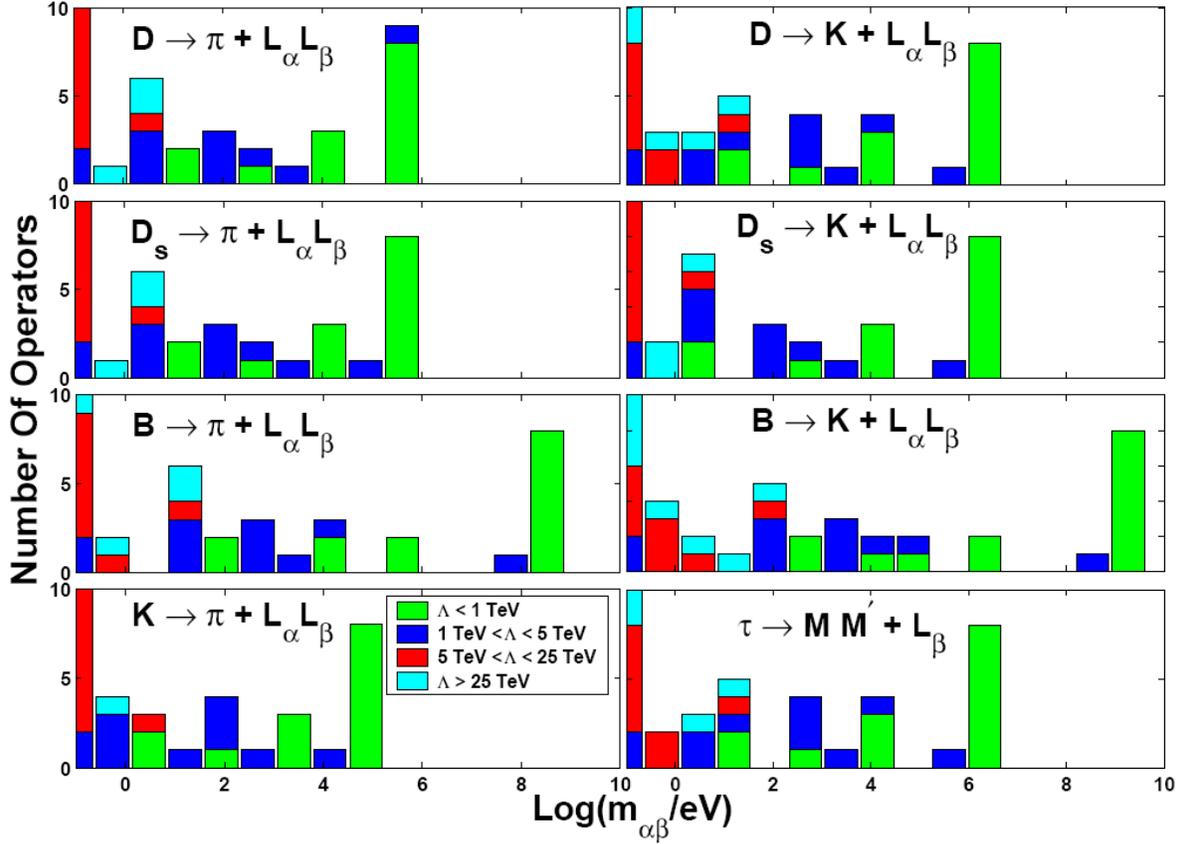}
\caption[$m_{\alpha\beta}^{\rm eff}$ distribution derived for
several rare LNV meson and $\tau$ decays]{$m_{\alpha\beta}^{\rm
eff}$ distribution for several rare LNV meson and $\tau$ decays.
Calculations assumed the charge lepton flavors $\ell _\alpha\ell
_\beta = \mu e$, while the $\tau$ decay histogram (lower right-hand
panel) was obtained assuming the final state mesons $M M^\prime =
KK$. The histogram bars are color-coded by suppression scale.
Current bounds on these processes are typically above $1~\rm{TeV}$
and are
 not visible at these small scales.} \label{fig:Meson}
%\end{center}
\end{figure}

Reference \cite{LNVUpperBounds} summarizes LNV upper bounds on all
of these processes in terms of the effective Majorana neutrino mass
matrix element $m_{\alpha\beta}^{\rm eff}$ that one would extract
from observation assuming that all decay rates are dominated by the
light neutrino exchange shown in $D_\nu$.  Hence, I can compare
operator expectations with current experimental limits in exactly
the same way as was done in Sec.~\ref{subsec:bb0nu}.  For a given
LNV meson decay, $m^{\rm eff}_{\alpha\beta}$ is defined from the
contribution of the different classes of diagrams to the rare meson
decay in question, exactly as $m_{ee}^{\rm eff}$ was defined in the
previous subsection (see Eqs.(\ref{eq:ad},\ref{eq:MeeEff})). Direct
estimates for  different  process reveal $m^{\rm eff}$ distributions
similar to that for $m_{ee}^{\rm eff}$ depicted Fig.~\ref{fig:BBEX},
up to ``rescalings'' that reflect the different kinematics and the
presence of small CKM mixing matrix elements. Results are summarized
in Fig.~\ref{fig:Meson} for a representative sample of charged meson
decays. Each histogram is labeled by its associated decay mode and
is color-coded to indicate the neutrino-mass constrained cutoff
scale $\Lambda_{\nu}$ of the different LNV effective operators.  For
simplicity, I refrain from listing operator names on the individual
histogram bars (as opposed to what was done in Fig.~\ref{fig:BBEX}).
The ``operator ordering'' is very similar to that of
Fig.~\ref{fig:BBEX}, especially in the low $\Lambda_{\nu}$ scale,
high effective mass regime where decay rate predictions are
particularly important. Note that the horizontal
 axes are relatively fixed for easy comparison and that the vertical direction
is truncated and does not reflect the true ``height'' of the lowest
mass bar (order one hundred operators).

Specifically, I present effective Majorana neutrino mass
distributions for the processes, reading down the panels from left
to right, $D\rightarrow \pi + \ell^{\pm}_{\alpha}
\ell^{\pm}_{\beta}$, $D\rightarrow K + \ell^{\pm}_{\alpha}
\ell^{\pm}_{\beta}$, $D_s \rightarrow \pi + \ell^{\pm}_{\alpha}
\ell^{\pm}_{\beta}$, $D_s \rightarrow K + \ell^{\pm}_\alpha
\ell^{\pm} _\beta$, $B\rightarrow \pi + \ell^{\pm}_{\alpha}
\ell^{\pm}_{\beta}$, $B\rightarrow K + \ell^{\pm}_{\alpha}
\ell^{\pm}_{\beta}$, $K\rightarrow \pi + \ell^{\pm}_{\alpha}
\ell^{\pm}_{\beta}$, as well as the rare $\tau$ decay, $\tau^{\pm}
\rightarrow M M^\prime + \ell^{\mp}_{\beta}$.\footnote{The actual
calculations displayed in Fig.~\ref{fig:Meson} assumed the charge
lepton flavors $\ell _\alpha\ell _\beta = \mu e$, while the $\tau$
decay histogram (lower right-hand panel) was produced assuming the
final state mesons $M M^\prime = KK$.}  Here the final state leptons
can be of any flavor allowed by energy conservation. Since, as
previously discussed and explicitly verified numerically, the
specific details of the distributions are mainly dictated by
kinematics and CKM matrix elements, these results are robust under
changes in the final state lepton flavors. The $\tau$ decay
distribution shown in the lower right panel is representative of all
possible decay products including first and second generation
charged leptons and light meson states. One should notice the
expected general operator trend within each histogram as the
characteristic cutoff scale is decreased, as well as the expected
peaks near $0.05~\rm{eV}$ dominated by light Majorana neutrino
exchange. Additionally, each distribution is much ``broader'' than
the one in Fig.~\ref{fig:BBEX}. This observation exemplifies the
fact that effective mass calculations depend critically on the
underlying process. Indeed, maximum $m_{ee}^{\rm eff}$ values can
reach nearly $10^{10}~\rm{eV}$ for the $B^+ \rightarrow K^- +
e^+e^+$ decay but only $10^3~\rm{eV}$ for $\beta\beta0\nu$. Current
upper bounds for $m^{\rm eff}$ from these processes, mostly well
above one TeV, are well beyond the largest operator predictions
here, ranging from $m_{e\mu}^{\rm eff} < 0.09~{\rm TeV}$ for the
case of $K^+ \rightarrow \pi^- e^+ \mu^+$ to $m_{\mu\mu}^{\rm eff} <
1800~{\rm TeV}$ for the case of $B^+ \rightarrow K^- \mu^+ \mu^+$
\cite{LNVUpperBounds}.  It is curious that the best meson decay
bounds come from rare LNV kaon process but, as can be seen in the
lower left panel of Fig.~\ref{fig:Meson}, these yield by far the
lowest predictions. Future experiments have the potential for
observing LNV for a select few operators only provided vast
improvements in meson production luminosities. Current and upgraded
B-factories \cite{BFactories} are expected to provide the most
significant improvements, considering the large derived $B$-meson
effective masses shown in Fig.~\ref{fig:Meson}.  Still, the best
cases from the figure yield only the tiny branching fraction
$1.8\times 10^{-17}$ for the case of the rare decay $B^+ \rightarrow
\pi^- e^+ \mu^+$, nearly eleven orders of magnitude below the
current experimental limit of $1.3\times 10^{-6}$ \cite{PDG}.

Another possible search mode involves the decay of the $Z$-boson
into LNV final states.  The dominant contributions to this process
are generally unrelated to the reactions summarized in
Eq.~(\ref{eq:GoldenCh}) and shown schematically in
Figs.~\ref{fig:Bb0nOps} and \ref{fig:MesonOps}.  While there is a
slight connection between them as one can always attach a $Z$-boson
to various fermion lines in each diagram, there are potentially
large lower order contributions arising within the operators
themselves. The latter, when present, can easily overtake the
associated ``golden mode'' counterparts.  In this context, such
processes can be thought of as the decay of the longitudinally
polarized $Z$-boson. Strict bounds exist on such decays from the
LEP-I \cite{LEP} and SLD \cite{SLD} experiments. Each element of the
operator set predicts decays into final state fermions with total
lepton number $L=2$.  The dilepton pair can be of any flavor and is
generally accompanied by two or four additional fermion states,
depending on the dimension of the operator.  I restrict our
discussion to the dimension-eleven operators comprising the majority
of the sample, as these are typically suppressed by lower cutoff
$\Lambda_{\nu}$ scales and, equally important, explicitly contain
Higgs doublets in their field content.  In this case, tree-level
decays result in a six-fermion final state which suffers from a
large phase space suppression and cumbersome multiplicities that are
likely to render even the most sophisticated search ineffective. The
only possibility of this type that yields a charged dilepton signal
is $Z \rightarrow \ell^{\pm}_\alpha\ell^{\pm}_\beta
q\bar{q}q\bar{q}$ (quarks of all allowed flavors implied), but many
other possibilities exist involving invisible final state neutrinos.
A little thought also reveals that closing fermion loops in an
attempt to obtain simpler final states and thus render the analysis
more tractable will necessarily result in final state neutrinos.
Therefore, the majority of the $Z$-boson LNV decay channels involve
invisible final states with practically undetermined total lepton
number.  The prospect of direct discovery by these means seems
dismal, but indirect constraints on LNV are still possible from
bounds on the $Z$-boson invisible decay width. There is currently a
statistically insignificant, but nonetheless captivating, $2\sigma$
deviation between the observed invisible decay width and SM
expectations assuming three light neutrino species \cite{PDG}. The
experimentally extracted branching ratio was found to be slightly
\emph{smaller} than its predicted value so that a new LNV
contribution of the form $Z \rightarrow \nu_\alpha \nu_\beta$ would
push the invisible branching ratio in the ``wrong'' direction. From
these bounds the decay width of any new contribution to the
$Z$-boson decay is constrained to be less than $2.0~\rm{MeV}$ at the
$95\%$ confidence level \cite{PDG,ZInvBound}. A quick estimate
reveals that this constrains the dominant LNV amplitudes $A_Z <
\sqrt{4\pi(2.0~{\rm MeV}/M_Z}) \sim 0.53$.  For the dimension-eleven
operators of interest, the largest possible amplitude is of order
$y^2/(16\pi^2)^2(v/\Lambda)^3$ where $y$ is an arbitrary fermion
Yukawa coupling and four powers of the cutoff scale $\Lambda$ are
removed by divergences in the closed diagram loops. The constraint
above translates into $y^2(v/\Lambda)^3 < 4.1\times 10^2$, which is
easily evaded by even the best case scenario of $y = y_{t} \approx
1$ and $\Lambda \approx v$.  Experimental bounds on $\Gamma_{inv}$
must be improved by a factor of a million before they start
significantly constraining LNV (under the assumptions made here).
This result holds for virtually all possible flavor structures. I
conclude that rare $Z$-boson decays are not practical discovery
modes for the LNV effects considered here, but look to future rare
$Z$-boson decay
 studies for more information.

In a similar way, one can also dismiss the case of rare $W$-boson
decays as promising probes of LNV.  As in the $Z$-boson case, the
$W$-boson can decay into a variety of $L = 2$ final states
proceeding either through couplings to left-handed fermion lines or
explicit operator content.  Here, however, there is no six-fermion,
same-sign dilepton final state with no neutrinos due to conservation
of charge and weak isospin, so the lowest order observable mode is
already loop suppressed to $W^- \rightarrow \ell ^-_\alpha\ell
^-_\beta + q\bar{q}$. Current $W$-boson decay bounds are far too
weak to constrain such suppressed LNV \cite{PDG} and are not likely
to improve to the level implied by the operators under
consideration, which predict the tiny decay rate $\Gamma_{LNV} \leq
m_W(4\pi)/(16\pi^2)^5(v/\Lambda)^{10} \approx 10^{-5}~\rm{MeV}$ in
the best case scenario of electroweak scale $\Lambda_{\nu}$.  I also
point out that, contrary to the $Z$-boson decay limits, there are no
robust, indirect bounds that can be used to constrain LNV in the
case of the $W$-boson.  Note that, despite dismal prospects for
gauge boson decay driven LNV discovery within the minimal framework
of ``natural'' effective operators, one can still construct
theoretically well-motivated models that will yield observable
signals. Particularly, in a weak-scale seesaw mechanism
($\mathcal{O}_1$), the new degrees of freedom, comprised mostly of
Majorana gauge singlet fermions (right-handed neutrinos), can
mediate visible, $\Delta L=2$, $W$-boson mediated processes
 with little or no scale/loop suppression.  This class of model
is analyzed in \cite{WDecay} and is exempt from the discussion
outlined here.

\subsection{Collider LNV Signatures}
\label{subsec:Collider}

If neutrino masses are a consequence of ultraviolet physics related
to cutoff scales around the TeV scale, I expect future high energy
collider searches to directly access the new LNV physics. For
example, the direct, resonant, production of new states could lead
to rather spectacular signals of these models. It would also
indicate the breakdown of the effective field theory approach
undertaken here. To pursue such possibilities, one must assume a
specific ultraviolet sector and study its signatures and
implications on a case by case basis.  In the looming shadow of the
LHC, \cite{LHC} and the more distant ILC \cite{ILC}, such an
analysis is highly warranted but will not be pursued here. Instead,
I assume that the masses of new ultraviolet degrees of freedom
remain out of the reach of next-generation accelerator experiments.
Such a situation can be easily accommodated within the context of
the preceding results, considering the order of magnitude nature of
the $\Lambda_{\nu}$ estimates.

I will concentrate on the process $e^-e^-\rightarrow
q\bar{q}q\bar{q}$ (which will usually manifest themselves as jets)
with no missing energy in an ILC-like environment \cite{ILC} with a
center-of-mass energy of $1~\rm{TeV}$ and an integrated luminosity
of $100~\rm{fb}^{-1}$.  I also make the oversimplifying  assumption
that the detector system has equal acceptance to all quark flavors,
and the ability to efficiently distinguish quarks, gluons and
$\tau$s.  By summing over all possible quark final states it is
simple to estimate the total LNV cross section for each effective
operator, assuming it is responsible for neutrino masses. Such
searches can be complemented  by looking at $e^-e^-\to W^-W^-$,
which have been discussed in detail in the literature
\cite{EmEm2WmWm}.  As discussed in Sec.~\ref{subsec:OtherProcesses},
the different  LNV operators couple to one or more gauge bosons via
an appropriately closed fermion loop or direct coupling to the Higgs
doublet field.

Charge and baryon number conservation dictate that the two quarks in
$e^-e^-\rightarrow q\bar{q}q\bar{q}$ are down-type quarks, while the
two antiquarks are up-type antiquarks. At the parton level, the
scattering process is similar to
 $\beta\beta0\nu$, which motivates exploiting simple
variations of the diagrams in Fig.~\ref{fig:Bb0nOps} in order to
calculate the relevant amplitudes, as was done in
Sec.~\ref{subsec:OtherProcesses}.  Here, the extensions are obvious:
use crossing symmetry to rotate all lepton lines into the initial
state and all quark lines to the final state taking special care to
insert appropriate CKM matrix elements where needed.  Due to the
large characteristic momentum transfer $Q$ of the $e^-e^-$
scattering, one must also ``expand'' the electroweak vertices and
account for gauge boson propagation.  With this in mind, the
amplitude calculations can be carried over directly from the
previous sections. Specific results are, however, quite distinct due
to the higher center-of-mass energies involved.  In the language of
the underlying diagrams mediating this reaction, for diagrams
characterized by TeV cutoff scales, diagram $D_9$, if allowed at
tree-level, will dominate the rates.  As in the previous cases, for
intermediate to high cutoff scales, general diagram dominance must
be addressed on a case by case basis.
 It is important to appreciate that, since these are non-renormalizable
effective interactions, cross-sections grow with center-of-mass
energy. For this reason, I expect many of the low cutoff scale
operators to yield observably large signals at the ILC.

\begin{figure}
\begin{center}
\includegraphics[angle=270,scale=.57]{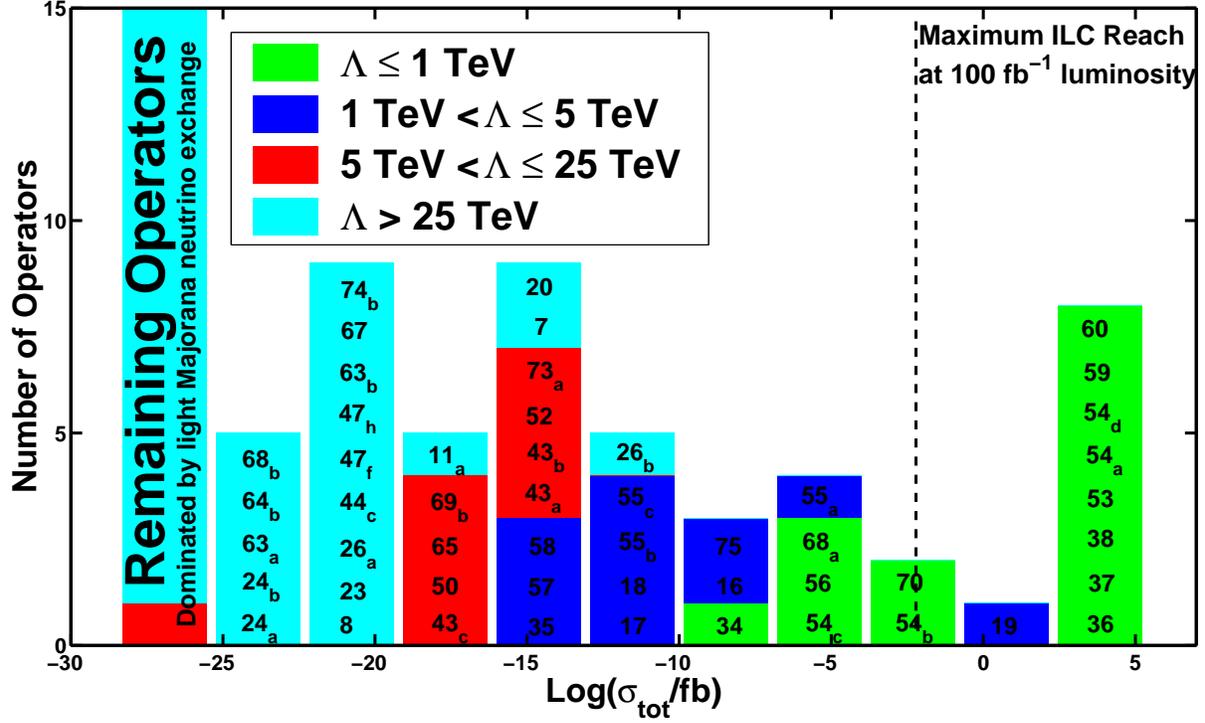}
\caption[Cross-section distribution for $e^-e^-\rightarrow
q\bar{q}q\bar{q}$ at an $e^-e^-$ collider]{Distribution of total
cross-section for the process $e^-e^-\rightarrow q\bar{q}q\bar{q}$
and no missing energy at an $e^-e^-$ collider with $1~\rm{TeV}$ of
center-of-mass energy. Estimates were obtained assuming the scales
$\Lambda_{\nu}$ derived in Sec.~\ref{sec:Scale}, as well as order
one coupling constants. The histogram bars are labeled with operator
names and color-coded by $\Lambda_{\nu}$ cutoff scale.  Also shown
(broken vertical line) is the reach of such an experiment assuming
$100~\rm{fb}^{-1}$ of integrated luminosity. The vertical axis is
truncated to best display the relevant features of the plot.}
\label{fig:ILC}
\end{center}
\end{figure}

Fig.~\ref{fig:ILC} shows the $e^-e^- \rightarrow q\bar{q}q\bar{q}$
cross-section distribution, in femtobarns, at the ILC, calculated
for all 129 of the analyzed LNV operators. Once again, the extracted
value of the cutoff energy scale $\Lambda_{\nu}$ assuming
constraints from neutrino masses are color-coded to indicate
operators associated with a low ($\Lambda_{\nu}\lesssim 10$~TeV) or
high ($\Lambda_{\nu}\gtrsim 10$~TeV)  ultraviolet cutoff. Each bar
is also labeled with the respective constituent operators, for
convenience. Note that the vertical axis is truncated at fifteen
operators (the left-most bin is over 60 operators high) to help
clearly display relevant features of the plot. I also highlight the
potential reach (defined as cross-section greater than the inverse
of the integrated luminosity) of the ILC with a broken vertical
line, assuming $100~\rm{fb}^{-1}$ of integrated luminosity. This
particular ILC luminosity value should be considered as a loose
lower bound, introduced to give a feeling for the observable scales
involved.  It has recently been argued, for example, that a
realistic machine should be able  to outperform this estimate by
over an order of magnitude \cite{ILC}.

A glance at Fig.~\ref{fig:ILC} reveals that it generally adheres to
the expected correlation of decreasing $\Lambda_{\nu}$ scales with
increasing LNV rates, similar to what is observed for other LNV
observable ({\it e.g.},~Fig.~\ref{fig:BBEX}). The similarities
between the different processes extend beyond mere trends to the
specific ordering of the operators within each histogram. This
reflects the common underlying interactions that drive these
processes.  The operators on the far right of the plot, topping off
the highest cross-sections, are exactly those operators with the
largest $m_{ee}^{\rm eff}$, now ``split'' into three different bars.
The large bar just below $10^5$~fb is composed of sub-TeV scale
operators with tree-level diagram $D_9$-like fermion content.
Slightly smaller are the expectations for $\mathcal{O}_{19}$, again
dominated by diagram $D_9$, but characterized by a slightly larger
$\Lambda_{\nu}$ scale (around one TeV). Moving down in
cross-section, this is followed by the low cutoff scale operators
$\mathcal{O}_{54_{b,c}}$ and $\mathcal{O}_{70}$, dominated by a
combination of diagrams $D_6$ and $D_7$. On the opposite end of the
plot I point out the large bar below $10^{-25}$~fb, composed mainly
of operators associated to high cutoff scales ($\Lambda_{\nu} >
25~\rm{TeV}$).  The contributions of these operators are dominated
by light Majorana neutrino exchange, but their histogram bar contain
far fewer models than their $\beta\beta0\nu$ counterpart, as many of
the latter have been driven up due to new diagram $D_4$ and $D_5$
contributions.  In general, the large center-of-mass energies tend
to magnify differences between interaction rates that were not
relevant in low-energy observables. This naively suggests that high
energy probes have a higher potential for distinguishing different
models.

There are eleven operators that lead to an observably  large (as
defined earlier) $e^-e^-\to q\bar{q}q\bar{q}$ cross-section at the
ILC. Note that all of these were already ``ruled out'' by current
$\beta\beta0\nu$ searches. As discussed in Sec.~\ref{subsec:bb0nu},
however,  these bounds only effectively limit the couplings of the
new physics to first generation of quarks and leptons, and hence, if
such a scenario is realized in nature, one should still expect large
contributions from decay modes that lead to second and third
generation final state quarks.  In fact, even one such heavy quark
is enough to bypass the constraints from $\beta\beta0\nu$ for
several effective operators.  Such reasoning implies that
constraints on the new physics flavor structure can be made quite
strong at a linear collider via analyzes of the flavor of the final
state quarks.  By identifying and comparing the outgoing quark
flavor one can extract individual limits on quark-lepton coupling
constants within the operators. Additionally, kinematics can be used
as a further operator probe.  For example, one can potentially
determine the dominant underlying LNV diagram (say $D_6$, $D_7$ or
$D_9$) by checking whether the  various kinematic distributions are
characteristic of $W$-boson exchange.

The ILC can cleanly select or discard some LNV scenarios.  This
characteristic is further enhanced by considerations of initial
electron polarization.  Planned linear colliders have the ability to
produce partially polarized beams (80\% polarization for $e^-$, 40\%
for $e^+$  \cite{ILC,ILCPol}). The power of a high energy polarized
$e^-e^-$ beam is in model identification and rejection. Of all
operators that yield observably large cross-sections, the
$e_L^-e_L^-$ mode can only probe $\mathcal{O}_{53}$, and therefore
any positive LNV signal cleanly identifies this as the operator
chosen by nature. In a similar way, the ILC running in its
$e_L^-e_R^-$ mode can easily observe LNV from $\mathcal{O}_{19}$,
$\mathcal{O}_{54_a}$, $\mathcal{O}_{54_d}$, $\mathcal{O}_{59}$ and
$\mathcal{O}_{60}$; and to a lesser extent, operators
$\mathcal{O}_{54_b}$ and $\mathcal{O}_{70}$, and possibly even
$\mathcal{O}_{54_c}$. Finally the $e_R^-e_R^-$ mode can probe
operators $\mathcal{O}_{36}$, $\mathcal{O}_{37}$ and
$\mathcal{O}_{38}$.  Within this framework, any LNV detected in one
ILC polarization mode will generally not be seen in the others. This
statement also applies to resonantly enhanced low scale operators
that lie outside the observability window.

While $e^-e^-$ collisions only probe effective operators that
``talk'' to first generation leptons, there are several lepton
collider processes that allow one to explore other members of the
charged lepton family. Future high energy muon colliders \cite{mumu}
could, in principle, also be used to study LNV.  In this case, all
of the preceding  discussions regarding the ILC are applicable.
Electron linear collider facilities can also be used to study
$\gamma e^-$ and $\gamma\gamma$ collisions \cite{gammagamma}.
$\gamma e^-$ collisions can be used to probe $\gamma e^-\to
\ell_{\alpha}^++X$ (and hence the ``$e\alpha$'' structure of
different LNV operators), while $\gamma\gamma\to
\ell_{\alpha}^{\pm}\ell_{\alpha}^{\pm}+X$ probes all the different
$\alpha,\beta$ charged lepton flavors. For $\gamma\gamma$
collisions, for example, considering projected ILC-like collider
parameters, one would expect the same operator distribution as
Fig.~\ref{fig:ILC}, shifted down in cross-section by, roughly,  a
factor of $\alpha^2 \sim 10^{-4}$.  Thus, a handful of operators
should be testable at a future $\gamma\gamma$ collider assuming
$100~\rm{fb}^{-1}$ of integrated luminosity.

\begin{figure}
\begin{center}
\includegraphics[scale=.7]{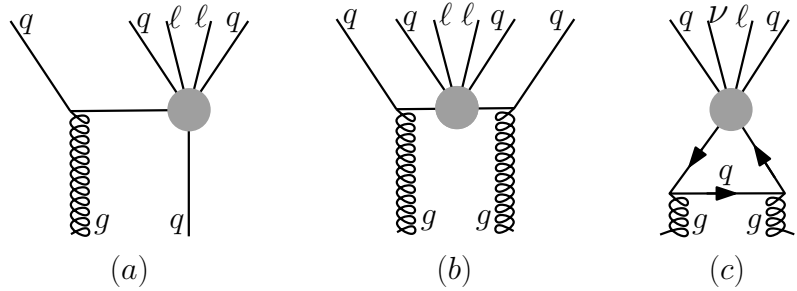}
\caption[Parton level gluon--gluon and gluon--quark LNV interaction
diagrams]{Parton level gluon--gluon and gluon--quark LNV
interactions relevant at high energy hadron colliders.  Each of
these yields a same sign dilepton signal with jets and no missing
energy.  Notice that the final state flavor structure is completely
arbitrary under the assumption of random order one coupling
constants.} \label{fig:LHC}
\end{center}
\end{figure}

The preceding analyses carry over to the case of  hadron colliders,
such as the LHC, in a relatively straightforward way.  The LHC, or
Large Hadron Collider, is a proton--proton machine that will operate
at a center-of-mass energy of $14~\rm{TeV}$ and a characteristic
integrated luminosity around $100~\rm{fb}^{-1}$ \cite{LHC} (in its
high luminosity mode). The relevant LNV variants of
Eq.~(\ref{eq:GoldenCh}) are $d d \rightarrow \ell_{\alpha} ^-
\ell_{\beta} ^- uu$ and $u u \rightarrow \ell_{\alpha} ^+ \ell_\beta
^+dd$ with no missing energy. Of course, at center-of-mass energies
well above a TeV, the proton--proton collisions are
 dominated by the gluon content of the proton, so most interactions at the LHC will
be initiated by gluon--gluon and gluon--quark scattering. The
dominant  LNV subprocesses are $q g \rightarrow
\ell^{\pm}_\alpha\ell^{\pm} _\beta q\bar{q}q$ and $gg \rightarrow
\ell^{\pm} _\alpha\ell^{\pm}_\beta q\bar{q}q\bar{q}$  and are
illustrated in diagrams $(a)$ and $(b)$ of Fig.~\ref{fig:LHC},
respectively. These are characterized by similar final states as the
quark--quark scattering reactions but, given that there is no
explicit gauge boson field content in the LNV operators in question
(Table \ref{tab:AllOps}), their amplitudes are  proportional to
unimportant order $\alpha_s$ and $\alpha_s^2$ coefficients,
respectively.  The parton level diagram $(c)$ shows the related
process $gg \rightarrow \ell _\alpha \nu _\beta + q\bar{q}$.  The
rate for this process can be estimated, relative to its four jet
cousins, by exchanging a final state phase space suppression for a
single loop suppression. In all three diagrams depicted in
Fig.~\ref{fig:LHC}, the LNV interaction regions represented by large
grey dots  contain all of the diagrams discussed earlier, meaning
that the operator amplitudes calculated for the ILC can be recycled
in this analysis. While all three bare diagrams are characterized by
rates of the same order of magnitude diagram $(c)$ leads to missing
transverse energy and potentially undetermined final-state lepton
number, rendering it a less than optimal experimental search mode.
Note that, in all of these cases, the external, and internal,
fermions outside of the LNV interaction region can be of any flavor.
Therefore, hadron collider experiments have, in principle,  access
to \emph{all} LNV operator parameters. Cleanly identifying and
constraining all said parameters should prove quite difficult for
all but the most obvious signatures.  The above statements regarding
signals at the LHC are also applicable at the Tevatron with some
minor, but important, modifications.  The Tevatron's $p\overline{p}$
collisions are at a much lower center-of-mass energy, roughly
$2~\rm{TeV}$, while the total expected integrated luminosity, less
than 10~fb$^{-1}$ per experiment, is orders of magnitude smaller.
These factors lead to much lower amplitudes, reduced by
approximately a factor of $\left(Q_{\rm{Tevatron}}/Q_{LHC}\right)^5
\approx 10^{-5}$.\footnote{Strictly speaking one must also account
for the proton's structure functions at the Tevatron's energy scale.
Unlike the LHC, where collisions are dominated by gluon--gluon
interactions, proton collisions at the Tevatron are dominated by
valence quark interactions. These considerations do not affect our
conclusions.} The smaller center-of-mass energy also limits the
Tevatron's ability to directly produce new physics states.  With
this in mind I conclude  that the Tevatron has little or no chance
of discovering LNV (within this minimal framework).

A detailed set of predictions for the LHC would require a much more
refined analysis, including the effects of parton structure
functions, flux distributions, and backgrounds, and as such is
beyond the scope of this general survey. I would, however, like to
point out that some of the reactions outlined here are subject to
large background rates.  While SM processes are lepton number
conserving, many can fake the LNV signals in the complicated
environment of a high energy hadronic interaction. The requirement
of no missing final state energy is particularly hard to accommodate
as some energy is always lost down the beampipe. As is typically
done, one must rely on the less restrictive conservation of
transverse momentum in order to constrain invisible states, such as
neutrinos.  SM same-sign dilepton production processes arising from,
say, $W$-boson pair production, are serious potential sources of
background. Furthermore, it is impossible to predict correlations
among final state jets without selecting a particular operator and
underlying model of new physics, making it difficult to impose
general cuts to reduce other hadronic backgrounds.  Of course, some
of the low scale LNV operators yield large enough total
cross-sections that even crude analyses may suffice to reveal their
existence.  I conclude by pointing out that  a large amount of
recent work has been dedicated to LNV searches at collider
facilities \cite{CollLNV}.  Most of these approach the subject from
the perspective of sub-TeV mass, mostly sterile Majorana neutrinos
that mix with the active neutrinos and are thus  related to light
neutrino masses via the seesaw mechanism \cite{SeeSaw}.  This
amounts to one example that leads to the dimension-five operator
$\mathcal{O}_1$, but where one assumes that the propagating degrees
of freedom are twelve or thirteen orders of magnitude lighter than
the ultraviolet cutoff scale $\Lambda_{\nu}$.\footnote{This can be
achieved in two different ways. Either the new physics is very
weakly coupled, or the new physics -- SM couplings are finely-tuned
\cite{fine_nus}. In order to observe right-handed neutrinos in
colliders, the latter must be realized.}  In this case, LNV
interactions are dominated by diagram $D_\nu$ of Fig.
\ref{fig:Bb0nOps} (where heavy (weak scale) neutrinos are also
exchanged),
 and as such one should make use of
specific kinematic cuts to reduce background rates. These cuts,
however, may also remove LNV signals resulting from many of the
scenarios explored here, particularly those whose rates are
dominated by $D_9$ at tree-level.  I urge experimentalists to
account for this possibility while analyzing future data sets.

\setcounter{footnote}{0} \setcounter{equation}{0}
\section{Neutrino mixing} \label{sec:Oscillations}

Table \ref{tab:AllOps} contains predictions for \emph{all} the
entries $m_{\alpha\beta}$, the Majorana neutrino mass matrix. These
are computed in the weak basis where the weak interactions and the
charged-lepton Yukawa couplings are diagonal, so that the
eigenvalues of the neutrino mass matrix are the neutrino masses
(bounded by oscillation experiments and, say, precision measurements
of tritium beta-decay  \cite{TritNuEffMass}), while its eigenvectors
determine the neutrino mixing matrix, constrained mostly by
oscillation experiments. Since different LNV effective operators
predict different flavor-structures for the neutrino mass matrix,
there is the possibility to constrain the different scenarios with
existing oscillation data \cite{OscBestFit}.  While I can only
predict the values of $m_{\alpha\beta}$ within, at best, an order of
magnitude, it is still possible to extract useful information from
the derived large scale structure of the expressions.  In particular
I can test the hypothesis of whether $\lambda$ values associated to
different lepton flavors are allowed to be of the same order of
magnitude. In order to obtain  more accurate predictions and further
probe the fine details of lepton mixing one must succumb to specific
models, beyond the scope and philosophy of this analysis.

The mass matrix for the three light Majorana neutrinos can be
reconstructed from nine observables: three masses $m_1,m_2,m_3$,
taken to be real and positive; three (real) mixing angles
$\theta_{12},\theta_{23},\theta_{13}$; and three CP-violating phases
$\delta,\phi_2,\phi_3$.  Here, $\delta$ is a so-called Dirac phase
that is generally present in the system regardless of the neutrino's
nature (Majorana or Dirac fermion), while $\phi_1,\phi_2$ are
so-called Majorana phases, only present if the neutrinos are
Majorana particles (which is the case of all scenarios under
consideration here). Oscillation data determine with relatively good
precision
 $\theta_{12}$, $\theta_{23}$,
$\Delta m^2_{12} \equiv m_2^2 - m_1^2$ and $|\Delta m^2_{13}| \equiv
|m_3^2 - m_2^2|$. I define neutrino masses such that $m_1<m_2$  and
$\Delta m^2_{12}<|\Delta m^2_{13}|$, so that the sign of $\Delta
m^2_{13}$ remains as an observables which characterizes the neutrino
mass hierarchy (``normal'' for $\Delta m^2_{13}>0$, ``inverted'' for
$\Delta m^2_{13}<0$). See, for example, \cite{NeutrinoReview} for
details.
 As for the third mixing angle, $\sin^2\theta_{13}$ is constrained
to be less than 0.025 (0.058) at $2\sigma$ $(4\sigma)$ from a three
neutrino global oscillation analysis \cite{OscBestFit}. A
considerable amount of uncertainty remains. In particular I have
only upper bounds on the absolute neutrino mass scale, from
kinematical measurements such as tritium beta decay
\cite{Mainz,Troitsk}, plus cosmological observations
\cite{CosmoSum,Fogli:2006yq,seljak,Hannestad:2006mi}.  Finally, the three CP violating phases are
completely unconstrained, and I have no information regarding the
neutrino mass hierarchy.

The above experimental results allow for several different
``textures'' for $m_{\alpha\beta}$ in our weak basis of choice (see,
for example, \cite{Frigerio:2002fb}). The purpose of this section is
to discuss whether any of the textures predicted by the different
LNV effective operators is ``ruled out'' by current observations.
Most of the analyzed operators imply ``anarchic'' \cite{anarchy}
neutrino masses.  This simply means that  all elements of the
neutrino mass matrix are uncorrelated and of the same order of
magnitude. This hypothesis is known to ``fit'' the current data very
well \cite{anarchy}. It will be further challenged by searches for
$\theta_{13}$ (the anarchic hypothesis favors large $\theta_{13}$
values) and probes that may reveal if the neutrino masses are
hierarchical or whether two or three of the masses are almost
degenerate (anarchy naively predicts the former). If future data
strongly points towards non-anarchic $m_{\alpha\beta}$, I will be
forced to conclude that there is nontrivial ``leptonic'' structure
in the dimensionless coefficients $\lambda$ of most of the LNV
operators considered here.

Many of the operators associated with a low neutrino-mass related
cutoff scale ($\Lambda_{\nu} \leq 10~\rm{TeV}$), on the other hand,
naively predict more structured neutrino mass matrices. Operators
\begin{equation}
\mathcal{O}_{7},\mathcal{O}_{8},\mathcal{O}_{19},\mathcal{O}_{20},\mathcal{O}_{34},\mathcal{O}_{35},\mathcal{O}_{54_{a,b,c,d}},\mathcal{O}_{55_{a,b,c}},\mathcal{O}_{56},\mathcal{O}_{57},\mathcal{O}_{58},\mathcal{O}_{59},\mathcal{O}_{60},\mathcal{O}_{70},\mathcal{O}_{75},
\label{eq:lalphaOps}
\end{equation}
which radiatively generate neutrino mass elements proportional to
distinct charged lepton Yukawa coupling ($y_{e}, y_{\mu},
y_{\tau}$),  yield mass matrices $m$ such that
\begin{equation}
m \propto \left(
  \begin{array}{ccc}
    y_{e} & y_{\mu} & y_{\tau} \\
    y_{\mu} & y_{\mu} & y_{\tau} \\
    y_{\tau} & y_{\tau} & y_{\tau} \\
  \end{array}
\right). \label{eq:texture1}
\end{equation}
Additionally, models described at low energies by
$\mathcal{O}_{36}$, $\mathcal{O}_{37}$ and $\mathcal{O}_{38}$
generate neutrino masses proportional to both associated charged
Yukawa couplings, such that
\begin{equation}
m \propto \left(
  \begin{array}{ccc}
    y_{e}y_{e} & y_{e}y_{\mu} & y_{e}y_{\tau} \\
    y_{e}y_{\mu} & y_{\mu}y_{\mu} & y_{\mu}y_{\tau} \\
    y_{e}y_{\tau} & y_{\mu}y_{\tau} & y_{\tau}y_{\tau} \\
  \end{array}
\right). \label{eq:texture2}
\end{equation}

The strongly hierarchial nature of the charged lepton masses ($y_e
\ll y_\mu \ll y_\tau$), implies that the $m_{\alpha\beta}$ elements
of Eqs.~(\ref{eq:texture1}) and (\ref{eq:texture2}) are expected to
be hierarchical as well. In particular,  the $ee$ matrix element,
$m_{ee}$, proportional to $y_e$ or $y_e^2$ is, for all practical
purposes, negligibly small\footnote{Quantitatively, in the scenarios
under investigation, $m_{ee}$ values are, respectively, up to order
one corrections, $y_e/y_\tau \sim 10^{-4}$ and $y_e^2/y_\tau^2 \sim
10^{-7}$ times the characteristic mass scale of the mass matrix.} in
both of these cases.  On the other hand, it is well known that only
a normal neutrino mass hierarchy is consistent with vanishing
$m_{ee}$ \cite{Normal0Mee}, so that both  Eqs.~(\ref{eq:texture1})
and (\ref{eq:texture2}) predict the neutrino mass ordering to be
normal. In the absence of extra structure, scenarios characterized
by the LNV operators listed in Eq.~(\ref{eq:lalphaOps}) plus
$\mathcal{O}_{36}$, $\mathcal{O}_{37}$ and $\mathcal{O}_{38}$ will
be ruled out if future  data favor an inverted mass hierarchy, or if
the neutrino masses end up  quasi-degenerate (regardless of the
hierarchy).  As will become clear shortly, Eqs.~(\ref{eq:texture1})
and (\ref{eq:texture2}) predict that the lightest neutrino mass
($m_1$ in this case) is  small ($\lesssim \sqrt{\Delta m_{12}^2}$).

A more detailed analysis reveals that naive expectations from
Eqs.~(\ref{eq:texture1}) are already disfavored, while those from
Eqs.~(\ref{eq:texture2}) are virtually excluded. Assuming the normal
hierarchy and very small $m_{ee}$, one can find a relation between
the neutrino mass eigenstates and the oscillation parameters, thus
reducing the number of free parameters in the mass matrix by one.
Consider the diagonalization of the neutrino mass matrix defined by
$m_{\alpha\beta} = UM^DU^T$ with $M^D = {\rm diag}(m_1,m_2e^{2
i\phi_2},m_3e^{2i\phi_3})$ and $U$ the neutrino mixing matrix,
expressed in the PDG parameterization. In this case,
\begin{equation}
m_{ee} = m_1\cos^2\theta_{12}\cos^2\theta_{13} +
m_2\sin^2\theta_{12}\cos^2\theta_{13}e^{2 i \phi_2} +
m_3\sin^2\theta_{13}e^{2 i\left(\phi_3-\delta\right)}.
\end{equation}
Setting $m_{ee}=0$, one can solve for $m_1$ and one of the Majorana
phases. Recalling that, for the normal mass hierarchy, $m_2 =
\sqrt{m_1^2 + \Delta m^2_{12}}$ and $m_3 = \sqrt{m_1^2 + \Delta
m^2_{13}}$,  and assuming small $\theta_{13}$ and $\eta \equiv
\sqrt{\Delta m^2_{12}/\Delta m^2_{13}}$,
\begin{eqnarray}
 \nonumber \frac{m_1}{\sqrt{\Delta m_{13}^2}} & \approx & \eta \frac{\sin^2\theta_S}{\cos^{1/2}2\theta_S}
 - \theta_{13}^2\frac{\cos^2\theta_S}{\cos
 2\theta_S}\cos [2(\phi_3 - \delta)],\\
 \phi_2 &\approx& \frac{\pi}{2} + \frac{1}{2}\arctan\left(\frac{4\theta_{13}^2}{\eta}\frac{\sqrt{\cos 2\theta_S}}{\sin^2 2\theta_S}\sin [2(\phi_3-\delta)]\right).
\label{eq:m1}
\end{eqnarray}
One can easily obtain approximate expressions for the other neutrino
masses ($m_2, m_3$) and hence all elements $m_{\alpha\beta}$. Upon
substituting the numeric best fit oscillation parameters to avoid
introducing a needlessly cumbersome expression, I get
\begin{eqnarray}
\frac{m_{\alpha\beta}}{\sqrt{\Delta m^2_{13}}} &=&
0.5e^{i2\phi_3}\left(
                      \begin{array}{ccc}
                        0 & 0 & 0 \\
                        0 & 1 & 1 \\
                        0 & 1& 1 \\
                      \end{array}
                    \right) + 0.71\theta_{13}e^{-i(\delta-2\phi_3)}\left(\begin{array}{ccc}
                        0 & 1 & 1 \\
                       1 & 0 & 0 \\
                       1 & 0 & 0 \\
                      \end{array}
                    \right) \\ \nonumber
                    &+& 0.45\eta\left(\begin{array}{ccc}
                        0 & -1.3 & 1 \\
                        -1.3 & -1 & 0.61 \\
                        1 & 0.61 & -0.36 \\
                      \end{array}
                    \right) \\ \nonumber
                    &+& 0.91\theta_{13}^2\cos [2(\delta-\phi_3)]\left(\begin{array}{ccc}
                        0 & 1 & -0.89 \\
                        1 & 0.12 & 0.02 \\
                        -0.89 & 0.02 & -0.12 \\
                      \end{array}
                    \right) \\ \nonumber
                     &+&
                    1.2i\theta_{13}^2\sin [2(\phi_3 - \delta)]\left(\begin{array}{ccc}
                        0 & 1 & -0.67 \\
                        1 & 1.2 & -0.83 \\
                        -0.67 & -0.83 & 0.56 \\
                      \end{array}
                      \right).
                    \label{eq:MassMatrix}
\end{eqnarray}

Eq.~(\ref{eq:MassMatrix})  suggests a clear hierarchy among the
mixing matrix elements. The four, lower box-diagonal $\mu-\tau$
elements dominate, followed by the off-diagonal $e\mu$ and $e\mu$
entries, and finally the vanishingly small $m_{ee}$. Except for the
vanishingly small $m_{ee}$, which was required  {\it a priori}, all
of the remaining properties follow directly from the experimentally
determined mixing parameters. Among the dominant $\mu-\tau$
submatrix, Eq.~(\ref{eq:MassMatrix}) predicts that all entries are
equal
 up to small order $\eta$ and $\theta_{13}$ corrections.
The magnitude, and sign, of these ``breaking terms'' can be tuned
with the phases $\phi_3$ and $\delta$, and to a lesser extent by
varying $\eta$ and $\theta_{13}$ within their allowed ranges. On the
other hand, the relative sizes of  $m_{e\mu}$ and $m_{e\tau}$ are
expected to be similar but not identical, {\it i.e.}, $m_{e\mu}\sim
m_{e\tau}\sim( m_{e\mu}-m_{e\tau})$.

While some of the gross features of  Eq.~(\ref{eq:MassMatrix}) are
shared by Eq.~(\ref{eq:texture1}) and Eq.~(\ref{eq:texture2}), a
finer analysis reveals several disagreements. The major discrepancy
lies in the required relations among the matrix elements.
Eq.~(\ref{eq:texture1}) predicts that all $m_{\alpha\tau}$ elements
are equal, while Eq.~(\ref{eq:texture2}) suggests $m_{e\tau} \ll
m_{\mu\tau} \ll m_{\tau\tau}$.  Both of these contradict, in
different ways, the experimental constraint $m_{e\tau} \ll
m_{\mu\tau} \approx m_{\tau\tau}$.  Additionally, both
Eq.~(\ref{eq:texture1}) and Eq.~(\ref{eq:texture2}) predict $m_{ee}
\ll m_{\mu\mu} \ll m_{\tau\tau}$, while observations require $m_{ee}
\ll m_{\mu\mu} \approx m_{\tau\tau}$.  Similarly, both sets of
operators suggest $m_{e\mu} \ll m_{e\tau}$ while, experimentally,
they are constrained to be similar.

\begin{figure}
\begin{center}
\includegraphics[scale=1]{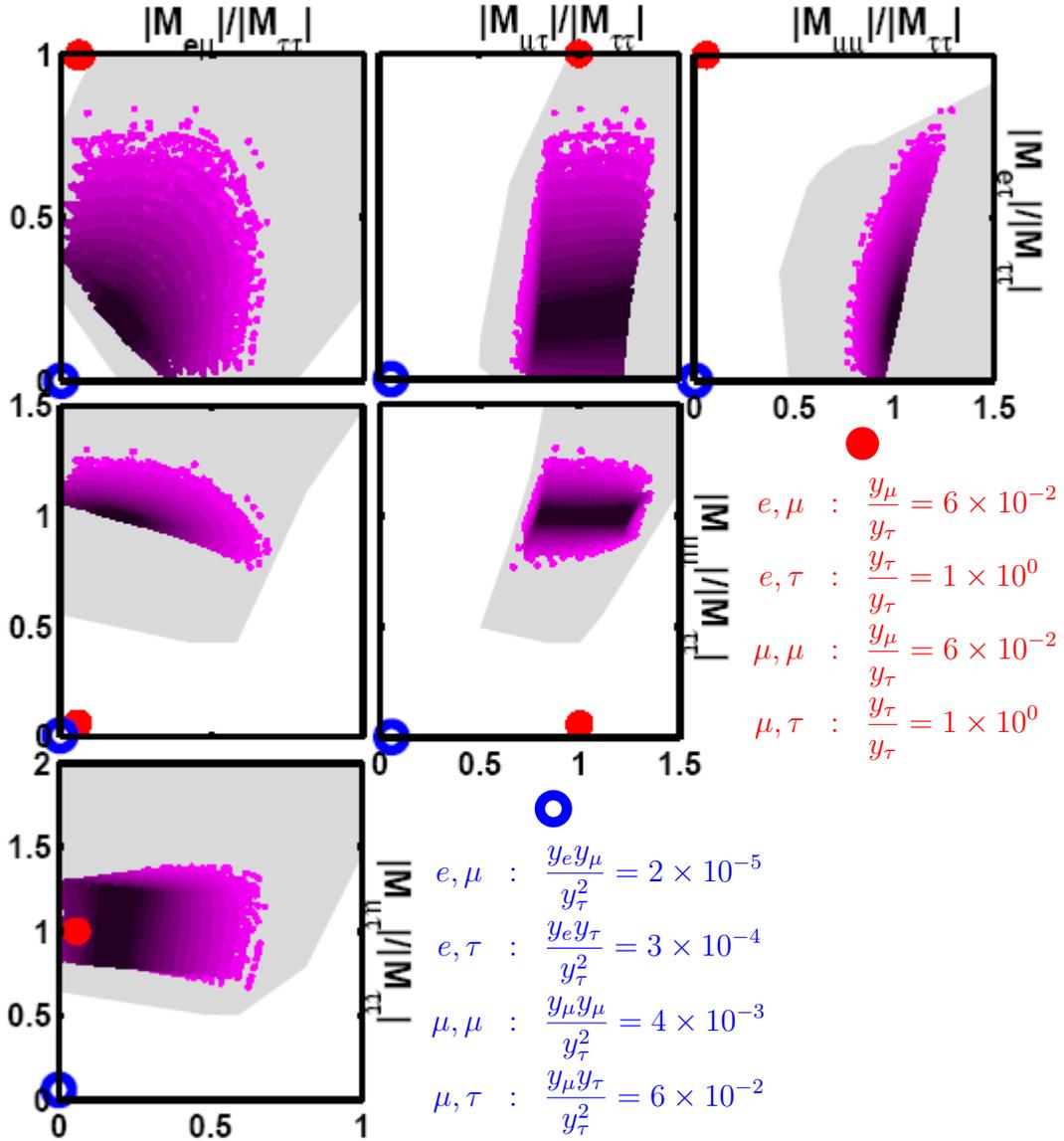}
\caption[Scatter plots of normalized Majorana neutrino mass matrix
elements]{Scatter plots of the symmetric Majorana neutrino mass
matrix elements normalized to $m_{\tau\tau}$. Each panel is produced
assuming the normal mass hierarchy and parameter constraints
insuring that $m_{ee} \leq 10^{-4}~\rm{eV}$. The light grey region
is calculated allowing all mixing parameters to vary within their
respective $95\%$ confidence intervals.  In the purple (darker)
regions, the solar and atmospheric parameters are held constant
while all phases are scanned within their physical ranges and
$\theta_{13}$ is varied between zero and its 4~$\sigma$ upper bound.
The $\sin^2\theta_{13}$ variation is illustrated by varying the
shading from dark to light. Also indicated by red (closed) and blue
(open) dots are the expectations derived from
Eqs.~(\ref{eq:texture1}) and (\ref{eq:texture2}), respectively,
along with a listing of their associated coordinate values.}
\label{fig:Mn_0ee}
\end{center}
\end{figure}

In order to quantify how much Eq.~(\ref{eq:texture1}) and
Eq.~(\ref{eq:texture2}) (dis)agree with our current understanding of
neutrino masses and lepton mixing, I numerically scanned the allowed
mass matrix parameter space assuming the normal neutrino mass
hierarchy and constraining $|m_{ee}| \leq y_e/y_\tau \times
1~\rm{eV} \approx 10^{-4}~\rm{eV}$. It should be noted that,
according to this relation, $m_{ee}$ is allowed to deviate by nearly
a factor of ten above naive expectations from mass matrix
Eq.~(\ref{eq:texture1}), thus accounting for the possible order of
magnitude uncertainties in operator scales and coupling constants.
This feature is only included for completeness, as one expects that
such $m_{ee}$ excursions from zero will generally have negligible
effect on the mass matrix due to the robust nature of
Eq.~(\ref{eq:MassMatrix}). Fig.~\ref{fig:Mn_0ee}, a scatter plot of
mixing matrix elements, depicts the result of such a scan.
 Note that  I plot the mass ratios with respect to
assumed-to-be-dominant  $m_{\tau\tau}$ element. The light grey
regions of the plot were produced allowing all oscillation
parameters to vary within their $95\%$ confidance bounds
\cite{OscBestFit} and phases to vary within their entire physical
range subject to the constraints discussed above.  In the purple
(dark) region, the phases and reactor mixing angle $\theta_{13}$ are
allowed to vary while all other mixing parameters are held fixed at
their best fit values. I depict the $\sin^2\theta_{13}$ variation
from zero to $0.06$ ($4\sigma$ upper bound \cite{OscBestFit}) by
varying the purple shading from dark to light.  It is easy to check
that the numeric (Fig.~\ref{fig:Mn_0ee}) and analytic results
(Eq.~(\ref{eq:MassMatrix})) are consistent both qualitatively and
quantitatively.

Fig.~\ref{fig:Mn_0ee} also depicts the predictions from
Eq.~(\ref{eq:texture1}) and Eq.~(\ref{eq:texture2}) with red
(closed) and blue (open) dots, respectively.    As expected, all the
predictions from Eq.~(\ref{eq:texture2})
 fall near the origin in each panel and
are safely excluded.  Because expectations  from
Eq.~(\ref{eq:texture2}) for all $m_{\alpha\beta}/m_{\tau\tau}$ are
much smaller than one, I also include the dot coordinate values for
both textures within the figure.  In order to render the neutrino
mass matrix predicted from $\mathcal{O}_{36}$, $\mathcal{O}_{37}$
and $\mathcal{O}_{38}$ consistent with experimental constraints on
neutrino masses and lepton mixing, one is required to choose very
hierarchical $\lambda$ coefficients. In more detail, one needs to
choose $\lambda$ values so that all mixing matrix elements are
enhanced relative to the dominant $m_{\tau\tau}\propto y_\tau
y_\tau$ by numerical factors that range -- for different entries --
from 100 to $10^{5}$. A possible mechanism for achieving this is to
suppress third generation couplings to new physics, thus driving up
the ratio $m_{\alpha\beta}/m_{\tau\tau}$ along with the required
cutoff scale $\Lambda_{\nu}$. This procedure would have to be
accompanied by a more modest reduction of the couplings of second
generation fermions. Basically, I need to impose a flavor structure
that ``destroys'' the naive flavor structure induced by the charged
lepton Yukawa coupling hierarchy. I can safely conclude that
$\mathcal{O}_{36}$, $\mathcal{O}_{37}$, and $\mathcal{O}_{38}$,
which suggest that the neutrino mass matrix has the form
Eq.~(\ref{eq:texture2}), are strongly disfavored by current neutrino
oscillation data and, if somehow realized in nature, must be
accompanied by a very nontrivial flavor structure.

On the other hand, the operators listed in Eq.~(\ref{eq:lalphaOps}),
which predict Eq.~(\ref{eq:texture1}), are not quite as disfavored.
In this case the hierarchies among different mass matrix elements
are softer, and one can ask whether the red dots in
Fig.~\ref{fig:Mn_0ee} can move toward the experimentally allowed
regions with order 1--10 relative shifts. Many of the predictions
are already in agreement with experimental constraints, or at least
close enough to be easily ``nudged'' toward acceptable levels with
order one coefficients. The figure reveals that only $m_{\mu\mu}$ is
predicted to be relatively too small. By enhancing it by a factor of
order $y_\tau/y_\mu \sim 20$ one obtains moderately good agreement
between Eq.~(\ref{eq:texture1}) and experimental requirements. I
therefore conclude that operators listed in Eq.~(\ref{eq:lalphaOps})
are at least marginally allowed by neutrino mixing phenomenology.

While essential for a complete understanding of neutrino masses and
mixing, improved measurements of the already determined mixing
angles and mass-squared differences will not
 help to further constrain/exclude any of  the LNV scenarios in question.
Considering our  parameter flexibility, only future neutrino
 experiments that provide qualitatively new results can aid in
this endeavor.  In particular, the experimental determination of the
neutrino mass hierarchy is essential in order to properly test  the
scenarios highlighted in this section, as they all predict, in the
absence of very non-trivial flavor structure in the LNV sector,  the
normal hierarchy. Next-generation neutrino oscillation experiments
are expected to provide non-trivial information regarding the
neutrino mass hierarchy. Most rely on a neutrino/anti-neutrino
oscillation asymmetry via Earth matter effects
\cite{NeutrinoReview,NuMassHier0t13}, and depend heavily on a
sufficiently large $\theta_{13}$ mixing angle.  The possibility that
$\theta_{13}$ is vanishingly small, where the standard approach is
ineffective, is addressed in \cite{NuMassHier0t13} considering both
oscillation and non-oscillation probes.  In that case, one can hope
to discern the neutrino mass spectrum in future neutrino factory
\cite{NuFact}/ Superbeam \cite{SupBeam} experiments coupled with
improved constraints on the effective masses extracted from tritium
beta decay \cite{TritNuEffMass} and cosmology
\cite{CosmoSum,Fogli:2006yq,seljak,Hannestad:2006mi}.\footnote{One traditionally includes the effective
$\beta\beta0\nu$ mass $m_{ee}$ given by Eq.~(\ref{eq:std_Mee}) in a
neutrino mass hierarchy analysis. However, as discussed in
Sec.~\ref{subsec:bb0nu}, $m_{ee}^{\rm eff}$ is a potentially
convoluted process-dependent quantity that generally has little
(directly) to do with neutrino masses. For this reason,
$\beta\beta0\nu$ constraints cannot be used to determine the
neutrino mass spectrum from the point of this analysis.} Note that
these non-oscillation probes  can be independently used to constrain
LNV models, as they provide information regarding the the magnitude
of the lightest mass eigenstate ($m_1$ [$m_3$] in the case of normal
[inverted] hierarchy). For example, if either cosmological
observations or tritium beta decay experiments see evidence for
non-zero neutrino masses (in more detail, they constrain $\Sigma =
\sum_i m_i$ and $m_{\nu_e}^2 = \sum_i m_i^2 |U_{ei}|^2$
respectively) such that $\Sigma\gg 0.05$~eV or $m_{\nu_e}\gg
0.01$~eV, one would conclude, assuming a normal mass hierarchy, that
$m_1 \gg \sqrt{\Delta m^2_{12}}$. This would destroy the possibility
of negligibly small $m_{ee}$, and hence disfavor the operators that
lead to mass matrices of the type Eq.~(\ref{eq:texture1}) and
(\ref{eq:texture2}).  Currently, $\Sigma$ and $m_{\nu_e}$ are
bounded to be below $0.94~{\rm eV}$ and $2.0~{\rm eV}$,
respectively, but the sensitivity to these observable is expected to
significantly improve with next-generation experiments to $0.1~{\rm
eV}$ \cite{FutureCosmo} and $0.2~{\rm eV}$ \cite{Katrin},
respectively.

\setcounter{footnote}{0} \setcounter{equation}{0}
\section{Phenomenologically interesting operators:  Sample Renormalizable Model} \label{sec:interesting}

Having superficially surveyed a large set of LNV operators, I am now
in a position to identify operators with ``interesting''
phenomenological features for further detailed study.  One  subset
of potentially interesting operators is characterized by those that,
when required to ``explain'' the observed neutrino masses, are
accompanied by a low cutoff scale of, say, less than several TeV.
Further requiring a small enough $m_{ee}^{\rm eff}$ in order to
evade current $\beta\beta 0 \nu$ constraints, this set contains only
seven elements:
$\mathcal{O}_{17},\mathcal{O}_{18},\mathcal{O}_{34},\mathcal{O}_{35},\mathcal{O}_{56},\mathcal{O}_{57},\mathcal{O}_{58}$.
Of these, all but operators $\mathcal{O}_{35}$ and
$\mathcal{O}_{58}$ (which lead to the zeroth-order neutrino mass
matrix Eq.~(\ref{eq:texture1}) and a suppressed $m_{ee}$) should
provided a positive LNV signal in the next round of double-beta
decay experiments, baring specific flavor symmetries or finely-tuned
couplings. Furthermore,  $\mathcal{O}_{56}$ leads to a
$\beta\beta0\nu$ rate that is higher than what is naively dictated
by the values of the neutrino masses. Finally, with the possible
exception of $\mathcal{O}_{56}$ which may mediate observable LNV
processes at high energy colliders, none of the seven operators
above are expect to mediate LNV violating phenomena (as defined
here) at accessible rates.

An ``orthogonal'' subset consists of the higher dimensional
operators already ``excluded'' by $\beta\beta0\nu$. Not including
those operators severely constrained by lepton mixing  in
Sec.~\ref{sec:Oscillations}, this list  contains 11 elements:
$\mathcal{O}_{16},\mathcal{O}_{19},\mathcal{O}_{53},\mathcal{O}_{54a,b,c,d}$,
$\mathcal{O}_{59},\mathcal{O}_{60},\mathcal{O}_{70},\mathcal{O}_{75}$.
Most of these are associated to cutoff scales of order the weak
scale, which are likely to already be constrained by different
searches for new degrees of freedom with masses around 100~GeV. Even
if those are considered to be excluded,
$\mathcal{O}_{16},\mathcal{O}_{19},\mathcal{O}_{75}$ are ``safely''
shielded from direct and indirect non-LNV searches,\footnote{Generic
new degrees of freedom at the weak scale are constrained by direct
and indirect searches at high energy colliders ({\it e.g.},
resonances and effective four-fermion interactions, respectively),
flavor-violating ({\it e.g.}, $\mu\to e\gamma$), and high precision
experiments ({\it e.g.}, measurements of the anomalous muon magnetic
moment).}  while still mediating potentially observable LNV effects
at colliders as long as the new physics does not couple, to zeroth
order, to first generation quarks (in order to evade the
$\beta\beta0\nu$ constraints).

Regardless of whether these different options for the LNV sector
lead to observable LNV phenomena, the low extracted cutoff scale of
\emph{all} the operators highlighted above  implies that new degrees
of freedom should be produced and, with a little luck, observed at
the LHC or, perhaps, the ILC. Furthermore, the TeV scale has already
been identified as an interesting candidate scale for new physics
for very different reasons, including the dark matter puzzle and the
gauge hierarchy problem. The fact that, perhaps, the physics
responsible for neutrino masses also ``lives'' at the TeV scale is
rather appealing.

In order to study this new physics, as already emphasized earlier,
ultraviolet complete manifestations of the physics that leads to the
effective operators are required. Here I discuss one concrete
example. Other examples (for different effective operators) were
discussed in \cite{Operators}. Given a specific LNV operator, it is
a simple matter to write down equivalent renormalizable Lagrangians.
I briefly illustrate this procedure by constructing a renormalizable
model that will lead to the  dimension-eleven operator
$\mathcal{O}_{56}$.  It is among the interesting LNV effective
operators of the sample highlighted above, since it is currently
unconstrained by  $\beta\beta0\nu$ searches regardless of the
quark-flavor structure of the operator, while $m_{ee}^{\rm eff}\gg
m_{ee}$ for $\beta\beta0\nu$. On the other hand, $\Lambda_{\nu}$ for
$\mathcal{O}_{56}$ is very low (below 500~GeV), so that the new
degrees of freedom may already be constrained by, for example,
Tevatron or LEP data. I will not worry about such constraints
henceforth, but will only comment on possible phenomenological
problems.

$\mathcal{O}_{56}$ can be accommodated by a wide variety of models,
as can be seen from its possible Lorentz structures. In terms of
scalar/tensor helicity-violating bilinears
$\Gamma_v=1,\sigma_{\mu\nu}$, and vector helicity-conserving
bilinears $\Gamma_c=\gamma_{\mu}$, these are
\begin{eqnarray}
\mathcal{O}_{56} &=&\nonumber \{(L^i\Gamma_v Q^j)(d^c\Gamma_v d^c)(\overline{d^c}\Gamma_v \overline{e^c}),(L^i\Gamma_v Q^j)(d^c\Gamma_c \overline{d^c})(d^c\Gamma_c \overline{e^c}),(L^i\Gamma_v d^c)(Q^j\Gamma_v d^c)(\overline{d^c}\Gamma_v \overline{e^c}),\\
& & \nonumber (L^i\Gamma_v d^c)(Q^j\Gamma_c \overline{d^c})(d^c\Gamma_c \overline{e^c}),(L^i\Gamma_v d^c)(Q^j\Gamma_c \overline{e^c})(d^c\Gamma_c \overline{d^c}),(L^i\Gamma_c \overline{d^c})(Q^j\Gamma_v d^c)(d^c\Gamma_c \overline{e^c}),\\
& & (L^i\Gamma_c \overline{d^c})(Q^j\Gamma_c
\overline{e^c})(d^c\Gamma_v d^c),(L^i\Gamma_c
\overline{e^c})(Q^j\Gamma_v d^c)(d^c\Gamma_c
\overline{d^c}),(L^i\Gamma_c \overline{e^c})(Q^j\Gamma_c
\overline{d^c})(d^c\Gamma_v d^c) \} \nonumber \\ 
 &\times& H^kH^l\epsilon_{ik}\epsilon_{jl}.
\end{eqnarray}

It is clear from the chiral field content that these operators
depend on combination of helicity-conserving and helicity-violating
interactions.  In particular, it is impossible to form any of the
operators in this long list with only the addition of vector boson
states: new heavy scalar and/or tensor particles are probably
required  if $\mathcal{O}_{56}$ is the proper tree-level
manifestation of the LNV physics at low-energies.\footnote{Other
possibilities include heavy vector-like fermions.} Furthermore, the
couplings of the new physics fields with one another must be
constrained in order to ``block'' the presence  of lower-dimensional
tree-level effective operators.  This usually implies the existence
of new exact (broken) symmetries to forbid (suppress) particular
interactions.

\begin{figure}
\begin{center}
\includegraphics[scale=.9]{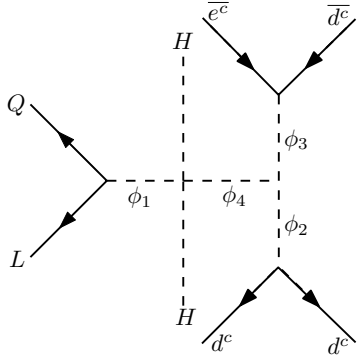}
\caption[Sample scalar interactions that lead to the effective
operator $\mathcal{O}_{56}$]{Sample scalar interactions that lead to
the ``interesting'' effective operator $\mathcal{O}_{56}$ with the
Lorentz structure
$(L^iQ^j)(d^cd^c)(\bar{e^c}\bar{d^c})H^kH^l\epsilon_{ik}\epsilon_{jl}$.}
\label{fig:Models}
\end{center}
\end{figure}

Certain Lorentz structures, those containing only $\Gamma_v$
bilinears, can be realized assuming that the LNV ultraviolet sector
contains only heavy \emph{scalar} fields and I concentrate, for
simplicity, on this possibility \cite{ScalarLNV}. Simple scalar
interactions that can lead to $\mathcal{O}_{56}$ are shown in the
diagram in Fig.~\ref{fig:Models}. Specifically, these  yield  the
effective operator Lorentz structure
$(L^iQ^j)(d^cd^c)(\bar{e^c}\bar{d^c})H^kH^l\epsilon_{ik}\epsilon_{jl}$
with the introduction of four charged scalar fields,
$\phi_1,\phi_2,\phi_3,\phi_4$.  The gauge structure is such that,
under $(SU(3)_c,SU(2)_L,U(1)_Y)$\footnote{In the case of $U(1)_Y$,
`transfoms as $X$' means `has hypercharge $X$'.}, $\phi_1$
transforms as a $(\bar{3},3,+1/3)$, $\phi_2$ as $(\bar{3},1,-2/3)$,
$\phi_3$ as $(3,1,-4/3)$, and $\phi_4$ as $(\bar{3},1,-2/3)$. While
$\phi_2$ and $\phi_4$ have identical gauge quantum numbers, they
have different baryon number ($2/3$ versus $-1/3$). $\phi_1$ has
baryon number $-1/3$, while $\phi_3$ has baryon number $1/3$. Lepton
number cannot be consistently assigned  as it is explicitly violated
by two units.

$\phi_4$, which does not couple to any of the SM fermions, plays an
essential role. It acts as a selective ``insulator'' that connects
the various interaction terms in such a way as to only alow certain
tree-level higher dimensional SM effective operators. All
renormalizable theories that lead to only very high dimensional
effective operators contain one or more of these ``hidden sector''
fields. Note that the new scalar fields should not acquire vacuum
expectation values in order to avoid the presence of lower
dimensional irrelevant operators that are likely to dominate
low-energy phenomenology and -- much more important -- to prevent
the spontaneous breaking of color or electromagnetic charge.

Given the scalar field content as well as its  transformation
properties under SM global and local symmetries, it is a simple
matter to write down the minimal interaction Lagrangian density for
the system.  A candidate renormalizable Lagrangian is
\begin{eqnarray}
\mathcal{L} &=& \mathcal{L}^{(SM)} + \sum_i \left(|D_\mu \phi_I|^2 +
M_i|\phi_i|^2\right) +y_1QL \phi_1 + y_2 d^cd^c \phi_2
 + y_3 e^cd^c \phi_3 + \lambda_{14}
\bar{\phi}_1\phi_4 H H \nonumber \\ 
 &+& \lambda_{234} M \phi_2\bar{\phi}_3\phi_4 +
h.c.~.
 \label{eq:sample}
 \end{eqnarray}
 Each term in Eq.~(\ref{eq:sample}), including those involving covariant derivatives $D_\mu$, is implicitly assumed to
 respect the gauge representations of the associated $\phi_i$
fields, as defined above.  The Yukawa-type couplings $y_i$,  as well
as the $\lambda_i$ scalar vertices are dimensionless, and assumed to
be of order one, while I assume all scalar masses $M_i$ to be of the
same order of magnitude. In this case, $\Lambda\sim M_i$. In the
$\lambda_{234}$ term, an overall mass scale $M$ has been ``factored
out'' and is assumed to be of the same order as the $M_i$. Note that
I neglect generation indicies, which are implied. In the case
$\lambda_{234}=0$, lepton number is a classical global symmetry of
Eq.~(\ref{eq:sample}), and one can view this three-scalar coupling
as the source of lepton number violation. One may even envision a
scenario where lepton number is spontaneously broken by the vacuum
expectation value of some SM singlet $\phi_5$ scalar field,
$\langle\phi_5\rangle=M$.

Provided all $M_i$ are around 0.5~TeV, as required if this
Lagrangian is to ``explain'' the observed light neutrino masses, LNV
is certainly \emph{not} the only (or even the main) consequence of
this model. The $y_1$ and $y_3$ terms for example, will mediate
$\mu\to e$-conversion in nuclei
 at very dangerous levels if their flavor structure is generic. $\phi_2$ can be resonantly
 produced in $dd$-collisions, while $\phi_1$ and $\phi_3$ qualify as scalar lepto-quarks, which
 are constrained by high energy collider experiments, including those at HERA \cite{HERA}, to weigh more than a few hundred GeV \cite{PDG}. For more details, I refer readers to, for example,
 the Particle Data Book \cite{PDG} and references therein.

 I will conclude this discussion by adding that several other effective operator can be realized
 in a very similar way. ${\mathcal O}_{19}$, for example, if it manifests itself with the Lorentz structure
 $(L^iQ^j)(d^cd^c)(\bar{e^c}\bar{u^c})\epsilon_{ij}$, can be realized by a Lagrangian very similar
 to Eq.~(\ref{eq:sample}) where the $d^c$ field in the $y_3$-coupling interaction is replaced by a $u^c$ field, and
 the $\phi_1$ field is replaced by an $SU(2)_L$ singlet (it is a triplet in Eq.~(\ref{eq:sample})). Of course, hypercharge assignments for the $\phi_i$ also need to be modified in a straight forward way. The
 associated non LNV phenomenology is similar, except for the fact that $\Lambda_{\nu}$ for  ${\mathcal O}_{19}$ (around 1~TeV) is larger than the one for ${\mathcal O}_{56}$ and hence ${\mathcal O}_{19}$ is less constrained by current experimental data. On the other hand, ${\mathcal O}_{19}$  predicts potentially much larger rates for LNV observables at colliders (see Fig,~\ref{fig:ILC}).

Our definition of ``interesting'' is  arbitrary and motivated only
by the fact that the physics of the ``interesting'' operators
highlighted earlier in this section will probably be explored at
next-generation collider and high-precision experiments.  One may
argue that many operators which lead to the observable neutrino
masses for high values of $\Lambda_{\nu}$ are interesting on their
own right, either due to the theoretically pleasing properties of
their associated potential ultraviolet completions, or by some
observational peculiarity. There are many examples of the first type
ranging from the different manifestations of the seesaw mechanism
\cite{SeeSaw,Bajc:2006ia,SeeSaw2,SeeSaw3} to the Zee model
\cite{ZeeModel} and the minimal supersymmetric SM with R-parity
violation \cite{R-parity}. Dedicated analysis of these cases have
been widely pursued in the literature and will not be discussed
here. I would also like to point out that some effective operators,
like  $\mathcal{O}_{7}$ and $\mathcal{O}_{8}$, are, according to our
criteria, very ``uninteresting.'' Both $\mathcal{O}_{7}$ and
$\mathcal{O}_{8}$  predict unobservably suppressed $\beta\beta0\nu$
rates (both predict small $m_{ee}$) and equally hopeless collider
prospects given that they are associated to very high cutoff scales,
$\Lambda_{\nu}\approx4\times 10^2~\rm{TeV}$ and
$\Lambda_{\nu}\approx6\times 10^3~\rm{TeV}$, respectively.  If
either of these operators are responsible for the observed tiny
neutrino masses, it is quite possible that I may never directly
detect LNV.  It is curious to consider possible means of indirect
detection or other observable consequences of the different
ultraviolet completions of such scenarios.\footnote{This is very
similar to the case of ${\mathcal O}_1$. The main redeeming feature
of ${\mathcal O}_1$, other than its simplicity, is the fact that
many of its ultraviolet completions allow one to explain the
matter--antimatter asymmetry of the universe \cite{Leptogenesis}.}
It would also be interesting to ask whether either of these elusive
models has any underlying theoretical motivation or whether they
allow one to solve other outstanding problems in particle physics.

\setcounter{footnote}{0} \setcounter{equation}{0}
\section{Conclusion}
\label{sec:conclusion}

If neutrino masses are a consequence of lepton-number violating
physics at a very high energy scale (higher than the scale of
electroweak symmetry breaking), new physics effects -- including the
generation of neutrino Majorana masses -- at low enough energies can
be parameterized in terms of irrelevant operators whose coefficients
are suppressed by inverse powers of an effective cutoff scale
$\Lambda$. As discussed before, $\Lambda$ is, roughly, the energy
scale above which new degrees of freedom must be observed if the new
ultraviolet physics is perturbative (if the new physics is very
weakly coupled, the masses of the new degrees of freedom can be much
smaller than $\Lambda$). I have explored a very large class of such
scenarios through 129 irrelevant operators of energy dimension less
than or equal to eleven that violate lepton number by 2 units. These
are tabulated in first two columns of Table~\ref{tab:AllOps}, along
with a summary of our results.

Analyzing each effective operator individually, I estimated the
predicted general form of the Majorana neutrino mass matrix. Our
results are listed in the third column of Table~\ref{tab:AllOps}. By
comparing each such estimate with our current understanding of
neutrino masses, I extracted the cutoff scale $\Lambda_{\nu}$ of
each effective operator, assuming that it provides the dominant
contribution to the observed neutrino masses. These results are
listed in the fourth column of Table~\ref{tab:AllOps} assuming light
neutrino masses equal to 0.05~eV (the square root of the atmospheric
mass-squared difference), and are summarized as follows. Depending
on the field content and dimension of the irrelevant operator, the
``lepton number breaking scale'' $\Lambda_{\nu}$ is predicted to be
anywhere from the weak scale ($\sim 0.1$~TeV) all the way up to
$10^{12}$~TeV (see Fig.~\ref{fig:OpSum}). This means that, depending
on how lepton number is violated and communicated to the SM, the
mass of the associated new degrees of freedom is predicted to be
anywhere between 100~GeV and $10^{12}$~TeV, \emph{even if all new
physics couplings are order one}. I note that in the case of all
variations of the seesaw mechanism (${\mathcal O}_1$),  neutrino
physics constrains $\Lambda_{\nu}=10^{12}$~TeV such that the new
degrees of freedom are either unobservably heavy, extremely weakly
coupled, or their couplings to the SM degrees of freedom are
finely-tuned. It is fair to say that this behavior is not
characteristic of all LNV ultraviolet physics. One sample
ultraviolet theory that leads to dimension-eleven LNV effective
operators was discussed in Sec.~\ref{sec:interesting}. Other
examples can be found in \cite{Operators}, and include supersymmetry
with trilinear R-parity violation and the Zee model.

Assuming that a particular operator is responsible for nonzero
neutrino masses, it is straight forward to ask whether it leads to
other observable consequences. Here, I concentrated on several LNV
observables, and included future LNV searches at the LHC and future
lepton machines (like the ILC), along with their ability to directly
produce (and hopefully observe) new physics states lighter than
several hundred GeV. In column five of Table~\ref{tab:AllOps}, I
list the most favorable modes of experimental observation for each
operator.  The different relevant probes are:  neutrinoless
double-beta decay ($\beta\beta0\nu$), neutrino oscillation and
mixing (mix), direct searches for new particles at the LHC (LHC) and
ILC (ILC), and virtual LNV effects at collider facilities (HElnv). I
find it  unlikely that other probes of LNV, including rare meson
decays, should yield a positive signal in the forseeable future.
This conclusion is strongly based on the fact that, for all of our
analysis, I assume that \emph{all} new physics degrees of freedom
are heavier than the weak scale. While the vast majority of
operators is most sensitive to searches for neutrinoless double-beta
decay, that is not true of all operators. Some lead to relatively
suppressed rates for $\beta\beta0\nu$ (mostly because they lead to
mass matrices with a very small $m_{ee}$) even if they are
associated to $\Lambda_{\nu}<1$~TeV, indicating that, for these
scenarios, we are more likely to observe the physics behind neutrino
masses directly at colliders than to see a finite lifetime for
$\beta\beta0\nu$. Other scenarios naively lead to $\beta\beta0\nu$
rates orders of magnitude higher than what is currently allowed by
data. If these are responsible for the generation of neutrino
masses, the new physics is constrained to be somewhat decoupled from
first generation quarks (for example). In this case, there is hope
that LNV phenomena at colliders, which are not restricted to first
generation quarks, occur with non-negligible rates.

The sixth column of Table~\ref{tab:AllOps} lists the current
``status'' of the operator as either experimentally unconstrained
(U), constrained (C), or disfavored (D).  Such labels are assigned
based only on the experimental probes reviewed in this work.  By
arbitrary convention, an `unconstrained' operator can safely
accommodate all existing data even if one assumes \emph{all} its
flavor-dependent dimensionless coefficients to be of order one. A
`constrained' operator can accommodate all existing data after one
allows some of the different flavor-dependent dimensionless
coefficients to be suppressed with respect to the dominant ones by a
factor of 100 or so (as described above).  `Disfavored' operators
can only accommodate all data  only if ``tuned'' much more severely
than the `constrained' ones, and are usually in trouble with more
than one ``type'' of constraint.  A glance at column six reveals
that 11 out of the 129 operators are disfavored by current data.
The most stringent constraints come from  $\beta\beta0\nu$, while
all `disfavored' operators are associated to cutoffs at or below
$1~\rm{TeV}$. Three of the `disfavored' operators,
$\mathcal{O}_{36}$, $\mathcal{O}_{37}$, and $\mathcal{O}_{38}$, are
also in disagreement with the  neutrino oscillation data (see
Sec.~\ref{sec:Oscillations}).

Our results illustrate that, as far as ``explaining'' neutrino
masses, the model-building scene is wide open even if one postulates
that neutrino masses arise as a consequence of lepton number
violating, ``heavy'' physics. Significant progress will only be
achieved once more experimental information becomes available. The
observation that neutrinoless double-beta decay occurs with a
nonzero rate will help point us in the right direction, but will
certainly not reveal much about the mechanism behind neutrino
masses. A more complete picture can only arise from combined
information from several observables, including other LNV
observables and the search for new physics at the electroweak scale.
Other important experimental searches, not discussed here,  include
all lepton-number conserving ``leptonic'' probes, such as precision
measurements of the anomalous magnetic moment of the muon, searches
for leptonic electric dipole moments, searches for charged-lepton
flavor violation (see \cite{LepFlavVio} for a model independent
discussion of this issue), and precision measurements of
neutrino--nucleon and neutrino--lepton scattering.

\chapter{Neutrino Phenomenology of Very Low-Energy Seesaws} \label{chap:SeeSawLSND}

The seesaw mechanism \cite{SeeSaw} is an appealing way to generate
the observed neutrino masses and lepton mixing matrix. The idea is
simple. Add an arbitrary number of singlet fermion states to the SM
matter content. The triviality of their quantum numbers allows them
to have Majorana masses of magnitude $M$, as well as couple to the
$SU(2)_L$ lepton and Higgs doublets. The latter vertices become
Dirac mass terms of magnitude $\mu$ after electroweak symmetry
breaking. The standard theoretical prejudice is that the Dirac
masses are of order the charged fermion masses, while the Majorana
masses are at some very high energy scale, $M\gg 100$~GeV. If this
is indeed the case, the resulting propagating neutrino degrees of
freedom separate into two quasi-decoupled groups: mostly active
states with very small masses $m\sim \mu^2/M$ suppressed by the new
physics scale, and mostly sterile states with very large masses $M$.
In this scenario, the mostly right-handed states are not directly
observable. Indeed, it is possible that if such a high-energy seesaw
is realized in Nature, its only observable consequence is that the
mostly active neutrinos have mass and mix.

 Of course, there is no direct evidence that $M$ -- which I refer to as the seesaw scale --
is large.  Large $M$ values are attractive for several reasons,
including the fact that one may relate $M$
 to the grand unified scale. On the other hand, all $M$ values are technically natural, given that when $M$ vanishes the global symmetry structure of the Lagrangian is enhanced: $U(1)_{B-L}$ is a symmetry of the Lagrangian if $M=0$, so that $M$ is often referred to as the lepton number breaking scale. This point was recently emphasized in \cite{SeeSaw_LSND}. Recent analyses have also revealed that there are several low-energy choices for the seesaw energy scale that allow one to address outstanding problems in particle physics and astrophysics. The main reason for this is that, unlike in the high-energy seesaw, in a low-energy seesaw the mostly right-handed states do not decouple but  remain as kinematically accessible sterile neutrinos.

 The data reported by the LSND short baseline neutrino oscillation experiment
\cite{LSND} can be explained by postulating the existence of light
($m\sim 1-10$~eV) sterile neutrino states.  This result is currently
being tested by the Fermilab MiniBooNE experiment \cite{minib} and,
if confirmed, will lead the community to seriously contemplate the
existence of light, SM singlet fermions.  It was pointed out in
\cite{SeeSaw_LSND} (see also \cite{light_rhn1}) that if $M\sim
1-10$~eV, the right-handed seesaw neutrinos can easily play the role
of the LSND sterile neutrinos. There is also evidence for mixed
sterile neutrinos at other energy scales: eV sterile neutrinos aid
in heavy element nucleosynthesis in supernovae, keV sterile
neutrinos can help explain the peculiar velocity of pulsars, and
remain viable warm dark matter candidates. In the past several
months, it has been shown that the seesaw right-handed neutrinos may
play the role of all these astrophysically/cosmologically inspired
sterile neutrinos \cite{nuSM_dark,nuSM_kicks}.

In this chapter, I consider the phenomenology of low-energy
($M\lesssim 1$~keV) seesaw scenarios, extending the analysis
performed in \cite{SeeSaw_LSND} in several ways. In
Sec.~\ref{sec:seesawMass}, I review the generation of neutrino mass
via the seesaw mechanism and apply it to relatively light
right-handed neutrino states. I pay special attention to the most
general active--sterile seesaw mixing matrix, whose parameters are
the main object of our study. In Sec.~\ref{sec:pheno}, I review the
several different ``evidences'' for sterile neutrinos, and discuss
whether these can be ``fit'' by the low-energy seesaw. I concentrate
on exploring solutions that can accommodate at the same time the
LSND anomaly and the astrophysical processes outlined above, but
also discuss different combinations of the seesaw parameters capable
of explaining only the astrophysics-related observables. In
Sec.~\ref{sec:future}, I examine other experimental probes that can
be used to explore low-energy seesaws -- regardless of their
relationship to the LSND anomaly, pulsar kicks, and warm dark
matter. I concentrate on the prospects of current/future tritium
beta-decay and neutrinoless double-beta decay experiments. I
conclude in Sec.~\ref{sec:SeeSawConclusion} by summarizing our results,
commenting on the plausibility of this scenario, and offering a
general outlook for the future.

\section{The Seesaw Mechanism and electron volt neutrino masses:
preliminaries} \label{sec:seesawMass}

In order to account for nonzero neutrino masses, I add to the SM
particle content
 three $SU(3)_c
\times SU(2)_L \times U(1)_Y$ gauge singlet fermion states $N_i$,
conventionally referred to as right-handed neutrinos. While sterile
under SM gauge interactions, right-handed neutrinos may still be
charged under new,  currently unknown gauge transformations.  Such
interactions, if at all present, are neglected in our analysis.

The most general renormalizable Lagrangian consistent with SM gauge
invariance is
\begin{equation}
\label{eq:seesaw} {\mathcal L}_{\nu}={\mathcal L}_{\rm old} -
\lambda_{\alpha
i}\bar{L}^{\alpha}HN^i-\sum_{i=1}^3\frac{M_{i}}{2}{\overline
N^c}^iN^i + H.c.,
\end{equation}
where ${\mathcal L}_{\rm old}$ is the traditional SM Lagrangian, $H$
is the Higgs weak doublet, $L^{\alpha}$,  $\alpha=e,\mu,\tau$, are
lepton weak doublets, $\lambda_{\alpha i}$ are neutrino Yukawa
couplings, and $M_i$ are Majorana masses for the $N_i$. I choose,
without loss of generality, the Majorana  mass matrix $M_R$ to be
diagonal and its eigenvalues $M_i$ to be real and positive. I also
choose the charged lepton Yukawa interactions and the charged weak
current interactions diagonal so that all physical mixing elements
are contained in the neutrino sector.

After electroweak symmetry breaking (when $H$ develops a vacuum
expectation value $v$), ${\mathcal L}_{\nu}$ will describe, aside
from all other SM degrees of freedom, six neutral massive Weyl
fermions
--- six neutrinos. The resulting mass terms can be expressed as:
\begin{equation}
\mathcal{L}_\nu \supset \frac{1}{2}\left(
                                \begin{array}{cc}
                                  \overline{\vec{\nu}} &
                                  \overline{\vec{N}^c}
                                \end{array}
                             \right)
                             \left(
                               \begin{array}{cc}
                                 0 & \mu \\
                                 \mu^T & M_R \\
                               \end{array}
                             \right)
                             \left(
                               \begin{array}{c}
                                 \vec{\nu}^c \\
                                 \vec{N} \\
                               \end{array}
                             \right)
\label{eq:lagrangian},
\end{equation}
where $\mu\equiv\lambda v$, and  $\vec{\nu}$ and $\vec{N}$ are
vectors of the three active neutrinos $(\nu_e,\nu_\mu,\nu_\tau)$ and
the three right-handed, sterile states, respectively.  Each entry in
the symmetric mass matrix of Eq~(\ref{eq:lagrangian}) is itself a
$3\times3$ matrix of mass parameters. Diagonalization of the mass
matrix yields eigenstates with Majorana masses that mix the
active--active states, related via the standard lepton mixing matrix
$V$, and the active--sterile states.  The physical neutrinos are
thus linear combinations of all active and sterile states.
Throughout I will work in the ``seesaw limit,'' defined by $\mu \ll
M_R$.  In this case, there are three mostly active light neutrinos
and three mostly sterile heavy neutrinos where `mostly' is
determined by the induced mixing parameters.

In the seesaw limit, the diagonalization is simple. Assuming, for
simplicity, that the mixing matrices are real:
\begin{eqnarray}
\left(
  \begin{array}{cc}
    0 & \mu \\
    \mu^T & M_R \\
  \end{array}
\right) &=& \left(
            \begin{array}{cc}
              1 & \Theta \\
              -\Theta^T & 1 \\
            \end{array}
          \right)
          \left(
            \begin{array}{cc}
              V & 0 \\
              0 & 1 \\
            \end{array}
          \right)
          \left(
            \begin{array}{cc}
              m & 0 \\
              0 & M_R \\
            \end{array}
          \right)
          \left(
            \begin{array}{cc}
              V^T & 0 \\
              0 & 1 \\
            \end{array}
          \right) \\ \nonumber 
 &~~~~~& \times
          \left(
            \begin{array}{cc}
              1 & -\Theta \\
              \Theta^T & 1 \\
            \end{array}
          \right) + \mathcal{O}(\Theta^2)
\label{eq:diag},
\end{eqnarray}
where $m$ is the diagonal matrix of light neutrino masses and
$\Theta$ is a matrix of active--sterile mixing angles found from the
relations
\begin{eqnarray}
\Theta M_R &=& \mu, \label{eq:mu} \\
 \Theta M_R \Theta^T &=& - V m V^T \label{eq:theta}.
\end{eqnarray}
The elements of $\Theta$ are small (${\mathcal O}(\mu/M_R)$), and
the standard seesaw relation ($VmV^T=-\mu{M}^{-1}_{R}\mu^T$) is
easily obtained by combining Eqs.~(\ref{eq:mu}) and
Eqs.~(\ref{eq:theta}). On the other hand, using
Eq.~(\ref{eq:theta}), I can relate the mixing parameters in $\Theta$
to the active--active mixing angles contained in $V$ and the
neutrino mass eigenvalues.  In the case of observably light sterile
neutrino masses, as considered in our analysis, this equation is
very useful, as it places testable constraints on observable
quantities. The general solution (first discussed in detail in
\cite{Casas:2001sr}) of Eq.~(\ref{eq:theta}) is
\begin{equation}
\Theta = -Vm^{1/2}OM_R^{-1/2} \label{eq:thetaM_constraint},
\end{equation}
where $O$ is an orthogonal matrix parameterized by three mixing
angles $\phi_{12},\phi_{13},\phi_{23}$.\footnote{In general, $O$ is
a \emph{complex} orthogonal matrix. Here, however, I will restrict
our analysis to \emph{real} neutrino mass matrices, unless otherwise
noted.} Physically, the mixing matrix $O$ is a consequence of our
freedom to choose the form of $M_R$.  An illustrative example is
found by considering the mass ordering of $M_R$ in its diagonal
form.  The reordered matrix $M_R(m_i\leftrightarrow m_j)$ is
equivalent to a $\pi/2$ rotation in the $i-j$ plane and therefore
represents the same physics as the original matrix, as it should. In
other words, $O M_R O^T$ is the physically relevant object, as
opposed to $O$ and $M_R$ separately. This object, when constrained
to be real, has six free parameters that I shall refer to as ``heavy
parameters:'' $\phi_{12},\phi_{13},\phi_{23},M_1,M_2,M_3$.

Using Eq.~(\ref{eq:thetaM_constraint}),  the $6\times6$ unitary
neutrino mixing matrix is
\begin{equation}
U = \left(
 \begin{array}{cc}
   V & \Theta \\
     -\Theta^T V & 1 \\
      \end{array}
       \right)
        \label{eq:U}.
\end{equation}
Note that, up to ${\mathcal O}(\Theta^2)$ corrections, $V$ is
unitary. $U$ is entirely described in terms of the three light
mixing angles, six mass eigenvalues and three angles $\phi_{ij}$.
Many combinations of these have been measured or constrained via
oscillation searches, cosmology and astrophysics.  In particular,
the two active neutrino mass-squared differences and mixing angles
have been measured \cite{neutrino_review,global_analy,PDG}, thus
leaving free the six heavy parameters and the absolute scale of
active neutrino masses.\footnote{The active neutrino mass hierarchy,
normal vs. inverted, is another (discrete) free parameter.}  With
$U$, the corresponding neutrino mass values and the SM couplings I
can calculate all observable quantities and compare them with data.

It is natural to wonder how well the seesaw approximation holds once
one starts to deal with $M_R$ values around  1~eV. From
Eq.~(\ref{eq:mu}), it is clear that one can choose for the expansion
parameter  $\sqrt{m/M_R}$.  In all scenarios considered here,
$\sqrt{m/M_R}<0.5$ (for $M\sim 1$~eV and $m \sim 0.3$~eV).  In the
worst case scenario, therefore, first order corrections are $55\%$
of the leading order terms, while second order corrections are near
$30\%$.   Corrections to most observables of interest are much
smaller than this because they are suppressed by larger right-handed
neutrino masses. The first non-trivial correction to
Eq.~(\ref{eq:thetaM_constraint}) occurs at second order and I find
that, for the ambitions of this chapter, all approximations are
under control. This simple argument has been verified numerically
for the most worrisome cases.

Before concluding this section, I wish to add that  operators that
lead, after electroweak symmetry breaking,  to Majorana masses for
the left--handed neutrino states ($M_L$) are also allowed if one
introduces $SU(2)_L$ Higgs boson triplets  or  nonrenormalizable
operators to the SM Lagrangian. While I neglect these ``active''
Majorana masses, I caution the reader that the existence of such
terms would alter our results significantly. In particular, assuming
the seesaw approximation holds, Eq.~(\ref{eq:theta}) would read
\begin{equation}
VmV^T+\Theta M_R \Theta^T=M_L, \label{eq:withML}
\end{equation}
which leads to a relationship between $\Theta$, $m$, and $M_R$
different from Eq.~(\ref{eq:thetaM_constraint}). If this were the
case, for example, it would no longer be true that the largest
$\Theta$ value (in absolute value) is constrained to be smaller than
$(m_{\rm max}/M_{\rm R,min})^{1/2}$, where $m_{\rm max}$ is the
largest element of $m$, while $M_{\rm R,min}$ is the smallest
element of $M_R$. On the other hand,  all objects on the left-hand
side of Eq.~(\ref{eq:withML}) are observables. Hence, in the case of
a low-energy seesaw, one can expect, in principle, to be able to
test whether there are contributions to the neutrino mass matrix
that are unrelated to the presence of right-handed neutrinos. By
measuring $V$, $m$, $M_R$, and $\Theta$, one can establish whether
$M_L$ is consistent with zero.

\section{Oscillation Phenomenology and Current Evidence for Low-Energy Seesaw}
\label{sec:pheno}

Here I examine a number of experimental and observational anomalies
that may be explained by light sterile neutrinos. More specifically,
I explore what these can teach us about the currently unknown
parameters of the seesaw Lagrangian, described in detail in
Sec.~\ref{sec:seesawMass}. In all cases I assume 3 mostly active and
3 mostly sterile neutrinos and, most of the time, will concentrate
on a $3+2+1$ picture of neutrino mass eigenstates, that is, three
mostly active sub-eV neutrinos, two mostly sterile eV neutrinos and
one almost completely sterile keV neutrino. The hope is that the
heavier state can  account for warm dark matter (section
\ref{subsec:WDM}) or pulsar kicks (Sec.~\ref{subsec:Kicks}), which
both require at least one keV neutrino, while the other two mostly
sterile states help ``explain'' the existing oscillation data where,
for all practical purposes, the heaviest neutrino decouples and I
are left with an effective $3+1$ or $3+2$ picture.  I remind readers
that a third possibility (2+2\footnote{It would have been rather
difficult to construct a 2+2 neutrino mass hierarchy using the
seesaw Lagrangian.}) is currently ruled out by solar and atmospheric
data \cite{CPTVorSterile,rule_out_4nu,Maltoni:2003yr} and will be
ignored. $3+1$ schemes that address the LSND anomaly are also
disfavored by global analysis of short baseline oscillation
experiments
\cite{CPTVorSterile,rule_out_4nu,Maltoni:2003yr,short_bl_analysis}
and, for this reason, I mostly concentrate on $3+2$ fits to the LSND
anomaly \cite{short_bl_analysis}.

Our analysis method is as follows:  For each experimental probe
considered I perform a $\chi^2$ ``fit'' of the mixing matrix $U$,
given by Eq.~(\ref{eq:U}), and neutrino masses to the ``data'', and
extract the region of parameter space that best explains the data.
In most cases I allow the light mixing angles and mass squared
differences to vary within their $1\sigma$ limits (according to
\cite{global_analy}),\footnote{In the case of 3+1 ``fits'' ({\it
cf.} Eqs.~(\ref{U_31_i},\ref{U31_n})), I kept the active  neutrino
parameters fixed at their best-fit values.} the angles $\phi_i$ to
vary unconstrained within their physical limits of $0-2\pi$, and the
lightest active neutrino mass eigenvalue $m_l$ to vary unconstrained
between $0-0.5$ eV. The quotation marks around  ``fit'' and ``data''
are meant to indicate that our methods are crude, in the sense that
I am fitting to previously processed experimental data, assuming a
diagonal correlation matrix with Gaussian uncertainties. In order to
avoid the subtleties involved in such a ``fit to a fit'', I hesitate
to mention actual confidence intervals, but are compelled to do so
for lack of a better measure of an allowed region.  I sometimes
present our best fit parameter points along with confidence
intervals, but warn the reader to avoid strict interpretations of
these numbers. While crude, our methodology of error analysis and
fitting provides a very useful  instrument for identifying whether
(and how) the low-energy seesaw can accommodate a particular
combination of data sets.

Before proceeding, it is useful to cement our notation. The neutrino
masses will be ordered in ascending order of magnitude from $m_1$ to
$m_6$ in the case of a normal active neutrino mass hierarchy
($m_2^2-m_1^2<m_3^2-m_1^2$), while in the case of an inverted mass
hierarchy they are ordered $m_3<m_1<m_2<m_4<m_5<m_6$ (in this case
$|m_3^2-m_1^2|>m_2^2-m_1^2$). The states with masses $m_{1,2,3}$ are
mostly active, while those with masses $m_{4,5,6}$ are mostly
sterile. Elements of the mixing matrix are referred to as $U_{\alpha
i}$, where $\alpha=e,\mu,\tau,s_1,s_2,s_3$ ($s$'s are the
right-handed neutrino degrees of freedom) and $i=1,2,3,4,5,6$. I
also define $\Delta m^2_{ji}=m_j^2-m_i^2$ and will refer to the
lightest active neutrino mass as $m_l$. In the case of normal
(inverted) active neutrino mass hierarchy $m_l=m_1$ ($m_l=m_3$).

\subsection{Short baseline oscillation constraints}
\label{subsec:LSND}

Here I analyze the constraints imposed on the unknown mixing
parameters by current neutrino oscillation data. I will assume that
all solar, reactor, long-baseline and atmospheric data are properly
fit with active--active oscillations, and that constraints on the
other seesaw parameters will be provided mostly by short-baseline
accelerator experiments. It is interesting to note that the
inclusion of the angles $\phi_{ij}$ introduces enormous freedom into
the system. Any one active--sterile mixing angle contained in
$\Theta$ can always be set to zero by an appropriate choice of $O$.
In fact, all but three elements may be set to zero simultaneously,
with only a single non-zero element in each row and column. In this
case, these are constrained to be around $\sqrt{m_l/M_{Ri}}$ where
$i$ is the column of the non-zero element. This is especially true
when the mostly active neutrino masses are quasi-degenerate. An
important ``sum rule of thumb'' is the following. For a given
right-handed neutrino mass $M_i$, the active--sterile mixing angle
squared is of order $m/M_i$, where $m$ is a typical active neutrino
mass. One can always choose parameters so that, for at most two
values of $\alpha=e,\mu,\tau$, $U_{\alpha i}$ are abnormally small.
In that case, however, the ``other'' $U_{\alpha i}$ is constrained
to saturate the bound $|U_{\alpha i}|^2\lesssim m_l/M_i$.

The most compelling evidence for light sterile neutrinos comes from
the short baseline oscillation experiment by the Liquid Scintillator
Neutrino Detector (LSND) collaboration at Los Alamos. Using a $\sim
30$ MeV $\overline{\nu}_\mu$ beam they observed a better than
$3\sigma$ excess of $\overline{\nu}_e$--like events above their
expected background at their detector some 30~m away
 from the production point \cite{LSND}. This evidence of $\overline{\nu}_\mu \leftrightarrow
\overline{\nu}_e$ oscillation requires a mass-squared difference
greater than $1~{\rm eV}^2$, clearly incompatible with the small
mass-squared differences observed between the active neutrinos.
Several mechanisms, such as CPT-violation \cite{CPTV,CPTVorSterile},
Lorentz invariance  violation \cite{lorentz_v}, quantum decoherence
\cite{decoherence}, sterile neutrino decay \cite{sterile_decay} and,
of course, oscillation into sterile neutrinos have been proposed to
explain this result.  Here I concentrate on the last possibility.

In order to take into account all short baseline data I ``fit'' our
mixing parameters and masses to the results of the $3+2$ performed
in \cite{short_bl_analysis}, which are summarized in Table
\ref{tab:LSND_fit} \cite{3p2fit}.  Here I assume that the heaviest,
mostly sterile state does not participate effectively in LSND
oscillations. This is guaranteed to happen if $m_6\gtrsim 10$~eV.
 On the other hand, $|U_{\alpha 6}|^2$ are partially constrained by our attempts to accommodate LSND data
 with seesaw sterile neutrinos, as will become clear in the next subsections.
\begin{table}[b]
\centering \caption[Parameter values used in analysis extracted from
global $3+2$ fit]{Parameter values used in our analysis. These were
extracted from a fit to all short baseline neutrino oscillation
experiments including LSND within the $3+2$ scenario
\cite{short_bl_analysis,3p2fit}. $1\sigma$ indicates a rough
estimate of the 1~sigma allowed range for the different parameters.}
\begin{tabular}{c|c c c c c c}
   & $U_{e4}$ & $U_{\mu 4}$ & $U_{e5}$ & $U_{e\mu 5}$ & $\Delta m^2_{41}~(\rm{eV}^2)$ & $\Delta m^2_{51}~(\rm{eV}^2)$ \\
   \hline
  Central Value & 0.121 & 0.204 & 0.036 & 0.224 & 0.92 & 22 \\
  $1\sigma$ & 0.015 & 0.027 & 0.034 & 0.018 & 0.08 & 2.4 \\
\end{tabular}
\label{tab:LSND_fit}
\end{table}

\begin{figure}
\begin{center}
\includegraphics[scale=0.6]{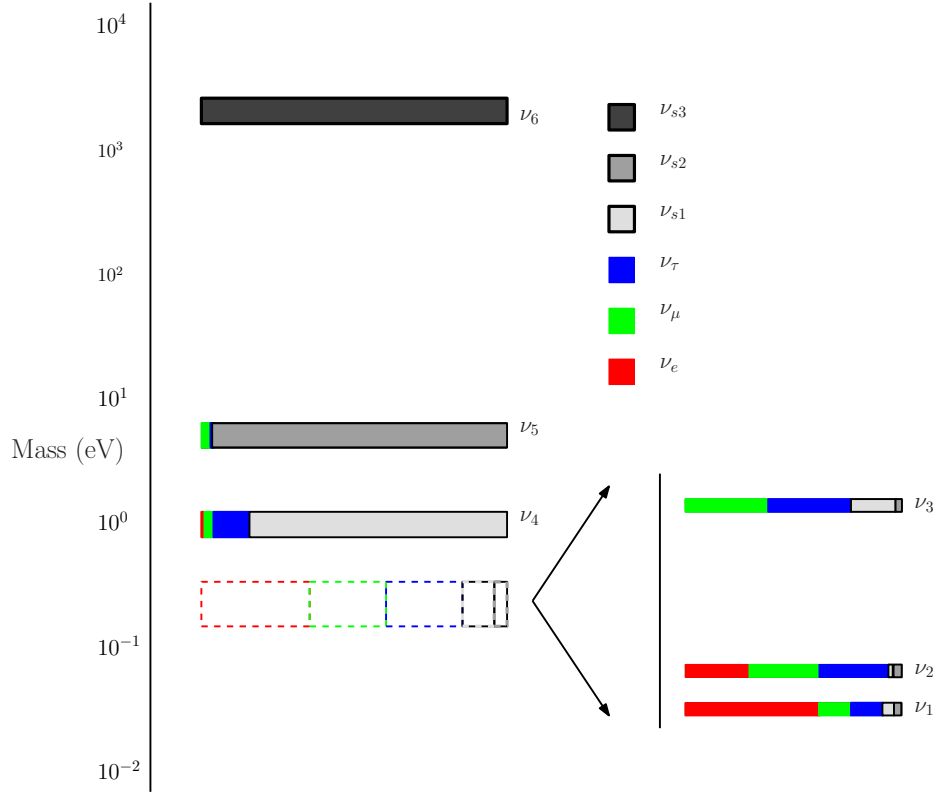}
\caption[Viable $3+2+1$ neutrino mass spectrum with flavor
composition]{Neutrino mass eigenstate spectrum, along with the
flavor composition of each state.  This case accommodates all
neutrino oscillation data, constraints from r-process
nucleosynthesis in supernovae, and may help explain anomalous pulsar
kicks (see text for details).  While I choose to depict a normal
hierarchy for the active neutrino states, an inverted active
neutrino mass hierarchy would have yielded exactly the same physics
(as far as the observables considered are concerned). }
\label{fig:spectrum}
\end{center}
\end{figure}

 I find that  $m_l$, the
lightest neutrino mass, is constrained to lie between
$(0.22-0.37)~\rm{eV}$, with a ``best fit'' value of $0.29~\rm{eV}$.
Thus  the active neutrino mass spectrum is predicted to be
quasi-degenerate.  A sample $6\times 6$ neutrino mixing matrix that
fits all data is 
\begin{equation}
U_{3+2} = \left(
  \begin{array}{cccccc}
   0.8301 &   0.5571 &   0.001365 &   0.1193 &  -0.009399 &
   -0.006513\\
  -0.3946 &   0.5866 &   0.7072 &   0.2016 &   0.2262 &
  0.0003363\\
   0.3932 &  -0.5879 &   0.7070 &   0.4760 &  -0.0949 &   0.001470\\
  -0.2067 &   0.09514 &  -0.4792 &   1 &                  0&
  0\\
   0.1343 &  -0.1832 &  -0.09284 &                  0&   1&
   0\\
   0.004963 &  0.004295 &  -0.001268 &                  0 &                 0&
   1\\
  \end{array}
\right), \label{U_32}
\end{equation}
while the associated masses are $m_1\simeq m_2\simeq m_3=0.28$~eV,
$m_4=1.0$~eV, $m_5=4.7$~eV, and $m_6=6.4$~keV. Note that the matrix
in Eq.~(\ref{U_32}) is only approximately unitary, up to corrections
of order 25\%. This result agrees qualitative with those obtained in
\cite{SeeSaw_LSND}. The neutrino masses and mixings obtained in this
``fit'' are depicted in Fig.~\ref{fig:spectrum}. I will use the
results of ``fits'' similar to this one throughout the chapter.

One can also  aim at a (currently disfavored) $3+1$ LSND
fit.\footnote{This is easily accomplished by requiring $m_5\gtrsim
10$~eV.}. In this case,  much lower $m_l$ values are also allowed,
extending well into the hierarchical spectrum range. In this case,
all $m_l$ values above $0.01~\rm{eV}$ and $0.03~\rm{eV}$ are
allowed, assuming an inverted and normal mass hierarchy,
respectively. This is to be compared with the results found in
\cite{SeeSaw_LSND}, where only trivial choices for $O$ were
considered. Examples that ``fit'' all oscillation data include, for
an inverted active mass hierarchy: $m_1\simeq m_2=0.066$~eV,
$m_3=0.043$~eV, $m_4=0.96$~eV, $m_5=5$~keV, and $m_6=10$~GeV,
together with
{\Small \begin{equation}
U_{3+1}^{\rm inverted} = \left(
  \begin{array}{cccccc}
   0.8305 &  0.5571 &  0 &  0.1359 &  -0.00009142 &  -0.000002198 \\
  -0.3939 &   0.5872 &   0.7071 &   0.2046 &   0.00005000 &
  0.000001202\\
   0.3939 &  -0.5872 &   0.7071 &  -0.04421 &  -0.003236 &
   -0.0000001857\\
  -0.01486 &  -0.2218 &  -0.1134 &   1 &                  0 &
  0\\
   0.001370 &  -0.001878 &   0.002253 &                  0 &  1         &
   0\\
   0.000002372 &   0.0000004094 &  -0.0000007187    &               0    &              0 &
   1\\
  \end{array}
\right). \label{U_31_i}
\end{equation}}
For a normal mass hierarchy, I find that $m_1=0.055$~eV,
$m_2=0.056$~eV , $m_3=0.0744$~eV, $m_4=0.96$~eV, $m_5=5$~keV,
$m_6=10$~GeV, and
{\Small \begin{equation}
U^{\rm normal}_{3+1} = \left(
  \begin{array}{cccccc}
   0.8305 &   0.5571 &                  0 &  0.1173 &  -0.002100 &
   0.000001418\\
  -0.3939 &   0.5872 &   0.7071 &   0.2176 &   0.0004625 &
  -0.000001364\\
   0.3939 &  -0.5872 &   0.7071 &   0.09802 &   0.002804 &
   0.000001283\\
  -0.05028 &  -0.1355 &  -0.2231 &   1 &                  0 &
  0\\
   0.0008214 &   0.002545 &  -0.002310 &                  0  & 1       &
   0\\
  -0.000002220 &   0.0000007646 &   0.00000005663 &                  0  &                0 &
  1\\
  \end{array}
\right), \label{U31_n}
\end{equation} }
``fit'' all oscillation data quite well.

Note that a null result from MiniBooNE is bound to place significant
limits on the seesaw energy scale. If all right-handed neutrino
masses are similar, the effective mixing angle that governs
$\nu_\mu\rightarrow\nu_e$ transitions is $\sin^22\theta_{\rm
MiniBooNE}\lesssim 4 m^2/M^2$. Hence, a null result at MiniBooNE
would rule out a seesaw energy scale $M$ lighter than 6~eV, assuming
all active neutrino masses $m$ are around 0.1~eV \cite{minib}. This
limit is sensitive to the lightest neutrino mass $m_l$ and can be
somewhat relaxed (similar to how I obtain a good 3+2 to all neutrino
data) by postulating a (mild) hierarchy of right-handed neutrino
masses and by assuming that sterile-electron and sterile-muon
neutrino mixing is suppressed with respect to naive expectations for
the lightest mostly sterile state(s). For larger values of $m_l$,
$M$ values around 10~eV are already constrained by
$\nu_{\mu}\to\nu_e$ searches at the NuTeV \cite{nutev} and NOMAD
\cite{nomad} experiments, and $\nu_{\mu}\to\nu_{\tau}$ searches at
CHORUS \cite{chorus}.

\setcounter{footnote}{0}
\subsection{Cosmological and Astrophysical Constraints, Warm Dark Matter}
\label{subsec:WDM}

Very light sterile neutrinos that mix with the active neutrinos are
constrained by several cosmological and astrophysical observables.
The ``seesaw'' right-handed neutrinos are no exception. Given that
active--sterile mixing angles $|U_{\alpha i}|^2\lesssim m_l/m_i$
($\alpha=e,\mu,\tau$, $i=4,5,6$), it turns out that for ``standard
cosmology,'' the right-handed neutrinos thermalize with the early
universe thermal bath of SM particles, as long as the reheat
temperature is higher than their Majorana masses. For the low seesaw
energy scales I am interested in, this is a problem. For the values
of $M_R$ under consideration here, thermal right-handed neutrinos
easily overclose the universe. Smaller $m_l$ values ($m_l\lesssim
10^{-5}$~eV) lead to the possibility that right-handed neutrinos are
the dark matter, as recently discussed in the literature
\cite{nuSM_dark,nuSM_kicks}. I comment on this and other
possibilities shortly.

\begin{figure}
\begin{center}
\includegraphics[angle = 270,scale=0.575]{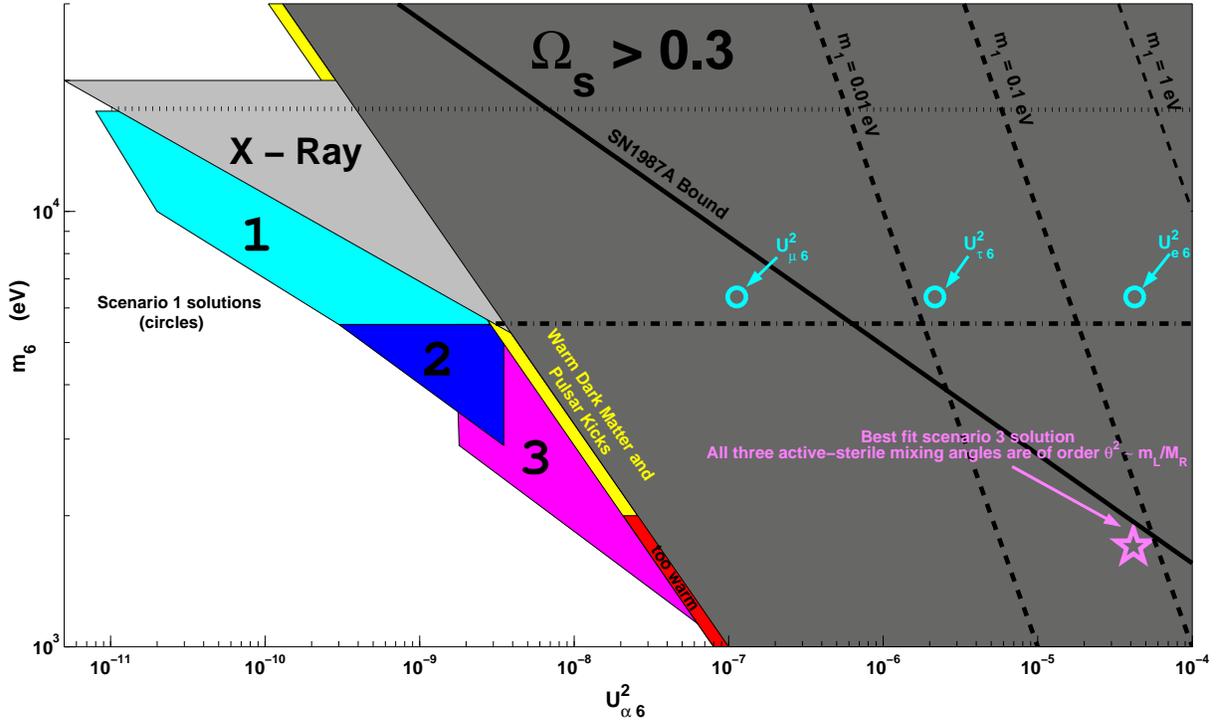}
\caption[Cosmological and astrophysical constraints on the
$|U_{\alpha 6}|^2\times m_6$-plane]{Adapted from \cite{kicks_gen}.
Cosmological and astrophysical constraints on the $|U_{\alpha
6}|^2\times m_6$-plane. In the large dark grey region, the density
of a thermal $\nu_6$ population is $\Omega_s>0.3$, while the light
grey `X-ray' region is disfavored by X-ray observations.  The
regions labeled 1,2,3 are preferred if one is to explain anomalous
pulsar kicks with active--sterile oscillations inside supernovae.
Regions 1 and 3 qualitatively extend inside the  $\Omega_s>0.3$ part
of the plane as indicated by the horizontal dotted and dash-dotted
lines, respectively. The regions `Warm Dark Matter'' and ``Too Warm
Dark Matter'' are meant to represent the region of parameter space
where thermal $\nu_6$ qualifies as a good (or bad) warm dark matter
candidate. The region above the solid diagonal line is disfavored by
the observation of electron (anti)neutrinos from SN1987A. The
diagonal dashed lines correspond to $U_{e6}^2=m_l/m_6$, for
different values of $m_l$.
 Also shown is our ``best fit'' sterile solution
for different  pulsar kick scenarios, assuming the $3+2$ LSND fit
for the lighter states.  The regions one and three best fit values
are represented by circles and a star respectively. See text for
details.} \label{fig:kicks}
\end{center}
\end{figure}

Fig.~\ref{fig:kicks} depicts the region of the $|U_{e6}|^2 \times
m_6$-plane in which the contribution of the heaviest neutrino
$\nu_6$ to $\Omega$ (the normalized energy density of the universe,
$\rho/\rho_c$) is larger than 0.3 (dark region). The same constraint
roughly applies for all $\alpha=e,\mu,\tau$ and $i=4,5,6$. The
dashed diagonal lines correspond to $|U_{\alpha 6}|^2=m_l/m_6$, for
different values of $m_l$. All lines lie deep within the dark
$\Omega_s>0.3$ region.

For smaller values of $M_R$, the situation is also constrained. For
$M_R$ values below tens of eV, thermal sterile neutrinos contribute
to the amount of hot dark matter in the universe
\cite{wmap,seljak,Hannestad:2006mi,Fogli:2006yq,Cirelli:2006kt}.
Right-handed neutrinos will thermalize as long as
$m_i\sin^2\theta_{i\alpha}\gtrsim 5\times 10^{-4}$~eV
\cite{thermalize,smirnov_renata}. In low energy seesaws, this
roughly translate into $m\gtrsim 10^{-3}$~eV, where $m$ is the
active neutrino mass scale. For the cases of interest here $m\gtrsim
\sqrt{\Delta m^2_{21}}\sim 10^{-2}$~eV, in which case $m_i$ values
above somewhere between\footnote{The upper bound on the sum of
neutrino masses from cosmology depends on several assumptions that
go into analyzing the different cosmological observations. These
include the issue of defining the values of the concordance
cosmological model and deciding which data sets to include in the
fit.}  0.2--2~eV are ruled out
\cite{wmap,Dodelson:2005tp,seljak,Hannestad:2006mi,Fogli:2006yq,Cirelli:2006kt}.
Note that in our ``best fit'' 3+2 solution to the LSND data, the sum
of all active neutrino masses violates slightly some of these
constraints.  This problem can be easily evaded if I choose $m_l$
values close to the lower end of the ``allowed region''. Other
``mild'' non-standard cosmology effects (see, for example,
\cite{Beacom:2004yd}) are also known to alleviate the hot dark
matter bound on neutrino masses. Note that any sterile neutrino
solution to the LSND anomaly faces a similar problem, which must be
resolved with non-standard cosmology.

Big-bang nucleosynthesis also proves to be a large obstacle when it
comes to the existence of light sterile neutrinos in thermal
equilibrium in the early Universe (at temperatures above several
MeV). In the absence of ``non-standard'' assumptions, big-bang
nucleosynthesis constrains the existence of new thermal relativistic
degrees of freedom (see, for example,
\cite{Cirelli:2006kt,Barger:2003zg}).

One way to avoid the bounds described above (see also, for example,
\cite{Abazajian:2004aj}) is to consider that the reheating
temperature $T_r$ of the universe is very low. This way,
right-handed neutrinos, in spite of their ``large'' mixing angles,
never reach thermal equilibrium in the early universe and neither
overclose the universe nor contribute to the amount of hot dark
matter. Quantitatively, $T_r\lesssim 5$~MeV is sufficient to avoid
eV-mass (or heavier) sterile neutrinos that are allowed to explain
the LSND anomaly \cite{low_reheat}. Unless otherwise noted, this is
the assumption I make here. Other possibilities include adding new
neutrino interactions to lighter scalars, so that neutrinos remain
in thermal equilibrium until they are non-relativistic
\cite{Beacom:2004yd}. According to \cite{Beacom:2004yd}, this can
even be accomplished for light neutrinos, as long as the
neutrino--scalar field coupling is finely tuned (see, however,
\cite{Cirelli:2006kt}). Yet another possibility is to consider that
the lepton asymmetry of the universe is large. The authors of
\cite{Chu:2006ua} have recently studied this issue in detailed, and
concluded that a lepton asymmetry of order $10^{-4}$ is required in
order to allow the existence of LSND sterile neutrinos to be in
agreement with data from large-scale structure and big-bang
nucleosynthesis (see also \cite{Abazajian:2004aj}).

Under these circumstances, it is interesting to consider whether
light seesaw right-handed neutrinos still qualify as good warm dark
matter. This could happen if their production in the early universe
was non-thermal.  One concern surrounding warm dark matter is
whether it is ruled out by large scale structure surveys. Here, I
will not add to this discussion but refer readers to the recent
literature on the subject \cite{warm_recent}. A brief summary of the
situation is as follows: constraints on warm dark matter can be
translated into a lower bound on the mostly sterile neutrino mass.
The lower bound has been computed by different groups, and lies
somewhere between 3 and 14~keV \cite{warm_recent}. Different lower
bounds depend on several issues, including which subset of Lyman
alpha-forest data is taken into account.

Another constraint on potential dark matter sterile neutrinos comes
from the observation of X-rays originating in galactic clusters.
Such regions of the universe should be overdense with warm dark
matter heavy neutrinos, which can be directly observed via their
radiative decay $\nu_6 \rightarrow \nu_i + \gamma$ \cite{Xray}.
Bounds from X-ray observations have been summarized very recently in
\cite{Kusenko:2006wa}. Combining the results of
\cite{Kusenko:2006wa} and Fig.~\ref{fig:kicks}, I find that for
lightest neutrino masses larger than $10^{-2}$~eV, such bounds can
only be avoided for $m_i\lesssim 100$~eV, where large scale
structure constraints on warm/hot dark matter are severe. This
qualitative analysis indicates that seesaw sterile neutrinos cannot,
simultaneously, fit the LSND data and serve as cold dark matter.

On the astrophysics side, the most severe constraint on light,
sterile neutrinos is provided by the observation of electron
(anti)neutrinos coming from SN1987A. The current analysis consists
of comparing the model-dependent neutrino flux at the surface of the
neutrinosphere with that detected on Earth.  Large sterile neutrino
mixing and mass would result in modification/depletion of the
detected neutrino signal (for a recent detailed discussion, see
\cite{sterile_probe}). Although only twenty neutrinos were observed
in this event, one can still place bounds on sterile-active neutrino
mixing. As far as ``LSND'' sterile neutrinos are concerned, these
bounds are still weaker than those obtained by the null short
baseline oscillation experiments \cite{SN1987_bound} and therefore
already accounted for in our analysis.  Heavier right-handed
neutrinos can, however, be excluded by SN1987A neutrino data.
Fig.~\ref{fig:kicks} depicts the region of parameter space excluded
by SN1987A data (region to the right of solid, diagonal line). This
bound is defined by $m_i\sqrt{2}\sqrt{U_{\alpha i}}>$0.22~keV
\cite{sn_bound,low_reheat}. See also \cite{smirnov_renata}.
According to Fig.~\ref{fig:kicks}, supernova bounds force the seesaw
scale to be below a few keV for $m_l$ values above $0.01$~eV.

%In the next subsection, we comment on a potential positive evidence for light sterile neutrinos coming from supernova explosions.

\subsection{Pulsar Kicks}
\label{subsec:Kicks}

 Pulsars are born from the
gravitational collapse of the iron core of a massive star.  These
core collapse supernova are an excellent source of neutrinos,
producing all (active) flavors copiously (see \cite{SN_review} for a
detailed review). Current observations point to the fact that some
pulsars move with peculiar velocities much greater than those
expected from an asymmetric supernova explosion mechanism.
Quantitatively, current three dimensional models yield velocities up
to $200~\rm{km/s}$ \cite{pulsar_disp} while pulsars moving at speeds
as high as $1600~\rm{km/s}$ have been observed. I note, however,
that some two-dimensional hydrodynamic studies \cite{Scheck:2006rw}
indicate that natural anisotropies generated during supernova
explosions can, in fact, yield large neutron star velocities
consistent with observations. However, more simulations seem to be
required in order to validate this claim. Here, I will operate under
the hypothesis that new physics, usually in the form of new neutrino
physics, is responsible for the large pulsar kicks.

Since roughly $99\%$ of the approximately $10^{53}~\rm{ergs}$ of
energy released in a core collapse supernova is in the form of
neutrinos, it is reasonable that neutrino physics provides a
solution to this anomaly. At these rates a small $(1-3) \%$
asymmetry in neutrino emission can account for the observed large
pulsar velocities.  Neutrinos are always \emph{produced}
asymmetrically in the polarized medium of the proto-neutron star,
due to the left--handed nature of their interactions. Unfortunately,
asymmetric production cannot solve this problem because the
associated medium densities are such that neutrinos undergo multiple
scattering within the star's interior, eventually diffusing out of
an effective surface, called the neutrinosphere, with all initial
asymmetries washed away. Several distinct mechanisms have been
formulated to sidestep this fact. Specifically, the existence of
large neutrino magnetic moments has been explored in
\cite{mmoment_kick}, and can be tested in next generation neutrino
scattering experiments \cite{nu_e_scattering,NuSOnGPRD,munu,texono}.  Proposed
solutions also exist which utilize standard three flavor neutrino
oscillations, where $\nu_\mu$ and $\nu_\tau$ appearing between their
neutrinosphere and the larger $\nu_e$ neutrinosphere can stream
unhindered out of the star \cite{active_kicks}.  This solution is,
however, currently disfavored by terrestrial oscillation
experiments.

I concentrate on  the case of oscillations into sterile neutrinos,
which can proceed in various ways, depending on the mass and
coupling of the relevant neutrinos as well as the properties of the
collapsing star, including its density and magnetic field. Following
\cite{kicks_gen}, I separate and analyze these within three distinct
categories.  Each one requires the existence of a keV-scale sterile
neutrino with very small couplings to the active flavors, of the
order $10^{-4} - 10^{-5}$, especially if light sterile neutrinos are
thermally produced in the early universe. Under these circumstances,
if seesaw neutrinos are to play the role of the sterile neutrinos
responsible for pulsar kicks,  $|U_{\alpha i}|^2\lesssim m/m_i$
($i=4,5,6$) must lie in the $10^{-9}$ range for $m_i\sim 10^{4}$~eV.
This implies $m\sim 10^{-5}$~eV and is only compatible with a
hierarchical active neutrino mass spectrum and very light $m_l$, as
identified in \cite{nuSM_dark,nuSM_kicks}.

Here, instead, I will concentrate on identifying solutions that will
address pulsar kicks and the LSND anomaly. According to the
discussion in the previous subsection, the mostly sterile neutrino
masses $m_4$ and $m_5$ are constrained to be less than 10~eV so that
a 3+2 solution to the LSND anomaly can be obtained from the seesaw
Lagrangian. The heaviest neutrino mass $m_6$ is unconstrained, so I
are free to vary it as needed in order to attack the pulsar peculiar
velocity issue.\footnote{I can neglect the lighter sterile neutrinos
($\nu_4$ and $\nu_5$) as they should not alter the kicking mechanism
significantly due to their small mass and non-resonant production.}
Naively, the  fraction of $\nu_{\alpha}$ ($\alpha=e,\mu,\tau$) in
$\nu_6$ is expected to be of the order $U_{\alpha 6} \sim
\sqrt{0.3~\rm{eV}/3\times 10^3~\rm{eV}} = 10^{-2}$, much too large
to satisfy the pulsar kick plus cosmology constraints summarized in
Fig.~\ref{fig:kicks}. On the other hand, once the $\Omega_S<0.3$
constraint is removed, the `pulsar kicks' allowed region of the
plane is significantly enlarged, as qualitatively  indicated by the
horizontal lines in Fig.~\ref{fig:kicks}. In this case, which I must
consider anyway if I am to have agreement between LSND and searches
for hot dark matter, one can envision explaining pulsar kicks and
the LSND data simultaneously. Note that once heavy sterile neutrinos
are ``removed'' (so that they do not overclose the universe),
constraints from X-ray observations (see Fig.~\ref{fig:kicks}) are
also removed.

In scenario 1, the pulsar kick is produced via an active--sterile
MSW resonance in the core of the proto-neutron star at large
densities, greater than $10^{14}~\rm{g/cm^3}$, and magnetic fields,
near $10^{16}~\rm{G}$ \cite{kicks_res}. The effective neutrino
matter potential in material polarized by a strong magnetic field
contains a term proportional to $\vec{k} \cdot \vec{B}/|\vec{k}|$
\cite{nu_Bfield1,nu_Bfield2}, where $\vec{k}$ is the neutrino's
three-momentum and $\vec{B}$ is the local magnetic field vector.
Clearly the MSW resonance occurs at a radius that depends on
$\vec{k} \cdot \vec{B}/|\vec{k}|$, the relative orientation of the
neutrino momentum and magnetic field.  Sterile neutrinos produced at
smaller radii (higher temperatures) carry greater average momentum
than those produced at larger radii (lower temperatures), yielding
an asymmetric momentum distribution of emitted neutrinos. This
asymmetry is capable of producing the observed pulsar kicks, in the
direction of the magnetic field, when the mass and coupling of the
sterile state is near $8~\rm{keV}$ and above $1.5\times 10^{-5}$,
respectively \cite{kicks_gen}. I found the  ``best fit'' to the LSND
data (using $\nu_4$ and $\nu_5$) and pulsar kicks (using $\nu_6$)
and $m_6>5$~keV. The $|U_{\alpha 6}|^2$ and $m_6$ ``best fit''
values are depicted in Fig.~\ref{fig:kicks}. This solution is
strongly disfavored by the observation of neutrinos from SN1987A.
The fact that $|U_{e6}|$ is much larger than the  other two
active--sterile mixing angles is due to the fact that $|U_{e4}|$ and
$|U_{e5}|$ are constrained by LSND data to be much smaller than
naive expectations (see Eq.~(\ref{U_32})). In order to reduce
$|U_{e6}|$, one would have to either reduce $m_l$ by an order of
magnitude -- which renders the 3+2 fit to oscillation data very poor
-- or increase $m_6$, which would only push  $|U_{e6}|$ deeper into
the region of parameter space ruled out by SN1987A. One can however,
find 3+1 solutions to LSND data where $\nu_5$ could pose as the
sterile neutrino that explains why pulsar peculiar velocities are so
large (see Eqs.~(\ref{U_31_i},\ref{U31_n})).

Scenario 2 also relies on a direction-dependent MSW resonance, this
time occurring outside the core where the matter density and
temperature are much lower.  Here, both the active and sterile
neutrinos are free to stream out of the star.  The departing active
flavors still have a small interaction cross-section, $\sigma \sim
G_F^2 E_\nu^2$, and can therefore deposit energy and momentum into
the star's gravitationally bound envelope proportional to the matter
it transverses.  Via the direction-dependent resonance, neutrinos
moving in the direction of the magnetic field remain active longer,
deposit more momentum, and thus kick the star forward.  The
observations can be explained in this case with a smaller sterile
neutrino mass and larger active--sterile coupling near $4~\rm{keV}$
and $4.5\times 10^{-5}$ respectively \cite{kicks_gen}.  In the case
of our LSND ``fit'' to the data, I can constrain one of $|U_{\mu
6}|^2$ or $|U_{\tau 6}|^2$ to lie inside region 2. The other
$|U_{\alpha6}|^2$ ($\alpha=e,\tau$ or $e,\mu$), however, are
constrained to be large, thus violating the SN1987A bound in much
the same way as the scenario 1 best fit results.�� Another
possibility is to choose all $|U_{\alpha6}|^2$ of the same order of
magnitude.  I do not explicitly consider these points as they reside
in the region of parameter space where region 2 and 3 overlap, and
behave in the same way as the point described below, under scenario
3.

Scenario 3 proceeds through off-resonance production of the sterile
neutrino in the proto-neutron star core \cite{kicks_nonres}.  The
amplitude for sterile neutrino production by a weak process is
proportional to $U_{\alpha 6}^m$, the effective mixing angle between
the heavy mass eigenstate and the flavor eigenstate. Initially, this
quantity is very small due to matter effects in the dense core. The
effective potential in the star's interior is quickly driven to zero
in the presence of sterile neutrino production by a negative
feedback mechanism.  If this occurs in a time less than the
diffusion time-scale for the active neutrinos, approximately
$(3-10)$~s, the mixing angle will reach its vacuum value
\cite{sterile_hotwarmcold}.  The sterile neutrinos will then be
produced and emitted asymmetrically and thus kick the pulsar to
large velocities.  Lower limits on the vacuum mass and mixing values
are derived by requiring that the off-resonance time scale
(inversely proportional to $m_6^4 \sin^2 2\theta_{\alpha 6}$) for
the evolution of the matter potential to zero be less than about ten
seconds. This places the sterile mass and mixing at approximately
$1~\rm{keV}$ and above $5\times 10^{-5}$, respectively. Since all
three active flavors are present in equal abundances, and all
contribute to the effective matter potential, the mixing angle in
question is not any particular $U_{\alpha 6}$. Rather it is the
angle, $\theta_6$ associated with the projection of $\nu_6$ onto the
space spanned by $\nu_e$, $\nu_\mu$ and $\nu_\tau$, that is
$\theta_6^2 \equiv U_{e 6}^2 + U_{\mu 6}^2 + U_{\tau 6}^2$.  From
Eq.~(\ref{eq:thetaM_constraint}) I see that $\theta_6 =
\sqrt{m_l/m_6}$ up to corrections due to the non-unitarity of $V$
and active neutrino mass differences. This is independent of  mixing
angles, and therefore cannot be tuned to be small. Our ``best fit''
region-3 solution is depicted in Fig.~\ref{fig:kicks} by a star. It
turns out that $U_{\alpha 6}$ have very similar values for
$\alpha=e,\mu,\tau$. In order to evade the SN1987A constraint, I
were forced to pick $m_l$ values close to lower bound of our 3+2
LSND ``fit'' ($m_l=0.22$~eV), so that $|U_{e4}|$, $|U_{\mu4}|$, and
$|U_{\mu5}|$ are close to the low-end of the allowed range in
Table~\ref{tab:LSND_fit}.

\subsection{Supernova Nucleosynthesis}
\label{subsec:Nucleosynthesis}

Core collapse supernova are believed to produce the observed heavy
element ($A \geq 100$) abundance through the r-process, or rapid
neutron capture process. Here I briefly review this mechanism (see
\cite{rprocess_nuwind} for a comprehensive review), as well as its
facilitation by the addition of active--sterile neutrino
oscillations \cite{nucleosynthesis1}.  This scenario begins in the
neutrino driven wind; that is, the wind of ejected nucleons driven
by neutrinos radiated from the cooling proto-neutron star. The
maintenance of equilibrium among neutrons, protons, and electron
(anti)neutrinos in neutrino capture processes leads to a
neutrino--rich environment. As the wind propagates, it cools enough
for all free protons to bind into $\alpha$ particles. In the ideal
r-process picture, as the wind cools further these $\alpha$
particles bind into intermediate size seed nuclei which later
undergo neutron capture to form the observed heavy r-process
elements.

This ideal scenario is dampened by the large number of electron
neutrinos present at the stage of $\alpha$ particle formation. These
will capture on the free neutrons, converting them to protons, which
in turn will fuse to make more $\alpha$ particles.  The end result
is a very small free neutron to $\alpha$ particle ratio, conditions
unfavorable for r-process element formation.  This is known as the
$\alpha$ effect and must be circumvented to produce the correct
distribution of heavy elements. A clear solution to this problem is
to reduce the number of electron neutrinos present at this stage,
which can be accomplished by resonant $\nu_e \rightarrow \nu_s$
conversion\footnote{In this mechanism the effective matter
potential, which depends on the number of electrons, positrons and
neutrinos, varies wildly as a function of distance from the core.
Along this radial direction there are three relevant MSW resonant
conversions that must be tracked and understood:  $\nu_e \rightarrow
\nu_s$, $\bar{\nu}_e \rightarrow \bar{\nu}_s$ and $\bar{\nu}_s
\rightarrow \bar{\nu}_e$. See \cite{nucleosynthesis1} for more
information. } \cite{nucleosynthesis1}.

The sterile neutrino solution to the r-process mechanism is modeled
and fit to the data in reference \cite{nucleosynthesis2} including
the effects of relevant nuclear physics and additional neutrino
oscillations in the star's envelope.  The analysis is expanded in
\cite{nucleosynthesis_fisscy} with the inclusion of fission cycling
of the produced heavy elements.  The analysis indicates the need for
an eV-scale sterile neutrino with an allowed parameter space much
larger than that constrained by LSND.  By itself, the requirement of
successful r-process in supernovae only weakly constrains the light
neutrino mass scale to be greater than $10^{-2}~\rm{eV}$ and
$10^{-3}~\rm{eV}$ for a $1~\rm{eV}$ and $10~\rm{eV}$ sterile
neutrino, respectively.  With regard to the LSND results, it has
been demonstrated that the $3+1$ oscillation scenario fits within
this parameter space \cite{nucleosynthesis_fisscy}. Considering that
the best fit mass-squared difference and mixing angles for the
fourth mass eigenstate, which makes up most of the lightest sterile
neutrino, is very similar between the $3+1$ and $3+2$ case
\cite{3p2fit}, it is reasonable to conclude that $\nu_e
\leftrightarrow \nu_{s4}$ resonant conversion will also fit within
this scenario.  Even oscillations into the heavier $\nu_{s5}$ state
can potentially solve this anomaly if the neutrino driven wind
expansion time-scale is sufficiently small, $ \leq 0.1~\rm{sec}$. To
conclude this section I note that, although the sterile neutrino
solution to the supernova nucleosynthesis problem fits well within
our seesaw framework, it adds no additional constraints, and
therefore does not increase the predictability of our scenario.

\section{Other Probes of the Seesaw Energy Scale}
\label{sec:future}

Here I survey other existing and future probes of light sterile
neutrinos.  As opposed to the previous cases, these probes are
perfectly consistent by themselves.  That is, extra heavy neutrinos
are not required to solve problems within the system. However, their
addition can lead to large modifications to the outcome of such
experiments, thus rendering the eV-scale seesaw scenario testable.
Specifically, I consider bounds from tritium beta-decay and
neutrinoless double-beta decay.  Observations in all of these areas
have already yielded useful constraints on sterile neutrinos, and
the situation is expected to improve in the next few years.

\subsection{Tritium Beta-Decay}

The endpoint of the electron energy spectrum in the beta-decay of
tritium is a powerful probe of nonzero neutrino masses.  This
results from the decay kinematics of the system which  is
necessarily modified by the presence of a massive neutrino. The
nonzero neutrino mass effect can be understood almost entirely from
the analysis of the phase space distribution of the emitted
electrons, and is therefore quite model independent. Existing beta
decay experiments extract limits on an effective electron neutrino
mass $m_{\nu_e}^2 = \sum_i |U_{ei}|^2 m_i^2$ \cite{TritNuEffMass},
provided that the neutrino masses are smaller than the detector
energy resolution. Currently the most stringent bounds on
$m_{\nu_e}^2$ are $(2.3)^2~\rm{eV}^2$ at $95\%$ confidence from the
Mainz experiment \cite{Mainz} and $(2.5)^2~\rm{eV}^2$ at $95\%$
confidence from the Troitsk experiment \cite{Troitsk}.  In the next
few years the Katrin experiment should exceed these limits by nearly
two orders of magnitude, probing down to $(0.2)^2~\rm{eV}^2$ at the
$90\%$ confidence level \cite{Katrin}.  One might naively compute
this effective mass for the ``best fit'' mixing parameters obtained
in the previous section. In this case, I expect a keV seesaw
neutrino to contribute to $m^2_{\nu_e}$ by a huge amount $\delta
m_{\nu_e}^2 = U_{e6}^2 m_6^2 \sim \frac{m_l}{m_6} m_6^2 = m_l m_6$
\cite{SeeSaw_LSND} so that it would be excluded by the current
precision measures of tritium beta-decay for $m_l\gtrsim
10^{-3}$~eV. This is clearly an incorrect treatment of the physics.
As pointed out in, for example,  \cite{Farzan:2001cj}, the existence
of a heavy neutrino state would produce a kink in the electron
energy spectrum of size $|U_{ei}|^2$ at an energy $E_0 - m_i$ as
well as a suppression of events at the endpoint of order $1 -
|U_{ei}|^2$. Here $U_{ei}$ ($i=4,5,6$) is the mixing between the
electron neutrino flavor eigenstate and the heavy mass eigenstate,
while $E_0 = 18.6~\rm{keV}$ is the endpoint energy of tritium
beta-decay.

\begin{figure}
\begin{center}
\includegraphics[scale=0.70]{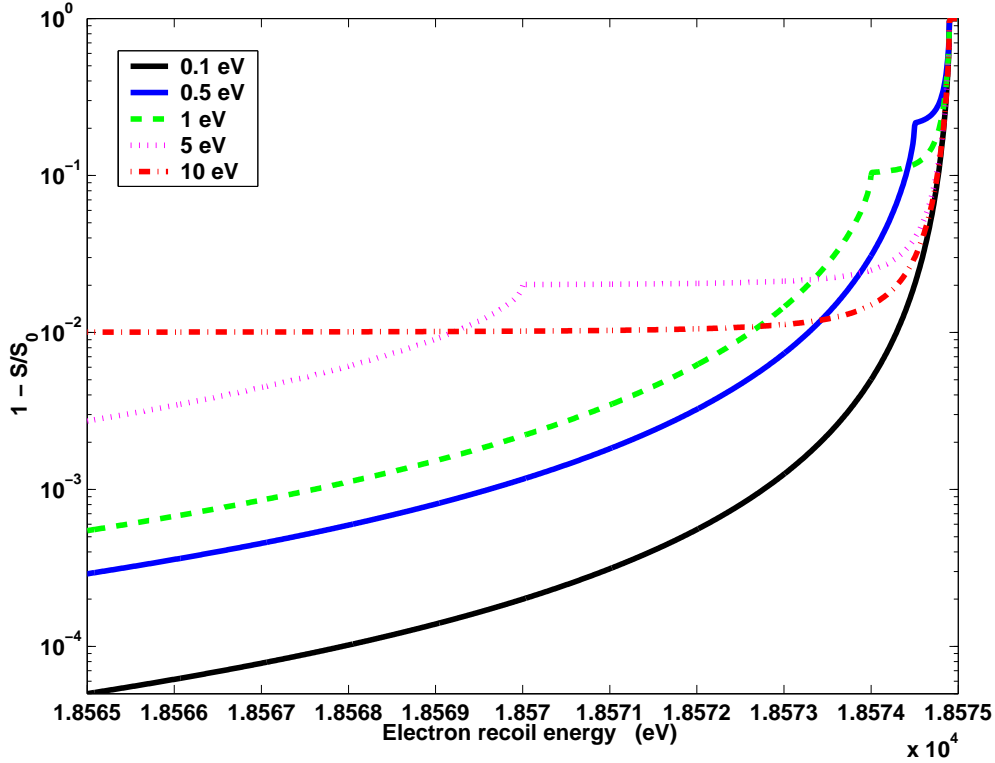}
\caption[$1-S/S_0$ as a function of the tritium
$\beta$-ray energy]{$1-S/S_0$ as a function of the $\beta$-ray
energy, where $S$ is the $\beta$-ray energy spectrum obtained
assuming three mostly active, degenerate neutrinos with mass
$m=0.1$~eV and one mostly sterile neutrino $\nu_i$ with
$m_i=0.1,0.5,1,5,$ and $10$~eV. The mixing angle is given by
$U_{ei}^2=m/m_i$. $S_0$ is the spectrum associated with massless
neutrinos. See text for details.} \label{fig:spectra}
\end{center}
\end{figure}

Fig.~\ref{fig:spectra} depicts $1-S/S_0$, where $S$ is the
$\beta$-ray energy spectrum obtained assuming three mostly active,
degenerate neutrinos with mass $m=0.1$~eV and one mostly sterile
neutrino $\nu_i$ with various masses $m_i$  and mixing  angle
$U_{ei}^2=m/m_i$, while $S_0$ is the spectrum associated with
massless neutrinos. One can readily observe ``kinks'' in the
spectrum above $m_i$. For $\beta$-energies above $E_0-m_i$, the
impact of the sterile state is to ``remove'' around $1 - |U_{ei}|^2$
of the $\beta$-rays from spectrum. This is most significant between
$E_0-m_i$ and $E_0$ minus the mass of the active neutrinos. For
energies below $E_0-m_i$, the spectrum agrees with that obtained
from the emission of one effective neutrino with mass-squared
$m_{\nu_e}^2$.

I estimate the sensitivity of future tritium beta-decay experiments
to the emission of one heavy state by considering the ratio between
the number of electrons with energies above $E_0-\Delta E$ in the
case of one heavy massive neutrino $\nu_i$ and in the case of
massless neutrinos
\begin{equation}
 R(U_{ei},M_R) = \frac{|U_{ei}|^2 \int_{E_0 - \Delta E}^{E_0}
 dE \frac{dN}{dE}(m_i) + \left(1 - |U_{ei}|^2 \right)\int_{E_0 - \Delta E}^{E_0} dE
 \frac{dN}{dE}(0)}{\int_{E_0 - \Delta E}^{E_0} dE
 \frac{dN}{dE}(0)},
 \label{equ:R}
 \end{equation}
 where $dN/dE$ is the energy distribution of $\beta$-rays, which depends on the neutrino mass $m_i$.
This expression can be easily generalized for more than one heavy
neutrino.  The advantage of using the ratio above is that potential
systematic uncertainties and normalization effects can be safely
ignored. An experiment is sensitive to a massive neutrino state if
it can distinguish $R$ from unity, a determination that should be
limited by statistics due to the very low $\beta$-ray flux in the
high-energy tail of the electron spectrum.

\begin{figure}
\begin{center}
\includegraphics[scale=0.70]{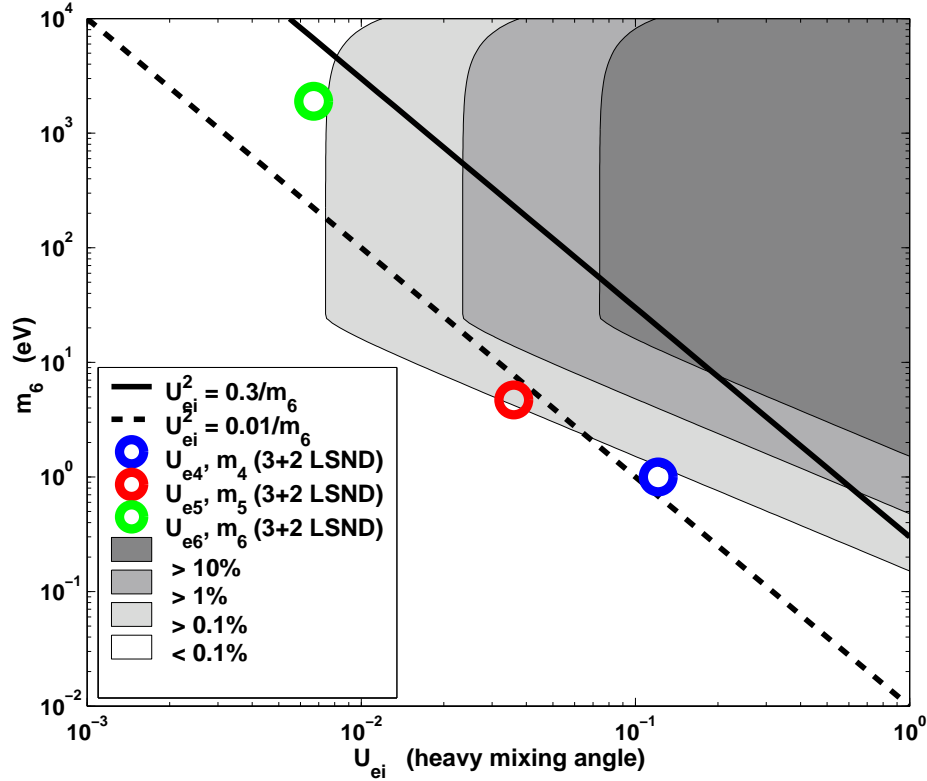}
\caption[Contour plot of constant $R$ assuming an energy window
$\Delta E = 25~\rm{eV}$]{Contour plot of constant $R$, as defined by
Eq.~(\ref{equ:R}), assuming an energy window $\Delta E =
25~\rm{eV}$.  The solid (dashed) line corresponds to
$\sqrt{m_l/m_i}$, a naive upper bound for $|U_{ei}|$, for
$m_l=0.3$~eV (0.01~eV). The circles correspond to $U_{ei}$ for the
three mostly sterile states obtained by our ``fit'' to other
neutrino data, Eq.~(\ref{U_32}). See text for details.}
\label{fig:Katrin}
\end{center}
\end{figure}

In order to compute $R$, I use an analytic expression for
Eq.~(\ref{equ:R}), which exists provided that one neglects nucleon
recoil in the decay. Fig.~\ref{fig:Katrin} depicts constant
$R$-contours in the $|U_{ei}|\times m_i$ plane. Contours were
computed for $\Delta E=25~\rm{eV}$, in order to allow one to easily
compare our results with the sensitivity estimates of the Katrin
experiment. After data-taking, Katrin is expected to measure $R$ at
the 0.1\%--1\% level (lightest grey region). Its sensitivity is
expected to be $\sqrt{m^2_{\nu_e}}>0.2$~eV at the 90\% confidence
level. This can be extracted from the plot by concentrating on the
$U_{ei}=1$ line.  Note that while the expected energy resolution for
Katrin is of order 1~eV, the expected number of signal events above
$E_0-1$~eV is minuscule (both in absolute terms and compared with
expected number of background events), so that most of the
sensitivity to nonzero neutrino masses comes from analyzing the
shape of the electron spectrum in the last tens of electron-volts. A
larger ``window'' would suffer from increased systematic
uncertainties, so that $\Delta E\sim 25$~eV is representative of
Katrin's optimal reach \cite{Katrin}.

The shape of the constant $R$ contours is easy to understand. As
already discussed, for $m_i>\Delta E$, the effect of the
right-handed neutrinos is to reduce the spectrum in an energy
independent way by $1-|U_{ei}|^2$, while for $m_i<\Delta E$, states
with the same effective mass-squared $m_i^2|U_{ei}|^2$ produce the
same effect in tritium beta-decay so that the diagonal lines
coincide with lines of constant $m_i^2|U_{ei}|^2$.

A more sensitive approach would be to ``bin'' the last tens of eV of the ``data'' into 1~eV bins, and fit the distribution to a massless neutrino hypothesis. For the values of the parameters in which I am interested, I find that one 25~eV bin yields roughly the same sensitivity to nonzero neutrino masses as twenty five 1~eV bins for large masses and small mixing angles. For smaller masses and larger mixing angles, a ``binned'' analysis should be sensitive to effects which are localized in individual bins (such as ``kinks'').  Another recent estimate of the sensitivity of tritium beta-decay experiments to heavy, sterile neutrinos can be found in  \cite{smirnov_renata}. Our results agree qualitatively. %Note that, unlike  \cite{smirnov_renata}, I provide a detailed explanation of our procedure to gauge the reach of tritium beta-decay experiments, along with a qualitative understanding of this reach.

Fig.~\ref{fig:Katrin} also depicts the loose upper bound for
$U_{ei}=\sqrt{m_l/m_i}$ as a function $m_i$, for $m_l=0.32$~eV and
$m_l=0.01$~eV. For $m_l\gtrsim 0.1$~eV, Katrin should be sensitive
to $M_R\lesssim1$~keV while for $m_l\gtrsim 0.01$~eV (the solar mass
scale) Katrin should be sensitive to $M_R\gtrsim10$~eV and
$M_R\lesssim 100$~eV, where here I assume that all right-handed
neutrino masses are of order $M_R$. In the case of seesaw parameters
that fit the LSND data with a 3+2 neutrino spectrum (see
Eq.~(\ref{U_32})), expectations are high as far as observing a
kinematical neutrino mass effect at Katrin, in spite of the fact
that the fit to LSND data requires $U_{e4}$ and $U_{e5}$ to be
``abnormally'' low. Fig.~\ref{fig:Katrin} depicts $U_{ei}$ and $m_i$
values for the heavy neutrinos (open circles). The contribution of
the heaviest of the two LSND-related sterile neutrinos is of order
the Katrin sensitivity, while the active contribution itself, which
leads to $m^2_{\nu_e}=\sum_{i=1,2,3}|U_{ei}^2m_i^2|\sim m_l^2$ is
already within the Katrin sensitivity range, given that large
$m_l>0.22$~eV values are required by our 3+2 LSND ``fit''. The
effect of $\nu_6$ is small if $m_6$ is larger than 1~keV (required
if one takes the ``pulsar kicks'' hint into account), but would be
very significant if $m_6$ were less than 1~keV.

\subsection{Neutrinoless Double-Beta Decay}

If the neutrinos are Majorana fermions -- as predicted in the case
of interest here -- lepton number is no longer a conserved quantity.
The best experimental probe of lepton number violation is the rate
for neutrinoless double-beta decay. This process, which violates
lepton number by two units, is currently the subject of intense
search \cite{bb0n_future,EVcommon}. If neutrino masses are the only source of
lepton number violation, the decay width for neutrinoless
double-beta decay is
\begin{equation}
\Gamma_{0\nu\beta\beta}\propto \left|\sum_i U_{ei}^2
\frac{m_i}{Q^2+m_i^2}{\mathcal M}(m_i^2,Q^2)\right|^2,
\label{eq:mbb}
\end{equation}
where $\mathcal M$ is the relevant nuclear matrix element and
$Q^2\sim 50^2$~MeV$^2$ is the relevant momentum transfer. In the
limit of very small neutrino masses ($m_i^2\ll Q^2$),
$\Gamma_{0\nu\beta\beta}$ is proportional to an effective neutrino
mass $|m_{ee}|$,
\begin{equation}
m_{ee} = \sum_i^n U_{ei}^2 m_i . \label{equ:m_ee}
\end{equation}
The sum is over all light neutrino mass eigenstates. In the case of
a low-energy seesaw, when all $m_i$, $i=1,\ldots,6$ are much smaller
than $Q^2$, it is easy to see that $m_{ee}$ vanishes
\cite{SeeSaw_LSND}. The reason for this is that, in the weak basis I
are working on (diagonal charged-lepton and charged weak-current),
$m_{ee}$ is the $ee$-element of the neutrino mass matrix, as defined
in Eq.~(\ref{eq:lagrangian}). One can trivially check that, by
assumption, not only does $m_{ee}$ vanish, but so do all other
$m_{\alpha\beta}$, $\alpha,\beta=e,\mu,\tau$. Note that this result
does not depend on the fact that I have been assuming all elements
of the neutrino mass matrix to be real \cite{Kayser_CP}.

For heavy $\nu_i$ neutrinos, $U_{ei}^2m_i$ no longer captures the
dependency of $\Gamma_{0\nu\beta\beta}$ on the exchange of $\nu_i$.
For $m_i^2\gg Q^2$, instead, the dependency on neutrino exchange is
proportional to $U_{ei}^2 /m_i$. If this is the case, the overall
contribution (including all heavy and light states) is no longer
proportional to $m_{ee}$ but, instead, can be qualitatively
expressed as a function of an effective $m^{\rm eff}_{ee}$,
\begin{equation}
m_{ee}^{\rm eff}\equiv Q^2\sum_i \frac{U_{ei}^2 m_i}{Q^2+m_i^2}.
\end{equation}
The approximation $\Gamma_{0\nu\beta\beta}\propto |m_{ee}^{\rm
eff}|$ is good as long as one can neglect the dependency of
${\mathcal M}$ on $m_i$ and is not expected to be a great
approximation when $m_i^2\sim Q^2$. Nonetheless, $m_{ee}^{\rm eff}$
still qualitatively captures the behavior of
$\Gamma_{0\nu\beta\beta}$ as a function of the sterile neutrino
masses and studying its behavior is sufficient for our ambitions in
this discussion.

\begin{figure}
\begin{center}
\includegraphics[angle = 270,scale=0.57]{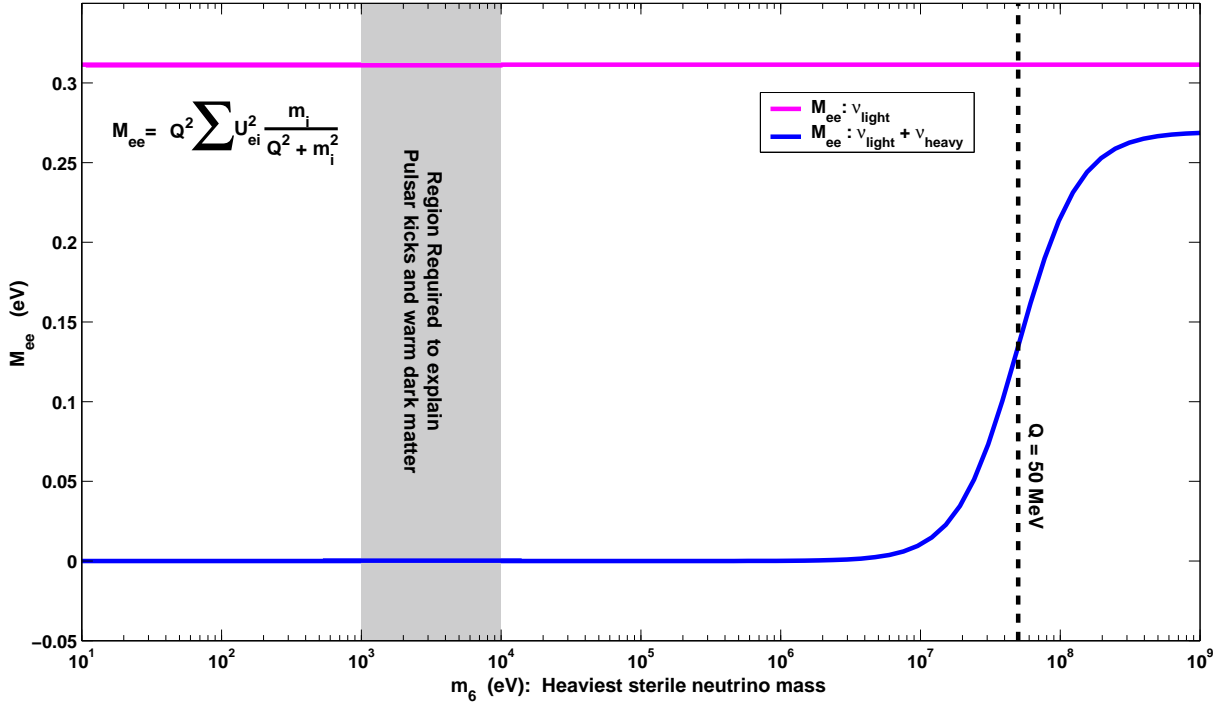}
\caption[Effective $m_{ee}$ for neutrinoless double-beta decay as a
function of $m_6$]{Effective $m_{ee}$ for neutrinoless double-beta
decay as a function of $m_6$, the heaviest right-handed neutrino
mass, assuming the existence of only light, active, neutrinos
(magenta curve), with a degenerate mass spectra, and for our ``best
fit'' $3+2$ LSND sterile neutrino solution (blue curve).  See text
for details. Also indicated is the parameter region preferred by
astrophysical hints of sterile neutrinos. I assume $Q =
50~\rm{MeV}$. In the case of a low-energy seesaw, $m_{ee}$ vanishes
as long as $m_6\ll Q$.} \label{fig:0nubb}
\end{center}
\end{figure}

Fig.~\ref{fig:0nubb} depicts $m_{ee}^{\rm eff}$ for our ``best fit''
$3+2$ LSND solution (see Sec.\ref{subsec:LSND}), as a function of
the unconstrained $m_6$. As advertised, $m_{ee}^{\rm eff}$ vanishes
for $m_6^2\ll Q^2$. The figure also depicts the ``active only''
value of $m_{ee}^{\rm active}=\sum_{i=1,2,3}U_{ei}^2m_i$. Even in
the limit $m_6^2\gg Q^2$, there is still partial cancellation
between the mostly active and mostly sterile ``LSND'' states.   This
is a feature of the Lagrangian I am exploring here, and is not in
general observed in other scenarios with light sterile neutrinos
tailor-made to solve the LSND anomaly.

Currently, the most stringent limits on this effective mass comes
from the Heidelberg-Moscow experiment \cite{bb0n_present} where they
find $m_{ee} < 0.91~\rm{eV}$ at $99\%$ confidence. In the near
future, experiments aim to reach down to $m_{ee}$ values close to
$10^{-2}$~eV \cite{bb0n_future,EVcommon}. A signal would rule out a seesaw
scale below tens of MeV. On the other hand, if I were to conclude
that the neutrino masses are quasi-degenerate (through, say, a
signal in tritium beta-decay) and if the LSND 3+2 solution were
experimentally confirmed, a vanishing result for $m_{ee}$ could be
considered strong evidence for a very small seesaw scale. On the
other hand, if this were the case ($m_{ee}$ zero, large active
neutrino masses), it would also be very reasonable to conclude that
neutrinos are Dirac fermions. Distinguishing between the two
possibilities would prove very challenging indeed. It is curious
(but unfortunate) that in a low energy seesaw model the neutrinos
are Majorana fermions, but all ``standard'' lepton-number violating
observables vanish, given that their rates are all effectively
proportional to $m_{\alpha\beta}$!

\section{Conclusions}
\label{sec:SeeSawConclusion}

The ``New Standard Model'' (equal to the ``old'' Standard Model plus
the addition of three gauge singlet Weyl fermions) is, arguably, the
simplest extension of the SM capable of accommodating neutrino
masses. This Lagrangian contains a new dimension-full parameter: the
right-handed neutrino mass scale $M$, which must be determined
experimentally. Unlike the Higgs mass-squared parameter, all $M$
values are technically natural given that the global symmetry of the
Lagrangian is enhanced in the limit $M\to 0$.

Very large $M$ values are \emph{theoretically} very intriguing, and
have received most of the attention of the particle physics
community. There are several strong hints that new phenomena are
expected at the electroweak breaking scale $\sim 10^3$~GeV, the
grand unified scale $\sim 10^{15-16}$~GeV, and the Planck scale
$\sim 10^{18-19}$~GeV, and it is tempting to associate $M$ to one of
these energy scales. Furthermore, large $M$ values provide an
elegant mechanism for generating the matter antimatter asymmetry of
the Universe \cite{leptogenesis}. Of course, large $M$ values are
experimentally very frustrating. It may ultimately prove impossible
to experimentally verify whether the New Standard Model is really
the correct way to describe Nature.

Here, I explore the opposite end of the $M$ spectrum, $M\lesssim
1$~keV. Such values are \emph{phenomenologically} very intriguing,
given that small $M$ values imply the existence of light sterile
neutrinos that mix significantly with the active neutrinos and can
potentially be directly observed. Furthermore, there are several
experimental and astrophysical phenomena that are best understood if
one postulates the existence of light, moderately mixed sterile
neutrinos. I find that by requiring all three right-handed neutrino
masses to be less than a few keV I can simultaneously explain all
neutrino oscillation data, including those from LSND,  explain the
large peculiar velocities of pulsars, and accommodate the production
of heavy elements in supernova environments. Our fit also provides
constraints for the ``active'' neutrino oscillation parameters, most
strongly to the lightest active neutrino mass. All successful
parameter choices that accommodate the LSND data require $m_l$ to be
``large'' ($m_l\gtrsim 0.1$~eV), and the ``best fit'' requires all
active neutrino masses to be quasi-degenerate. It is important to
emphasize that the presence of light sterile neutrinos that mix
relatively strongly with the active neutrinos is only in agreement
with cosmological data (especially large scale structure and
big-bang nucleosynthesis) if non-standard cosmological ingredients
are present.

Fig.~\ref{fig:spectrum} depicts such a scenario. This six neutrino
mass spectrum (including mixing angles) fits all neutrino
oscillation data (including those from LSND), provides a sterile
neutrino solution to the pulsar kick puzzle, and contains all the
necessary ingredients for heavy element nucleosynthesis in
supernovae. The heaviest of the neutrinos does not qualify as
thermal warm dark matter (in the absence of new cosmological
ingredients,  its presence would overclose the Universe). Note that
even if it were non-thermally produced, constraints from the
observation of X-rays from the center of the galaxy would rule out
$\nu_6$ as a good dark matter candidate. Lighter $\nu_6$ masses will
evade X-ray constraints, but would render $\nu_6$ too ``hot,'' and
hence not a good dark matter candidate.

On the negative side, low $M$ values are, theoretically, rather
puzzling. In order to obtain the observed light neutrino masses,
neutrino Yukawa couplings are required to be much smaller than the
electron Yukawa coupling, and it is tempting to believe that such
small numbers are proof that a more satisfying understanding of
fermion masses must exist. Furhermore, thermal leptogenesis is no
longer an option (see, however, \cite{Shaposhnikov:2006xi}).
Finally, the fact that $M$ is naively unrelated to other mass scales
can also be perceived as disheartening, but, in our opinion, should
be interpreted as evidence that there is more to the lepton sector
than meets the eye.

Regardless of one's preference for a high or low seesaw energy
scale, and independent of whether the data from LSND and the
astronomical observables discussed above have anything to do with
sterile neutrinos, our main point is that the determination of  $M$
is an \emph{experimental} issue. In the near/intermediate future,
low energy seesaw scales will be probed by several experiments, most
importantly measurements of the end-point of tritium beta-decay, the
MiniBooNE experiment, searches for neutrinoless double-beta decat
and, if we get lucky, the detection of neutrinos from a nearby
supernova explosion. I find, for example, that Katrin should be
sensitive to seesaw energy scales below tens of keV if all
right-handed neutrino masses are similar, while null results from
MiniBooNE would severely constrain right-handed neutrino masses
below several eV. I conclude by pointing out that larger (but still
``small'') values of $M$ are much harder to constrain. For GeV
sterile neutrinos, typical active--sterile mixing angles are
$U_{\alpha i}^2\lesssim 10^{-10}$, probably too small to observe in
particle physics processes.  It is frustrating (and, I hope,
ultimately false) that we seem to be unable to experimentally
distinguish $M\sim 1$~GeV from $M\sim 10^{14}$~GeV\ldots

% Uncomment the 'singlespace' environment and '\bibsep' command
% if needed - some bibliographic styles overide the definition
% of 'thebibliography' in nuthesis.cls
%
%\begin{singlespace}
%\bibsep 12pt

%\end{singlespace}

\appendices

\chapter{Standard Model of Particle Physics}\label{app:SM}
The Standard Model (SM) is a local Lorentz invariant quantum field
theory defined by its gauge symmetry, field content and symmetry
breaking scale. The gauge symmetry is
\begin{equation}
SU(3)_C \times SU(2)_L \times U(1)_Y.
\end{equation}
Each group has its own distinct gauge coupling and is necessarily
associated with spin one gauge boson fields transforming under the
adjoint group representations.  These are the eight gluons (G),
three weak isobosons (W), and single hypercharge boson (B),
respectively.  The fermion field content, along with its associated
quantum numbers, is listed in Table \ref{tab:FieldContent}.  Here,
$i$ is a flavor label that runs over the three known generations. In
other words, there are at least three copies of each fermion
generation identical in interactions, which are only
mass scale.  Additionally, there is a single complex Lorentz
scalar, the Higgs boson H, with quantum numbers $({\bf 1},{\bf 2},-1/2)$.
%\begin{center}
\begin{table} 
\begin{tabular}{|c|ccc|}
\hline
Field & $SU(3)_c$ & $SU(2)_L$ & $U(1)_y$ \\ 
\hline
$L_i$ & {\bf 1} & {\bf 2} & $-\frac{1}{2}$ \\ 
$\overline{e_i}$ & {\bf 1} & {\bf 1} & 1\\
$Q_i$ & {\bf 3} & {\bf 2}& $ \frac{1}{6}$\\
$\overline{u_i}$ & $\overline{{\bf 3}}$ &{\bf 1} & $-\frac{2}{3}$\\
$\overline{d_i}$ & $\overline{{\bf 3}}$ &{\bf 1} & $\frac{1}{3}$\\
\hline 
\end{tabular} \label{tab:FieldContent} \caption[Standard Model fermion content with quantum numbers]{Summary table of the Standard Model fermion field content along with their associated gauge quantum numbers.  The fields are written in terms of their chiral left handed projections which may be conjugated to yield the right handed projections.  The boldface entries describe the representaion of the field under the nonabelian symmetry groups.}
\end{table}
%\end{center}

From here it is a simple matter to write down the most general
Lagrangian incorporating these principles.  This consists of three
parts; the gauge sector
\begin{equation}
\mathcal{L}_{Gauge} = \imath\sum_i \overline{L_i}\gamma_\mu D^\mu L_i + \overline{Q_i}\gamma_\mu D^\mu Q_i + \overline{e_i}\gamma_\mu D^\mu e_i + \overline{d_i}\gamma_\mu D^\mu d_i + \overline{u_i}\gamma_\mu D^\mu u_i + |D^\mu H|^2,
\end{equation}
the Yukawa sector
\begin{equation}
\mathcal{L}_{Yukawa}= y_\ell^{ij} \overline{L_i}H e_j + y_d^{ij} \overline{Q_i} H d_j + y_u^{ij} \overline{Q_i} H^\dagger u_j + h.c.\footnote{The simple multiplicative notation employed here is schematic.  Since all terms are weak isosinglets it is clear that some spinor products between doublets are nontrivial involving the antisymmetric tensor $\epsilon_{\alpha \beta}$. },
\end{equation}
and the Higgs sector
\begin{equation}
\mathcal{L}_{Higgs} = \mu^2 H^\dagger H + \eta (H^\dagger H)^2. \label{eq:Higgs}
\end{equation}
Here, $D_\mu = \partial_\mu - ig_sG_\mu \lambda - i g W_\mu \tau - i g^\prime B_\mu Y$ is the covariant derivative containing the gauge fields with the matricies $\lambda$ and $\tau$, the generators of $SU(3)_c$ and $SU(2)_L$ in the appropriate representation.
Notice that all fermions and gauge bosons are strictly massless in
this theory. This follows from the chiral nature of the fermions
where the left and right handed fields transform differently under
$SU(2)_L \times U(1)_Y$ thus forbidding the required
$\overline{\psi_L}\psi_R$ mass term.

The standard treatment now postulates that the $\mu^2 < 0$ in
Eq.~(\ref{eq:Higgs}) so that H acquires a vacuum expectation value
(vev).  This breaks electroweak symmetry
\begin{equation}
SU(2)_L \times U(1)_Y \rightarrow U(1)_Q
\end{equation}
at the energy scale characterized by $v$.  Once done, only one
abelian gauge symmetry remains to mediate electromagnetic
interactions.  Upon expanding the SM Lagrangian about $v$, one finds
that the charged fermions (now under Q) have acquired masses as well
as three of the electroweak gauge bosons.  These extra longitudinal
degrees of freedom arise from the Goldstone Higgs bosons ``eaten''
by the gauge fields.  One can easily rewrite the Lagrangian terms
expanded about the new vacuum and in the basis of definite gauge
boson mass, but this is not needed for what follows.  For
completeness, I do point out that orthogonal linear combinations of
$W_1$ and $W_2$ combine to form the massive $W^\pm$ bosons, and
similarly, combinations of $W_3$ and $B$ combine to form the massive
$Z$ and (photon) $A$ bosons.  The $SO(2)$ transformation that
facilitates the rotation to this mass basis is defined by $\theta_w$,
the weak mixing angle.  One should also notice that the massless
photon is now the field associated with the remaining unbroken
symmetry.

From the point of view of this thesis, the most important feature of
the SM Lagrangian is its accidental symmetries.  The gauge and Higgs
sectors contain a huge global symmetry $SU(3)^5_f$ corresponding to
the freedom to rotate each of the five field families in generation
space.  The Yukawa terms substantially break this but still leave
some residual symmetries.  In particular, I find that if I assign
identical U(1) charges, called baryon number (B), to $Q_i$, $u_i$
and $d_i$ while holding the rest of the field content neutral, B is
conserved.  Similarly, I see that Lepton number (L) is conserved
provided that I assign identical charges to both $L_i$ and $e_i$
with the remaining content neutral\footnote{Of course, in the SM
with no means of generating neutrino mass, a separately conserved
lepton number may be assigned to each lepton family.  This is broken
by the observed neutrino mixing phenomena induced by BSM physics}.
This is not the whole story, as both B and L are broken by
nonperturbative instaton effects.  At the end of the day, only one
non-anomalous symmetry remains, conserving the quantity $B-L$,
$U(1)_{B-L}$.  It is important to understand that this symmetry is
completely accidental within the SM and that any additional new
physics is likely to break it.  This implies that searches for B and
L violation are ideal laboratories for studying physics BSM.  The
search for Lepton Number Violation (LNV) is the major unifying theme
of this work, as it is intimately related to the Majorana nature of
neutrinos.

\chapter{Extended $SU(2)$ neutrino mixing matrix notation}
\label{app:notation}

 Neutrino mixing and all of its invariance may be
expressed as products of three distinct matrix classes in addition
to the identity ${\bf I}$ \footnote{It is also true that one may use
the full $SU(n)$ algebra to parameterize $n$ neutrino mixing in a
more elegant way than in my extended $SU(2)$ parameterization.
These algebras are well known, but their use would require new
commutator relations for each separate case, which is not conducive
to my global approach.}. Two are defined in terms of continuous
variables. These are real orthogonal rotations in the $a-b$ plane
${\bf R^{ab}}(\theta)$ and single phase rotations ${\bf
P^{a}}(\phi)$ defined by
\begin{equation}
{\bf R^{ab}}(\theta) \equiv \left\{\begin{aligned}
~[{\bf R^{ab}}(\theta)]_{aa} = [{\bf R^{ab}}(\theta)]_{bb} = \cos\theta ~~~~~\\
~[{\bf R^{ab}}(\theta)]_{ab} = -[{\bf R^{ab}}(\theta)]_{ba} =-\sin\theta\\
~[{\bf R^{ab}}(\theta)]_{ij} = \delta_{ij} ~~~~~~~ (ij\neq a,b)~~~~~
\end{aligned}
\right. \label{eq:Rab}
\end{equation}
and
\begin{equation}
[{\bf P^{a}}(\phi)]_{ij} = \delta_{ij} e^{i\phi \delta_{ia}},
\label{eq:Pa}
\end{equation}
respectively.  The discrete transformation
 \begin{equation}
 {\bf A^{ab}} = {\bf A^{ba}} = \frac{1}{2} {\bf
 R^{ab}}(\frac{\pi}{2})
 \left\{ {\bf P^b}(\pi) - {\bf P^a}(\pi) \right\}
\end{equation}
completes the set.  It is a simple matter to write an arbitrary
mixing matrix with such components. For example, the complex
rotation matrix typically assigned to the $1-3$ plane in standard
three neutrino mixing analysis is ${ \bf
\widetilde{R}^{ab}}(\theta,\delta) = {\bf P^a}(-\delta){\bf
P^b}(\delta){\bf R^{ab}}(\theta){\bf P^a}(\delta){\bf
P^b}(-\delta)$. Additionally, discrete permutations of $ab$ vector
elements are accomplished with ${\bf S^{ab}} = {\bf
R^{ab}}(\pi/2){\bf P^b}(\pi)$. Manipulations of these matrices
utilize their commutation relations defined by

\begin{eqnarray}
\left[{\bf R^{ab}}(\theta),{\bf R^{bc}}(\theta^\prime) \right] &=&
\sin\theta\left(1-\cos\theta^\prime\right){\bf A^{ab}} +
\sin\theta^\prime\left(1-\cos\theta\right){\bf A^{bc}} \\ \nonumber
 &~~~~~~& +
\frac{1}{2}\sin\theta \sin\theta^\prime \left\{ {\bf
R^{ac}}(-\frac{\pi}{2}) - {\bf
R^{ac}}(\frac{\pi}{2}) \right\}\\
\left[ {\bf R^{ab}}(\theta),{\bf P^a}(\phi) \right] &=& \sin\theta
\left(e^{i\phi}-1\right) {\bf A^{ab}}\\
\left[ {\bf R^{ab}}(\theta),{\bf A^{ab}} \right] &=& -\sin\theta
\left\{ {\bf P^b}(\pi) - {\bf P^a}(\pi) \right\}\\
\left[ {\bf R^{ab}}(\theta),{\bf A^{ac}} \right] &=& \sin\theta {\bf
A^{bc}} + \frac{1}{2}\left( \cos\theta - 1\right)\left\{ {\bf
R^{ac}}(-\frac{\pi}{2}) - {\bf R^{ac}}(\frac{\pi}{2}) \right\}\\
\left[ {\bf A^{ab}},{\bf P^{a}}(\phi) \right] &=&
\frac{1}{2}\left(e^{i\phi} - 1\right)\left\{ {\bf
R^{ab}}(-\frac{\pi}{2}) - {\bf R^{ab}}(\frac{\pi}{2}) \right\}\\
\left[ {\bf A^{ab}},{\bf A^{ac}} \right] &=& \frac{1}{2}\left\{ {\bf
R^{bc}}(-\frac{\pi}{2}) - {\bf R^{bc}}(\frac{\pi}{2}) \right\}.
\end{eqnarray}
All other commutators vanish.  This matrix set closes upon itself
facilitating a simple platform to perform manipulations.  In that
spirit, the following identities are easy to prove and useful in
calculations:
\begin{eqnarray}
{\bf P^a}(\phi){\bf P^a}(-\phi) &=& I\\
{\bf R^{ab}}(\theta){\bf R^{ab}}(-\theta) &=& {\bf R^{ab}}(\theta){\bf R^{ba}}(\theta) = I\\
{\bf R^{ab}}(\theta + \theta^\prime) &=& {\bf R^{ab}}(\theta){\bf R^{ab}}(\theta^\prime)\\
{\bf P^{a}}(\phi + \phi^\prime) &=& {\bf P^{a}}(\phi){\bf P^{a}}(\phi^\prime)\\
 {\bf P^a}(\pi){\bf P^b}(\pi) &=& {\bf R^{ab}}(\pi)\\
 {\bf P^b}(\pi) {\bf R^{ab}}(\theta) &=& {\bf R^{ab}}(-\theta) {\bf P^b}(\pi)\\
 {\bf R^{ab}}(\theta){\bf P^a}(\phi) - {\bf P^b}(\phi){\bf
R^{ab}}(\theta) &=& \cos\theta \left(e^{i\phi}-1\right) {\bf
A^{ab}}\\
{\bf A^{ab}P^a}(\phi) &=& {\bf P^b}(\phi){\bf A^{ab}} \\
{\bf R^{ab}}(\theta){\bf R^{ca}}(\frac{\pi}{2}) &=& {\bf
R^{ca}}(\frac{\pi}{2}){\bf R^{cb}}(\theta).
\end{eqnarray}


\begin{thebibliography}{xxx}


%\cite{de Gouvea:2007xp}
\bibitem{de Gouvea:2007xp}
  A.~de Gouvea and J.~Jenkins,
  %``A Survey of Lepton Number Violation Via Effective Operators,''
  Phys.\ Rev.\  D {\bf 77}, 013008 (2008)
  [arXiv:0708.1344 [hep-ph]].
  %%CITATION = PHRVA,D77,013008;%%

\bibitem{MySeeSaw}
A.~de Gouv\^ea, J.~Jenkins and N.~Vasudevan,
 %``Neutrino phenomenology of very low-energy seesaws,''
 Phys.\ Rev.\  D {\bf 75}, 013003 (2007)
 [arXiv:hep-ph/0608147];
 %%CITATION = PHRVA,D75,013003;%%

\bibitem{MyParameterPaper}
A.~de Gouv\^ea, J.~Jenkins, arXiv:0804.3627 [hep-ph]


%\cite{Mahbubani:2006kq}
\bibitem{Mahbubani:2006kq}
For a review of the gauge hierarchy problem and some possible solutions
  R.~Mahbubani (thesis),
  ``Beyond the Standard Model: The Pragmatic approach to the gauge hierarchy
  problem,''
  %%CITATION = UMI-32-17817;%%


%\cite{Langacker:1980js}
\bibitem{Langacker:1980js}
  P.~Langacker,
  %``Grand Unified Theories And Proton Decay,''
  Phys.\ Rept.\  {\bf 72}, 185 (1981).
  %%CITATION = PRPLC,72,185;%%


%\cite{Trodden:2004st}
\bibitem{Trodden:2004st}
  M.~Trodden and S.~M.~Carroll,
  %``TASI lectures: Introduction to cosmology,''
  arXiv:astro-ph/0401547.
  %%CITATION = ASTRO-PH/0401547;%%

 
\bibitem{SeeSaw} P.~Minkowiski, Phys.\ Lett.\ B {\bf 67}, 421 (1977);
M. Gell-Mann, P. Ramond and R. Slansky in {\it Supergravity}, eds.
D. Freedman and P. Van Niuenhuizen (North Holland, Amsterdam, 1979),
p.~315; T. Yanagida in {\it Proceedings of the Workshop on Unified
Theory and Baryon Number in the Universe}, eds. O.~Sawada and
A.~Sugamoto (KEK, Tsukuba, Japan, 1979); S.L.~Glashow, {\it 1979
Carg\`ese Lectures in Physics --- Quarks and Leptons}, eds.
M.~L\'evy {\it et al.} (Plenum, New York, 1980), p.~707. See also
R.N. Mohapatra and G. Senjanovi\'c, Phys.\ Rev.\ Lett.\ {\bf 44},
912 (1980) and J.~Schechter and J.W.F.~Valle,
   Phys.\ Rev.\  D {\bf 22}, 2227 (1980).


%\cite{de Gouvea:2000cq}
\bibitem{de Gouvea:2000cq}
  A.~de Gouvea, A.~Friedland and H.~Murayama,
  %``The dark side of the solar neutrino parameter space,''
  Phys.\ Lett.\  B {\bf 490}, 125 (2000)
  [arXiv:hep-ph/0002064].
  %%CITATION = PHLTA,B490,125;%%


%\cite{deGouvea:2002gf}
\bibitem{deGouvea:2002gf}
  A.~de Gouvea, B.~Kayser and R.~N.~Mohapatra,
  %``Manifest CP violation from Majorana phases,''
  Phys.\ Rev.\  D {\bf 67}, 053004 (2003)
  [arXiv:hep-ph/0211394].
  %%CITATION = PHRVA,D67,053004;%%
  
  
%\cite{Langacker:1998pv}
\bibitem{Langacker:1998pv}
  P.~Langacker and J.~Wang,
  %``Neutrino anti-neutrino transitions,''
  Phys.\ Rev.\  D {\bf 58}, 093004 (1998)
  [arXiv:hep-ph/9802383].
  %%CITATION = PHRVA,D58,093004;%%


%\cite{Petcov:2001sy}
\bibitem{Petcov:2001sy}
  S.~T.~Petcov and M.~Piai,
  %``The LMA MSW solution of the solar neutrino problem, inverted neutrino  mass
  %hierarchy and reactor neutrino experiments,''
  Phys.\ Lett.\  B {\bf 533}, 94 (2002)
  [arXiv:hep-ph/0112074].
  %%CITATION = PHLTA,B533,94;%%


%\cite{Fritzsch:1986gv}
\bibitem{Fritzsch:1986gv}
  H.~Fritzsch and J.~Plankl,
  %``The Mixing of Quark Flavors,''
  Phys.\ Rev.\  D {\bf 35}, 1732 (1987).
  %%CITATION = PHRVA,D35,1732;%%



%\cite{Jenkins:2007ip}
\bibitem{Jenkins:2007ip}
  E.~Jenkins and A.~V.~Manohar,
  %``Rephasing Invariants of Quark and Lepton Mixing Matrices,''
  arXiv:0706.4313 [hep-ph].
  %%CITATION = ARXIV:0706.4313;%%

  
%\cite{de Gouvea:2007xp}
%\bibitem{de Gouvea:2007xp}
 % A.~de Gouvea and J.~Jenkins,
  %``A Survey of Lepton Number Violation Via Effective Operators,''
  %arXiv:0708.1344 [hep-ph].
  %%CITATION = ARXIV:0708.1344;%%

  
\bibitem{NeutrinoReview}
See, for example, M.~C.~Gonzalez-Garcia and M.~Maltoni,
  %``Phenomenology with Massive Neutrinos,''
  arXiv:0704.1800 [hep-ph];
  %%CITATION = ARXIV:0704.1800;%%
A.~Strumia and F.~Vissani,
  %``Neutrino masses and mixings and.,''
  arXiv:hep-ph/0606054;
  %%CITATION = HEP-PH/0606054;%%
 R.~N.~Mohapatra {\it et al.},
 %``Theory of neutrinos: A white paper,''
 arXiv:hep-ph/0510213;
 %%CITATION = HEP-PH/0510213;%%
 A.~de Gouv\^ea,
 %``Neutrinos Have Mass - So What?,''
 Mod.\ Phys.\ Lett.\  A {\bf 19}, 2799 (2004)
 [arXiv:hep-ph/0503086];
 %%CITATION = MPLAE,A19,2799;%%
 A.~de Gouv\^ea,
 %``Neutrino physics,''
 arXiv:hep-ph/0411274.
 %%CITATION = HEP-PH/0411274;%%

 
\bibitem{Operators}
 K.~S.~Babu and C.~N.~Leung,
 %``Classification of effective neutrino mass operators,''
 Nucl.\ Phys.\  B {\bf 619}, 667 (2001)
 [arXiv:hep-ph/0106054].
 %%CITATION = NUPHA,B619,667;%%


\bibitem{OscBestFit}
 M.~Maltoni, T.~Schwetz, M.~A.~Tortola and J.~W.~F.~Valle,
 %``Status of global fits to neutrino oscillations,''
 New J.\ Phys.\  {\bf 6}, 122 (2004)
 [arXiv:hep-ph/0405172].
 %%CITATION = NJOPF,6,122;%%

 
\bibitem{Mainz}
 C.~Kraus {\it et al.},
 %``Final results from phase II of the Mainz neutrino mass search in  tritium
 %beta decay,''
 Eur.\ Phys.\ J.\  C {\bf 40}, 447 (2005)
 [arXiv:hep-ex/0412056].
 %%CITATION = EPHJA,C40,447;%%

 
\bibitem{Troitsk}
 V.~M.~Lobashev {\it et al.},
 %``Direct search for neutrino mass and anomaly in the tritium  beta-spectrum:
 %Status of 'Troitsk neutrino mass' experiment,''
 Nucl.\ Phys.\ Proc.\ Suppl.\  {\bf 91}, 280 (2001).
 %%CITATION = NUPHZ,91,280;%%

  
\bibitem{CosmoSum}
For recent estimates see and comprehensive review, see J.~Lesgourgues and S.~Pastor,
  %``Massive neutrinos and cosmology,''
  Phys.\ Rept.\  {\bf 429}, 307 (2006)
  [arXiv:astro-ph/0603494].
  %%CITATION = PRPLC,429,307;%%

%CosmoSum,Fogli:2006yq,seljak,Hannestad:2006mi
%%%%%%%%%%%%%%%%%%%%%%%%%%%%%%%%%%%%%%%%%%%%%%%%%%%%%%%%%%%%%%%%%%%%%%%%%%%%%%%%%%%
 \bibitem{Fogli:2006yq}
  G.~L.~Fogli {\it et al.},
  %``Observables sensitive to absolute neutrino masses: A reappraisal after
  %WMAP-3y and first MINOS results,''
  arXiv:hep-ph/0608060.
  %%CITATION = HEP-PH 0608060;%%

\bibitem{seljak} U.~Seljak, A.~Slosar and P.~McDonald,
  %``Cosmological parameters from combining the Lyman-alpha forest with CMB,
  %galaxy clustering and SN constraints,''
  JCAP {\bf 0610}, 014 (2006).
  %%CITATION = ASTRO-PH 0604335;%%

%\cite{Hannestad:2006mi}
\bibitem{Hannestad:2006mi}
S.~Hannestad and G.~G.~Raffelt,
  %``Neutrino masses and cosmic radiation density: Combined analysis,''
  astro-ph/0607101.
%%%%%%%%%%%%%%%%%%%%%%%%%%%%%%%%%%%%%%%%%%%%%%%%%%%%%%%%%%%%%%%%%%%%%%%%%%%%%%%%%%%%%



\bibitem{TritNuEffMass}
For a recent detailed discussion see
 Y.~Farzan and A.~Yu.~Smirnov,
 %``On the effective mass of the electron neutrino in beta decay,''
 Phys.\ Lett.\  B {\bf 557}, 224 (2003)
 [arXiv:hep-ph/0211341].
 %%CITATION = PHLTA,B557,224;%%


  %\cite{anarchy}
\bibitem{anarchy} L.J.~Hall, H.~Murayama and N.~Weiner,
  %``Neutrino mass anarchy,''
  Phys.\ Rev.\ Lett.\  {\bf 84}, 2572 (2000)
  [arXiv:hep-ph/9911341];
  %%CITATION = PRLTA,84,2572;%%
A.~de Gouv\^ea and H.~Murayama,
  %``Statistical test of anarchy,''
  Phys.\ Lett.\  B {\bf 573}, 94 (2003)
  [arXiv:hep-ph/0301050].
  %%CITATION = PHLTA,B573,94;%%


\bibitem{Bajc:2006ia}
E.~Ma,
 %``Pathways to naturally small neutrino masses,''
 Phys.\ Rev.\ Lett.\  {\bf 81}, 1171 (1998)
 [arXiv:hep-ph/9805219];
 %%CITATION = PRLTA,81,1171;%%
 B.~Bajc and G.~Senjanovic,
 %``Seesaw at LHC,''
 arXiv:hep-ph/0612029.
 %%CITATION = HEP-PH/0612029;%%

 
\bibitem{SeeSaw2}
 R.~N.~Mohapatra and G.~Senjanovic,
 %``Neutrino Masses And Mixings In Gauge Models With Spontaneous Parity
 %Violation,''
 Phys.\ Rev.\  D {\bf 23}, 165 (1981);
 %%CITATION = PHRVA,D23,165;%%
 G.~B.~Gelmini and M.~Roncadelli,
  %``Left-Handed Neutrino Mass Scale And Spontaneously Broken Lepton Number,''
  Phys.\ Lett.\  B {\bf 99}, 411 (1981).
  %%CITATION = PHLTA,B99,411;%%

  
 \bibitem{SeeSaw3}
 R.~Foot, H.~Lew, X.~G.~He and G.~C.~Joshi,
 %``SEESAW NEUTRINO MASSES INDUCED BY A TRIPLET OF LEPTONS,''
 Z.\ Phys.\  C {\bf 44}, 441 (1989).
 %%CITATION = ZEPYA,C44,441;%%

 
\bibitem{Leptogenesis}
M.~Fukugita and T.~Yanagida,
  %``Baryogenesis Without Grand Unification,''
  Phys.\ Lett.\  B {\bf 174}, 45 (1986).
  %%CITATION = PHLTA,B174,45;%%
For a recent review, see
 W.~Buchmuller, R.~D.~Peccei and T.~Yanagida,
 %``Leptogenesis as the origin of matter,''
 Ann.\ Rev.\ Nucl.\ Part.\ Sci.\  {\bf 55}, 311 (2005)
 [arXiv:hep-ph/0502169].
 %%CITATION = ARNUA,55,311;%%

  
\bibitem{fine_nus}
A.~Pilaftsis,
  %``Radiatively induced neutrino masses and large Higgs neutrino couplings in
  %the standard model with Majorana fields,''
  Z.\ Phys.\  C {\bf 55}, 275 (1992);
  %%CITATION = ZEPYA,C55,275;%%
J.~Bernab\'eu, A.~Santamaria, J.~Vidal, A.~Mendez and J.W.F.~Valle,
  %``Lepton Flavor Nonconservation at High-Energies in a Superstring Inspired
  %Standard Model,''
  Phys.\ Lett.\  B {\bf 187}, 303 (1987);
  %%CITATION = PHLTA,B187,303;%%
W.~Buchmuller and D.~Wyler,
  %``Dilatons and majorana neutrinos,''
  Phys.\ Lett.\  B {\bf 249}, 458 (1990);
  %%CITATION = PHLTA,B249,458;%%
W.~Buchmuller and C.~Greub,
  %``Heavy Majorana neutrinos in electron - positron and electron - proton
  %collisions,''
  Nucl.\ Phys.\  B {\bf 363}, 345 (1991);
  %%CITATION = NUPHA,B363,345;%%
A.~Datta and A.~Pilaftsis,
  %``Revealing the Majorana nature of heavy neutrinos via a heavy Higgs boson,''
  Phys.\ Lett.\  B {\bf 278}, 162 (1992);
  %%CITATION = PHLTA,B278,162;%%
G.~Ingelman and J.~Rathsman,
  %``Heavy Majorana neutrinos at e p colliders,''
  Z.\ Phys.\  C {\bf 60}, 243 (1993);
  %%CITATION = ZEPYA,C60,243;%%
C.A.~Heusch and P.~Minkowski,
  %``Lepton flavor violation induced by heavy Majorana neutrinos,''
  Nucl.\ Phys.\  B {\bf 416}, 3 (1994).
  %%CITATION = NUPHA,B416,3;%%
For recent discussions see J.~Kersten and A.~Yu.~Smirnov,
  %``Right-Handed Neutrinos at LHC and the Mechanism of Neutrino Mass
  %Generation,''
  arXiv:0705.3221 [hep-ph];
  %%CITATION = ARXIV:0705.3221;%%
A.~de Gouv\^ea,
  %``GeV Seesaw, Accidentally Small Neutrino Masses, and Higgs Decays to
  %Neutrinos,''
  arXiv:0706.1732 [hep-ph].
  %%CITATION = ARXIV:0706.1732;%%

    
\bibitem{LowScaleSeeSaw}
D.~Gorbunov and M.~Shaposhnikov,
  %``How to find neutral leptons of the nuMSM?,''
  arXiv:0705.1729 [hep-ph].
  %%CITATION = ARXIV:0705.1729;%%



%%%%%%%%%%%%%%%%%%%%%%%%%%%%%%%%%%%%%%%%%%%%%%%%%%%%%%%%%%%%%%%%%
\bibitem{nuSM_dark}
  T.~Asaka, S.~Blanchet and M.~Shaposhnikov,
  %``The nuMSM, dark matter and neutrino masses,''
  Phys.\ Lett.\ B {\bf 631}, 151 (2005);
  %%CITATION = HEP-PH 0503065;%%

%\bibitem{MySeeSaw}
%A.~de Gouv\^ea, J.~Jenkins and N.~Vasudevan,
 %``Neutrino phenomenology of very low-energy seesaws,''
 %Phys.\ Rev.\  D {\bf 75}, 013003 (2007)
 %[arXiv:hep-ph/0608147];
 %%CITATION = PHRVA,D75,013003;%%

%\cite{SeeSaw_LSND}
\bibitem{SeeSaw_LSND}
  A.~de Gouv\^ea,
  %``Seesaw energy scale and the LSND anomaly,''
  Phys.\ Rev.\ D {\bf 72}, 033005 (2005).
  %%CITATION = HEP-PH 0501039;%%

%%%%%%%%%%%%%%%%%%%%%%%%%%%%%%%%%%%%%%%%%%%%%%%%%%%%%%%%%%%%%%%%%

  
\bibitem{LNVUpperBounds}
 A.~Atre, V.~Barger and T.~Han,
 %``Upper bounds on lepton-number violating processes,''
 Phys.\ Rev.\  D {\bf 71}, 113014 (2005)
 [arXiv:hep-ph/0502163].
 %%CITATION = PHRVA,D71,113014;%%

 
\bibitem{PDG}
 W.~M.~Yao {\it et al.}  [Particle Data Group],
 %``Review of particle physics,''
 J.\ Phys.\ G {\bf 33}, 1 (2006).
 %%CITATION = JPHGB,G33,1;%%

 
\bibitem{SupBeam}
 W.~T.~Weng {\it et al.},
 %``The neutrino superbeam from the AGS,''
 J.\ Phys.\ G {\bf 29}, 1735 (2003);
 %%CITATION = JPHGB,G29,1735;%%
 Y.~Oyama,
 %``Results from K2K and status of T2K,''
 arXiv:hep-ex/0512041;
 %%CITATION = HEP-EX/0512041;%%
 M.~G.~Albrow {\it et al.},
 %``Physics at a Fermilab proton driver,''
 arXiv:hep-ex/0509019.
 %%CITATION = HEP-EX/0509019;%%

  
 \bibitem{NuFact}
See, for example,
 C.~H.~Albright {\it et al.}  [Neutrino Factory/Muon Collider
                 Collaboration],
 %``The neutrino factory and beta beam experiments and development,''
 arXiv:physics/0411123;
 %%CITATION = PHYSICS/0411123;%%
 S.~Geer,
 %``Neutrino beams from muon storage rings: Characteristics and physics
 %potential,''
 Phys.\ Rev.\  D {\bf 57}, 6989 (1998)
 [Erratum-ibid.\  D {\bf 59}, 039903 (1999)]
 [arXiv:hep-ph/9712290].
 %%CITATION = PHRVA,D57,6989;%%

  
 \bibitem{BetaBeam}
 P.~Zucchelli,
 %``A novel concept for a anti-nu/e / nu/e neutrino factory: The beta beam,''
 Phys.\ Lett.\  B {\bf 532}, 166 (2002);
 %%CITATION = PHLTA,B532,166;%%

 
\bibitem{EmEm2munu}
 J.~Maalampi and N.~Romanenko,
 %``Testing lepton number violation with the reaction e- e- --> mu nu q
 %anti-q,''
 Phys.\ Lett.\  B {\bf 474}, 347 (2000)
 [arXiv:hep-ph/9909416].
 %%CITATION = PHLTA,B474,347;%%

 
\bibitem{BB0n_review}
See, for example,
 M.~Doi, T.~Kotani and E.~Takasugi,
 %``Double Beta Decay And Majorana Neutrino,''
 Prog.\ Theor.\ Phys.\ Suppl.\  {\bf 83}, 1 (1985);
 %%CITATION = PTPSA,83,1;%%
 A.~S.~Barabash,
 %``Double beta decay experiments,''
 JINST {\bf 1}, P07002 (2006)
 [arXiv:hep-ex/0602037].
 %%CITATION = JINST,1,P07002;%%



\bibitem{EVcommon}
S.R.~Elliott and P.~Vogel,
  %``Double beta decay,''
  Ann.\ Rev.\ Nucl.\ Part.\ Sci.\  {\bf 52}, 115 (2002);
  [arXiv:hep-ph/0202264];
  %%CITATION = HEP-PH 0202264;%%

  
\bibitem{NuMatrixEl}
 For a recent assessment, see
 V.~A.~Rodin, A.~Faessler, F.~Simkovic and P.~Vogel,
  %``Assessment of uncertainties in QRPA 0nu beta beta-decay nuclear matrix
  %elements,''
  Nucl.\ Phys.\  A {\bf 766}, 107 (2006) [arXiv:nucl-th/0503063].
  %%CITATION = NUPHA,A766,107;%%
%``Erratum: Assessment of uncertainties in QRPA $0\nu\beta\beta$-decay nuclear
  %matrix elements [Nucl. Phys. A 766, 107 (2006)],''
  Erratum arXiv:0706.4304 [nucl-th].

     
\bibitem{BB0nPosSig}
 H.~V.~Klapdor-Kleingrothaus, I.~V.~Krivosheina, A.~Dietz and O.~Chkvorets,
 %``Search for neutrinoless double beta decay with enriched Ge-76 in Gran
 %Sasso 1990-2003,''
 Phys.\ Lett.\  B {\bf 586}, 198 (2004)
 [arXiv:hep-ph/0404088].
 %%CITATION = PHLTA,B586,198;%%

 
\bibitem{HMoscowCurrent}
 H.~V.~Klapdor-Kleingrothaus {\it et al.},
 %``Latest results from the Heidelberg-Moscow double-beta-decay experiment,''
 Eur.\ Phys.\ J.\  A {\bf 12}, 147 (2001)
 [arXiv:hep-ph/0103062].
 %%CITATION = EPHJA,A12,147;%%
 
% No SPIRES record found for cite request IGEX
\bibitem{IGEX}
  C.~E.~Aalseth {\it et al.}  [IGEX Collaboration],
  %``The IGEX Ge-76 neutrinoless double-beta decay experiment: Prospects for
  %next generation experiments,''
  Phys.\ Rev.\  D {\bf 65}, 092007 (2002)
  [arXiv:hep-ex/0202026].
  %%CITATION = PHRVA,D65,092007;%%
  

\bibitem{BB0nFuture}
 S.~M.~Bilenky, C.~Giunti, J.~A.~Grifols and E.~Masso,
 %``Absolute values of neutrino masses: Status and prospects,''
 Phys.\ Rept.\  {\bf 379}, 69 (2003)
 [arXiv:hep-ph/0211462];
 %%CITATION = PRPLC,379,69;%%
C.~Aalseth {\it et al.},
  %``Neutrinoless double beta decay and direct searches for neutrino mass,''
  arXiv:hep-ph/0412300.
  %%CITATION = HEP-PH/0412300;%%

  
%\cite{Flanz:1999ah}
\bibitem{Flanz:1999ah}
  M.~Flanz, W.~Rodejohann and K.~Zuber,
  %``Bounds on effective Majorana neutrino masses at HERA,''
  Phys.\ Lett.\  B {\bf 473}, 324 (2000)
  [Erratum-ibid.\  B {\bf 480}, 418 (2000)]
  [arXiv:hep-ph/9911298].
  %%CITATION = PHLTA,B473,324;%%

    
\bibitem{BFactories}
 A.~G.~Akeroyd {\it et al.}  [SuperKEKB Physics Working Group],
 %``Physics at super B factory,''
 arXiv:hep-ex/0406071;
 %%CITATION = HEP-EX/0406071;%%
 I.~I.~Bigi and A.~I.~Sanda,
 %``Is super-B sufficiently superb? On the motivation for a super-B  factory,''
 arXiv:hep-ph/0401003.
 %%CITATION = HEP-PH/0401003;%%

 
\bibitem{LEP}
 See, for example, J.~Drees,
 %``Review of final LEP results or a tribute to LEP,''
 Int.\ J.\ Mod.\ Phys.\  A {\bf 17}, 3259 (2002)
 [arXiv:hep-ex/0110077].
 %%CITATION = IMPAE,A17,3259;%%

 
\bibitem{SLD}
 See, for example, M.~Breidenbach,
 %``Overview Of The Sld,''
 IEEE Trans.\ Nucl.\ Sci.\  {\bf 33}, 46 (1986).
 %%CITATION = IETNA,33,46;%%

 
\bibitem{ZInvBound}
 A.~Abbaneo {\it et al.} [LEP Electroweak Working Group, SLD Electroweak and Heavy Flavor Groups],
  %``A combination of preliminary electroweak measurements and constraints  on
  %the standard model,''
  arXiv:hep-ex/0312023. See also
  %%CITATION = HEP-EX/0312023;%%
 M.~Carena, A.~de Gouv\^ea, A.~Freitas and M.~Schmitt,
 %``Invisible Z-boson decays at e+ e- colliders,''
 Phys.\ Rev.\  D {\bf 68}, 113007 (2003)
 [arXiv:hep-ph/0308053].
 %%CITATION = PHRVA,D68,113007;%%

   
\bibitem{WDecay}
  S.~Bar-Shalom, N.~G.~Deshpande, G.~Eilam, J.~Jiang and A.~Soni,
  %``Majorana neutrinos and lepton-number-violating signals in top-quark and
  %W-boson rare decays,''
  Phys.\ Lett.\  B {\bf 643}, 342 (2006)
  [arXiv:hep-ph/0608309].
  %%CITATION = PHLTA,B643,342;%%

     
\bibitem{LHC}
 See, for example, G.~Rolandi,
 %``The LHC machine and experiments,''
 Int.\ J.\ Mod.\ Phys.\  A {\bf 21}, 1654 (2006).
 %%CITATION = IMPAE,A21,1654;%%

 
\bibitem{ILC}
 See, for example,
 M.~Alabau Pons, P.~Bambade, O.~Dadoun, R.~Appleby and A.~Faus-Golfe,
 %``Optimization of the e- e- option for the ILC,''
 arXiv:physics/0609043;
 %%CITATION = PHYSICS/0609043;%%
 R.~D.~Heuer,
 %``The International Linear Collider Ilc: A Status Report,''
 Nucl.\ Phys.\ Proc.\ Suppl.\  {\bf 154}, 131 (2006).
 %%CITATION = NUPHZ,154,131;%%
See also \url{http://www.linearcollider.org/cms/}.


\bibitem{EmEm2WmWm}
 D.~London, G.~Belanger and J.~N.~Ng,
 %``New Tests Of Lepton Number Violation At Electron - Electron Colliders,''
 Phys.\ Lett.\  B {\bf 188}, 155 (1987).
 %%CITATION = PHLTA,B188,155;%%

 
\bibitem{ILCPol}
 K.~Moffeit, M.~Woods, P.~Schuler, K.~Moenig and P.~Bambade,
% ``Spin rotation schemes at the ILC for two interaction regions and  positron
% polarization with both helicities,''
 SLAC-TN-05-045, LCC-0159, IPBI-TN-2005-2.
 %%CITATION = IPBI-TN-2005-2;%%

 
  \bibitem{mumu}
 C.~M.~Ankenbrandt {\it et al.},
  %``Status of muon collider research and development and future plans,''
  Phys.\ Rev.\ ST Accel.\ Beams {\bf 2}, 081001 (1999)
  [arXiv:physics/9901022];
  %%CITATION = PRSTA,2,081001;%%
M.~M.~Alsharoa {\it et al.}  [Muon Collider/Neutrino Factory
Collaboration],
  %``Recent progress in neutrino factory and muon collider research within the
  %Muon collaboration,''
  Phys.\ Rev.\ ST Accel.\ Beams {\bf 6}, 081001 (2003)
  [arXiv:hep-ex/0207031].
  %%CITATION = PRSTA,6,081001;%%
See also \url{http://www.fnal.gov/projects/muon_collider/}.

 
\bibitem{gammagamma}
See, for example, V.~I.~Telnov,
  %``The photon collider at ILC: Status, parameters and technical problems,''
  Acta Phys.\ Polon.\  B {\bf 37}, 1049 (2006)
  [arXiv:physics/0604108];
  %%CITATION = APPOA,B37,1049;%%
E.~Accomando {\it et al.}  [CLIC Physics Working Group],
  %``Physics at the CLIC multi-TeV linear collider,''
  arXiv:hep-ph/0412251.
  %%CITATION = HEP-PH/0412251;%%

% No SPIRES record found for cite request CollLNV

 
\bibitem{CollLNV}
See, for example,
  F.~del Aguila, J.~A.~Aguilar-Saavedra and R.~Pittau,
  %``Heavy neutrino signals at large hadron colliders,''
  arXiv:hep-ph/0703261;
  %%CITATION = HEP-PH/0703261;%%
  A.~Datta, M.~Guchait and A.~Pilaftsis,
  %``Probing lepton number violation via majorana neutrinos at hadron
  %supercolliders,''
  Phys.\ Rev.\  D {\bf 50}, 3195 (1994)
  [arXiv:hep-ph/9311257];
  %%CITATION = PHRVA,D50,3195;%%
    F.~M.~L.~de Almeida, Y.~D.~A.~Coutinho, J.~A.~Martins Sim\~oes, A.~J.~Ramalho, S.~Wulck and M.~A.~B.~do Vale,
  %``Discriminating among the theoretical origins of new heavy Majorana
  %neutrinos at the CERN LHC,''
  Phys.\ Rev.\  D {\bf 75}, 075002 (2007)
  [arXiv:hep-ph/0703094];
  %%CITATION = PHRVA,D75,075002;%%
    T.~Han and B.~Zhang,
  %``Signatures for Majorana neutrinos at hadron colliders,''
  Phys.\ Rev.\ Lett.\  {\bf 97}, 171804 (2006)
  [arXiv:hep-ph/0604064];
  %%CITATION = PRLTA,97,171804;%%
  O.~Panella, M.~Cannoni, C.~Carimalo and Y.~N.~Srivastava,
  %``Signals of heavy Majorana neutrinos at hadron colliders,''
  Phys.\ Rev.\  D {\bf 65}, 035005 (2002)
  [arXiv:hep-ph/0107308].;
  %%CITATION = PHRVA,D65,035005;%%
    A.~Ali, A.~V.~Borisov and N.~B.~Zamorin,
  %``Majorana neutrinos and same-sign dilepton production at LHC and in  rare
  %meson decays,''
  Eur.\ Phys.\ J.\  C {\bf 21}, 123 (2001)
  [arXiv:hep-ph/0104123];
  %%CITATION = EPHJA,C21,123;%%
    A.~Ali, A.~V.~Borisov and N.~B.~Zamorin,
  %``Same-sign dilepton production via heavy Majorana neutrinos in  proton
  %proton collisions,''
  arXiv:hep-ph/0112043.
  %%CITATION = HEP-PH/0112043;%%

    
%\cite{Frigerio:2002fb}
\bibitem{Frigerio:2002fb}
  M.~Frigerio and A.~Yu.~Smirnov,
  %``Neutrino mass matrix: Inverted hierarchy and CP violation,''
  Phys.\ Rev.\  D {\bf 67}, 013007 (2003)
  [arXiv:hep-ph/0207366];
  %%CITATION = PHRVA,D67,013007;%%
  %M.~Frigerio and A.~Yu.~Smirnov,
  %``Structure of neutrino mass matrix and CP violation,''
  Nucl.\ Phys.\  B {\bf 640}, 233 (2002)
  [arXiv:hep-ph/0202247];
  %%CITATION = NUPHA,B640,233;%%
  A.~Merle and W.~Rodejohann,
  %``The elements of the neutrino mass matrix: Allowed ranges and  implications
  %of texture zeros,''
  Phys.\ Rev.\  D {\bf 73}, 073012 (2006)
  [arXiv:hep-ph/0603111].
  %%CITATION = PHRVA,D73,073012;%%

  
\bibitem{Normal0Mee}
  See, for example, S.~M.~Bilenky, C.~Giunti, C.~W.~Kim and S.~T.~Petcov,
  %``Short-baseline neutrino oscillations and neutrinoless double-beta decay in
  %schemes with an inverted mass spectrum,''
  Phys.\ Rev.\  D {\bf 54}, 4432 (1996)
  [arXiv:hep-ph/9604364].
  %%CITATION = PHRVA,D54,4432;%%

   
\bibitem{NuMassHier0t13}
  A.~de Gouv\^ea and W.~Winter,
  %``What would it take to determine the neutrino mass hierarchy if  theta(13)
  %were too small?,''
  Phys.\ Rev.\  D {\bf 73}, 033003 (2006)
  [arXiv:hep-ph/0509359];
  %%CITATION = PHRVA,D73,033003;%%
  A.~de Gouv\^ea and J.~Jenkins,
  %``Non-oscillation probes of the neutrino mass hierarchy and vanishing
  %|U(e3)|,''
  arXiv:hep-ph/0507021;
  %%CITATION = HEP-PH/0507021;%%
  A.~de Gouv\^ea, J.~Jenkins and B.~Kayser,
  %``Neutrino mass hierarchy, vacuum oscillations, and vanishing U|e3|,''
  Phys.\ Rev.\  D {\bf 71}, 113009 (2005)
  [arXiv:hep-ph/0503079].
  %%CITATION = PHRVA,D71,113009;%%

  
  
\bibitem{FutureCosmo}
See, for example, J.~Lesgourgues and S.~Pastor in \cite{CosmoSum,Fogli:2006yq,seljak,Hannestad:2006mi};
K.~N.~Abazajian and S.~Dodelson,
  %``Neutrino mass and dark energy from weak lensing,''
  Phys.\ Rev.\ Lett.\  {\bf 91}, 041301 (2003)
  [arXiv:astro-ph/0212216].
  %%CITATION = PRLTA,91,041301;%%

  
\bibitem{Katrin}
A.~Osipowicz {\it et al.}  [KATRIN Collaboration],
  %``KATRIN: A next generation tritium beta decay experiment with sub-eV
  %sensitivity for the electron neutrino mass,''
  arXiv:hep-ex/0109033.
  %%CITATION = HEP-EX 0109033;%%
  See also \url{http://www-ik.fzk.de/~katrin/index.html}.

    
\bibitem{ScalarLNV}
 For a scenario that also pursues lepton-number violation from new exotic scalar fields, see
 G.~K.~Leontaris, K.~Tamvakis and J.~D.~Vergados,
 %``Lepton And Family Number Violation From Exotic Scalars,''
 Phys.\ Lett.\  B {\bf 162}, 153 (1985).
 %%CITATION = PHLTA,B162,153;%%
%\cite{Adloff:1999tp}

   
  \bibitem{HERA}
 See, for example,
 C.~Adloff {\it et al.}  [H1 Collaboration],
  %``A search for leptoquark bosons and lepton flavor violation in e+ p
  %collisions at HERA,''
  Eur.\ Phys.\ J.\  C {\bf 11}, 447 (1999)
  [Erratum-ibid.\  C {\bf 14}, 553 (2000)]
  [arXiv:hep-ex/9907002].
  %%CITATION = EPHJA,C11,447;%%

  
\bibitem{ZeeModel}
 A.~Zee,
 %``Quantum Numbers Of Majorana Neutrino Masses,''
 Nucl.\ Phys.\  B {\bf 264}, 99 (1986).
 %%CITATION = NUPHA,B264,99;%%

  %\cite{Dreiner:1997uz}
\bibitem{R-parity}
For overviews, see
  R.~Barbier {\it et al.},
  %``R-parity violating supersymmetry,''
  Phys.\ Rept.\  {\bf 420}, 1 (2005)
  [arXiv:hep-ph/0406039];
  %%CITATION = PRPLC,420,1;%%
  H.~K.~Dreiner,
  %``An introduction to explicit R-parity violation,''
  arXiv:hep-ph/9707435.
  %%CITATION = HEP-PH/9707435;%%


  
%%%%%%%%%%%%%%%%%%%%%%%%%%%%%%%%%%%%%%%%%%%%%%%%%check this%%%%%%%%%%%%%%%%%%%%%%%
  %\cite{leptogenesis}
\bibitem{leptogenesis}
For a recent review, see
  W.~Buchmuller, R.D.~Peccei and T.~Yanagida,
  %``Leptogenesis as the origin of matter,''
  Ann.\ Rev.\ Nucl.\ Part.\ Sci.\  {\bf 55}, 311 (2005).
  %[arXiv:hep-ph/0502169].
  %%CITATION = HEP-PH 0502169;%%



\bibitem{LepFlavVio}
  D.~Black, T.~Han, H.~J.~He and M.~Sher,
  %``tau - mu flavor violation as a probe of the scale of new physics,''
  Phys.\ Rev.\  D {\bf 66}, 053002 (2002)
  [arXiv:hep-ph/0206056].
  %%CITATION = PHRVA,D66,053002;%%



%  \cite{LSND}
\bibitem{LSND}
  A.~Aguilar {\it et al.}  [LSND Collaboration],
  %``Evidence for neutrino oscillations from the observation of anti-nu/e
  %appearance in a anti-nu/mu beam,''
  Phys.\ Rev.\ D {\bf 64}, 112007 (2001).
  %%CITATION = HEP-EX 0104049;%%

  

%   \cite{minib}
\bibitem{minib}
  A.~Bazarko  [MiniBooNE Collaboration],
  %``MiniBooNE: Status of the booster neutrino experiment,''
  Nucl.\ Phys.\ Proc.\ Suppl.\  {\bf 91}, 210 (2001).
  %%CITATION = HEP-EX 0009056;%%

  
%\cite{light_rhn1}
\bibitem{light_rhn1}
  W.~Krolikowski,
  %``Can one of three righthanded neutrinos be light enough to produce a  small
  %LSND effect?,''
  Acta Phys.\ Polon.\ B {\bf 35}, 2241 (2004).
  %[arXiv:hep-ph/0404118].
  %%CITATION = HEP-PH 0404118;%%

    
\bibitem{nuSM_kicks}
T.~Asaka, A.~Kusenko and M.~Shaposhnikov,
  %``Opening a new window for warm dark matter,''
  Phys.\ Lett.\ B {\bf 638}, 401 (2006).
  %%CITATION = HEP-PH 0602150;%%

  
  %\cite{Casas:2001sr}
\bibitem{Casas:2001sr}
  J.A.~Casas and A.~Ibarra,
  %``Oscillating neutrinos and mu --> e, gamma,''
  Nucl.\ Phys.\ B {\bf 618}, 171 (2001).
  %[arXiv:hep-ph/0103065].
  %%CITATION = HEP-PH 0103065;%%


 %\cite{neutrino_review}
\bibitem{neutrino_review}
A.~Strumia and F.~Vissani,
  %``Neutrino masses and mixings and.,''
  hep-ph/0606054;
  %%CITATION = HEP-PH 0606054;%%
R.N.~Mohapatra and A.Yu.~Smirnov,
  %``Neutrino mass and new physics,''
  hep-ph/0603118;
  %%CITATION = HEP-PH 0603118;%%
R.N.~Mohapatra {\it et al.},
  %``Theory of neutrinos: A white paper,''
  hep-ph/0510213;
  %%CITATION = HEP-PH 0510213;%%
A.~de Gouv\^ea,
  %``2004 TASI lectures on neutrino physics,''
  hep-ph/0411274.
  A.~de Gouv\^ea,
  %``Neutrinos Have Mass - So What?''
  Mod.\ Phys.\ Lett.\ A {\bf 19}, 2799 (2004).
  %%CITATION = HEP-PH 0503086;%

  
 %\cite{global_analy}
\bibitem{global_analy}
  M.~Maltoni, T.~Schwetz, M.~A.~Tortola and J.~W.~F.~Valle,
  %``Status of global fits to neutrino oscillations,''
  New J.\ Phys.\  {\bf 6}, 122 (2004).
  %[arXiv:hep-ph/0405172].
  %%CITATION = HEP-PH 0405172;%%


   
%\cite{CPTVorSterile}
\bibitem{CPTVorSterile}
  A.~Strumia,
  %``Interpreting the LSND anomaly: Sterile neutrinos or CPT-violation
  %or...?,''
  Phys.\ Lett.\ B {\bf 539}, 91 (2002).
  %[arXiv:hep-ph/0201134].
  %%CITATION = HEP-PH 0201134;%%




%\cite{rule_out_4nu}
\bibitem{rule_out_4nu}
  M.~Maltoni, T.~Schwetz, M.~A.~Tortola and J.~W.~F.~Valle,
  %``Ruling out four-neutrino oscillation interpretations of the LSND
  %anomaly?,''
  Nucl.\ Phys.\ B {\bf 643}, 321 (2002).
 % [arXiv:hep-ph/0207157].
  %%CITATION = HEP-PH 0207157;%%


    
%\cite{Maltoni:2003yr}
\bibitem{Maltoni:2003yr}
  M.~Maltoni, T.~Schwetz, M.~A.~Tortola and J.~W.~F.~Valle,
  %``Can four neutrinos explain global oscillation data including LSND \&
  %cosmology?,''
  hep-ph/0305312.
  %%CITATION = HEP-PH 0305312;%%


    
  %\cite{short_bl_analysis}
\bibitem{short_bl_analysis}
  M.~Sorel, J.~M.~Conrad and M.~Shaevitz,
  %``A combined analysis of short-baseline neutrino experiments in the (3+1)
  %and (3+2) sterile neutrino oscillation hypotheses,''
  Phys.\ Rev.\ D {\bf 70}, 073004 (2004).
  %[arXiv:hep-ph/0305255].
  %%CITATION = HEP-PH 0305255;%%

  
  
  %\cite{CPTV}
\bibitem{CPTV}
H.~Murayama and T.~Yanagida,
  %``LSND, SN1987A, and CPT violation,''
  Phys.\ Lett.\ B {\bf 520}, 263 (2001);
  %[arXiv:hep-ph/0010178];
  %%CITATION = HEP-PH 0010178;%%
G.~Barenboim, L.~Borissov, J.D.~Lykken and A.Yu.~Smirnov,
  %``Neutrinos as the messengers of CPT violation,''
  JHEP {\bf 0210}, 001 (2002);
  %[arXiv:hep-ph/0108199];
  M.~C.~Gonzalez-Garcia, M.~Maltoni and T.~Schwetz,
  %``Status of the CPT violating interpretations of the LSND signal,''
  Phys.\ Rev.\ D {\bf 68}, 053007 (2003).
  %[arXiv:hep-ph/0306226].
  %%CITATION = HEP-PH 0306226;%%

  
  
  %\cite{lorentz_v}
\bibitem{lorentz_v}
 V.~A.~Kostelecky and M.~Mewes,
  %``Lorentz violation and short-baseline neutrino experiments,''
  Phys.\ Rev.\ D {\bf 70}, 076002 (2004);
  %%CITATION = HEP-PH 0406255;%%
   V.~A.~Kostelecky and M.~Mewes,
  %``Lorentz and CPT violation in neutrinos,''
  Phys.\ Rev.\ D {\bf 69}, 016005 (2004);
  %%CITATION = HEP-PH 0309025;%%
   L.~B.~Auerbach {\it et al.}  [LSND Collaboration],
  %``Tests of Lorentz violation in anti-nu/mu $\to$ anti-nu/e oscillations,''
  Phys.\ Rev.\ D {\bf 72}, 076004 (2005);
  %[arXiv:hep-ex/0506067];
  A.~de Gouv\^ea and Y.~Grossman,
  %``A three-flavor, Lorentz-violating solution to the LSND anomaly,''
  hep-ph/0602237;
  %%CITATION = HEP-PH 0602237;%%
T.~Katori, A.~Kostelecky and R.~Tayloe,
  % ``Global three-parameter model for neutrino oscillations using Lorentz
  %violation,''
  hep-ph/0606154.
  %%CITATION = HEP-PH 0606154;%%


    
%\cite{decoherence}
\bibitem{decoherence}
  G.~Barenboim and N.~E.~Mavromatos,
  %``CPT violating decoherence and LSND: A possible window to Planck scale
  %physics,''
  JHEP {\bf 0501}, 034 (2005).
  %[arXiv:hep-ph/0404014].
  %%CITATION = HEP-PH 0404014;%%

  
  
  %\cite{sterile_decay}
\bibitem{sterile_decay}
  S.~Palomares-Ruiz, S.~Pascoli and T.~Schwetz,
  %``Explaining LSND by a decaying sterile neutrino,''
  JHEP {\bf 0509}, 048 (2005).
  %[arXiv:hep-ph/0505216].
  %%CITATION = HEP-PH 0505216;%%


  
%\cite{3p2fit}
\bibitem{3p2fit}
  M.~Sorel, private communication.

  
  
\bibitem{nutev} S.~Avvakumov {\it et al.},
  % ``A search for nu/mu --> nu/e and anti-nu/mu --> anti-nu/e oscillations  at
  %NuTeV,''
  Phys.\ Rev.\ Lett.\  {\bf 89}, 011804 (2002).
%[arXiv:hep-ex/0203018].
  %%CITATION = HEP-EX 0203018;%%

    
\bibitem{nomad}
P.~Astier {\it et al.}  [NOMAD Collaboration],
  %``Search for nu/mu --> nu/e oscillations in the NOMAD experiment,''
  Phys.\ Lett.\ B {\bf 570}, 19 (2003)
  %[arXiv:hep-ex/0306037].
  %%CITATION = HEP-EX 0306037;%%


    
\bibitem{chorus}
E.~Eskut {\it et al.}  [CHORUS Collaboration],
   %``Search for nu/mu --> nu/tau oscillation using the tau decay modes into  a
  %single charged particle,''
  Phys.\ Lett.\ B {\bf 434}, 205 (1998);
  %%CITATION = PHLTA,B434,205;%%
Phys.\ Lett.\ B {\bf 424}, 202 (1998).
  %%CITATION = PHLTA,B424,202;%%


  %\cite{kicks_gen}
\bibitem{kicks_gen}
  A.~Kusenko,
  %``Pulsar kicks from neutrino oscillations,''
  Int.\ J.\ Mod.\ Phys.\ D {\bf 13}, 2065 (2004).
  %[arXiv:astro-ph/0409521].
  %%CITATION = ASTRO-PH 0409521;%%

  
  
\bibitem{wmap} D.N.~Spergel {\it et al.},
   %``Wilkinson Microwave Anisotropy Probe (WMAP) three year results:
  %Implications for cosmology,''
  astro-ph/0603449.

  
 % \bibitem{seljak} U.~Seljak, A.~Slosar and P.~McDonald,
  %%``Cosmological parameters from combining the Lyman-alpha forest with CMB,
  %%galaxy clustering and SN constraints,''
  %JCAP {\bf 0610}, 014 (2006).
  %%%CITATION = ASTRO-PH 0604335;%%

%%\cite{Hannestad:2006mi}
%\bibitem{Hannestad:2006mi}
%S.~Hannestad and G.~G.~Raffelt,
  %%``Neutrino masses and cosmic radiation density: Combined analysis,''
 % astro-ph/0607101.




  

\bibitem{Cirelli:2006kt}
  M.~Cirelli and A.~Strumia,
  %``Cosmology of neutrinos and extra light particles after WMAP3,''
  astro-ph/0607086.
  %%CITATION = ASTRO-PH 0607086;%%






%\cite{thermalize}
\bibitem{thermalize}
K.~Enqvist, K.~Kainulainen and M.~J.~Thomson,
  %``Stringent cosmological bounds on inert neutrino mixing,''
  Nucl.\ Phys.\ B {\bf 373}, 498 (1992);
X.~Shi, D.~N.~Schramm and B.~D.~Fields,
  %``Constraints on neutrino oscillations from big bang nucleosynthesis,''
  Phys.\ Rev.\ D {\bf 48}, 2563 (1993).



  

%\cite{smirnov_renata}
\bibitem{smirnov_renata}
A.Yu.~Smirnov and R.Z.~Funchal,
   %``Sterile neutrinos: Direct mixing effects versus induced mass matrix of
  %active neutrinos,''
  Phys.\ Rev.\ D {\bf 74}, 013001 (2006).
  %[arXiv:hep-ph/0603009].


%\cite{Dodelson:2005tp}
\bibitem{Dodelson:2005tp}
  S.~Dodelson, A.~Melchiorri and A.~Slosar,
  %``Is cosmology compatible with sterile neutrinos?,''
  Phys.\ Rev.\ Lett.\  {\bf 97}, 04301 (2006).
  %%CITATION = ASTRO-PH 0511500;%%

   

%\cite{Beacom:2004yd}
\bibitem{Beacom:2004yd} See, for example,
J.F.~Beacom, N.F.~Bell and S.~Dodelson,
  %``Neutrinoless universe,''
  Phys.\ Rev.\ Lett.\  {\bf 93}, 121302 (2004).
  %%CITATION = ASTRO-PH 0404585;%%


%\cite{Barger:2003zg}
\bibitem{Barger:2003zg}
  See, for example, V.~Barger, J.~P.~Kneller, H.~S.~Lee, D.~Marfatia and G.~Steigman,
  %``Effective number of neutrinos and baryon asymmetry from BBN and WMAP,''
  Phys.\ Lett.\ B {\bf 566}, 8 (2003).
    %%CITATION = HEP-PH 0305075;%%


        
%\cite{Abazajian:2004aj}
\bibitem{Abazajian:2004aj}
  K.~Abazajian, N.~F.~Bell, G.~M.~Fuller and Y.~Y.~Y.~Wong,
   %``Cosmological lepton asymmetry, primordial nucleosynthesis, and sterile
  %neutrinos,''
  Phys.\ Rev.\ D {\bf 72}, 063004 (2005).
  %[arXiv:astro-ph/0410175].
  %%CITATION = ASTRO-PH 0410175;%%

  
 %\cite{low_reheat}
\bibitem{low_reheat}
  G.~Gelmini, S.~Palomares-Ruiz and S.~Pascoli,
  %``Low reheating temperature and the visible sterile neutrino,''
  Phys.\ Rev.\ Lett.\  {\bf 93}, 081302 (2004).
  %[arXiv:astro-ph/0403323].
  %%CITATION = ASTRO-PH 0403323;%%

 
%\cite{Chu:2006ua}
\bibitem{Chu:2006ua}
For a recent analysis, see Y.~Z.~Chu and M.~Cirelli,
  %``Sterile neutrinos, lepton asymmetries, primordial elements: How much of
  %each?,''
  Phys.\ Rev.\ D {\bf 74}, 085015 (2006).
  %%CITATION = ASTRO-PH 0608206;%%

  
    
\bibitem{warm_recent} For very recent discussions, see
K.~Abazajian,
  %``Linear cosmological structure limits on warm dark matter,''
  Phys.\ Rev.\ D {\bf 73}, 063513 (2006);
  %[arXiv:astro-ph/0512631].
  %%CITATION = ASTRO-PH 0512631;%%
U.~Seljak, A.~Makarov, P.~McDonald and H.~Trac,
  %``Can sterile neutrinos be the dark matter?,''
  astro-ph/0602430;
  %%CITATION = ASTRO-PH 0602430;%%
K.~Abazajian and S.M.~Koushiappas,
  %``Constraints on sterile neutrino dark matter,''
  Phys.\ Rev.\ D {\bf 74}, 023527 (2006);
  %[arXiv:astro-ph/0605271].
  %%CITATION = ASTRO-PH 0605271;%%
M.~Viel, J.~Lesgourgues, M.G.~Haehnelt, S.~Matarrese and A.~Riotto,
  %``Can sterile neutrinos be ruled out as warm dark matter candidates?,''
  astro-ph/0605706.
  %%CITATION = ASTRO-PH 0605706;%%

  
    
%\cite{Xray}
\bibitem{Xray}
  K.~Abazajian, G.~M.~Fuller and W.~H.~Tucker,
  %``Direct detection of warm dark matter in the X-ray,''
  Astrophys.\ J.\  {\bf 562}, 593 (2001).
  %[arXiv:astro-ph/0106002].
  %%CITATION = ASTRO-PH 0106002;%%
  Recent discussions can be found in
K.~Abazajian and S.M.~Koushiappas in \cite{warm_recent};
C.R.~Watson, J.F.~Beacom, H.~Yuksel and T.P.~Walker,
  %``Direct X-ray constraints on sterile neutrino warm dark matter,''
  astro-ph/0605424.
  %%CITATION = ASTRO-PH 0605424;%%

  
%\cite{Kusenko:2006wa}
\bibitem{Kusenko:2006wa}
  A.~Kusenko,
  %``Detecting sterile dark matter in space,''
  astro-ph/0608096.
  %%CITATION = ASTRO-PH 0608096;%%

  
%\cite{sterile_probe}
\bibitem{sterile_probe}
  M.~Cirelli, G.~Marandella, A.~Strumia and F.~Vissani,
  %``Probing oscillations into sterile neutrinos with cosmology,  astrophysics
  %and experiments,''
  Nucl.\ Phys.\ B {\bf 708}, 215 (2005).
  %[arXiv:hep-ph/0403158].
  %%CITATION = HEP-PH 0403158;%%

    
%\cite{SN1987_bound}
\bibitem{SN1987_bound}
  M.~Sorel and J.~M.~Conrad,
  %``Supernova neutrinos and LSND,''
  Phys.\ Rev.\ D {\bf 66}, 033009 (2002).
  %[arXiv:hep-ph/0112214].
  %%CITATION = HEP-PH 0112214;%%

  
\bibitem{sn_bound} K.~Kainulainen, J.~Maalampi and J.~T.~Peltoniemi,
  %``Inert neutrinos in supernovae,''
  Nucl.\ Phys.\ B {\bf 358}, 435 (1991).
  %%CITATION = NUPHA,B358,435;%%

  
%\cite{SN_review}
\bibitem{SN_review}
  H.~A.~Bethe,
  %``Supernova Mechanisms,''
  Rev.\ Mod.\ Phys.\  {\bf 62}, 801 (1990).
  %%CITATION = RMPHA,62,801;%%

  
  
%\cite{pulsar_disp}
\bibitem{pulsar_disp}
  Z.~Arzoumanian, D.~F.~Chernoffs and J.~M.~Cordes,
  %``The Velocity Distribution of Isolated Radio Pulsars s,''
  Astrophys.\ J.\  {\bf 568}, 289 (2002).
  %[arXiv:astro-ph/0106159].
  %%CITATION = ASTRO-PH 0106159;%%

    
%\cite{Scheck:2006rw}
\bibitem{Scheck:2006rw}
L.~Scheck, T.~Plewa, H.~T.~Janka, K.~Kifonidis and E.~Mueller,
  %``Pulsar Recoil by Large-Scale Anisotropies in Supernova Explosions,''
  Phys.\ Rev.\ Lett.\  {\bf 92}, 011103 (2004).
  %%CITATION = ASTRO-PH 0307352;%%
  L.~Scheck, K.~Kifonidis, H.~T.~Janka and E.~Mueller,
  %``Multidimensional Supernova Simulations with Approximative Neutrino
  %Transport I. Neutron Star Kicks and the Anisotropy of Neutrino-Driven
  %Explosions in Two Spatial Dimensions,''
  astro-ph/0601302.
  %%CITATION = ASTRO-PH 0601302;%%

  
  %\cite{mmoment_kick}
\bibitem{mmoment_kick}
  M.~B.~Voloshin,
  %``Resonant Helicity Flip Of Electron-Neutrino Due To Magnetic Moment And
  %Dynamics Of Supernova,''
  Phys.\ Lett.\ B {\bf 209}, 360 (1988).
  %%CITATION = PHLTA,B209,360;%%

  
%\cite{nu_e_scattering}
\bibitem{nu_e_scattering}
  A.~de Gouv\^ea and J.~Jenkins,
  %``What can we learn from neutrino electron scattering?,''
  Phys.\ Rev.\ D {\bf 74}, 033004 (2006).
  %%CITATION = HEP-PH 0603036;%%

\bibitem{NuSOnGPRD}
T.~Adams {\it et al.} (J~Jenkins), arXiv:0803.0354 [hep-ph]
  
  \bibitem{munu}
  Z.~Daraktchieva {\it et al.}  [MUNU Collaboration],
  %``Final results on the neutrino magnetic moment from the MUNU experiment,''
  Phys.\ Lett.\ B {\bf 615}, 153 (2005).
  %[arXiv:hep-ex/0502037].
  %%CITATION = HEP-EX 0502037;%%

  
\bibitem{texono}
  H.~B.~Li {\it et al.}  [TEXONO Collaboration],
  %``New limits on neutrino magnetic moments from the Kuo-Sheng reactor neutrino
  %experiment,''
  Phys.\ Rev.\ Lett.\  {\bf 90}, 131802 (2003).
  %[arXiv:hep-ex/0212003].
  %%CITATION = HEP-EX 0212003;%%

  
%\cite{active_kicks}
\bibitem{active_kicks}
  A.~Kusenko and G.~Segre,
  %``Velocities of pulsars and neutrino oscillations,''
  Phys.\ Rev.\ Lett.\  {\bf 77}, 4872 (1996).
  %[arXiv:hep-ph/9606428].
  %%CITATION = HEP-PH 9606428;%%

    
  %\cite{kicks_res}
\bibitem{kicks_res}
  M.~Barkovich, J.~C.~D'Olivo and R.~Montemayor,
  %``Neutrinospheres, resonant neutrino oscillations, and pulsar kicks,''
  hep-ph/0503113.
  %%CITATION = HEP-PH 0503113;%%

    
%\cite{nu_Bfield1}
\bibitem{nu_Bfield1}
  H.~Nunokawa, V.~B.~Semikoz, A.Yu.~Smirnov and J.W.F.~Valle,
  %``Neutrino conversions in a polarized medium,''
  Nucl.\ Phys.\ B {\bf 501}, 17 (1997).
  %[arXiv:hep-ph/9701420].
  %%CITATION = HEP-PH 9701420;%%


%\cite{nu_Bfield2}
\bibitem{nu_Bfield2}
  S.~Esposito and G.~Capone,
  %``Neutrino propagation in a medium with a magnetic field,''
  Z.\ Phys.\ C {\bf 70}, 55 (1996).
  %[arXiv:hep-ph/9511417].
  %%CITATION = HEP-PH 9511417;%%

  
  
%\cite{kicks_nonres}
\bibitem{kicks_nonres}
  G.~M.~Fuller, A.~Kusenko, I.~Mocioiu and S.~Pascoli,
  %``Pulsar kicks from a dark-matter sterile neutrino,''
  Phys.\ Rev.\ D {\bf 68}, 103002 (2003).
  %[arXiv:astro-ph/0307267].
  %%CITATION = ASTRO-PH 0307267;%%

  
  
  %\cite{sterile_hotwarmcold}
\bibitem{sterile_hotwarmcold}
  K.~Abazajian, G.~M.~Fuller and M.~Patel,
  %``Sterile neutrino hot, warm, and cold dark matter,''
  Phys.\ Rev.\ D {\bf 64}, 023501 (2001).
  %[arXiv:astro-ph/0101524].
  %%CITATION = ASTRO-PH 0101524;%%

  
    
%\cite{rprocess_nuwind}
\bibitem{rprocess_nuwind}
  S.~Wanajo, T.~Kajino, G.~J.~Mathews and K.~Otsuki,
  %``The r-Process in Neutrino-Driven Winds from Nascent, Compact Neutron Stars
  %of Core-Collapse Supernovae,''
  astro-ph/0102261.
  %%CITATION = ASTRO-PH 0102261;%%

  
%\cite{nucleosynthesis1}
\bibitem{nucleosynthesis1}
  G.~C.~McLaughlin, J.~M.~Fetter, A.~B.~Balantekin and G.~M.~Fuller,
  %``An Active-Sterile Neutrino Transformation Solution for r-Process
  %Nucleosynthesis,''
  Phys.\ Rev.\ C {\bf 59}, 2873 (1999).
  %[arXiv:astro-ph/9902106].
  %%CITATION = ASTRO-PH 9902106;%%

      
  %\cite{nucleosynthesis2}
\bibitem{nucleosynthesis2}
  J.~Fetter, G.~C.~McLaughlin, A.~B.~Balantekin and G.~M.~Fuller,
  %``Active-sterile neutrino conversion: Consequences for the r-process and
  %supernova neutrino detection,''
  Astropart.\ Phys.\  {\bf 18}, 433 (2003).
  %[arXiv:hep-ph/0205029].
  %%CITATION = HEP-PH 0205029;%%


%\cite{nucleosynthesis_fisscy}
\bibitem{nucleosynthesis_fisscy}
  J.~Beun, G.~C.~McLaughlin, R.~Surman and W.~R.~Hix,
  %``Fission cycling in supernova nucleosynthesis: Active-sterile neutrino
  %oscillations,''
  hep-ph/0602012.
  %%CITATION = HEP-PH 0602012;%%

  
%\cite{Farzan:2001cj}
\bibitem{Farzan:2001cj}
  Y.~Farzan, O.L.G.~Peres and A.Yu.~Smirnov,
  %``Neutrino mass spectrum and future beta decay experiments,''
  Nucl.\ Phys.\ B {\bf 612}, 59 (2001).
  %[arXiv:hep-ph/0105105].
  %%CITATION = HEP-PH 0105105;%%


  %\cite{bb0n_future}
\bibitem{bb0n_future}
  F.~T.~Avignone,
  %``Strategies for next generation neutrinoless double-beta decay
  %experiments,''
  Nucl.\ Phys.\ Proc.\ Suppl.\  {\bf 143}, 233 (2005).
  %%CITATION = NUPHZ,143,233;%%

    
  %\cite{Kayser_CP}
\bibitem{Kayser_CP}
  B.~Kayser,
  %``Cpt, Cp, And C Phases And Their Effects In Majorana Particle Processes,''
  Phys.\ Rev.\ D {\bf 30}, 1023 (1984).
  %%CITATION = PHRVA,D30,1023;%%

    
%\cite{bb0n_present}
\bibitem{bb0n_present}
  A.M.~Bakalyarov, A.Y.~Balysh, S.T.~Belyaev, V.I.~Lebedev and S.V.~Zhukov
                  [C03-06-23.1 Collaboration],
  %``Results of the experiment on investigation of Germanium-76 double beta
  %decay,''
  Phys.\ Part.\ Nucl.\ Lett.\  {\bf 2}, 77 (2005)
  [Pisma Fiz.\ Elem.\ Chast.\ Atom.\ Yadra {\bf 2}, 21 (2005)].
  %[arXiv:hep-ex/0309016].
  %%CITATION = HEP-EX 0309016;%%

    
%\cite{Shaposhnikov:2006xi}
\bibitem{Shaposhnikov:2006xi}
  M.~Shaposhnikov and I.~Tkachev,
  %``The nuMSM, inflation, and dark matter,''
  Phys.\ Lett.\ B {\bf 639}, 414 (2006).
  %[arXiv:hep-ph/0604236].
  %%CITATION = HEP-PH 0604236;%%



\end{thebibliography}
\end{document}